%% file: TBE.tex
\documentclass[twoside,12pt]{article}

\usepackage{epsfig}
\usepackage{graphics}
\usepackage{graphicx}
\usepackage{amsmath}
\usepackage{bm}
\usepackage{latexsym}
\usepackage{footnote}

\input{definitions.tex}

\begin{document}

\input{frontmatter.tex} 

\input{section1.tex} 

\input{section2.tex} 

\input{section3.tex} 

\input{section4.tex} 

\input{section5.tex} 

\input{section6.tex} 

\input{section7.tex} 

\input{section8.tex} 

\input{references.tex}

%% file: definitions.tex
\def\Journal#1#2#3#4{{#1} {#2} (#4) #3 }

\def\NPA{{\em Nucl. Phys.} A}

\def\NPB{{\em Nucl. Phys.} B}
\def\PLB{{\em Phys. Lett.} B}

\def\PRL{{\em Phys. Rev. Lett.}}
\def\PREV{\em Phys. Rev.}
\def\PREP{\em Phys. Rep.}
\def\PRD{{\em Phys. Rev.} D}
\def\PRC{{\em Phys. Rev.} C}
\def\EPJA{{\em Eur. Phys. J.} A}
\def\EPJC{{\em Eur. Phys. J.} C}

\def\JPG{{\em J. Phys.} G}
\def\ANNP{\em Ann. Phys.} 
\def\RMP{{\em Rev. Mod. Phys.}}

\def\PPNP{\em Prog. Part. Nucl. Phys.}
\def\CPC{\em Comp. Phys. Comm.}
\def\SJNP{\em Sov. J. Nucl. Phys.}
\def\RPJ{\em Russ. Phys. Jour.}

\newcommand{\be}{\begin{equation}}
\newcommand{\ee}{\end{equation}}
\newcommand{\bea}{\begin{eqnarray}}
\newcommand{\eea}{\end{eqnarray}}
\newcommand{\nn}{\nonumber}

\def\Re{{\cal R}e}
\def\Im{{\cal I}m}
\def\ubar{\bar{u}}

\def\gtorder{\mathrel{\raise.3ex\hbox{$>$}\mkern-14mu
 \lower0.6ex\hbox{$\sim$}}}
\def\ltorder{\mathrel{\raise.3ex\hbox{$<$}\mkern-14mu
 \lower0.6ex\hbox{$\sim$}}}
\def\mugegm{\mu_p G_E / G_M}
\def\gegm{G_E / G_M} 
\def\ge{G_E}
\def\gm{G_M}

\def\gep{G_E^p}

\def\mugegmn{\mu_n G_E^n / G_M^n}
\def\gegmn{G_E^n / G_M^n} 
\def\gen{G_E^n}
\def\gmn{G_M^n}

\newcommand{\eps}{\varepsilon}
\newcommand{\sw}{\sin^2\theta_W}
\newcommand{\cw}{\cos^2\theta_W}
\newcommand{\vslash}[1]{#1 \hspace{-0.42 em} /}

\newcommand{\fslash}[1]{\mbox{$\!\not\!#1$}}
 
\def\eqref#1{Eq.~(\ref{eq:#1})}

\def\deg{^\circ}

\def\f3{\widetilde{F}_3}
\def\y2g{Y_{2\gamma}}
\def\etal{{\em et al.}}
\def\ie{{\em i.e.}}

%% file: frontmatter.tex
\title{ \vspace{1cm}
	Review of two-photon exchange in electron scattering%
	\footnote{Dedicated to the memory of John A. Tjon.}}

\author{J. Arrington,$^1$
	P. G. Blunden,$^2$
	W. Melnitchouk$^3$	\\ \\
$^1$Physics Division, Argonne National Laboratory, Argonne,
	Illinois 60439, USA	 \\
$^2$University of Manitoba, Winnipeg, Manitoba, Canada R3T 2N2	\\
$^3$Jefferson Lab, Newport News, Virginia 23606, USA}

\date{}

\maketitle

\begin{abstract}
We review the role of two-photon exchange (TPE) in electron--hadron
scattering, focusing in particular on hadronic frameworks suitable
for describing the low and moderate $Q^2$ region relevant to most
experimental studies.
We discuss the effects of TPE on the extraction of nucleon form
factors and their role in the resolution of the proton electric
to magnetic form factor ratio puzzle.
The implications of TPE on various other observables, including
neutron form factors, electroproduction of resonances and pions,
and nuclear form factors, are summarized.
Measurements seeking to directly identify TPE effects, such as through
the angular dependence of polarization observables, nonlinear $\eps$
contributions to the cross sections, and via $e^+ p$ to $e^- p$ cross
section ratios, are also outlined.
In the weak sector, we describe the role of TPE and $\gamma Z$
interference in parity-violating electron scattering, and assess their
impact on the extraction of the strange form factors of the nucleon
and the weak charge of the proton.
\end{abstract}

\newpage
\tableofcontents

%% file: section1.tex
\section{Introduction} \label{sec:intro}

The electromagnetic probe has been a primary experimental tool in the
study of hadron physics for many decades.  Electromagnetic interactions
are extremely well understood, and the pointlike nature of electrons
and muons make them ideal probes of the internal structure of hadrons.
Lepton scattering experiments provide the cleanest information available
on fundamental quantities such as hadron form factors and parton
distributions.  In addition, the relatively small value of the
electromagnetic coupling means that measurements on nuclei
probe the entire nuclear volume, without the significant attenuation
of the beam over the length of the nucleus that yields a dominance of
surface effects in some hadron beam measurements.

Because of the power of the electromagnetic probe, a great deal of our
information on the structure of the nucleon comes from unpolarized
measurements of the inclusive lepton-nucleon cross section.
More recently, polarized beams have been used to provide additional
information on the spin structure of the nucleon and to improve our
knowledge of the nucleon form factors.
As one of the most fundamental observables characterizing the composite
nature of the nucleon, electromagnetic form factors have over the past
few decades provided considerable insight into the nucleon internal
structure, with the electric ($\ge$) and the magnetic ($\gm$) form
factors encoding the (transverse) spatial distributions of the nucleon's
charge and magnetization
(for reviews see Refs.~\cite{Per07,Arr07a,Arr11}).
While these can in principle be extracted from the unpolarized cross
sections, polarization measurements have played a critical role in 
studies of the nucleon form factors over the last decade.

The polarization measurements have led to a renaissance in studies
of the structure of the proton and neutron, providing significantly
improved measurements of their form factors over a wide range of
momentum transfer $Q^2$ (which, loosely speaking, reflects the inverse
resolution at which the structure is probed).
However, while these measurements have considerably improved the
precision with which the form factors could be extracted, they
also revealed a significant discrepancy with extractions from the
unpolarized cross sections in kinematic regions where both techniques
provide precise measurements.  Because these, and essentially all other
electron scattering measurements, are analyzed in the one-photon
exchange or Born approximation, this discrepancy led to a serious
reexamination of the possible role played by two-photon exchange
(TPE) corrections.
Early measurements and calculations suggested that TPE effects were
small, although recent studies have provided convincing evidence that
these corrections can nonetheless be extremely important in specific
observables.

In the 10 years since the form factor discrepancy was confirmed, a great
deal of progress has been made in understanding TPE contributions.
There have been several approaches used to calculate TPE corrections
for a variety of reactions and observables, as well as a significant
effort aimed at constraining these experimentally.
At present there are calculations of TPE that are consistent with all
existing experimental constraints, which can explain the form factor
discrepancy, and which allow extraction of the proton form factors
without yielding a significant theoretical uncertainty in the extraction.
In addition, a set of experiments is underway which will allow for
direct verification of these calculations, and thus provide a final
resolution to the issue.
With extensive experimental checks of the TPE calculations, reliable
estimates can be made for other reactions, and it will be possible to
identify other cases where TPE effects may be large enough that they
pose concerns for the interpretation of precision experiments.

It is therefore timely to review the recent experimental and theoretical
efforts dedicated to studying TPE in electromagnetic processes, as well
as the impact of two-boson ($\gamma$ or $Z$) exchange corrections which
enter into weak processes induced by electromagnetic probes.
In Sec.~\ref{sec:elastic} we provide an overview of the relevant
electron scattering formalism, including definitions of kinematics
and scattering amplitudes.
Section~\ref{sec:experiments} summarizes the initial evidence
for TPE effects, as well as constraints from early measurements.
The formalism and calculations of TPE corrections are introduced
in Sec.~\ref{sec:TPE}, and their impact on elastic scattering
measurements examined in Sec.~\ref{sec:impact}.
Implications of TPE for observables in other electron--hadron
scattering reactions are discussed in Sec.~\ref{sec:other}.
Finally, in Sec.~\ref{sec:z} we review the role of two-photon and
$\gamma Z$ interference in parity-violating electron scattering
and their impact on the extraction of the strange form factors
of the nucleon and the weak charge of the proton.
We end with conclusions and the outlook for future studies of
two-photon exchange in Sec.~\ref{sec:conc}.

%% file: section2.tex
\section{Elastic electron--nucleon scattering}
\label{sec:elastic}

In this section we define the general kinematics of elastic
electron--nucleon scattering (Sec.~\ref{ssec:kinematics}),
and present amplitudes and cross sections in the one-photon
exchange or Born approximation (Sec.~\ref{ssec:Born}).
Following this we discuss the extraction of the electromagnetic form
factors in the Born approximation using the Rosenbluth separation
and polarization transfer methods (Sec.~\ref{ssec:BornFF}).

\subsection{\it Kinematics}
\label{ssec:kinematics}

For the elastic scattering process $e N \to e N$ the four-momenta of
the initial and final electrons are labeled by $k$ and $k'$, with
corresponding energies $E$ and $E'$, and of the initial and final
nucleons by $p$ and $p'$, respectively.
The four-momentum transfer from the electron to the nucleon is given
by $q = p'-p = k-k'$, with $Q^2 \equiv -q^2 > 0$.
Conventionally the scattering cross section is defined in terms of
$Q^2$ and the electron scattering angle $\theta$, or equivalently
the dimensionless quantities
\bea
\tau\ =\ {Q^2 \over 4M^2}\, ,\ \ \ \ & &
\eps\ =\ {\nu^2 - \tau (1+\tau) \over \nu^2 + \tau (1+\tau)}\, ,
\label{eq:kin}
\eea
where $\nu = k \cdot p/M^2 - \tau$.
%
In the target rest frame the variable $\eps$ is related to the
scattering angle $\theta$ by
\be
\eps\ =\ \left(1 + 2 (1+\tau) \tan^2{{\theta\over2}}\right)^{-1},
\ee
and is identified with the relative flux of longitudinal virtual photons.
In terms of $\tau$ and $\eps$ the incident electron energy is
\be
E\ =\ M\left( \tau + \sqrt{\tau (1+\tau)(1+\eps)/(1-\eps)} \right),
\ee
and the scattered electron energy is $E'=E-2 M \tau$.

One can also express the elastic cross section in terms of any two of the
Mandelstam variables $s$ (total electron--nucleon invariant mass squared),
$t$, and $u$, where
\be
s=(k+p)^2=(p'+k')^2,\ \ \ \ \
t=(k-k')^2=-Q^2,\ \ \ \ \
u=(p-k')^2=(p'-k)^2,
\label{eq:Mandelstam}
\ee
with the constraint $s + t + u = 2 M^2 + 2 m_e^2$.
Furthermore, the variable $\nu$ is related to the Mandelstam
variables by $\nu = (s-u)/(4M^2)$.  The electron mass $m_e$ can
generally be ignored at the kinematics of interest here.  In particular,
there are no mass singularities in the limit $m_e \to 0$ in either the
one-photon exchange amplitude or the {\em total} TPE amplitude.

\begin{figure}[htb]
\begin{center}
\begin{minipage}{8cm}
\includegraphics[angle=270,scale=0.6,clip=true,bb=60 0 320 400]{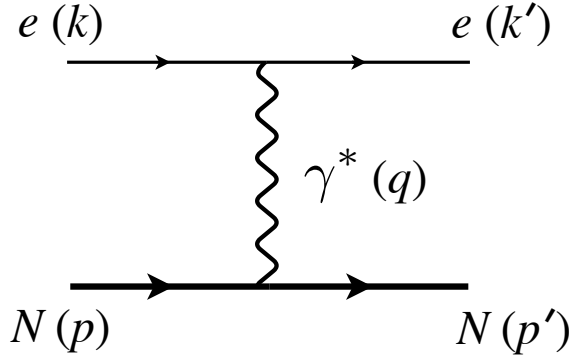}
\end{minipage}
\begin{minipage}{16.5cm}
\caption{Elastic electron--nucleon scattering in the one-photon
	exchange (Born) approximation.  Particle momenta are
	indicated in parentheses.}
\label{fig:OPE}
\end{minipage}
\end{center}
\end{figure}

\subsection{\it Born approximation}
\label{ssec:Born}

In the Born approximation (see Fig.~\ref{fig:OPE}), the
electron--nucleon scattering invariant amplitude can be written as
\bea
{\cal M}_\gamma &=& -{e^2 \over q^2}\, j_{\gamma \mu}\, J_\gamma^\mu ,
\label{eq:Mg}
\eea
where $e$ is the electric charge, and the matrix elements of the
electromagnetic leptonic and hadronic currents are given in terms
of the lepton ($u_e$) and nucleon ($u_N$) spinors by
\be
j_{\gamma \mu}\
=\ \ubar_e(k')\, \gamma_\mu\, u_e(k)\ , \ \ \ \ \ \
J_\gamma^\mu\
=\ \ubar_N(p')\, \Gamma_\gamma^\mu(q)\, u_N(p)\, .
\ee
Our metric and other unstated conventions follow Ref.~\cite{Pes95}.
Note that other conventions for amplitudes have also been used in
the TPE literature \cite{Blu03, Blu05, Kon05, Tjo08, Blu10}.
The electromagnetic hadron current operator $\Gamma_\gamma^\mu$ is
parametrized by the Dirac ($F_1$) and Pauli ($F_2$) form factors as
\be
\Gamma_\gamma^\mu(q)\
=\ \gamma^\mu\, F_1(Q^2)\
+\ {i \sigma^{\mu\nu} q_\nu \over 2 M}\, F_2(Q^2)\, ,
\label{eq:Jg}
\ee
where $M$ is the nucleon mass.
In terms of the amplitude ${\cal M}_\gamma$, the differential Born
cross section is given by
\bea
{ d\sigma \over d\Omega }
&=& \left( { \alpha \over 4 M Q^2 } {E' \over E} \right)^2
    \left| {\cal M}_\gamma \right|^2\
 =\ { \sigma_{\rm Mott} \over \eps (1+\tau) }\, \sigma_R\, ,
\label{eq:sigma0}
\eea
where $\alpha = e^2/4\pi$ is the electromagnetic fine structure constant,
and the Mott cross section for the scattering from a point particle is
\bea
\sigma_{\rm Mott}\
=\ { \alpha^2 E' \cos^2(\theta/2) \over 4 E^3 \sin^4(\theta/2) }\, .
\label{eq:Mott}
\eea
The reduced Born cross section $\sigma_R$ is given by
\be
\sigma_R\ =\ \eps\, G_E^2(Q^2)\ + \tau\, G_M^2(Q^2)\, ,
\label{eq:sigmaR}
\ee
where the Sachs electric and magnetic form factors $G_{E,M}(Q^2)$
are defined in terms of the Dirac and Pauli form factors as
\be
G_E(Q^2)\ =\ F_1(Q^2) - \tau F_2(Q^2)\, ,\ \ \ \
G_M(Q^2)\ =\ F_1(Q^2) + F_2(Q^2)\, .
\label{eq:GEMdef}
\ee
The form factors are normalized such that $G_E^{p\,(n)}(0)=1\,(0)$
and $G_M^{p\,(n)}(0)=\mu_{p\,(n)}$ for the proton (neutron), where
$\mu_{p\,(n)} = 2.793\,(-1.913)$ is the proton (neutron) magnetic moment.

\subsection{\it Form factors in the Born approximation}
\label{ssec:BornFF}

The standard technique to extract the electric and magnetic form factors
of the proton has been the Rosenbluth, or longitudinal-transverse (LT),
separation method \cite{Ros50}.  Using the fact that the Born level
form factors in Eq.~(\ref{eq:sigmaR}) are functions of $Q^2$ only,
analyzing the cross section as a function of the longitudinal photon
polarization $\eps$ at fixed $Q^2$ allows one to extract $G_M^2$
from the $\eps$-intercept, and $G_E^2$ from the slope in $\eps$,
once standard radiative corrections have been applied.
The cross section at $\theta=180\deg$ ($\eps \to 0$) depends only on the
magnetic form factor $G_M$, while the cross section at smaller angles
is a combination of magnetic and electric contributions.
Because of the $\eps/\tau$ weighting of $G_E^2$ relative to $G_M^2$,
the contribution from the electric form factor to the cross section
is suppressed at large $Q^2$.
The proton form factor ratios extracted via the Rosenbluth technique have
generally been consistent with $Q^2$ scaling, $|G_E| \approx |G_M/\mu_p|$
\cite{Wal94, Arr03, Chr04, Qat05}.
Note that because the cross sections are sensitive to the squares of the
form factors, the signs on the form factors cannot be determined from
Rosenbluth separations alone.

An alternative method of extracting the ratio $R$ utilizes polarization
degrees of freedom to increase the sensitivity to the electric form
factor at large $Q^2$.  Here, longitudinally polarized electrons are
scattered from an unpolarized proton target, with the polarization
of the recoiling proton detected, $\vec{e} p \to e \vec{p}$.
The polarization of the incident electron (or recoil proton) is
characterized by the spin four-vector \cite{Blu05, Max00P}
\be
s^\mu\
=\ \left( { \bm{\zeta} \cdot \bm{k} \over m };
            \bm{\zeta} + \bm{k} { \bm{\zeta} \cdot \bm{k} \over m (m+E) }
   \right),
\label{eq:s}
\ee
where $m$ and $E$ are the particle's mass and energy, and the
three-dimensional spin vector $\zeta$ specifies the spin direction in
the rest frame.  In the limit $\bm{k} \to 0$, the spin four-vector
$s^\mu \to (0; \bm{\zeta})$.  
Since $\zeta$ is a unit vector, $\bm{\zeta}^2 = 1$, from 
Eq.~(\ref{eq:s}) one has $s^2 = -1$ and $k \cdot s = 0$.
For incident electron energies $E \gg m_e$, the electron spin
four-vector $s_e^\mu$ can be related to the electron helicity
$h = \bm{\zeta_e} \cdot \hat{\bm{k}}$ by
\be
s_e^\mu\ \approx\ h\, { k^\mu \over m_e }\, .
\label{eq:s_e}
\ee
The coordinate axes are chosen so that the recoil proton momentum
$\bm{p}'$ defines the $z$ axis, in which case for longitudinally
polarized protons one has $\bm{\zeta_p} = \hat{\bm{p}}'$.
In the Born approximation the elastic cross section for scattering
a longitudinally polarized electron with a recoil proton polarized
longitudinally is then given by
\bea
{ d\sigma^{(L)} \over d\Omega }
&=& h\ \sigma_{\rm Mott}\
   {E + E' \over M} \sqrt{\tau \over 1+\tau} \tan^2{\theta\over 2}\
    G_M^2\, .
\label{eq:sigL}
\eea
For a proton detected with transverse polarization the $x$ axis
is defined to be in the scattering plane,
$\hat{\bm{x}} = \hat{\bm{y}}\ \times \hat{\bm{z}}$,
where $\hat{\bm{y}} = \hat{\bm{k}} \times \hat{\bm{k}}'$ defines
the direction perpendicular, or normal, to the scattering plane.
The cross section for producing a transversely polarized proton,
$\bm{\zeta_p} \cdot \bm{p}' = 0$, is given by
\bea
{ d\sigma^{(T)} \over d\Omega }
&=& h\ \sigma_{\rm Mott}\
    2 \sqrt{\tau \over 1+\tau} \tan{\theta\over 2}\
    G_E\, G_M\, ,
\label{eq:sigT}
\eea
while in the Born approximation the normal polarization is
identically zero.
Taking the ratio of the transverse to longitudinal proton cross
sections then yields the ratio of the electric to magnetic proton
form factors,
\be
-\mu_p \sqrt{\tau (1+\eps)\over 2 \eps}\, {P_T \over P_L }\
=\ -\mu_p { E+E'\over 2 M } \tan{\theta\over 2}\ { P_T \over P_L }\
=\ \mu_p {G_E \over G_M}\, ,
\label{eq:poltrans}
\ee
where $P_L$ and $P_T$ are the polarizations of the recoil proton
longitudinal and transverse to the proton momentum in the scattering
plane, proportional to the longitudinal and transverse cross sections
in Eqs.~(\ref{eq:sigL}) and (\ref{eq:sigT}), respectively.

\begin{figure}[ht]
\begin{center}
\begin{minipage}{8cm}
\epsfig{file=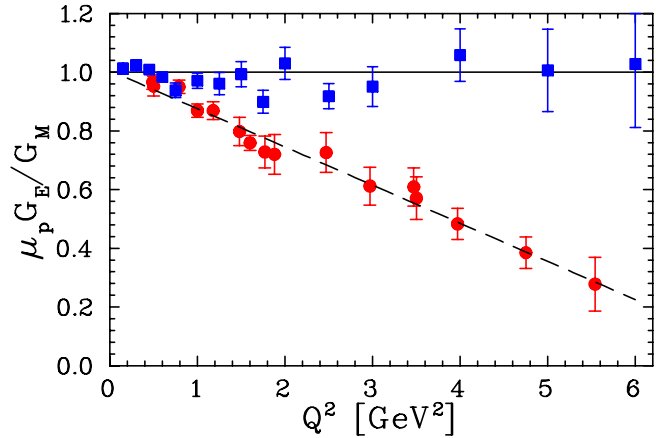,scale=0.5}
\end{minipage}
\begin{minipage}{16.5cm}
\caption{Ratio of proton electric to magnetic form factors as extracted
	using Rosenbluth (LT) separation \cite{Wal94} (squares) and
	polarization transfer measurements \cite{Jon00, Gay02} (circles).
	Figure adapted from Ref.~\cite{Arr03}.}
\label{fig:GEGM}
\end{minipage}
\end{center}
\end{figure}

In a series of recent experiments at Jefferson Lab \cite{Jon00, Gay01,
Gay02, Pun05, Mac06, Ron07, Puc10, Zha11, Puc11, Ron11}, the polarization
transfer (PT) technique has been used to accurately determine the ratio
$\gegm$ up to $Q^2 = 8.5$~GeV$^2$.  In addition, there have been
complementary measurements using polarized targets at MIT-Bates
\cite{Cra06} and Jefferson Lab \cite{Jon06}.  The results, illustrated
in Fig.~\ref{fig:GEGM}, are in striking contrast to the ratio obtained
via LT or Rosenbluth separations, showing an approximately linear
decrease of $R$ with $Q^2$ which is in strong violation of the $Q^2$
scaling behavior (see also Refs.~\cite{Per07, Arr07a, Jai03, Bel03}).

The discrepancy between the LT and PT measurements of $\gegm$ has
stimulated considerable activity, both theoretically and experimentally,
over the past decade.  Attempts to reconcile the measurements have
mostly focused on improved treatments of radiative corrections,
particularly those associated with two-photon exchange, which can lead
to additional angular (and thus $\eps$) dependence of the cross section.
In the following sections we discuss experimental efforts to better 
understand the discrepancy, and then describe theoretical attempts to
compute TPE corrections and assess their impact on various observables.

%% file: section3.tex
\section{Experimental observables and measurements}
\label{sec:experiments}

\subsection{\it Verification of the discrepancy}

\begin{figure}[htb]
\begin{center}
\epsfig{file=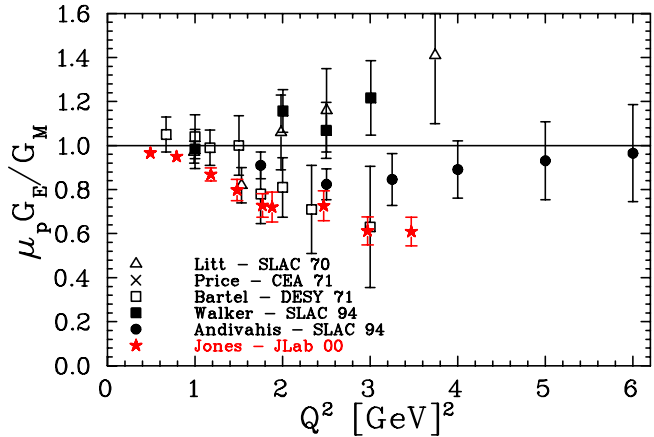,scale=0.5}
\begin{minipage}{16.5cm}
\caption{Ratio of proton electric to magnetic form factors as extracted
	from the initial high $Q^2$ polarization transfer measurement 
	\cite{Jon00} (stars) and previous Rosenbluth (LT) 
	separations~\cite{Wal94, Lit70, Pri71, Bar73, And94}.}
\label{fig:GEGM_raw}
\end{minipage}
\end{center}
\end{figure}

The striking difference between Rosenbluth~\cite{Bos95} and the early
polarization transfer~\cite{Jon00, Gay02} measurements of the proton
electromagnetic form factors shown in Fig.~\ref{fig:GEGM} led to 
significant activity aimed at understanding and resolving this 
discrepancy.
It was noted early on~\cite{Jon00} that there was significant scatter
between the results of different Rosenbluth extractions
\cite{Wal94, Lit70, Pri71, Bar73, And94}, as illustrated in
Fig.~\ref{fig:GEGM_raw}, suggesting that the problem was related
to the cross section measurements.
At high $Q^2$, $\ge$ yields only a small, angle-dependent correction
to the cross section, leading to the possibility that a systematic 
difference between small- and large-angle measurements could yield
large corrections to $\gegm$ that would increase in importance
with increasing $Q^2$.
Thus, it was initially suggested that the observed difference may
be due to some experimental error in one or more of the cross section
measurements, rather than a true discrepancy between the techniques.

Because most extractions combine multiple data sets covering different
angular regions, uncertainty in the relative normalization of the data 
sets leads to large uncertainties in the form factors that can be 
strongly correlated between points at different $Q^2$ values.
Many extractions allow the normalization of each experiment to vary as 
part of the fit, to improve consistency in the overlap regions, but the
uncertainty in the determination of the relative normalization factors 
was usually neglected.  At high $Q^2$, a change in the normalization 
between large- and small-angle measurements can systematically shift 
all of the high $Q^2$ results within the extraction.

A detailed reanalysis of the world's cross section data for moderate
to high $Q^2$ values \cite{Arr03} showed that the cross sections were
consistent when accounting for the quoted normalization uncertainties
in the different data sets, except for some of the small-angle data 
from Ref.~\cite{Wal94}.  Excluding these data and performing a global
extraction of $\ge$ and $\gm$ yielded results that were still in
significant disagreement with the polarization transfer extractions.
Furthermore, it was not possible to reproduce the polarization results
by varying the normalization of the data sets within their quoted
uncertainties, or by simply excluding one of the 20 data sets included
in the analysis.  However, it was noted that a systematic correction
with an $\eps$ dependence of $\sim$~5\% for $Q^2 \gtorder 2$~GeV$^2$
would be able to explain the discrepancy between the polarization
measurements and the Rosenbluth extractions \cite{Arr03}.

\begin{figure}[thb]
\begin{center}
\includegraphics[width=8cm]{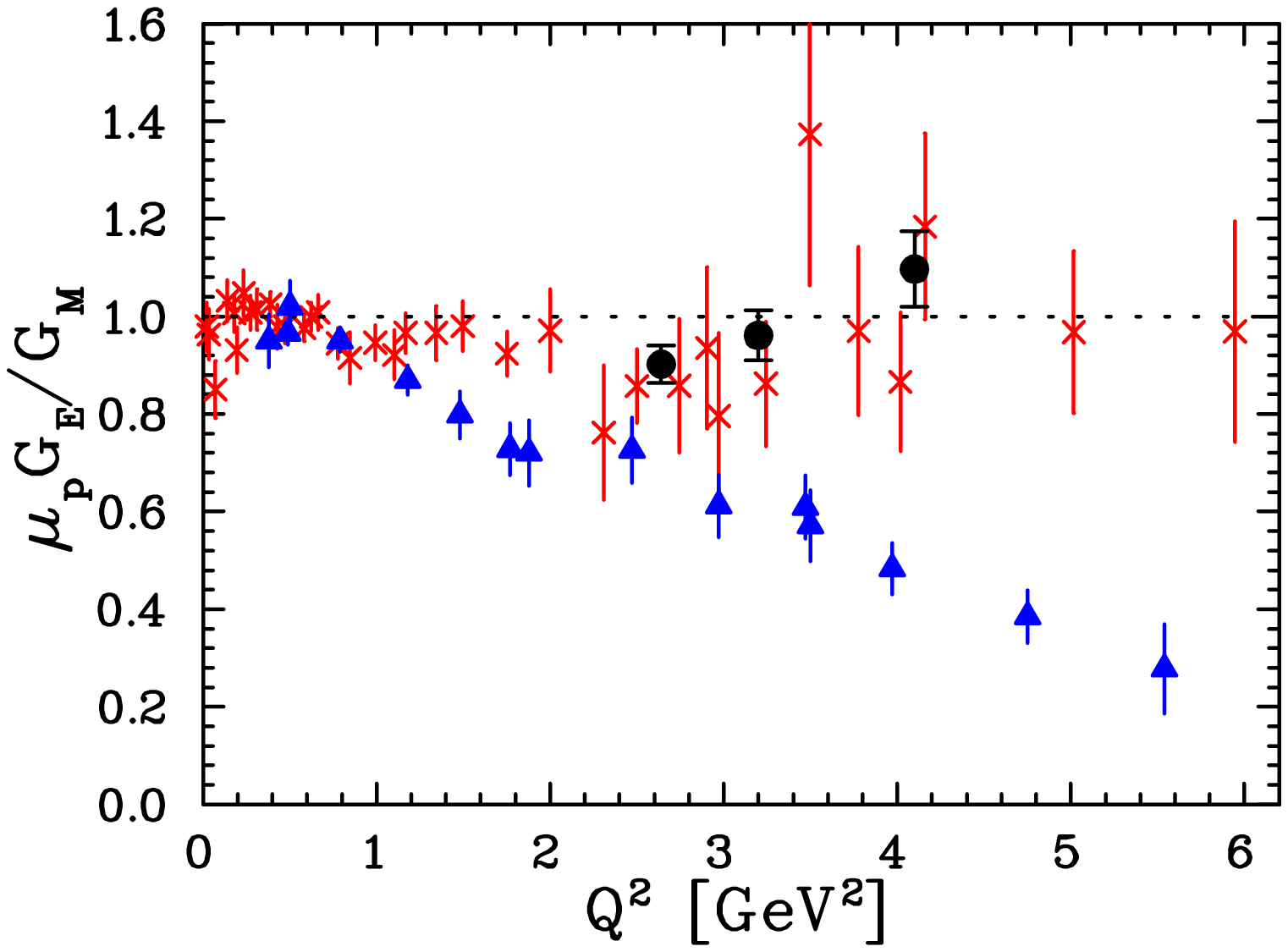}
\hspace*{0.5cm}
\includegraphics[width=8cm]{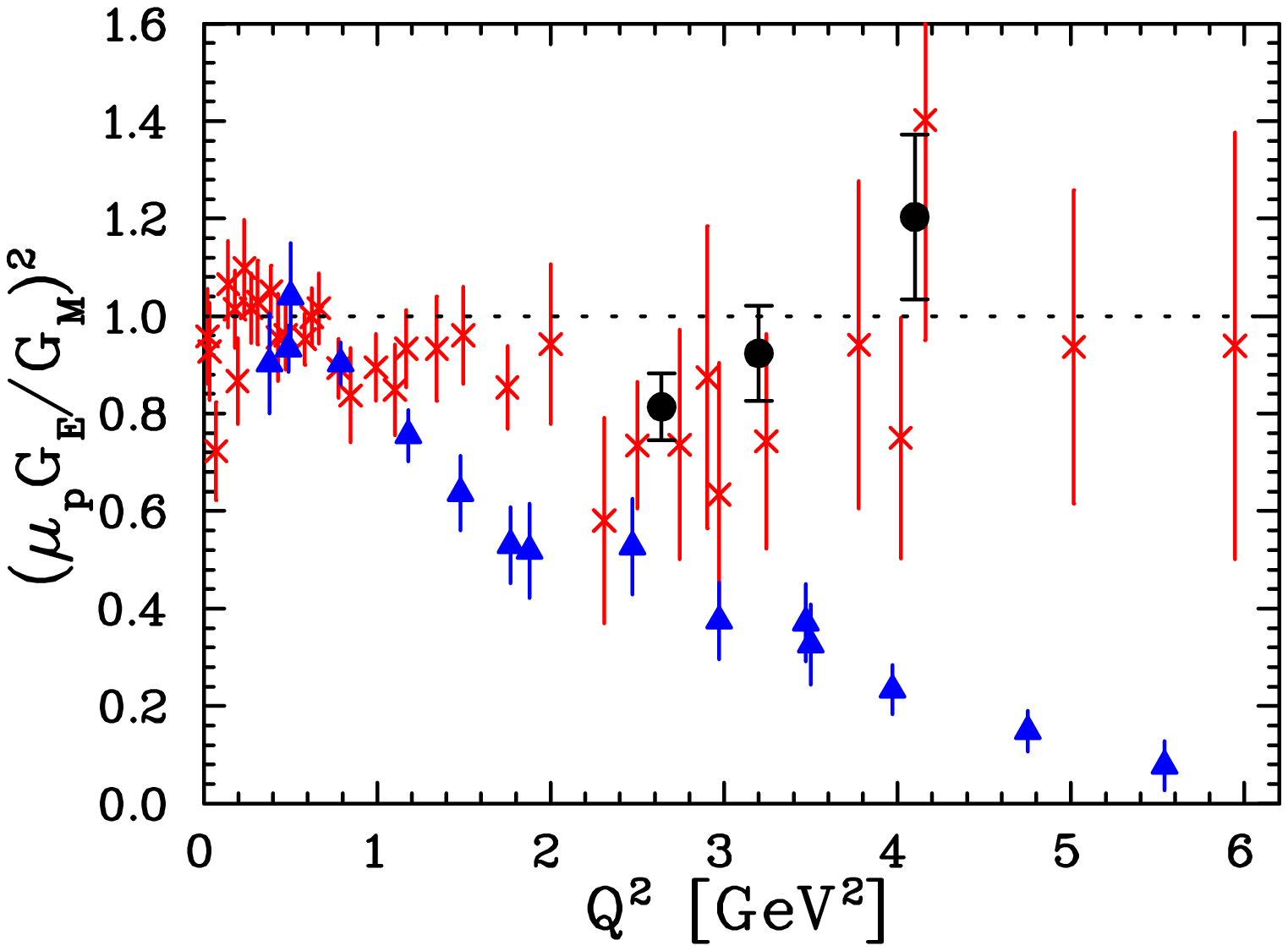}
\begin{minipage}{16.5cm}
\caption{Proton electric to magnetic form factor ratio $\mugegm$
	{\bf (left)} and $(\mugegm)^2$ {\bf (right)} extracted
	from the high-precision Super-Rosenbluth experiment \cite{Qat05}
	(filled circles), compared to polarization extractions
	\cite{Jon00, Gay02} (triangles) and a global analysis of
	previous cross section measurements~\cite{Arr04a} (crosses).
	Note that the slope of the reduced cross section in the
	Rosenbluth measurements is directly sensitive to $(\mugegm)^2$,
	so the right figure best shows the significance of the
	discrepancy at the cross section level.  Figure adapted from
	Ref.~\cite{Qat05}.}
\label{fig:e01001}
\end{minipage}
\end{center}
\end{figure}

Shortly thereafter, several extremely high precision measurements of
$\gegm$ were made using a modified Rosenbluth technique \cite{Qat05,
Qat06}.  Because the extraction is very sensitive to angle-dependent
corrections, the measurement used detection of the recoil proton,
rather than the scattered electron.  In the traditional Rosenbluth
separation, varying $\eps$ while keeping $Q^2$ fixed means that the
scattered electron energy and angle vary with $\eps$, as does the
scattering cross section.  This means that any momentum- or
rate-dependent efficiencies or systematic corrections yield
$\eps$-dependent effects.  With proton detection, the momentum of
the detected particle does not vary for measurements at a fixed $Q^2$
value, and the cross section for detection of the recoil proton is only
very weakly dependent on angle.  These and several other corrections and
systematic uncertainties are thus strongly suppressed in the extraction
of $\gegm$ \cite{Qat05, Qat06}.  In this measurement, care was taken
to ensure that the \textit{relative} cross section uncertainties were
minimized, even though this could increase the absolute uncertainties,
because only the relative uncertainties enter into the ratio $\gegm$.
The experiment yielded measurements for $Q^2$=2.64, 3.2 and 4.1~GeV$^2$ 
with precision comparable to the recoil polarization results, as shown
in Fig.~\ref{fig:e01001}.  This confirmed a clear and significant
discrepancy with the polarization measurements, which ruled out the
possibility that the difference was simply caused by a small systematic
error in some of the earlier measurements.

It is worth noting that the radiative corrections applied to the
elastic scattering cross sections have evolved somewhat over time.
While the early work of Mo and Tsai \cite{Tsa61, Mo69} provides
the standard basis for applying radiative corrections, the SLAC
measurements in the 1990s \cite{Wal94, And94} accounted for additional
radiative correction terms such as vacuum polarization contributions
from higher mass leptons, which have also been included in other
recent extractions.  In the global analysis~\cite{Arr04a}, these
corrections were also included for older measurements to minimize the
difference in the radiative correction procedures over the full body
of data.  However, the additional terms have essentially no dependence
on the electron scattering angle, and therefore have minimal impact
on the extraction of $\ge$ at large $Q^2$.

\subsection{\it Early estimates of two-photon exchange corrections}
\label{sec:estimates}

Once it was clear that there was a systematic discrepancy between the
Rosenbluth and polarization transfer techniques, it was important to
determine how large of an effect would be needed to explain the
discrepancy, and determine if TPE contributions of this size could be
ruled out by existing measurements.  Early attempts to calculate the
TPE contributions in the 1950s and 1960s, using only the proton in
the intermediate state~\cite{Mck48,Lew56}, or including some excited
intermediate states~\cite{Dre57, Dre59, Wer61, Cam67, Cam69a, Gre69,
Bro70}, yielded extremely small TPE contributions, typically well
below one percent of the Born cross section\footnote{The TPE diagram
also contains an infrared-divergent contribution, which is canceled by
a corresponding divergent term in bremsstrahlung emission, as discussed
in more detail in Sec.~\ref{ssec:general}.}.
It was difficult, however, to determine how reliable these estimates
were, especially at large $Q^2$ values.
Even before the form factor discrepancy became clear, there were some
estimates examining the radiative corrections~\cite{Ent01} and TPE
contributions~\cite{Max00T}.  Coulomb distortion, corresponding to
the soft-photon contribution to TPE, had also been investigated.
This was mainly of interest at low $Q^2$~\cite{Ros00}, where it
contributes a significant portion of the full TPE correction.
After the observation of the discrepancy, it was examined at higher
$Q^2$~\cite{Arr04c}, but found to have a relatively small effect
compared to the observed discrepancy.  For the most part,
investigations have focused on the effect of TPE corrections
\cite{Blu03, Max00T, Gui03, Rek04, Afa03, Che04} beyond the
soft-photon approximation.  Calculations of TPE corrections will
be presented in detail in Sec.~\ref{sec:TPE}, and their impact
on the measurements in Sec.~\ref{sec:impact}.

\begin{figure}[thb]
\begin{center}
\includegraphics[height=6.0cm,width=8.5cm]{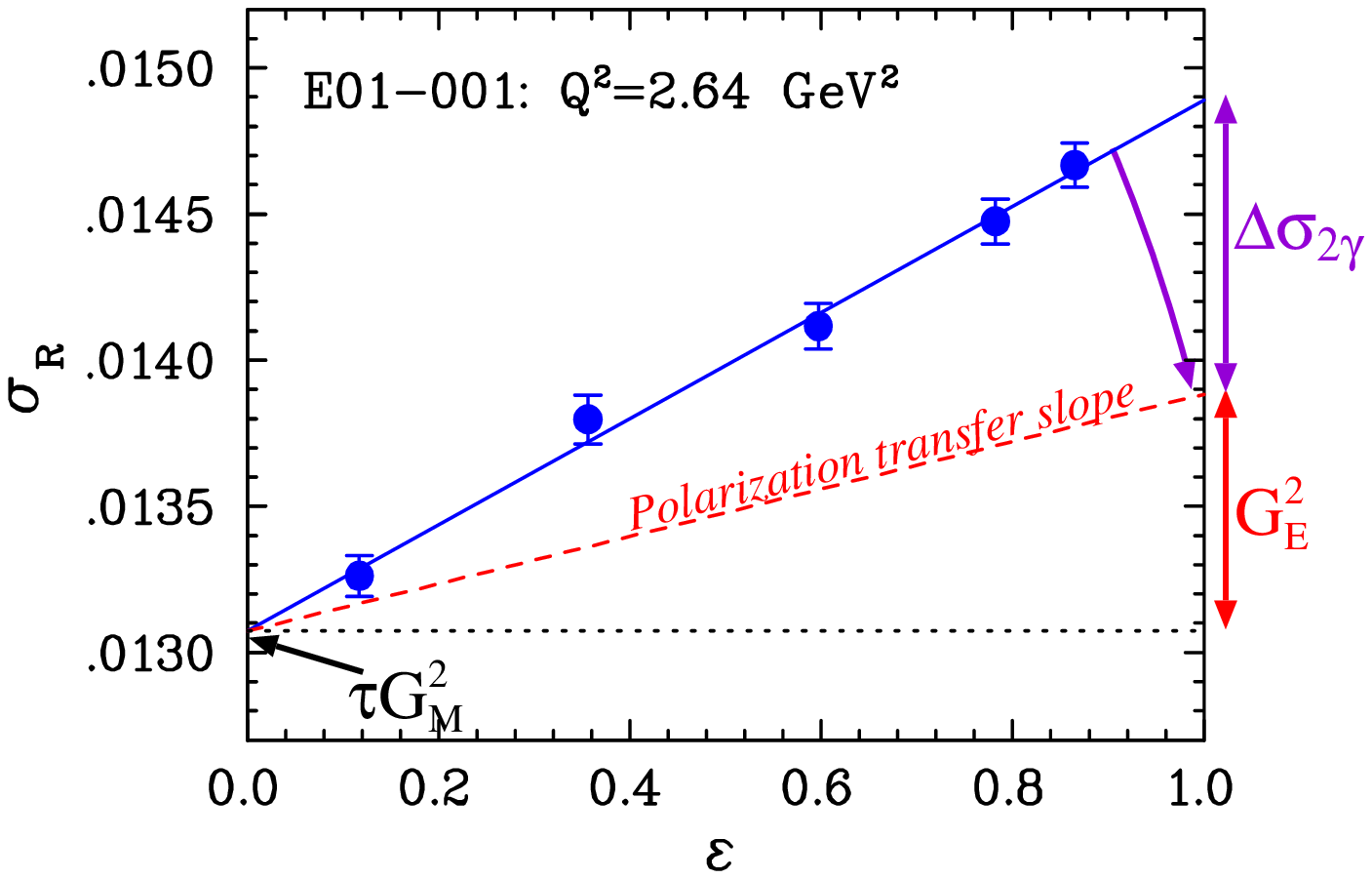}
\includegraphics[height=6.0cm,width=8.5cm]{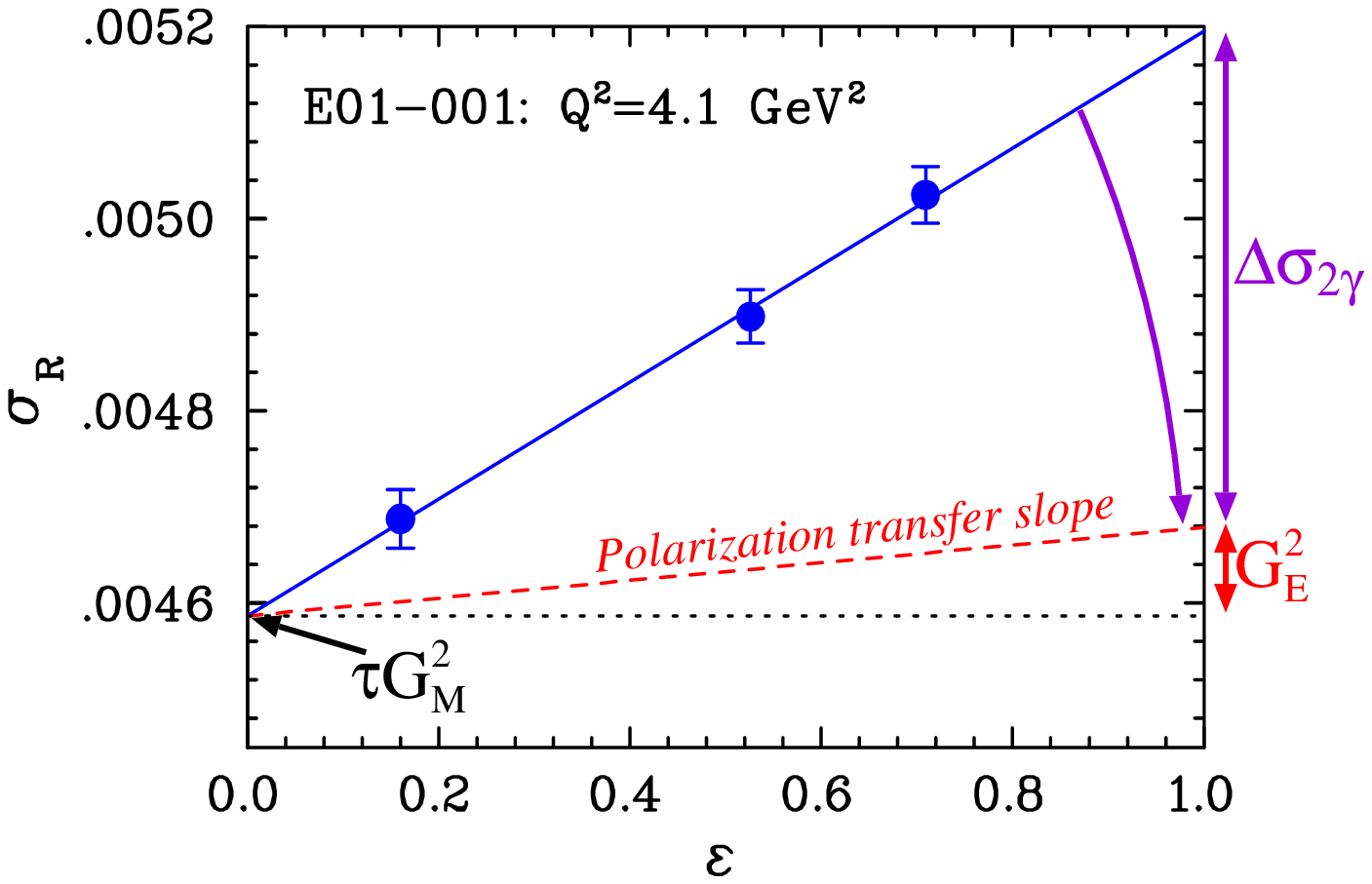}
\begin{minipage}{16.5cm}
\caption{The $\eps$ dependence of the reduced cross section as
	predicted from the polarization transfer results for $\gegm$
	(red dashed line), and as measured by Jefferson Lab experiment
	E01-001 \cite{Qat05} (circles; blue solid line is best fit).
	If the polarization transfer represents the true form factors,
	TPE yields more than half of the $\eps$ dependence at
	2.64~GeV$^2$ {\bf (left)}, and 85\% at 4.1~GeV$^2$ {\bf (right)}.
	The PT measurements give only the slope of $\sigma_R$, and the
	curve has been arbitrarily normalized to match the LT separations
	at $\eps=0$.}
\label{fig:delta_sigma}
\end{minipage}
\end{center}
\end{figure}

Initial investigations focused on using measurements of the discrepancy
to determine the nature of the corrections required to explain the data.
Analyses of the discrepancy that assume it is due to missing strength
in the cross section measurements~\cite{Arr03, Arr04a} indicated that
the difference could be explained by an error in the $\eps$ dependence
of the cross section of approximately $5-8\%$ for $Q^2 > 2-3$~GeV$^2$.
Similar results are obtained by comparing global analyses of the form
factors, with and without the high-$Q^2$ Rosenbluth data included
\cite{Bel08, Alb09}, under the assumption that the difference will be
entirely due to the TPE contributions.
This approach gives a less direct extraction of the TPE contribution to
the cross section and is sensitive to other assumptions made in the fit,
but it allows the extraction to be performed in the context of a more
global analysis of form factor data.

Figure~\ref{fig:delta_sigma} shows the $\eps$ dependence observed in
the Super-Rosenbluth experiment~\cite{Qat05} and as expected from the
polarization transfer measurements~\cite{Jon00, Gay02}, assuming the Born
approximation.  The polarization transfer measurement constrains only the
slope, not the value of the cross section, and has been arbitrarily
normalized to agree with the Rosenbluth extraction at $\eps=0$.
If some systematic correction to the cross section explains the
difference, then the $\eps$ dependence shown by the dashed line
is related to the contribution from the electric form factor,
and the remaining slope must come from the missing correction.
This correction should be close enough to linear such that it does
not spoil the linearity expected from the Rosenbluth formula, since the
reduced cross section is consistent with a linear $\eps$ dependence
within the current uncertainties.

If the discrepancy is entirely related to TPE contributions, then there
are additional constraints that can be included in such extractions.
For example, TPE corrections must be zero at $\eps=1$, as discussed
in Sec.~\ref{sec:TPE}, so one would expect the two lines in
Fig.~\ref{fig:delta_sigma} to meet at $\eps=1$ rather than $\eps=0$.
Such constraints were included in later attempts to extract the TPE
contribution to the cross section.
Two such extractions~\cite{Tom05, Che07} use symmetry arguments to choose
functional forms for the TPE corrections and then fit the difference
between Rosenbluth and polarization data to extract the TPE contributions.
While the symmetry arguments do not give a unique $\eps$ dependence,
they suggest that it may be more natural to take the correction to be
a simple function of $x=\sqrt{(1+\eps)/(1-\eps)}$ rather than a function
of $\theta$ or $\eps$.  Note, however, that the specific form chosen in
Ref.~\cite{Tom05} is divergent for $\theta \to 0\deg$, and so does not
yield the correct limit for $\eps=1$.  Similar extractions are performed
in Refs.~\cite{Arr04a, Bor07a, Qat11, Gra11}, again fitting only the TPE
contributions to the cross section, but making somewhat different
assumptions about the functional form.  A detailed comparison of the
assumptions and procedures of many of these extractions can be found
in Ref.~\cite{Qat11}.

It is worth noting that many of these TPE extractions are only reliable
at high $Q^2$ values.  All of the extractions discussed thus far assume
that the TPE impact on polarization measurements is negligible.
While this assumption is reasonable at high $Q^2$ values, where the
Rosenbluth extraction of $\ge$ is extremely sensitive to TPE, it is not
as good at lower $Q^2$ values where both the polarization and cross
section measurements may have small but similar TPE contributions.
This can introduce additional uncertainty in the form factors at low
$Q^2$ in cases where both the TPE contribution and form factors are
extracted.  A clear example is Ref.~\cite{Arr04a}, where the TPE
contribution extracted at large $Q^2$ is applied as a correction
at all $Q^2$ values in extracting the form factors, leading to an
overestimate of TPE effects at low $Q^2$.

The first attempt to use the discrepancy to extract the TPE amplitudes
was made in Ref.~\cite{Gui03}, which took into account the impact of TPE
on the polarization measurements as well as on the cross section data.
Because there are three TPE amplitudes, each of which depends on $Q^2$
and $\eps$, significant assumptions have to be made about their $\eps$
dependence.  Thus, different extractions~\cite{Gui03, Arr05, Bor07}
yielded very different amplitudes, depending on the constraints or
approximations used to select the amplitudes and functional forms
used in the analysis.
However, because the dominant effect at high $Q^2$ is related to the
TPE contribution to the cross section, all of these analyses yield
similar results in this region: an $\eps$ dependence of 5--8\%
at high $Q^2$, although with different overall normalizations.
Later analyses~\cite{Bor11a, Gut11} were able to make less
model-dependent extractions with the inclusion of new information
on TPE in polarization measurements~\cite{Mez11}.  These results
will be discussed in more detail in Sec.~\ref{ssec:impact_PT}.

\begin{figure}[thb]
\begin{center}
\includegraphics[width=9.0cm]{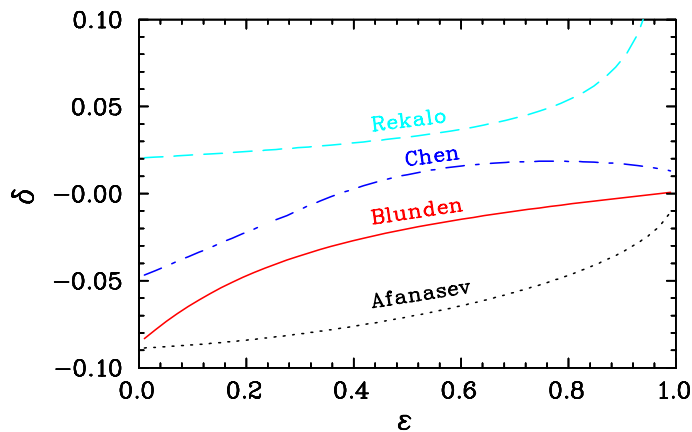}
\begin{minipage}{16.5cm}
\caption{The TPE correction $\delta$, defined via
	$\sigma = \sigma_0 (1+\delta)$, where $\sigma_0$ is the
	Born cross section, to the elastic $ep$ cross section
	from the calculations of
	Rekalo~\etal~\cite{Rek04} (dashed),
	Chen~\etal~\cite{Che04} (dot-dashed),
	Blunden~\etal~\cite{Blu03} (solid), and 
	Afanasev~\etal~\cite{Afa03} (dotted)
	for $Q^2 \approx 5$~GeV$^2$, after scaling the calculations
	to yield an $\eps$ dependence of approximately 6\% over the
	$\eps$ range of existing data.}
\label{fig:calcs_e01001}
\end{minipage}
\end{center}
\end{figure}

Early calculations were performed in hadronic~\cite{Blu03} and
partonic~\cite{Che04} frameworks, yielding corrections which could
explain roughly half of the discrepancy at large $Q^2$ values.
Calculations at the parton level in the double logarithm approximation
\cite{Afa03} yielded a different form for the $\eps$ dependence, with
nonlinearities appearing at large $\eps$.  Finally, invariance under
C-parity and crossing symmetry were used \cite{Rek04} to argue that
the TPE corrections should depend on $x=\sqrt{(1+\eps)/(1-\eps)}$.
The range of predictions of these models is shown in 
Fig.~\ref{fig:calcs_e01001} for $Q^2 \approx 2-3$~GeV$^2$, where each
model has been scaled to give an overall $\eps$ dependence of the
size needed to resolve the discrepancy between the Rosenbluth and
polarization measurements.   For the partonic model \cite{Che04}, the
calculation is not expected to be valid at low $Q^2$ or $\eps$ values.
Many of these calculations have been updated since, and the most recent
results will be discussed in Sec.~\ref{sec:TPE}, but these were the
models available at the time that measurements to further examine the
TPE corrections were being considered.

While scaled to yield approximately the same $\eps$ dependence, the
different models would nonetheless yield significantly different results
for the extracted form factors.  The Chen~\etal~calculation \cite{Che04}
has little $\eps$ dependence for $\eps>0.5$, meaning that it would have
little impact on Rosenbluth extractions if they did not have data at low
$\eps$.  The form of the correction in Ref.~\cite{Rek04} yields large
effects for $\eps \to 1$, but small corrections to $\gm$ for measurements
at low $\eps$.  The other calculations have the largest corrections at
low $\eps$, and thus larger corrections to direct measurements of $\gm$.
In addition, some calculations predict large deviations from the linear
$\eps$ dependence of the Born approximation at very high $\eps$, while
others predict the greatest deviations at low $\eps$.  Based on this
large range of predictions, several experimental tests, using existing
data or new measurements, were performed to provide independent
constraints on any possible TPE contributions.

\subsection{\it Experimental signatures of two-photon exchange contributions}
\label{sec:exp_overview}

Two-photon exchange contributions to elastic electron--proton scattering
manifest themselves in several different ways.  The real part of the TPE
amplitude modifies both the unpolarized cross section and the
polarization transfer components used to extract $\gegm$, and thus
measurements sensitive to the real part of TPE amplitudes yield       
constraints that are directly relevant to the form factor discrepancy.
These contributions, however, must be disentangled from the dominant  
single-photon contributions.
The imaginary part of the amplitudes, on the other hand, leads to  
non-zero values for the Born-forbidden normal asymmetries       
(Sec.~\ref{ssec:normal}), which allow two-photon effects to be directly
isolated.  However, since these are not directly connected to the form
factor discrepancy, they were not the initial focus of experimental
investigations.  We summarize here some of the early experiments that
provided constraints on the TPE contributions.

\subsubsection{Comparisons of positron--proton and electron--proton scattering}
\label{sec:positron}

\begin{figure}[h]
\begin{center}
\includegraphics[height=5.5cm,width=8.0cm]{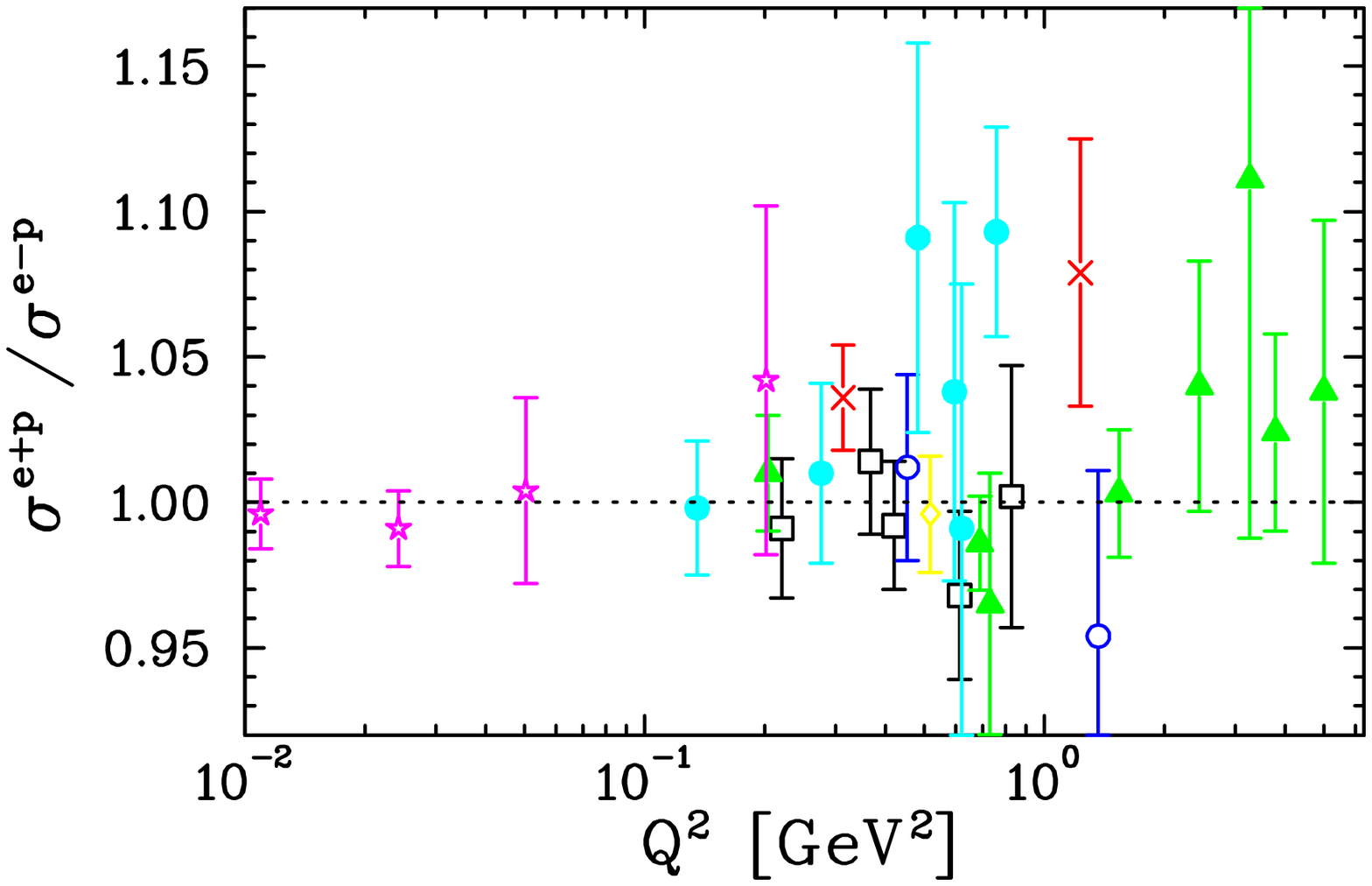}
\hspace*{0.5cm}
\includegraphics[height=5.5cm,width=8.0cm]{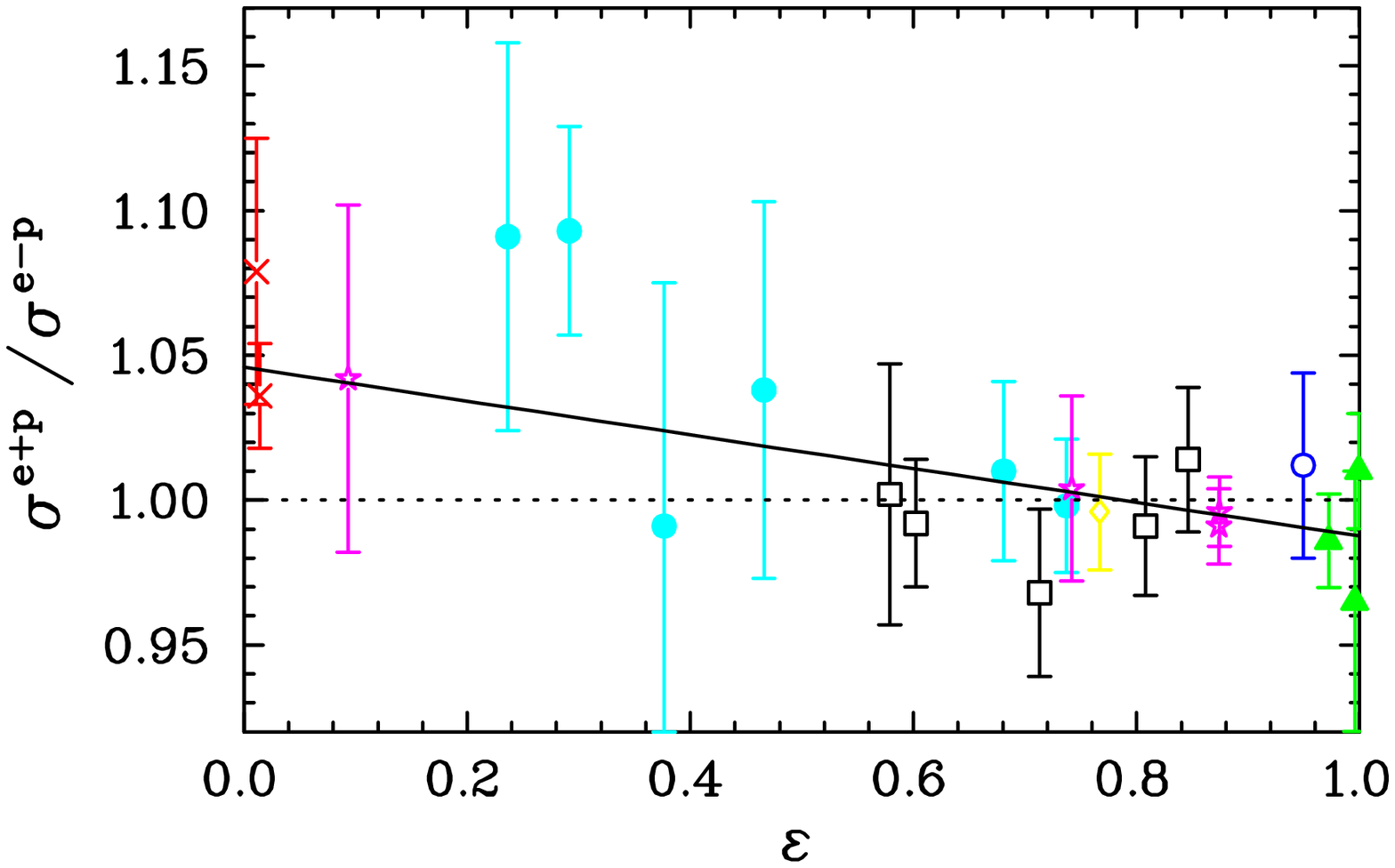}
\begin{minipage}{16.5cm}
\caption{The ratio of positron--proton to electron--proton scattering
	as a function of $Q^2$ {\bf (left)} and $\eps$ {\bf (right)}.
	The $\eps$ dependence plot includes only measurements below
	$Q^2=2$~GeV$^2$. The solid curve is a linear fit to the data
	yielding a slope of $-0.057(18)$.  Figure adapted from
	Ref.~\cite{Arr04b}.}
\label{fig:epluseminus}
\end{minipage}
\end{center}
\end{figure}

In unpolarized electron--proton scattering the TPE contributions cannot
in practice be separated from the Born cross section empirically.
The cleanest means of identifying TPE effects in unpolarized $ep$
scattering is to compare positron--proton and electron--proton cross
sections, where the interference between one- and two-photon exchange
has the opposite sign for positron and electron beams
(Sec.~\ref{ssec:general}).
While experimentally difficult, these were considered crucial tests
of the electron scattering technique, and several experiments were
performed in the 1960s to study possible TPE effects.
The comparisons of $e^+ p$ and $e^- p$~\cite{You62, Bro65, And66, Bar67,
And68, Bou68, Mar68} and $\mu^+ p$ and $\mu^- p$~\cite{Cam69b} scattering
were interpreted as supporting the conclusions of the early calculations
that two-photon corrections were extremely small ($< 1\%$).
Figure~\ref{fig:epluseminus}~(left panel) shows a compilation of all
such measurements for elastic scattering as a function of $Q^2$.
However, the low intensity of $e^+$ ($\mu^+$) beams made precise
measurements nearly impossible for large $Q^2$ or small $\eps$ values.

The data were reexamined~\cite{Arr04b} in light of the form factor
discrepancy, which suggested an $\eps$-dependent effect.
Measurements at high $Q^2$ were limited to small scattering angle
(large $\eps$) and so could not set a meaningful limit on angle-dependent
TPE contributions if they were small at forward scattering angles.
Figure~\ref{fig:epluseminus}~(right panel) shows the ratio of positron
to electron cross section as a function of $\eps$ for measurements at
$Q^2 \leq 2$~GeV$^2$.  There is some evidence of a charge-dependent term
in the $e^\pm p$ elastic cross section ratio at small values of $\eps$,
although the data at low $\eps$ are not very precise.  A linear fit
yields a 6\% $\eps$ dependence in the positron to electron ratio,
implying a 3\% $\eps$ dependence for the electron--proton cross section.
This is roughly half of the $\eps$ dependence needed to explain the
discrepancy at high $Q^2$, but the average $Q^2$ value of the points
showing nonzero TPE is less than 0.5~GeV$^2$, well below the region
of the observed discrepancy.  In addition, these measurements need
to apply a model-dependent correction for the charge-dependent
bremsstrahlung contributions to isolate TPE effects, and such corrections
have not been applied in a consistent fashion for all of the measurements.

Because of the limited statistics, the low $Q^2$ values of the large
angle measurements, and the model dependence in extracting the TPE
contributions, these data are insufficient to make a strong conclusion
about the presence of TPE effects.  There are three experiments which
aim to improve the precision and kinematic coverage of $e^\pm p$
comparisons~\cite{VEPP, e07005, olympus}.  Two of the measurements
\cite{VEPP, e07005} have completed data taking, and the third is
expected to begin in the near future.  While all of these will be
limited to $Q^2 \ltorder 2$~GeV$^2$, they will make precise measurements
over a significant $\eps$ range, and allow direct tests of calculations
of the TPE effects.  Additional details on the planned measurements are
presented in Sec.~\ref{sec:impact_positron}.

\subsubsection{Improved measurements of the Rosenbluth--polarization
	transfer discrepancy}

Given the difficulty of making precise comparisons of positron and
electron scattering at high $Q^2$ and large scattering angle, one must
complement these direct measurements with other studies that can be
used to constrain TPE contributions.  The most compelling evidence to
date is the discrepancy between Rosenbluth and polarization measurements
of the proton form factor ratio $\gegm$ which can be used to constrain
the overall size of the $\eps$-dependent TPE effects under the assumption
that TPE fully explains the discrepancy and that all other corrections
are accounted for.

Even at high $Q^2$ values, the discrepancy is only at the 2--3~$\sigma$
level when examined as a function of $(\mugegm)^2$,
Fig.~\ref{fig:e01001}~(right panel), which is directly related to
the measured slope of the reduced cross section.  While new or
updated high $Q^2$ polarization measurements are now available
\cite{Puc10,Puc11}, the comparison is clearly limited by the
uncertainties in the Rosenbluth separation measurements.
Jefferson Lab experiment E05-017~\cite{e05017} made an extended set
of Super-Rosenbluth measurements, using the same technique as
Ref.~\cite{Qat05}, but covering a much larger $Q^2$ range, and the
data are currently being analyzed.

Below $Q^2=1$~GeV$^2$ the sensitivity to TPE corrections is smaller,
and significant improvements in both the polarization and Rosenbluth
measurements are necessary to search for indication of TPE
contributions.  Extended Rosenbluth measurements from Mainz~\cite{Ber10}
and high-precision polarization measurements from Jefferson Lab
\cite{Zha11, Ron11, Pao10} are now available.
The new data show reasonable agreement between Rosenbluth and
polarization measurements below $Q^2=1$~GeV$^2$, suggesting that
TPE contributions may be small in this region.
However, because the impact of TPE corrections is reduced at low
$Q^2$, and the Rosenbluth extractions~\cite{Ber10} did include a
partial implementation of TPE (using the Coulomb distortion correction
in the $Q^2=0$ limit from Ref.~\cite{Mck48}), it is not yet clear to
what extent the agreement sets significant limits on TPE.

\subsubsection{Experimental limits on nonlinearities}

In the Born approximation, the reduced cross section at fixed $Q^2$
depends linearly on $\eps$.  Any deviation from linearity must come
from terms that are not included in the standard radiative correction
procedures.  Thus, the difference between polarization and Rosenbluth
measurements of the form factors is related to the average linear
contribution of TPE, while any deviation from linearity is a clear 
indicator of effects beyond the Born approximation.  At lower $Q^2$ 
values, observing such a deviation would be a clear signature of 
effects beyond one-photon exchange and would provide quantitative 
information on the \textit{nonlinear} component of the such effects.
At large $Q^2$, the contribution from $\ge$ becomes small enough
that almost all of the $\eps$ dependence comes from TPE (or other
corrections), as shown in Fig.~\ref{fig:delta_sigma}.  In this region,
the $\eps$ dependence of the reduced cross section allows one to
isolate the full $\eps$ dependence of the contributions beyond the
Born approximation.

A detailed examination of nonlinear contributions was performed by
fitting the reduced cross section to the form
\be
\sigma_R\ =\ P_0\, \left[ 1 + P_1 (\eps-0.5) + P_2 (\eps-0.5)^2 \right] .
\label{eq:curvature}
\ee
Note that the quadratic term is expanded around $\eps=0.5$ such that
$P_2$ represents the relative magnitude of the nonlinear term compared
to the average cross section, rather than the $\eps=0$ cross section, 
which becomes extremely small at low $Q^2$.  The coefficient $P_2$
provides a simple measure of the relative size of nonlinear terms,
and the uncertainty $\delta P_2$ can be used to set limits on
${\cal O}(\eps^2)$ terms.
Conventional Rosenbluth separation measurements have found $P_2$ to be 
consistent with zero, and the best constraint \cite{And94} yields 
$\delta P_2 \approx 10.5$\%.  The recent Jefferson Lab E01-001 data
\cite{Qat05} obtained improved limits on $P_2$ ($\delta P_2 =4.4$\%) by
detecting the struck proton rather than the scattered electron.

\begin{figure}[thb]
\begin{center}
\includegraphics[height=5cm,width=8.1cm]{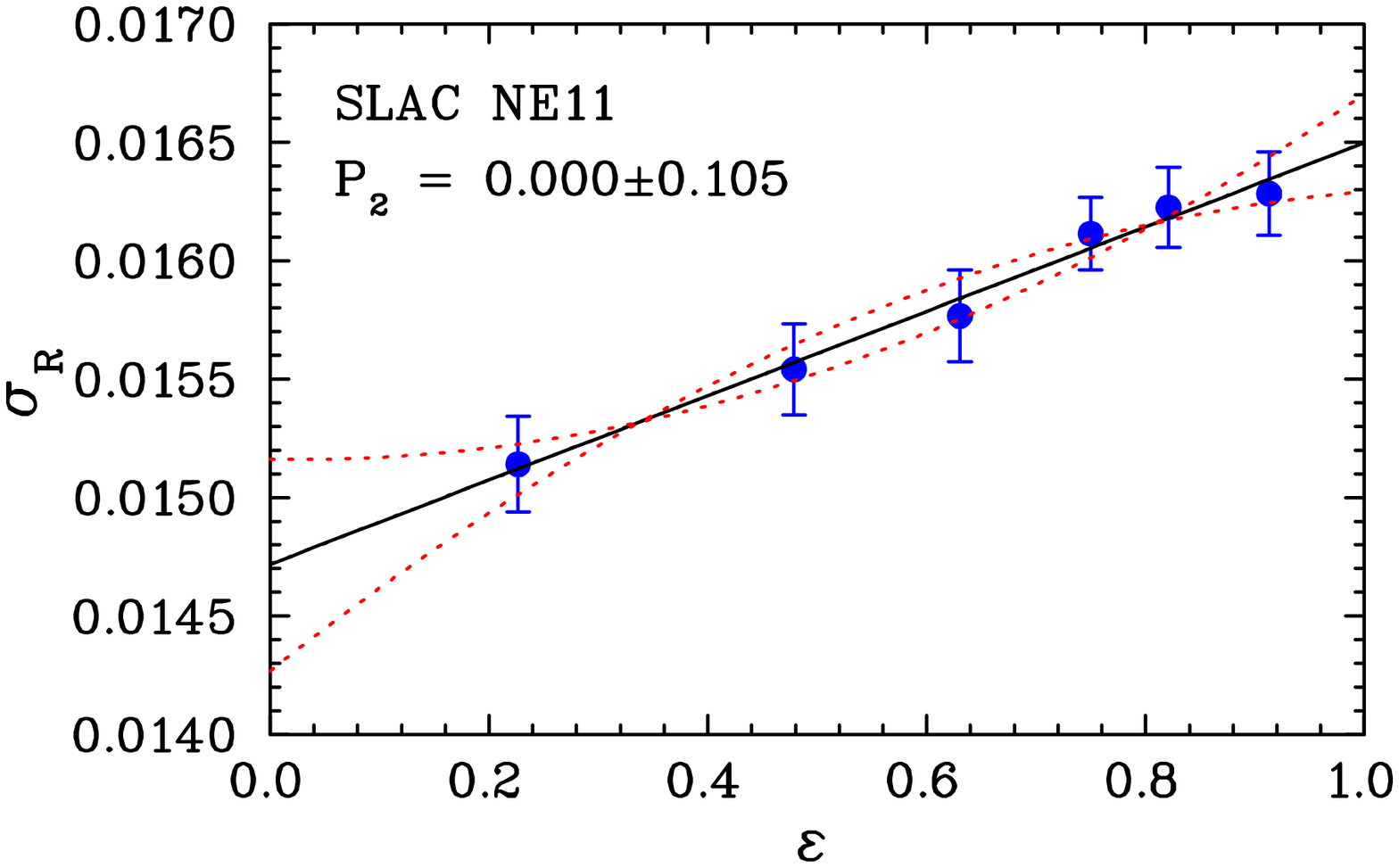}
\hspace*{0.4cm}\includegraphics[height=5cm,width=8.1cm]{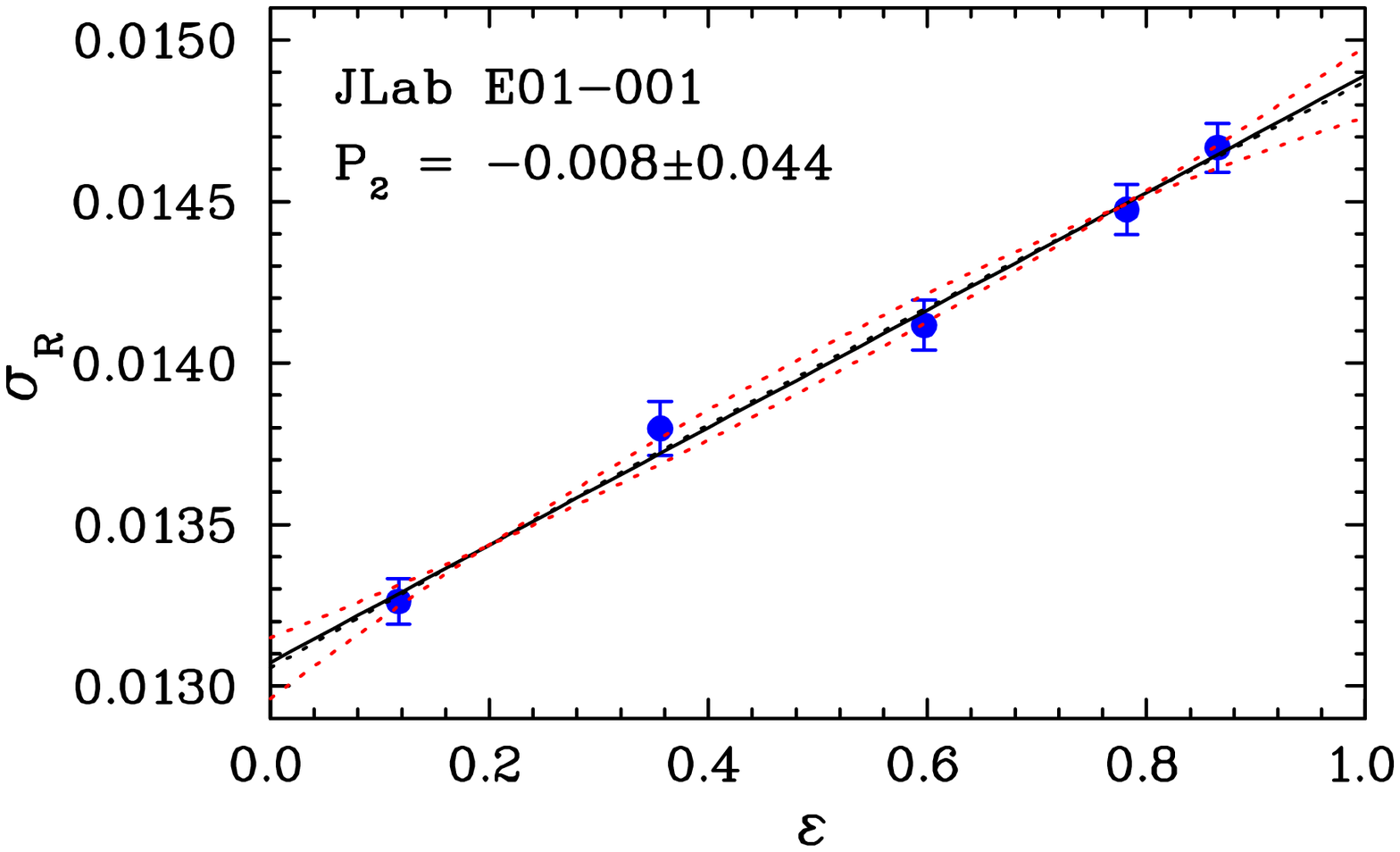}
\begin{minipage}{16.5cm}
\caption{The $\eps$ dependence of the reduced cross section from
	SLAC NE11~\cite{And94} (using only data from the 8~GeV
	spectrometer) {\bf (left)} and for Jefferson Lab experiment
	E01-001 \cite{Qat05} {\bf (right)}.  The solid line is a
	linear fit, while the dotted lines are quadratic fits with
	$P_2 = \pm 0.105$ for NE11, and $P_2 = -0.008 \pm 0.044$ for
	E01-001 ({\it i.e.} $1\sigma$ variations on the central value).}
\label{fig:ne11}
\end{minipage}
\end{center}
\end{figure}

\begin{figure}[thb]
\begin{center}
\includegraphics[width=7.8cm]{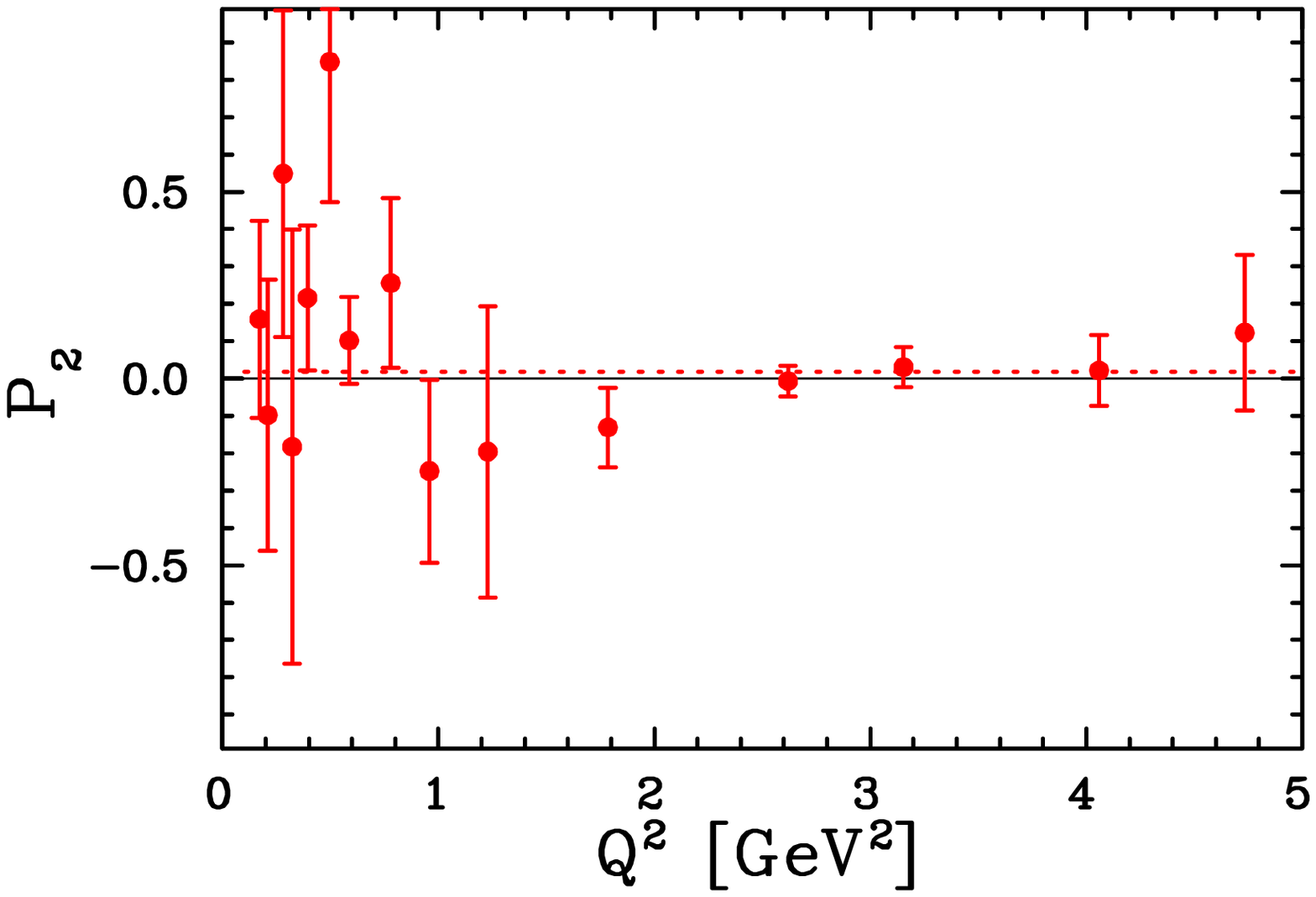}
\hspace*{0.4cm}
\includegraphics[width=8.2cm]{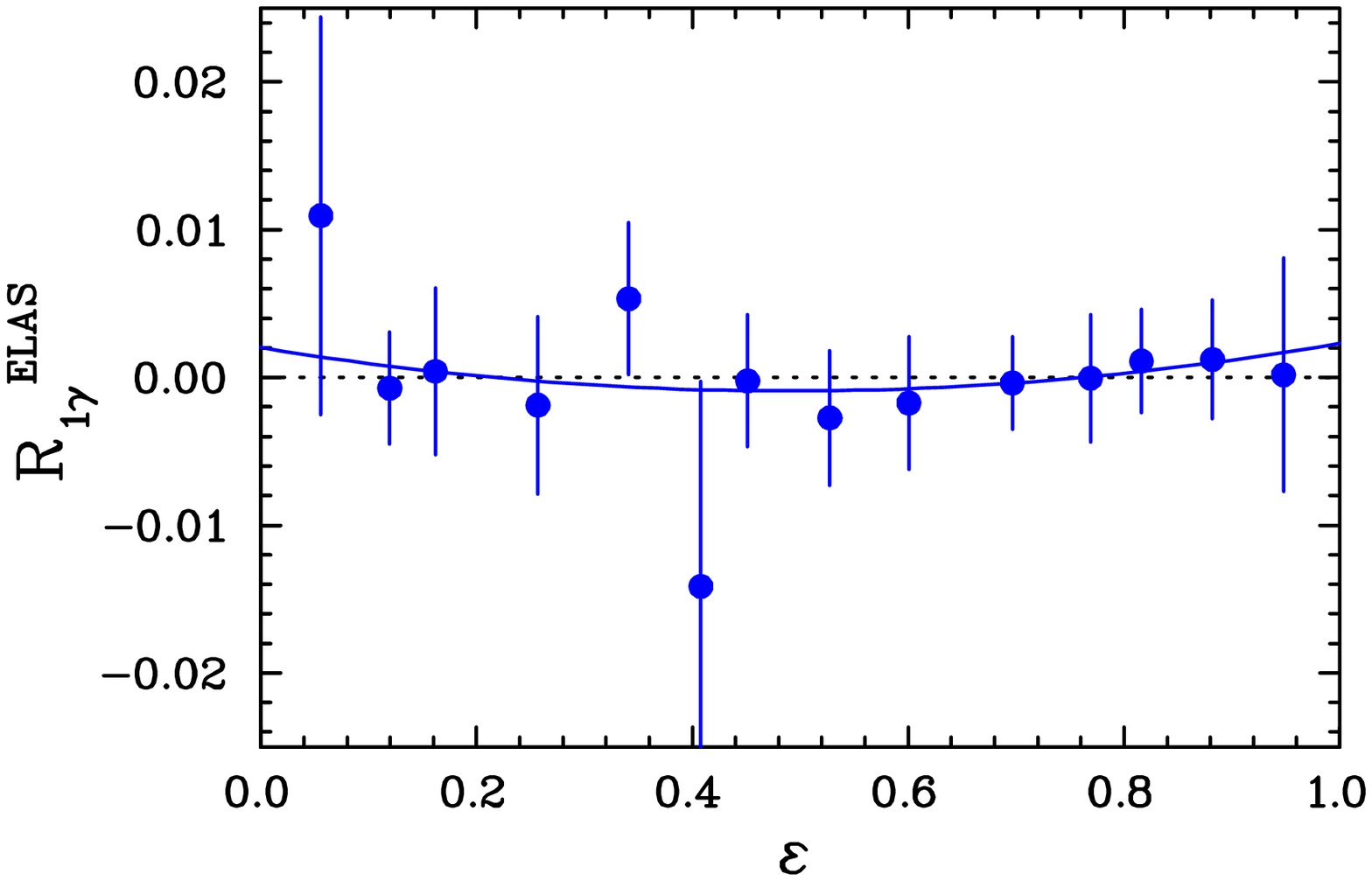}
\begin{minipage}{16.5cm}
\caption{
{\bf (Left)}
	Nonlinearity parameter $P_2$ (see Eq.~(\ref{fig:ne11})) from
	a global analysis~\cite{Tva06} of elastic cross section data. 
	The dotted line indicates the global average,
	$\langle P_2 \rangle = 0.019 \pm 0.027$.
{\bf (Right)}
	Weighted average of the deviation of the cross section
	measurements from linear fits,
	$R_{1\gamma} = (\sigma_{\rm data}-\sigma_{\rm fit})
		     / \sigma_{\rm fit}$.
	The curve is a quadratic fit to the residual deviation yielding
	a quadratic term of $0.9 \pm 2.0$\%.
	Figure adapted from Ref.~\cite{Tva06}.}
\label{fig:allp2}
\end{minipage}
\end{center}
\end{figure}

Figure~\ref{fig:ne11} shows the reduced cross section as a function
of $\eps$ for the best conventional Rosenbluth separation (SLAC NE11
\cite{And94}), and from the Super-Rosenbluth extraction in Jefferson
Lab experiment E01-001~\cite{Qat05}.  The data are consistent with
no nonlinear contributions.  A global analysis of nonlinearities,
including both elastic scattering, resonance region measurements,
and deep-inelastic scattering was performed in Ref.~\cite{Tva06}.
This analysis concluded that all of the measurements are consistent
with a linear $\eps$ dependence.  For the elastic measurements,
the limits on nonlinear contributions, shown in the left panel
of Fig.~\ref{fig:allp2}, are not very significant except for
$Q^2 = 2-4$~GeV$^2$, where there are recent high precision
measurements \cite{Qat05}.  The right panel shows the difference
between the cross section measurements and the linear fits, averaged
over all $Q^2$, which gives an indication of how well the possible
deviations from linearity are constrained as a function of $\eps$.
The improved measurements of Jefferson Lab experiment E05-017 should
allow for improved constraints over a range in $Q^2$.

Similar analyses were performed by other groups for $ep$ \cite{Tom05}
and $ed$ elastic scattering~\cite{Rek99}.  In Ref.~\cite{Tom05}, the
nonlinear contributions were parametrized using constraints from
C-parity and crossing symmetry.  The parametrization of the TPE
contributions there (relative to the Born cross section) was taken
to be proportional to $x=\sqrt{(1+\eps)/(1-\eps)}$, which diverges for
$\eps \to 1$.  Thus, any data with strong constraints at high $\eps$
yields a very tight constraint on the coefficient of the divergent TPE
contributions, but still yields large nonlinear contributions at higher
$\eps$ values.  While this parametrization is consistent with the
symmetries being considered by the authors, it is inconsistent with
TPE contributions, as discussed in Ref.~\cite{Che07}, and does not
provide meaningful constraints on nonlinear TPE effects.

Based on similar considerations, the analysis of TPE effects in $ed$
scattering by Rekalo~\etal~\cite{Rek99} expands the TPE contributions
as a Taylor series in $x$.  However, since $x$ is always larger than
unity, and in fact diverges as $\eps \to 1$ ($\theta \to 0$), the
expansion of the TPE contribution in powers of $x$ does not provide
a meaningful way to examine the data.

In Ref.~\cite{Abi07} estimates were provided for $P_2$ based on the
partonic TPE calculations~\cite{Che04, Afa05} (see Sec.~\ref{ssec:highQ}).
This work found that the calculated nonlinearities were of approximately
the same size as the limits set by the global analysis~\cite{Tva06},
but had a large dependence on the model used for the calculation.

\subsection{\it Conclusions from early experiments}

While examinations of previous data and the early measurements aimed at
understanding the form factor discrepancy did not provide clear answers,
they were consistent with the idea that larger than expected TPE
contributions could explain the observed effects.  The effects had to be
small at forward scattering angles, to be consistent with comparisons
of positron and electron scattering, and yield an angle dependence of
$\sim 5\%$ at high $Q^2$, approximately linear in $\eps$, to resolve the
discrepancy.  While these studies did not provide definitive evidence
that TPE were responsible, they still provide some of the most
significant constraints on certain aspects of the TPE contributions.

Because these corrections were larger than previous estimates of TPE,
significant effort was put into understanding them theoretically
and placing experimental constraints through new measurements.
These efforts were not just limited to examining TPE corrections
to the unpolarized cross sections; if TPE yields effects on the cross
section on the order of a few percent, it may also enter into other
observables at the same level and contribute to reactions beyond
elastic $ep$ scattering.  While experimental tests of TPE are for the
most part limited to elastic $ep$ scattering, measurements of several
observables can be performed.  With a range of measurements, one hopes
to constrain models well enough that they can provide reliable,
semi-quantitative estimates for TPE contributions in other reactions.
This will allow one to determine where residual TPE effects may have
a significant impact in other reactions.

As mentioned in Sec.~\ref{sec:exp_overview}, the initial experimental
investigations focused on observables directly related to the cross 
section measurements.  The initial work focused on existing data, while
new measurements were being performed to improve the precision and
kinematical coverage of Rosenbluth separations, polarization
measurements, and comparisons of positron and electron scattering.
Additional measurements were proposed to study the $\eps$ dependence
of the polarization observables~\cite{Mez11} (Sec.~\ref{ssec:impact_PT}),
which are $\eps$ independent in the Born approximation, as well as
observables which are forbidden in the Born approximation
(Sec.~\ref{ssec:normal}).  Before addressing these additional
measurements, in the following sections we discuss calculations of TPE
corrections within several theoretical frameworks (Sec.~\ref{sec:TPE}),
comparisons of these calculations to additional measurements
(Sec.~\ref{sec:impact}), and estimates of TPE contributions in
reactions beyond elastic $ep$ scattering (Sec.~\ref{sec:other}).

%% file: section4.tex
\section{Two-photon exchange}
\label{sec:TPE}

\begin{figure}[htb]
\begin{center}
\begin{minipage}{8cm}
\hspace*{-3.5cm}
\includegraphics[angle=270,scale=0.7,clip=true,bb=60 0 200 600]{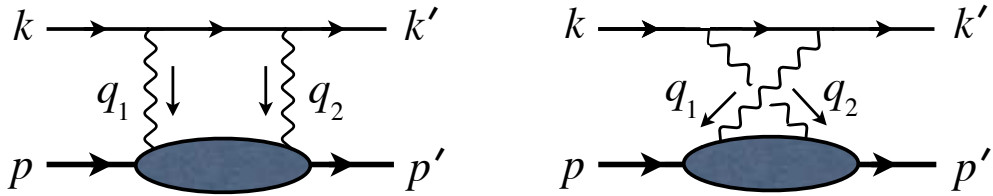}
\end{minipage}
\begin{minipage}{16.5cm}
\caption{Box and crossed-box two-photon exchange contributions to elastic
	electron--nucleon scattering.  The overall four-momentum transfer
	to the nucleon is $q = k - k' = q_1 + q_2$.}
\label{fig:TPE}
\end{minipage}
\end{center}
\end{figure}

In view of the failure to understand the discrepancy between the 
Rosenbluth and polarization transfer measurements of $\gegm$ in terms
of standard radiative corrections, the focus soon turned to revisiting  
the methodologies used in computing the box and crossed-box two-photon  
exchange corrections, illustrated in Fig.~\ref{fig:TPE}.
An experimental--theoretical working group was established at Jefferson 
Lab, with the goal of identifying possible directions for resolving the 
discrepancy.  The first quantitative calculation resulting from this 
renewed focus was made by Blunden \etal~\cite{Blu03}, who computed the
effect on $\gegm$ from TPE, incorporating explicitly the nucleon's 
substructure.  Thereafter followed a number of other studies, examining 
TPE in a variety of frameworks, and exploring reactions beyond elastic 
$ep$ scattering.
In a parallel effort, Guichon and Vanderhaeghen \cite{Gui03} provided a
generalized formalism for elastic scattering, allowing for possible TPE
contributions, and demonstrated that it was natural to have TPE 
contributions which could significantly change the LT extraction of 
$\ge$ with minimal impact on the PT measurements.

In this section we review these efforts, paying particular attention to 
the conventional hadronic-level calculations which are most applicable 
to data analysis at low to moderate $Q^2$ values.  Before turning to the
(model-dependent) TPE contributions, we first discuss some general 
properties of radiative corrections in order to set the stage for the 
recent improvements.

\subsection{\it General properties of radiative corrections}
\label{ssec:general}

At order $\alpha^2$, radiative corrections to the one-photon exchange 
cross section $\sigma_R^0$, Eq.~(\ref{eq:sigma0}), include processes 
arising from the exchange of a second virtual photon
($\delta_{\rm virt}$), and inelastic scattering processes involving
the emission of a real bremsstrahlung photon ($\delta_{\rm brem}$),
so that
\be
\sigma_R\
=\ \sigma_R^0 \left( 1 + \delta_{\rm virt} + \delta_{\rm brem} \right).
\label{eq:sigR}
\ee
In analyzing the virtual corrections for a $ep$ scattering, it is
convenient to separate terms into ``soft'' parts, which are independent
of hadronic structure, and ``hard'' parts, which are model dependent.
Soft here implies that any interaction of the second virtual photon
with the proton occurs with vanishingly small momentum transfer.
The soft parts are therefore the same for protons as they are for
scattering from pointlike particles ({\it e.g.} in $e^- \mu^+$
scattering).  All of the infrared (IR) divergences for the virtual
diagrams are contained in the soft parts, and cancel in the total
amplitude.

\begin{figure}[t]
\begin{center}
\includegraphics[height=6cm]{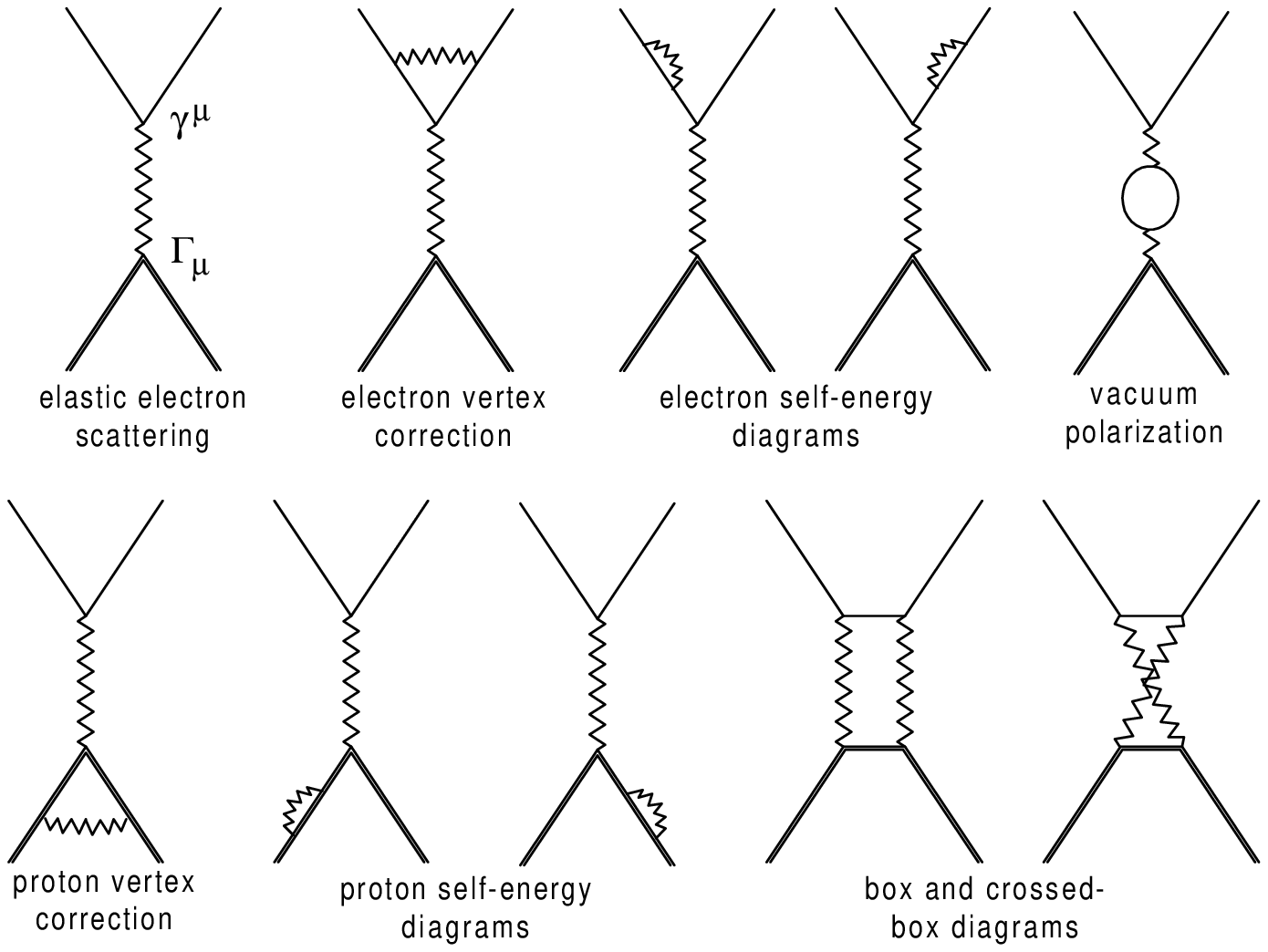}
\hspace*{1cm}\includegraphics[height=3cm]{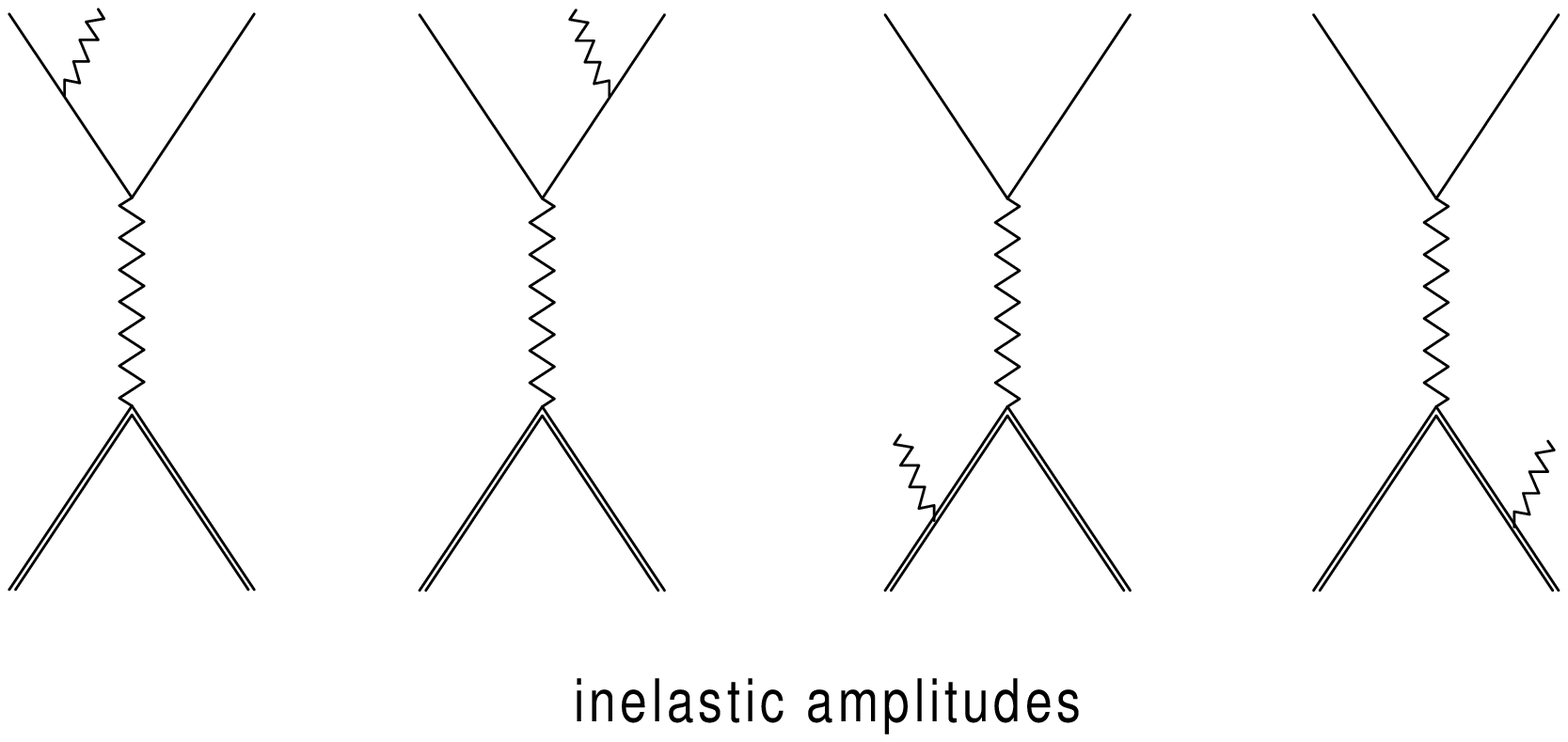}
\begin{minipage}{16.5cm}
\caption{The complete set of diagrams to order $\alpha^2$,
	including the virtual {\bf (left)} and bremsstrahlung
	(or inelastic) {\bf (right)} contributions.
	Figure taken from Ref.~\cite{Max00T}.}
\label{fig:RCall}
\end{minipage}
\end{center}
\end{figure}

If we denote the amplitude for all one-loop virtual corrections, 
illustrated in Fig.~\ref{fig:RCall}, by ${\cal M}_{\rm 1-loop}$, then 
${\cal M}_{\rm 1-loop}$ can be written as the sum of a ``factorizable'' 
soft term, proportional to the Born amplitude ${\cal M}_\gamma$, and a 
non-factorizable hard part ${\cal M}_{\rm hard}$,
\be
{\cal M}_{\rm 1-loop}\
=\ f(Q^2,\eps)\, {\cal M}_\gamma\ +\ {\cal M}_{\rm hard}\, ,
\label{eq:M1}
\ee
with $f(Q^2,\eps)$ a purely kinematic factor.
Therefore $\delta_{\rm virt}$ is given by
\be
\delta_{\rm virt}\
=\ 2 f(Q^2,\eps)\
+\  { 2 \Re\{ {\cal M}_\gamma^*\, {\cal M}_{\rm hard} \}
    \over |{\cal M}_\gamma|^2}\
\equiv\ \delta_{\rm soft}\ +\ \delta_{\rm hard}\ .
\label{eq:dvirt}
\ee
All of the virtual processes in Fig.~\ref{fig:RCall} contribute to the 
soft terms in $\delta_{\rm soft}$.  In practice these terms dominate, 
and are the ones accounted for in the standard radiative corrections 
of Mo and Tsai~\cite{Tsa61, Mo69}.  Furthermore, the functions 
$f(Q^2,\eps)$ for the vacuum polarization, self-energy, and vertex 
corrections are $\eps$-independent (although the vertex terms are 
IR-divergent), and therefore have no relevance for the LT separation 
aside from an overall normalization factor.  Hence, of the factorizable 
terms, only the TPE can contribute to the $\eps$-dependence of the
cross section (Fig.~\ref{fig:TPE}).  It should be noted that the
decomposition of the TPE effect into soft and hard parts is not
unique, as discussed below.

The terms which depend on hadronic structure are contained in
${\cal M}_{\rm hard}$, and arise from the proton vertex and TPE 
corrections.  For the proton, the hadronic vertex correction was 
analyzed by Maximon and Tjon~\cite{Max00T}, and found to be $< 0.5\%$ 
in magnitude for $Q^2 < 6$~GeV$^2$.  Since the proton vertex correction 
does not have a strong $\eps$-dependence, it will not affect the LT 
analysis, and can be safely neglected in examining the form factor
discrepancy.

For the inelastic bremsstrahlung contribution, illustrated in 
Fig.~\ref{fig:RCall}, the amplitude for real photon emission can also
be written in the form of Eq.~(\ref{eq:M1}). In the soft photon 
approximation (SPA) one again keeps only the factorizable terms that 
are independent of hadron structure, in which case the cross section
is a simple kinematic factor times the Born cross section, 
Eq.~(\ref{eq:sigma0}).  A significant $\eps$-dependence arises
from this contribution due to the frame dependence of the angular
distribution of the emitted photon.

Typically an infinitesimal photon mass $\lambda$ is introduced in the
photon propagator as a bookkeeping device to regulate the IR divergences.
For the box and crossed-box TPE diagrams in Fig.~\ref{fig:TPE}, this
$\lambda$ dependence is canceled by the bremsstrahlung interference
contribution with a soft-photon emitted from the electron and proton
(\ie~by cutting one of the (soft) photon propagators).  This produces
a correction to the cross section depending on the proton charge $Z$.
A similar cancellation occurs for the IR divergences in the electron 
vertex and proton vertex corrections, which depend on $Z^0$ and $Z^2$, 
respectively.  For positron-proton scattering one can take $Z=-1$,
so that only the TPE (and bremsstrahlung interference) terms are 
relevant in comparison with the electron-proton case.

Although the standard radiative corrections are model independent,
there have been significant improvements to the original work of Mo and
Tsai~\cite{Tsa61, Mo69}, removing many of the mathematical and other
approximations used in that work.  An important contribution was the 
calculation of Maximon and Tjon \cite{Max00T}, whose analysis differs 
from that of Refs.~\cite{Tsa61, Mo69} in two substantive aspects.
First, they evaluated the inelastic bremsstrahlung cross section
without any approximation, using integrals given in closed form by 
't~Hooft and Veltman~\cite{Hoo79}.  The exact expressions are simpler
in form than the approximate ones given in Ref.~\cite{Mo69}.
In the limit $M \to \infty$, corresponding to a static Coulomb potential, 
they reproduce exactly the result first given by Schwinger \cite{Sch49}.  
Second, in the evaluation of the contribution of the TPE diagrams, they 
make a less drastic approximation than that made in \cite{Tsa61}. 
Specifically, in the integrands corresponding to the relevant 
IR-divergent amplitudes, they make a soft-photon approximation for the 
matrix elements of the current appearing in the numerator (as in 
\cite{Tsa61}), but not for the propagators, which appear in the
denominator.  The required integrals (scalar four-point functions) have 
also been given in \cite{Hoo79, Pas79}; the resulting expressions are 
again considerably simpler than those obtained in \cite{Tsa61}, where 
the soft-photon approximation is also made in the denominators.

Other improvements have also been made to the work of Mo and Tsai 
\cite{Mo69}.  These include higher mass vacuum polarizations terms
(not just $e^+e^-$) from Ref.~\cite{Wal94}, improved implementation
of multi-photon exchange and angle-dependent bremsstrahlung for
coincidence reactions~\cite{Ent01, Wei09}, and improved treatment of
multi-photon exchange in the structure function method~\cite{Bys07}.
The first three were simply improvements to the Mo-Tsai framework,
made independently of the form factor discrepancy.  The last \cite{Bys07} 
was an attempt to explain the discrepancy.  While they found differences
between their results and Mo-Tsai based radiative corrections, the
comparison was not made for identical conditions, and more work is
required to determine whether there is really any substantial difference.


\subsection{\it Soft two-photon exchange effects}
\label{ssec:softTPE}

The improvements in the computation of the radiative corrections made by 
Maximon and Tjon \cite{Max00T} have a number of important effects on the 
corrections of order $Z$. The form of the bremsstrahlung correction 
depends on whether the scattered electron or the recoil proton is 
detected. For the detection of the scattered electron, the bremsstrahlung 
correction in the SPA is found to be
\be
\delta_{\rm brem}({\rm MTj})\
=\ {2\alpha Z\over \pi}
   \left[
     \ln\eta\, \ln\left( \frac{(2\eta \Delta E)^2}{y \lambda^2} \right)
   + {\rm Li}_2 \left( 1-\frac{\eta}{y} \right)
   - {\rm Li}_2 \left( 1-\frac{1}{\eta y} \right)
   \right],
\label{eq:dbremMTj}
\ee
with $\eta=E/E'$ the ratio of incident to final electron energies,
${\rm Li}_2$ is the dilogarithm function,
and $y=1+2 \tau+2 \sqrt{\tau (1+\tau)}$.
Here $\Delta E$ is the maximum allowable energy loss in the lab frame 
due to detector acceptance or experimental cuts, below which one cannot  
determine that a soft photon has been emitted.  This exact expression is 
simpler in form than the approximate one of Mo and Tsai \cite{Tsa61,Mo69}.  
Their expression is too long to reproduce here, but it takes the form
\be
\delta_{\rm brem}({\rm MoT})\
=\ {2\alpha Z \over \pi}
   \ln\eta\,
   \ln\left(\Delta E^2 \over \lambda^2\right)\
+\ \text{finite terms}.
\label{eq:dbremMoT}
\ee
From this it is clear that the logarithmic dependence on both $\Delta E$ 
and $\lambda$ in Eqs.~(\ref{eq:dbremMTj}) and (\ref{eq:dbremMoT}) is the 
same, and so the difference between the two treatments is a finite 
kinematic factor.

The IR-divergent part of the box amplitude can be separated from the 
IR-finite part by analyzing the structure of the photon propagators in  
the integrand.  The two poles, where the photons are soft, occur at  
$q_1=0$ ($q_2=q$) and at $q_1=q$ ($q_2=0$).
Evaluating the {\em numerator} of the integral at either value gives a 
contribution to the TPE amplitude for the box diagram as
\bea
{\cal M}_{\gamma\gamma}^{\rm box}
&=& -{Z\alpha \over 2\pi} (s-M^2)\, q^2\,
    {\cal M}_\gamma\ {1 \over i\pi^2}\int\,d^4q_1\,
    {1 \over \left[q_1^2-\lambda^2\right]
	     \left[(q-q_1)^2-\lambda^2\right]
	     \left[(k-q_1)^2-m_e^2\right]
	     \left[(p+q_1)^2-M^2\right]}	\nn\\
&=& -{Z\alpha \over 2\pi} (s-M^2)\, q^2\,
    {\cal M}_\gamma\ D_0(s;m_e,\lambda,M,\lambda),
\label{eq:MIRMTj}
\eea
where $s = M^2 + 2 M E$ is the Mandelstam variable in the laboratory
frame.  The integral over the product of four propagators is expressed
in terms of the four-point Passarino-Veltman function
$D_0(s)$~\cite{Pas79}, which can be evaluated numerically using
the program LoopTools~\cite{Hah99}.
General expressions for the asymptotic expansion of all four-point 
IR-divergent integrals have been given by Beenakker and Denner 
\cite{Bee90}, who find
\be
D_0(s;m_e,\lambda,M,\lambda)\
=\ {2 \over (s-M^2) q^2}
   \ln\left(M^2-s\over m_e M\right)
   \ln\left(-q^2\over \lambda^2\right),
\ee
in the limit $(s-M^2) \gg m_e^2, m_e M$. Here one is only interested in 
the real part of this expression.  The convention for the logarithm is 
$\ln(-z)=\ln(z) - i\pi$ when $z<0$, which is achieved by setting
$z \to z+i 0^+$.  For the crossed-box one can make use of crossing 
symmetry.  This requires that the crossed-box amplitude
${\cal M}_{\gamma\gamma}^{\rm xbox}$ obey the relation
\be
{\cal M}_{\gamma\gamma}^{\rm xbox}(u,t)\
=\ +\left. {\cal M}_{\gamma\gamma}^{\rm box}(s,t) \right|_{s\to u}.
\label{eq:cross}
\ee
where $t=q^2$.  Thus the total box plus crossed-box amplitude must
be even under the interchange $s \leftrightarrow u$, or equivalently
$E \leftrightarrow -E'$.  Note that since $M^2-u>0$, the crossed-box 
amplitude has no imaginary part.  Combining these expressions, and 
taking  the real part only, the final result for the soft virtual 
correction is \cite{Max00T}
\be
\delta_{\rm IR}({\rm MTj})
= -{2\alpha Z\over \pi} \ln{\eta}\, \ln{Q^2\over\lambda^2}.
\label{eq:deltaIRMTj}
\ee
Note that the dependence on the electron mass $m_e$ has dropped out in
the final expression, and the logarithmic IR singularity in $\lambda$ is 
exactly canceled when added to Eq.~(\ref{eq:dbremMTj}).

By contrast, in earlier treatments \cite{Tsa61, Cam69a, Mei63} the 
SPA is also applied to one of the propagators, for example replacing 
$1/(q-q_1)^2$ by $1/q^2$ when $q_1\to 0$, and $1/q_1^2$ by $1/q^2$ 
when $q_1\to q$.  Hence the IR-divergent contribution is
\bea
{\cal M}_{\gamma\gamma}^{\rm box}
&=& -2 {Z\alpha \over 2\pi} (s-M^2)\,
    {\cal M}_\gamma\ {1 \over i\pi^2}\int\,d^4q_1\,
    {1 \over \left[q_1^2-\lambda^2\right]
	     \left[(k-q_1)^2-m_e^2\right]
	     \left[(p+q_1)^2-M^2\right]}	\nn\\
&=& -2 {Z\alpha \over 2\pi} (s-M^2)\,
    {\cal M}_\gamma\ C_0(s;m_e,\lambda,M),
\label{eq:MIRMoT}
\eea
where $C_0(s)$ is the three-point Passarino-Veltman function
\cite{Pas79}, and a factor of 2 accounts for the contribution from
both poles.  To facilitate comparison, note that Tsai \cite{Tsa61} 
introduces the function $K(-k,p)$ in evaluating this integral,
which is equal to $(s-M^2) C_0(s)$, with $s=(k+p)^2$.  In this 
approximation one therefore expects
\bea
\delta_{\rm IR}
&=& -{2\alpha Z\over \pi} \left[ (s-M^2) C_0(s) - (u-M^2) C_0(u) \right],
							\nn\\
&=& {2\alpha Z \over \pi}
   \left[ {\rm Li}_2 \left( 1 + {M \over 2 E} \right)
        - {\rm Li}_2 \left( 1 - {M \over 2 E'} \right)
        - \frac{1}{2} \ln(-\eta)\,
		      \ln\left( -4 M^2 E E' \over \lambda^4 \right)
   \right],
   \label{eq:deltaIRMY}
\eea
for which the crossing symmetry property is manifest.  In writing the
last line we have again used the asymptotic expansion of three-point 
IR-divergent functions given in Beenakker and Denner \cite{Bee90}.
Equation~(\ref{eq:deltaIRMY}) is equivalent to the form given by Meister 
and Yennie~\cite{Mei63}, which was used in Refs.~\cite{Che04, Afa05}.
Equation~(\ref{eq:deltaIRMY}) is larger than Eq.~(\ref{eq:deltaIRMTj}) 
by a finite amount of approximately $\alpha \pi$ -- a point that was 
also noted in Refs.~\cite{Che04, Afa05}.

Tsai \cite{Tsa61} makes a further mathematical approximation, to
replace $p \to -p$ in the {\em box contribution only}, or equivalently
$s \to M^2-2 M E$.  Although this additional approximation spoils the 
crossing symmetry, it eliminates the term $(i\pi)^2$ that arises in 
Eq.~(\ref{eq:deltaIRMY}) from the product of logarithms with negative 
argument, and is therefore closer to the result of Maximon and Tjon, 
Eq.~(\ref{eq:deltaIRMTj}).  At the time, Mo and Tsai argued that the 
resulting expression is closer to the exact calculations of $ee$
scattering \cite{Mo69}.  The final result for the Mo-Tsai soft virtual 
correction can then be expressed as
\be
\delta_{\rm IR}({\rm MoT})\
=\ {2\alpha Z \over \pi}
   \left[ {\rm Li}_2 \left( 1 - {M \over 2 E} \right)
        - {\rm Li}_2 \left( 1 - {M \over 2 E'} \right)
        - \frac{1}{2} \ln\eta\,
		      \ln\left( 4 M^2 E E' \over \lambda^4 \right)
   \right],
\label{eq:deltaIRMoT}
\ee
which is clearly no longer asymmetric under $E \leftrightarrow -E'$.

\begin{figure}[t]
\vspace*{1cm}
\begin{center}
\begin{minipage}{8cm}
\hspace*{-1cm}\includegraphics[height=6cm]{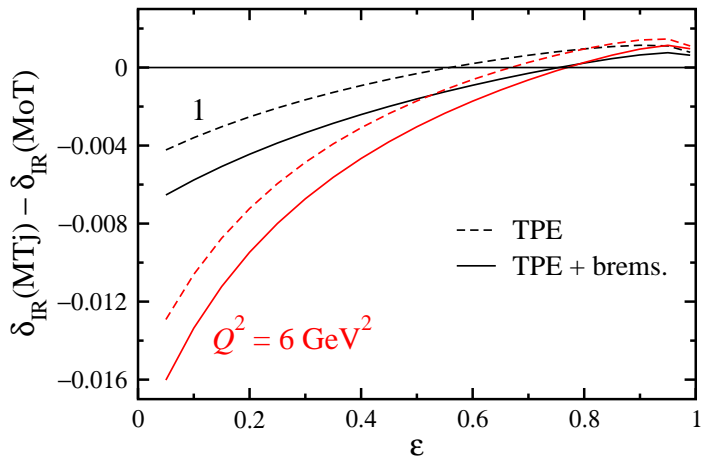}%
\end{minipage}
\begin{minipage}{16.5cm}
\caption{Difference between the model-independent corrections of
	Maximon-Tjon \cite{Max00T} and Mo-Tsai \cite{Tsa61, Mo69}
	for virtual TPE (dashed curves) and virtual+real (solid).
	The upper (lower) dashed and solid curves correspond to
	$Q^2=1\, (6)$~GeV$^2$.}
\label{fig:dMTjMoT}
\end{minipage}
\end{center}
\end{figure}

Because the Mo-Tsai result is the one generally used in existing 
experimental analyses calculations of the full TPE contribution
presented here will be taken with respect to 
$\delta_{\rm IR}({\rm MoT})$ rather than $\delta_{\rm IR}({\rm MTj})$
(except where indicated).
It is useful to compare the $\eps$-dependence of these two treatments
of the soft virtual corrections.  The difference
$\delta_{\rm IR}({\rm MTj}) - \delta_{\rm IR}({\rm MoT})$ is independent
of $\lambda$, and is shown in Fig.~\ref{fig:dMTjMoT} (dashed curves)
as a function of $\eps$ for $Q^2=1$ and 6~GeV$^2$.
Note that the difference vanishes as $Q^2 \to 0$.
The different treatments of the IR-divergent terms already have
significance for the LT separation, resulting in roughly a 1\%
change in the cross section over the range of $\eps$.
Also shown in Fig.~\ref{fig:dMTjMoT} is a comparison for the total
correction linear in $Z$ (solid curves), which includes the improvements
to the bremsstrahlung correction, {\em viz.}
$\left[\delta_{\rm IR}({\rm MTj}) + \delta_{\rm brem}({\rm MTj})\right]
- \left[\delta_{\rm IR}({\rm MoT}) + \delta_{\rm brem}({\rm MoT})\right]$.
This is appropriate when the scattered electron is detected, and is the
procedure adopted in the generalized parton distribution calculations of
Refs.~\cite{Che04, Afa05}.  From these curves it is clear that most of
the difference arises from the treatment of the TPE diagrams rather than
the bremsstrahlung correction.

\subsection{\it Hadron structure effects: elastic intermediate states}
\label{ssec:nucleon}

Going beyond the Mo-Tsai \cite{Tsa61, Mo69} and Maximon-Tjon
\cite{Max00T} approximations, Blunden {\it et al.} \cite{Blu03}
proceeded to evaluate the box and crossed-box TPE diagrams without
resorting to any of the soft-photon approximations discussed above.
In particular, they considered explicitly the effects of
incorporating the hadronic structure of the nucleon, parametrized
through hadronic electromagnetic form factors.
%
%
%
Specifically, Blunden {\it et al.} computed~\cite{Blu03, Blu05, Tjo09}
the total TPE amplitude
\be
{\cal M}_{\gamma\gamma}\
=\ {\cal M}_{\gamma\gamma}^{\rm box}\
+\ {\cal M}_{\gamma\gamma}^{\rm xbox},
\label{eq:Mgg}
\ee
where
\bea
{\cal M}_{\gamma\gamma}^{\rm box\, (xbox)}
&=& -ie^4 \int {d^4 q_1 \over (2\pi)^4}\
    L_{\mu\nu}^{\rm box\, (xbox)} H^{\mu\nu}_N\, 
    \Delta_F(q_1,\lambda)\ \Delta_F(q_2,\lambda),
\label{eq:Mggbox}
\eea
with the box and crossed-box leptonic tensors given by
\be
L_{\mu\nu}^{\rm box}\
=\ \ubar_e(k')\, \gamma_\mu\, S_F(k-q_1,m_e)\, \gamma_\nu\, u_e(k),
\ \ \ \ \ \
L_{\mu\nu}^{\rm xbox}\
=\ \ubar_e(k')\, \gamma_\nu\, S_F(k-q_2,m_e)\, \gamma_\mu\, u_e(k).
\label{eq:Lmunu_bc}
\ee
Alternatively, one can obtain the crossed-box term directly from the box
term by applying the crossing symmetry relation Eq.~(\ref{eq:cross}).
The electromagnetic nucleon elastic hadronic tensor
$H^{\mu\nu}_N$ is given by
\be
H^{\mu\nu}_N\
=\ \ubar_N(p')\, \Gamma_\gamma^\mu(q_2)\, S_F(p+q_1,M)\,
		 \Gamma_\gamma^\nu(q_1)\, u_N(p).
\label{eq:Hmunu}
\ee
Here the electromagnetic current operator $\Gamma_\gamma^\mu$ is
given in Eq.~(\ref{eq:Jg}), and the fermion (electron and nucleon)
and gauge boson (photon) propagators are given by
\be
S_F(k,m)
= { (\fslash{k} + m) \over k^2 - m^2 + i\epsilon }\, ,\ \ \ \ \ \
\Delta_F(k,\lambda)
= { 1 \over k^2 - \lambda^2 + i\epsilon }\, ,
\ee
respectively.  The infinitesimal photon mass $\lambda$ is introduced
to regulate the infrared divergences.
One can verify explicitly that the integrals in Eq.~(\ref{eq:Mggbox})
satisfy the crossing symmetry constraint in Eq.~(\ref{eq:cross}).

The relative correction to the elastic Born cross section,
Eq.~(\ref{eq:sigma0}), due to the interference of the one- and two-photon
exchange amplitudes (Figs.~\ref{fig:OPE} and \ref{fig:TPE}) is 
given by
\be
\delta_{\gamma\gamma}\  
=\ { 2 \Re \left( {\cal M}_\gamma^* {\cal M}_{\gamma\gamma} \right)
     \over  \left| {\cal M}_\gamma \right|^2 }.
\label{eq:delta_gg}
\ee
Typically, experimental analyses of form factor data apply radiative
corrections based on the Mo-Tsai prescription~\cite{Tsa61, Mo69} or
modified approaches based on the same general framework 
\cite{Wal94, Ent01}, which include approximating the TPE
contribution by the IR prescription $\delta_{\rm IR}{\rm (MoT)}$
in Eq.~(\ref{eq:deltaIRMoT}).  To determine the effect of the full, 
hadron-structure dependent correction $\delta_{\gamma\gamma}$ on the 
data, one must therefore compare the $\eps$-dependence of the full 
calculation with that of $\delta_{\rm IR}({\rm MoT})$. 
A meaningful comparison can be made by considering the difference
\be
\overline\delta\
\equiv\ \delta_{\gamma\gamma} - \delta_{\rm IR}({\rm MoT}),
\label{eq:Delta_dif}
\ee
in which the IR divergences cancel, and which is independent of   
$\lambda$.

\begin{figure}[tb]
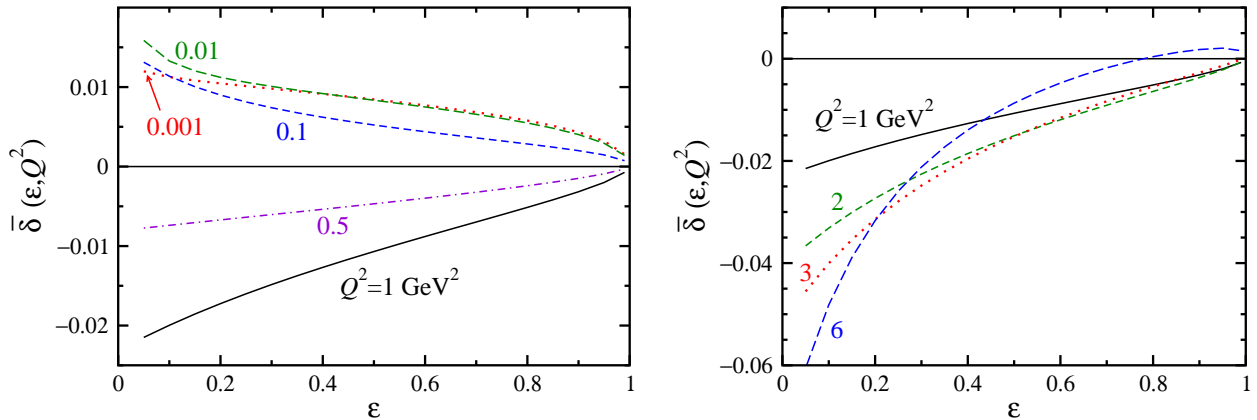

\begin{center}
\begin{minipage}{8cm}
\hspace*{-4.2cm}\includegraphics[height=5.5cm,angle=0]{Figs/fig13a.eps}%
\hspace*{0.5cm}\includegraphics[height=5.5cm,angle=0]{Figs/fig13b.eps}
\end{minipage}
\begin{minipage}{16.5cm}
\caption{Difference $\overline\delta$ between the full TPE correction
	to the elastic cross section \cite{Blu03} and the commonly used 
	Mo-Tsai approximation (\ref{eq:deltaIRMoT}) \cite{Tsa61, Mo69},
	for $Q^2=0.001$--1~GeV$^2$ {\bf (left)}, and
	for $Q^2=1$--6~GeV$^2$ {\bf (right)}.}
\label{fig:del_gg}
\end{minipage}
\end{center}
\end{figure}

The results for the difference $\overline\delta$ between the full
calculation and the MT approximation are shown in Fig.~\ref{fig:del_gg} 
as a function of $\eps$ for several values of $Q^2$ from
0.001 to 1~GeV$^2$ (left panel) and 1 to 6~GeV$^2$ (right panel).
The hadron structure dependent corrections are most significant at
low $\eps$, where they range from 2\% to $\sim 6\%$ over this $Q^2$ 
range.  At the lower $Q^2$ values, $\overline\delta$ is approximately
linear in $\eps$, but significant deviations from linearity are
observed with increasing $Q^2$, especially at small $\eps$.

Note that assuming that the current operator $\Gamma_\gamma^\mu$
in ${\cal M}_{\gamma\gamma}$ has a similar structure off-shell
as on-shell, with phenomenological form factors at the $\gamma NN$
vertices, introduces model dependence into the calculation, as the
radiative corrections are used to determine the experimental form
factors in the first place.  In principle this dependence could be
removed by iteratively extracting the form factors with the inclusion
of TPE corrections and feeding those form factors into subsequent
TPE calculations.
Even without iterating, however, because $\delta_{\gamma\gamma}$
is a ratio, the model dependence mostly cancels, provided the same 
phenomenological form factors are used for both ${\cal M}_\gamma$ and 
${\cal M}_{\gamma\gamma}$.  This was demonstrated in Ref.~\cite{Blu05} 
by comparing the results with those obtained using a dipole form
\be
G_D(Q^2)\ =\ \left( 1 + {Q^2 \over \Lambda_D^2} \right)^{-2},
\label{eq:GD}
\ee
with mass $\Lambda_D = 0.84$~GeV.
The model dependence is very weak at the lower $Q^2$ value, with
virtually no effect on the $\eps$ slope.  At larger $Q^2$ values
the differences increase, but the general trend of the correction
remains unchanged, so that one can conclude that the model dependence
of the calculation is quite modest.  At $Q^2 \gtorder 6$~GeV$^2$, which
will be accessible in future experiments at Jefferson Lab, significant 
deviations from linearity are expected over the entire $\eps$ range.

The TPE correction $\overline\delta$ is also found to be insensitive
to the high-$Q^2$ behavior of the $\gegm$ ratio.  Using form factor
inputs from parametrizations obtained by fitting only LT-separated
data \cite{Arr04a, Mer96} and those in which $\ge$ is constrained by
the polarization transfer data \cite{Arr04a, Bra02}, the differences
are almost indistinguishable up to $Q^2 = 6$~GeV$^2$ \cite{Blu05}.
(Note that the $\gm$ form factor itself also differs by a few
percent between the various parametrizations.)

\subsection{\it Inelastic contributions}
\label{ssec:inel}

While the TPE diagrams in Fig.~\ref{fig:TPE} contain contributions
from intermediate states containing all possible excitations of the
nucleon, the above discussion has thus far been restricted to the
nucleon elastic component.
In view of the prominent role of the $\Delta(1232)$ resonance, for
example, in many hadronic reactions, it is important to evaluate its
contribution to the TPE amplitude, as well as the possible role of
other inelastic intermediate states.

\subsubsection{$\Delta$ intermediate states}
\label{sssec:Delta}

The first estimate of the possible role played by excitations of the
nucleon in elastic $ep$ scattering was by Drell and Fubini \cite{Dre59}.
Using dispersion theory methods within a nonrelativistic approximation,
they found that for energies up to $E \sim 1$~GeV, the $\Delta(1232)$
resonance contribution increased the Born cross section by $\sim 1\%$.
Later, Campbell \cite{Cam67, Cam69a} used relativistic Rarita-Schwinger
spinors for the $\Delta$, but retained the soft-photon approximation
\cite{Mo69}, together with the assumption of $M1$ transition dominance.
For large scattering angles, the correction from the $\Delta$ was found
to be non-negligible for $Q^2 \sim 1-5$~GeV$^2$, but the estimate was
limited by the uncertainty in the $\gamma N \Delta$ form factor.
Partial cancellation between the nucleon and $\Delta$ contributions
for center of mass energies above 500~MeV was observed by Greenhut
\cite{Gre69}, who used the second Born approximation, neglecting recoil,
and assuming magnetic only coupling.  The conclusion of this work was
that the combined $N$ and $\Delta$ channels were not expected to exceed
$\sim 1\%$ of the cross section, although the approximations inherent
in the calculation rendered it less reliable for energies above 1~GeV
\cite{Gre69}.

More recently, Kondratyuk {\it et al.} \cite{Kon05} computed the
$\Delta(1232)$ resonance contribution within the same relativistic
framework that was used to compute the nucleon intermediate states
in Sec.~\ref{ssec:nucleon}, without invoking any of the approximations
of the earlier studies \cite{Dre59, Cam67, Cam69a, Gre69}.
Following Refs.~\cite{Jon73, Kon01}, the $\gamma N \Delta$ vertex is
defined as
\bea
\Gamma_{\gamma N \to \Delta}^{\alpha\mu}(p_\Delta,q)
&=& {1 \over 2 M_\Delta^2}\sqrt{2\over 3}
\Big\{
  g_1(Q^2)
  \left[ g^{\alpha \mu} \fslash{q}\fslash{p}_\Delta
    - \fslash{q} \gamma^\alpha p_\Delta^\mu 
    - \gamma^\alpha \gamma^\mu q\cdot p_\Delta
    + \fslash{p}_\Delta\,\gamma^\mu q^\alpha 
  \right]				\nn\\
& & \hspace*{-2.5cm}
+\ g_2(Q^2)
  \left[ q^\alpha p_\Delta^\mu - g^{\alpha\mu} q\cdot p_\Delta
  \right]\
+\ {g_3(Q^2) \over M_\Delta}
  \left[ q^2 \left( \gamma^\alpha p_\Delta^\mu 
		  - g^{\alpha\mu} \fslash{p}_\Delta
	     \right)
       + q^\mu \left( q^\alpha \fslash{p}_\Delta
		    - \gamma^\alpha q\cdot p_\Delta
	       \right)
  \right]
\Big\} \gamma_5,
\label{eq:gDN}
\eea
where $p_\Delta$ and $q$ are the momenta of the {\em outgoing} $\Delta$
and {\em incoming} photon, with corresponding Lorentz indices $\alpha$
and $\mu$, respectively.
The factor $\sqrt{2/3}$ is the $N\to\Delta$ isospin transition factor.
Electromagnetic gauge invariance requires that
$q_\mu \Gamma_{\gamma N \to \Delta}^{\alpha\mu}(p_\Delta,q) = 0$.
The coupling constants $g_i \equiv g_i(Q^2=0)$ for $i=1,2,3$ can
be related to the magnetic, electric and Coulomb components of the
$\gamma N\Delta$ vertex by $g_1 = g_M$, $g_2 = g_M+g_E$, $g_3 = g_C$
\cite{Kon05, Tjo09}.
The corresponding conjugate vertex can be obtained from the relation
\cite{Kon05, Zho10}
\be
\Gamma_{\Delta \to \gamma N}^{\mu\alpha}(p_\Delta,q)
= \gamma_0 \left[ \Gamma_{\gamma N \to \Delta}^{\alpha\mu}(p_\Delta,q)
           \right]^\dagger \gamma_0,
\label{eq:Gconj}
\ee
where $p_\Delta$ and $q$ are the momenta of the {\em incoming} $\Delta$
and {\em outgoing} photon.
%

The amplitude of the box diagram with a $\Delta$ intermediate state
can then be written as \cite{Kon05, Tjo09}
\bea
{\cal M}_{\gamma\gamma}^{(\Delta)}  
&=& - ie^4 \int {d^4 q_1 \over (2\pi)^4}\
    L_{\mu\nu} H^{\mu\nu}_\Delta\,
    \Delta_F(q_1,0)\, \Delta_F(q_2,0),
\eea
where $L_{\mu\nu}$ is the leptonic box tensor of Eq.~(\ref{eq:Lmunu_bc}),
and the crossed-box is obtained by crossing symmetry,
Eq.~(\ref{eq:cross}).  The $\Delta$ hadronic tensor is given by
\bea
H^{\mu\nu}_\Delta
&=& \ubar_N(p')\,
    \Gamma_{\Delta\to \gamma N}^{\mu\alpha}(p+q_1,-q_2)\,
     S_{\alpha\beta}(p+q_1,M_\Delta)\,
    \Gamma_{\gamma N\to\Delta}^{\beta\nu}(p+q_1,q_1)\,
    u_N(p),
\label{eq:MgDg}
\eea
where the $\Delta$ propagator is given by
\be
S_{\alpha\beta}(p_\Delta,M_\Delta)\
=\ -S_F(p_\Delta,M_\Delta)\, {\cal P}^{3/2}_{\alpha\beta}(p_\Delta),\quad
\ee
where the projection operator
\be
{\cal P}^{3/2}_{\alpha\beta}(p_\Delta)\
=\ g_{\alpha\beta}
-  {1 \over 3} \gamma_\alpha \gamma_\beta
-  {1 \over 3 p_\Delta^2}
   \left( \fslash{p}_\Delta \gamma_\alpha p_{\Delta\beta}
	+ \gamma_\beta \fslash{p}_\Delta p_{\Delta\alpha}
   \right)
\ee
ensures that only spin-3/2 components are present.  Suppression of the
unphysical spin-1/2 contributions also leads to the condition on the
vertex
$p_{\Delta\alpha} \Gamma_{\gamma N \to \Delta}^{\alpha\mu}(p_\Delta,q)
= 0$.
Note that the mass difference $M_\Delta - M$ between the initial
and intermediate hadronic states renders the integral (\ref{eq:MgDg})
IR-finite, so that in contrast to Eq.~(\ref{eq:Mgg}), a photon
mass parameter $\lambda$ is not needed to regulate divergences.

\begin{figure}[htb]
\begin{center}
\begin{minipage}{10cm}
\includegraphics[scale=0.5]{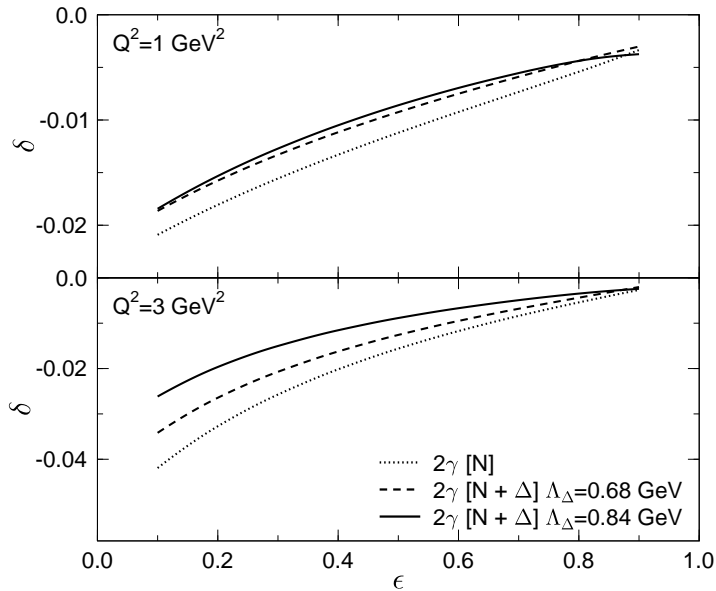}%
\end{minipage}
\begin{minipage}{16.5cm}
\vspace*{-0.5cm}
\caption{Sum of the nucleon and $\Delta(1232)$ contributions
	to the TPE correction to the $ep$ elastic cross section
	for two values of the cut-off mass $\Lambda_\Delta$.
	Figure taken from Ref.~\cite{Kon05}.}
\label{fig:del_ggD}
\end{minipage}
\end{center}
\end{figure}

In Refs.~\cite{Kon05, Tjo09} the three $\gamma N\Delta$ transition
form factors $g_i(Q^2) \equiv g_i\, F_V^\Delta(Q^2)$ ($i=1,2,3$)
were assumed to have dipole shapes,
   $F_V^\Delta(Q^2) = (1+Q^2/\Lambda_\Delta^2)^{-2}$,
with a dipole mass $\Lambda_\Delta = 0.84$~GeV for each.
The electric and magnetic couplings were taken to have the values
$g_M=7$ and $g_E=9$ \cite{Kon05}, obtained from a K-matrix analysis
of pion photoproduction data \cite{Kon01}.
A more realistic $\pi N$ coupled channel quasi-potential study
\cite{Pas04} gives similar values, $g_M=6.3$ and $g_E=3.4$.
The Coulomb coupling $g_C$ is not as well constrained as the electric
and magnetic couplings, and Kondratyuk~\etal~\cite{Kon05} considered
the range between $g_C = -2$ to 0.  Tjon~\etal~\cite{Tjo09} used an
estimate from the nucleon--$\Delta$ E2/M1 transition strength,
$g_C=5.8$, but found the sensitivity to variations of $g_C$ small.
Similar values for the couplings were also used by
Zhou~\etal~\cite{Zho10}, although defined with different
normalizations.  The sensitivity to $g_C$ was found to be weak
at low $Q^2$ and $\eps$, becoming stronger at forward angles and
large $Q^2$ ($Q^2 \gtorder 3$~GeV$^2$, $\eps \sim 1$) \cite{Zho10}.

The $\eps$ dependence of the relative $\Delta$ and nucleon TPE
contributions to the elastic cross section from Ref.~\cite{Kon05}, for
$g_C=0$, is shown in Fig.~\ref{fig:del_ggD} at $Q^2=1$ and 3~GeV$^2$.
The most striking feature of $\Delta$ corrections is its positive sign,
in agreement with the early estimates \cite{Dre59, Cam67, Cam69a, Gre69},
which is opposite to that of the nucleon, and the corresponding
{\it negative} slope in $\eps$.  This has the effect of attenuating
the (negative) nucleon TPE correction, making it somewhat smaller
especially at backward angles.  The magnitude of the $\Delta$ correction
is considerably smaller than the nucleon, however, so that the general
features of the TPE correction in Fig.~\ref{fig:del_gg} are not affected
by the $\Delta$.

These features are also largely insensitive to the details of the
$\gamma N \Delta$ form factor, particularly at low $Q^2$, as the
comparison of the results in Fig.~\ref{fig:del_ggD} with cut-offs
$\Lambda_\Delta=0.84$ and 0.68~GeV illustrates \cite{Kon05}.
At higher $Q^2$, on the other hand, both the magnitude of the
$\Delta$ correction and its dependence on the form factor model
increase, although for very large $Q^2$ the pure resonance
calculation itself becomes more questionable.

\subsubsection{Nucleon resonances}
\label{sssec:resonance}

If the $\Delta(1232)$ resonance makes a non-negligible contribution
to the TPE correction, at least in some kinematic regions, the
question naturally arises whether other, higher-mass resonances
could also play some role.  Kondratyuk and Blunden \cite{Kon07}
extended the formalism of Refs.~\cite{Blu03, Kon05}, generalizing it
to include the full spectrum of the most important hadron resonances
as intermediate states involving spin $1/2$ and $3/2$ resonances.
The masses of the resonances and their nucleon-photon coupling
constants are based on dynamical multichannel calculations
\cite{Kon01, Kor00, Pen02} of nucleon Compton scattering at low and
intermediate energies.  The resonance TPE effects turn out to
be not too sensitive to the details of these models.

In general the contributions of all the heavier resonances are much 
smaller than those of the nucleon and $\Delta$ ($P_{33}$)~\cite{Kon05}.
However, there is an interesting interplay between the contributions
of the spin $1/2$ and spin $3/2$ resonances, which is analogous to
the partial cancellation of the two-photon exchange effects of the
nucleon and $\Delta$ intermediate states, found in Ref.~\cite{Kon05}.
Notwithstanding the smallness of the resonance contributions, their
inclusion in the TPE diagrams leads to a better agreement between
the Rosenbluth and polarization transfer data analyses, especially
at higher values of the momentum-transfer squared $Q^2$.

\begin{figure}[tb]
\begin{center}
\begin{minipage}{10cm}
\includegraphics[scale=0.5]{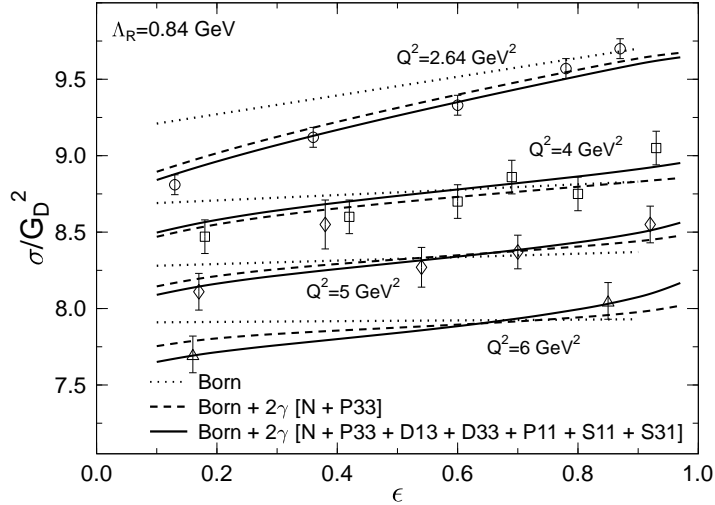}%
\end{minipage}
\begin{minipage}{16.5cm}
\vspace*{-0.5cm}
\caption{Reduced cross section, scaled by the dipole form factor squared,
	showing the effect of adding TPE corrections to the Born cross
	section.  The intermediate state includes a nucleon and the
	indicated hadron resonances.  The curves for $Q^2=2.64$, 4 and
	6~GeV$^2$ have been shifted vertically by $-0.04$, $+0.04$ and
	$+0.09$, respectively, for clarity, and are compared with data
	from Refs.~\cite{Wal94,Qat05}.
	The nucleon-only result is slightly larger than the
	$N+\Delta(P_{33})$, lying close to the full calculation.
	Figure taken from Ref.~\cite{Kon07}.}
\label{fig:del_res}
\end{minipage}
\end{center}
\end{figure}

The total TPE correction is given by the sum of the separate
hadron contributions,
\be
\delta\ =\ \delta^N + \delta^\Delta + \delta^{D_{13}} + \delta^{D_{33}}
	 + \delta^{P_{11}} + \delta^{S_{11}} + \delta^{S_{31}}.
\label{eq:delres}
\ee
The coupling constants in the vertices are taken from the Dressed K-Matrix
Model, whose essential ingredients are described in Ref.~\cite{Kon01}.
The calculated two-photon corrections to the reduced cross section are
displayed in Fig.~\ref{fig:del_res}.  The one-photon exchange cross
sections are shown by the dotted lines.  The cross sections including
additional TPE corrections are shown by the dashed lines for the sum of
the nucleon and $\Delta$ contributions, and by the solid lines for the 
full result with all resonances.  In general, each resonance two-photon 
correction is proportional to a sum of squares of the nucleon-photon 
coupling constants of that resonance.  This sets the scale of the 
magnitude of the resonance contributions.

As an example at the moderately high value $Q^2=4$~GeV$^2$, the TPE 
corrections from the included resonances can be classified by their 
signs and orders of magnitude as follows.   For $0 < \eps < 1$, the 
corrections change smoothly between the values
\bea
& & -5.0\  \ltorder\ \delta^N\        \ltorder\ 0\%,\ \ \ \ \ \ \ \ \
     1.9\  \gtorder\ \delta^\Delta\   \gtorder\ 0\%,\ \ \ \ \,
    -0.7\  \ltorder\ \delta^{D_{13}}\ \ltorder\ 0\%,	\nn\\
& & -0.3\  \ltorder\ \delta^{D_{33}}\ \ltorder\ 0\%,\ \ \
    -0.15\ \ltorder\ \delta^{P_{11}}\ \ltorder\ 0\%,\ \ \ \
     0.06\ \gtorder\ \delta^{S_{11}}\ \gtorder\ 0\%,\ \ \ 
     0.01\ \gtorder\ \delta^{S_{31}}\ \gtorder\ 0\%,	\nn
\eea
listed in order of decreasing magnitude.  Inclusion of the excited
state resonance contributions therefore effectively reduces the
nucleon elastic TPE correction by $\sim 15\%$ at this $Q^2$ value.
At lower $Q^2$ the resonance contributions are even less important
relative to the nucleon-only contribution.

Figure~\ref{fig:del_res} shows that at low to moderate $Q^2$ the
total TPE corrections as a function of $\eps$ lie between those of
the nucleon and $\Delta$ intermediate states.  In addition to the
dominant nucleon and $\Delta$ contributions, the $D_{13}$ gives the
most important correction among the remaining resonances.  This is
consistent with the well-known prominence of the $D_{13}$ in the
second resonance region of the Compton scattering cross section
({\em e.g.} see Ref.~\cite{Kor00, Pen02} and references therein).

Inclusion of contributions of intermediate states with masses larger
than $\sim 2$~GeV becomes impractical within a hadronic approach when
one moves into the deep-inelastic continuum.  Here it becomes more
efficient to use either partonic degrees of freedom, discussed in
Sec.~\ref{ssec:highQ}, or dispersion relations, discussed in
Sec.~\ref{ssec:DRgg}

\subsection{\it Scattering from a point target}
\label{ssec:emu}

It is useful to examine the TPE corrections to elastic scattering
from a structureless pointlike target, such as a $\mu^+$.
or in the limit of hard scattering from quarks in the nucleon.
The pointlike limit is realized by setting $F_1=1$ and $F_2=0$ in the 
electromagnetic current operator appearing in Eq.~(\ref{eq:Hmunu}),
or equivalently by replacing $\Gamma_\gamma^\mu(q) \to \gamma^\mu$.
The reduced cross section is then $\sigma_R^0 = \tau + \eps$.
In this case, we take the IR divergent part of the TPE correction
to be the model-independent Maximon-Tjon result of
Eq.~(\ref{eq:deltaIRMTj}); the remainder is denoted as
$\delta_{\rm hard}$, {\em viz.}
$\delta_{\gamma\gamma} = \delta_{\rm IR}({\rm MTj}) + \delta_{\rm hard}$.

\begin{figure}[htb]
\vspace*{0.6cm}
\begin{center}
\begin{minipage}{10cm}
\hspace*{1cm}\includegraphics[height=6cm]{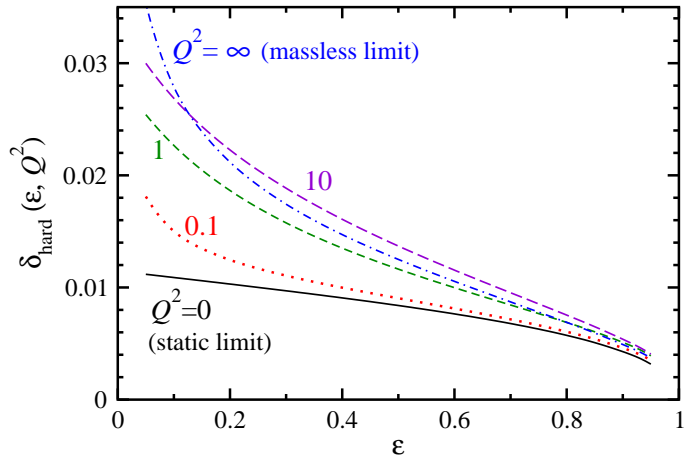}%
\end{minipage}
\begin{minipage}{16.5cm}
\caption{TPE corrections to scattering from a (positively charged)
	particle of mass
	$M=0.939$~GeV for various values of $Q^2$ (in GeV$^2$).
	Note that the limit $Q^2 \to 0$ also corresponds to the
	static limit $M \to\infty$ for any finite $Q^2$, and the
	limit $Q^2 \to\infty$ also corresponds to the massless
	limit $M \to 0$ for any finite $Q^2$.}
\label{fig:emu}
\end{minipage}
\end{center}
\end{figure}

Figure~\ref{fig:emu} shows the result for $\delta_{\rm hard}$ as
a function of $\eps$ for various values of $Q^2$ ranging from 0 to
$\infty$.  Clearly $\delta_{\rm hard}$ is a positive quantity which 
generally increases with $Q^2$ between the two limiting cases.
In both limits simple expressions for $\delta_{\rm hard}$ exist in
terms of a single variable.  For a pointlike target the results should
exhibit a scale invariance under the appropriate dimensionless variable,
which can be taken to be $\tau$ and $\eps$.

\begin{enumerate}
\item[(i)]
The limit $\tau \to 0$ is the same as either $Q^2 \to 0$ or the
static limit $M \to \infty$ with $Q^2$ finite.
In this case $\delta_{\rm hard}$ is a function of a single variable. 
Making use of the asymptotic expansions of the Passarino-Veltman
functions, a particularly simple expression exists in this limit, 
namely
\be
\delta_{\rm hard} = {\alpha \pi\over x+1},
\label{eq:del_hard0}
\ee
with $x=\sqrt{(1+\eps)/(1-\eps)}$.  This static limit is also realized
in the second Born approximation, which we discuss below. It agrees with
the expression given in Ref.~\cite{Mck48}, provided one identifies
$\sin(\theta/2)=1/x$ in this limit.
\item[(ii)]
The limit $\tau \to \infty$ is the same as either $Q^2 \to \infty$ or
the limit $M \to 0$ with $Q^2$ finite.  Once again $\delta_{\rm hard}$
is a function of a single variable only,  and can be expressed as
\be
\delta_{\rm hard}\
=\ {\alpha \over \pi\left(x^2+1\right)}
   \left\{ \ln \left(x+1\over x-1\right)
	 + x \left[\pi ^2 + \ln^2 \left( x+1 \over 2 \right)
		 + \ln^2 \left(x-1\over 2\right)
		 - \ln \left(x^2-1\over 4\right)
	     \right]
   \right\}.
\label{eq:del_hardInf}
\ee
This expression, plus the soft IR contribution
$\delta_{\rm IR}({\rm MTj})$, agrees with that for scattering from
massless quarks given in Eq.~(27) of Ref.~\cite{Afa05}, and also in
earlier work.  However, those results were not expressed in terms
of a single variable.
\end{enumerate}

The behavior of TPE for hard scattering from a pointlike target stands
in stark contrast to the $ep$ elastic scattering results, shown in
Fig.~\ref{fig:del_gg}, which is larger and has the opposite sign at
high $Q^2$.  However, as one decreases $Q^2$, the $ep$ TPE correction
becomes smaller, eventually switching sign from negative to positive
as $Q^2$ decreases.  For $Q^2 \to 0$, the hadronic result reproduces
the pointlike limit shown in Fig.~\ref{fig:emu} (and given in
Eq.~(\ref{eq:del_hard0})) -- independent of the anomalous magnetic
form factor $F_2$.
(Note that   
$\delta_{\rm IR}({\rm MTj}) \to \delta_{\rm IR}({\rm MoT})$ 
in the limit $Q^2 \to 0$, so that $\bar\delta \to \delta_{\rm hard}$).
This suggests that the magnetic interaction is unimportant in this limit.
This is consistent with the low-$Q^2$ analysis of Ref.~\cite{Bor07}.

Kuraev and Tomasi-Gustafsson~\cite{Kur10} have suggested that $e\mu$ 
scattering should be a function of $x$, and that it represents an 
``upper limit'' for the hadronic TPE corrections.  In the context of
the discussion here, while this may be a positive upper bound, at high 
$Q^2$ the full calculation gives a correction larger in magnitude but 
opposite in sign.

To fully understand the underlying reasons for this behavior, one can
invoke the second Born approximation.  In the static limit, equivalent
to the case where the proton has infinite mass, the scattering amplitude
for TPE coincides with the amplitude for electron scattering in a Coulomb
potential in second Born approximation.  This limit was first considered
by Dalitz~\cite{Dal51}.

\begin{figure}[tb]
\vspace*{2cm}
\begin{center}
\vspace*{-2cm}
\hspace*{5cm}\includegraphics[height=6.5cm]{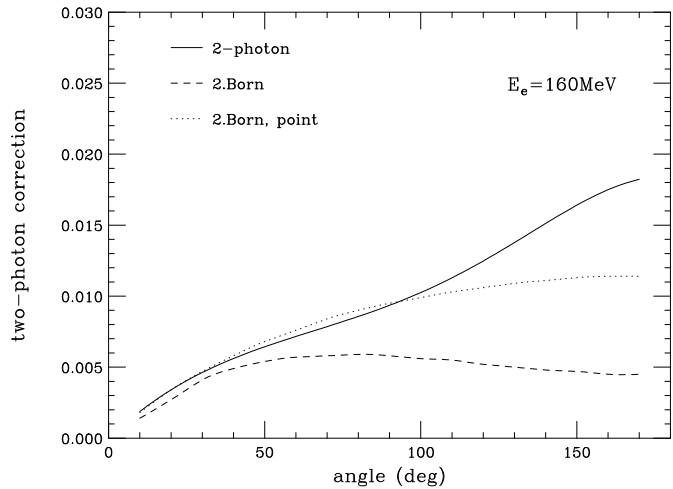}%
\hspace*{-13cm} 
\begin{minipage}{16.5cm}
\vspace*{2cm}
\caption{Relative contribution of TPE to elastic $ep$ scattering at
	$E=160$~MeV. The results in second Born approximation account for
	the Coulomb distortion (exchange of second soft photon) only.
        Figure taken from Ref.~\cite{Blu05S}.}
\label{fig:2born}
\end{minipage}
\end{center}
\end{figure}

In Fig.~\ref{fig:2born} the TPE correction is shown as a function
of $\theta$ for a typical electron energy involving a range of low
$Q^2$ values \cite{Blu05S}.
The TPE contribution is compared to the contribution involving
only the Coulomb distortion of the electron, arising from a second
{\it soft} photon, calculated according to Ref.~\cite{Sic98} in
second Born approximation.
The same contribution, but for a point nucleus, is given for
comparison.  Remarkably good agreement between the two approaches
is obtained at forward angles.
Figure~\ref{fig:2born} shows that at forward angles the TPE
contribution is dominated by the Coulomb distortion, while at backward
angles the exchange of two hard photons contributes appreciably.
This is consistent with the observations about the pointlike limit,
and the independence of this limit from $F_2$.

The behavior shown in Fig.~\ref{fig:del_gg}~(left) can also be understood
from the second Born approximation \cite{Isk93}.  The second interaction
provides a ``focusing effect'' --- accelerating electrons towards the
target.  For a pointlike target this should increase scattering at
backward angles.  However, because of an increased momentum transfer
there is a competing effect from a reduction in cross section due to
the proton electric and magnetic form factors.  At larger $Q^2$ this
reduction wins out, and the total cross section is reduced at backward
angles.  For positrons, the opposite effect is expected.

\subsection{\it High $Q^2$ partonic models}
\label{ssec:highQ}

In the regime of high $Q^2$, two approaches to the TPE effect on elastic
scattering have been taken by different groups.  In Refs.~\cite{Che04, 
Afa05}, the hard scattering part of TPE was studied in a partonic 
approach using different models for generalized parton distributions 
(GPDs).  In this approach it is assumed that both photons interact with 
the same quark.  The results of these calculations on the cross sections 
and other observables have been examined in detail in Ref.~\cite{Car07}, 
and the details will not be duplicated here.

\begin{figure}[tbh]
\vspace*{0.5cm}
\begin{center}
\def\www{0.16}
\centering
(a)
\includegraphics[width=\www\textwidth]{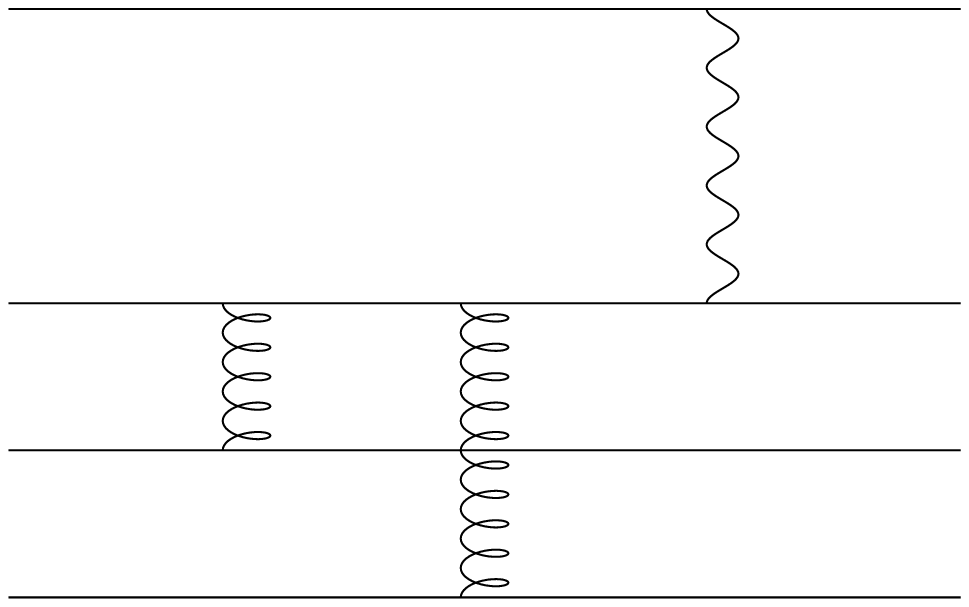}\hspace*{.5cm}
(b)
\includegraphics[width=\www\textwidth]{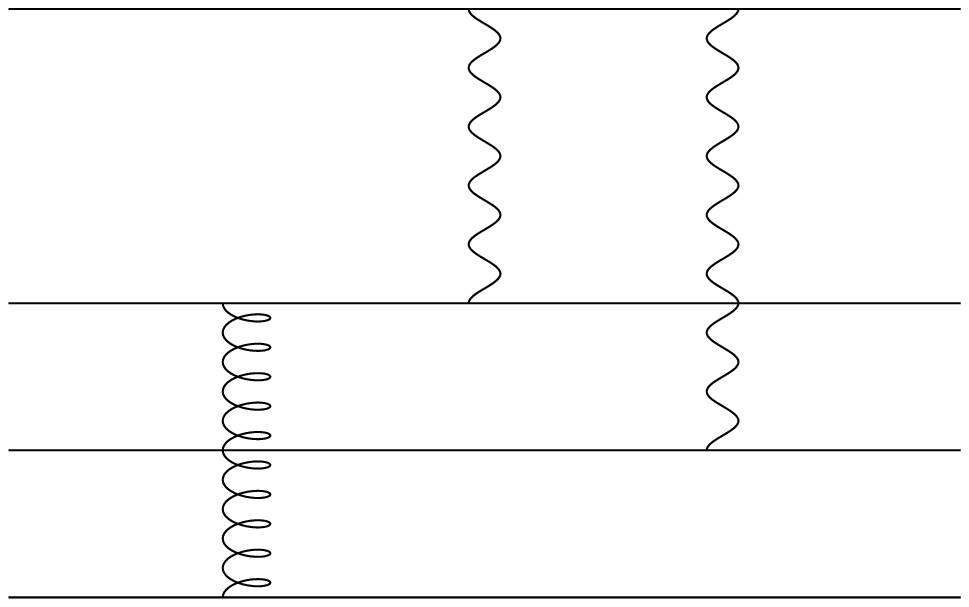}\hspace*{.5cm}
(c)
\includegraphics[width=\www\textwidth]{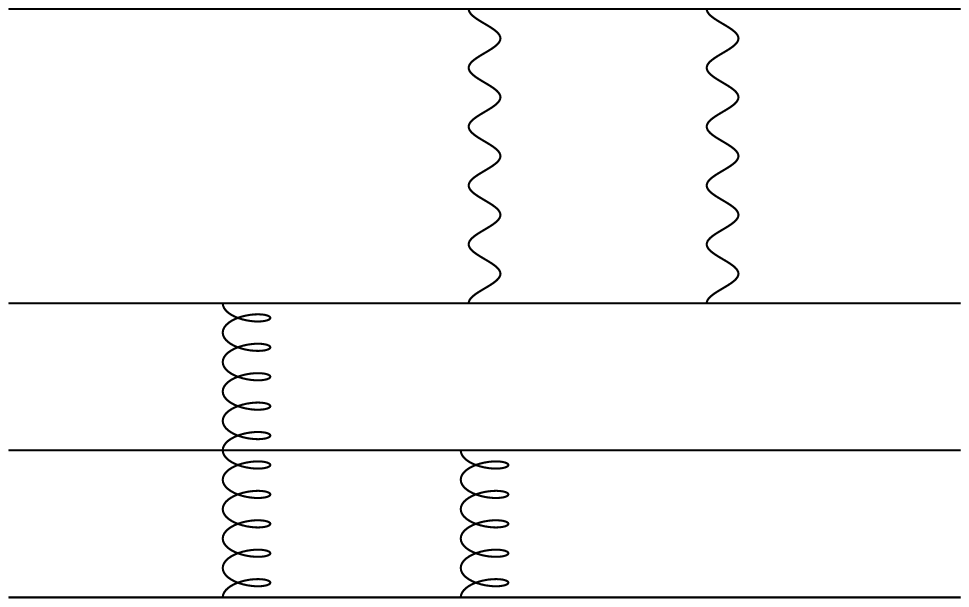}
\begin{minipage}{16.5cm}
\caption{Typical pQCD diagrams for
	{\bf (a)} one-photon exchange,
	{\bf (b)} leading order TPE involving 1 hard gluon, and
	{\bf (c)} subleading order TPE involving 2 hard	gluons.
	Figure adapted from Ref.~\cite{Bor09}.}
\label{fig:pQCD}
\end{minipage}
\end{center}
\end{figure}

Our observations about the pointlike limit in Sec.~\ref{ssec:emu} suggest
that hard TPE corrections involving both photons interacting with the same
particle have the opposite sign to the hadronic calculations at backward
angles.  Borisyuk and Kobushkin \cite{Bor09} studied TPE in the framework
of perturbative QCD (pQCD).  In this approach, it turns out that the most 
important diagrams for backward angle scattering are the ones where the
two photons interact with different quarks.  This allows the possibility
to generate a hard TPE correction that is negative at backward angles.
The argument given by Borisyuk and Kobushkin is as follows.

Referring to the diagrams in Fig.~\ref{fig:pQCD}, one-photon exchange
diagrams like (a) need 2 hard gluons to turn the momentum of all 3
quarks, and are therefore of order $\alpha \alpha_s^2/Q^6$, where
$\alpha_s$ is the strong coupling.  The TPE diagrams like (b),
involving 2 quarks, need only 1 hard gluon, and are of order
$\alpha^2\alpha_s/Q^6$.  Therefore the ratio of two-photon to
one-photon exchange is not of order $\alpha$, as one might expect,
but of order $\alpha/\alpha_s$.  By contrast, diagrams like (c),
where both photons interact with the same quark, require 2 hard gluons,
and are therefore suppressed by an additional factor of $\alpha_s$.

\begin{figure}[bth]
\begin{center}
\includegraphics[width=7cm]{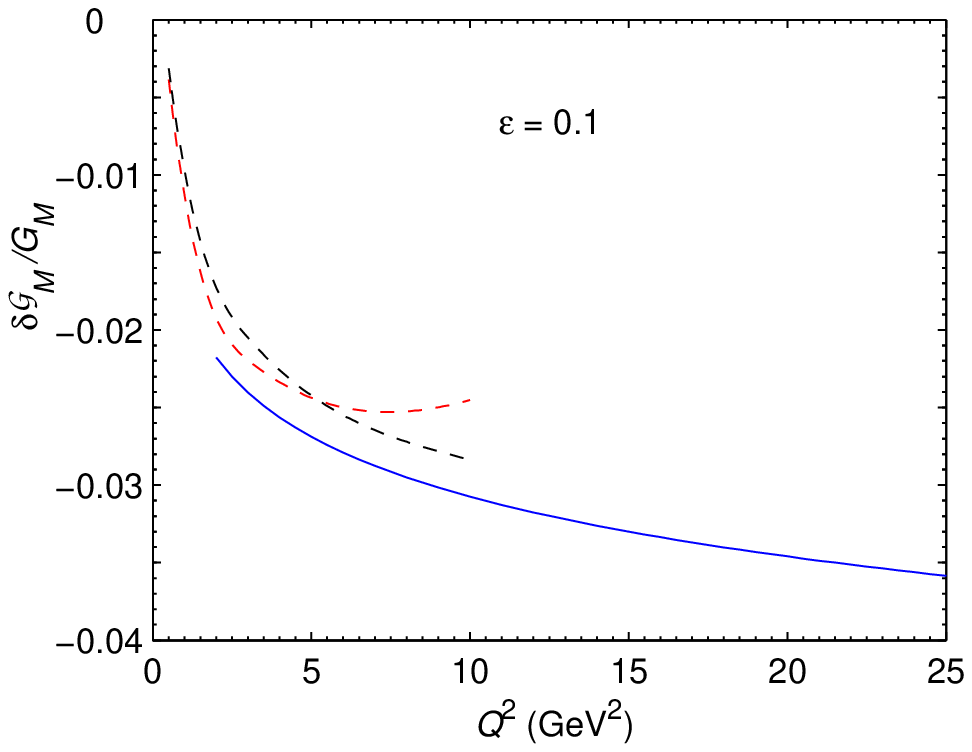}
\hspace*{0.5cm}\includegraphics[width=7cm]{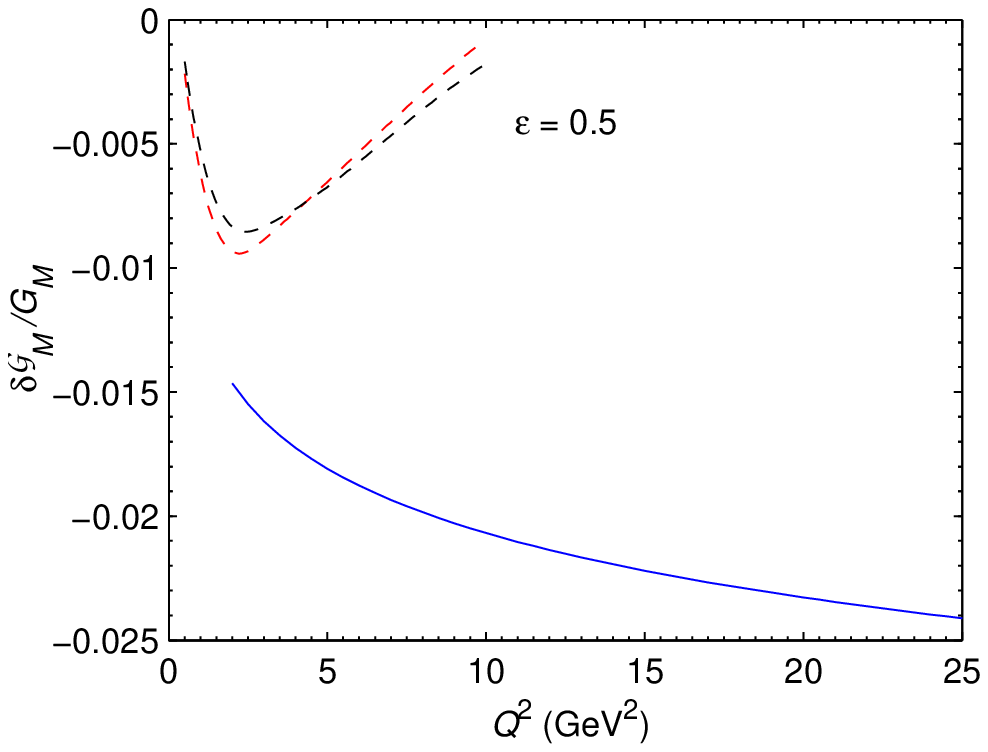}
\begin{minipage}{16.5cm}
\caption{TPE amplitude vs. $Q^2$ for two values of $\eps$.
	The dashed curves are the hadronic calculation using two
	different form factors, and solid curves are the pQCD results.
	Figure taken from Ref.~\cite{Bor09}.}
\label{fig:BKpQCD}
\end{minipage}
\end{center}
\end{figure}

Numerically, Borisyuk and Kobushkin \cite{Bor09} express the cross
section correction in terms of the relative TPE amplitude
$\delta G_M/G_M$, where $G_M = G_M^{\rm Born} + \delta G_M$, which has
linear $\eps$ dependence.  This amplitude grows logarithmically with
$Q^2$, reaching 3.5\% at $Q^2=30$~GeV$^2$.  This may offer an interesting 
avenue to merge the hadronic approach, valid at low to moderate $Q^2$, 
with the pQCD approach at high $Q^2$.  Figure~\ref{fig:BKpQCD} shows
how the hadronic and pQCD results may connect approximately at around 
$Q^2=3$~GeV$^2$ for two values of $\eps$. A complementary pQCD analysis
by Kivel and Vanderhaeghen \cite{Kiv09} also finds a comparable effect.

\begin{figure}[h]
\begin{center}
\includegraphics[width=8.0cm]{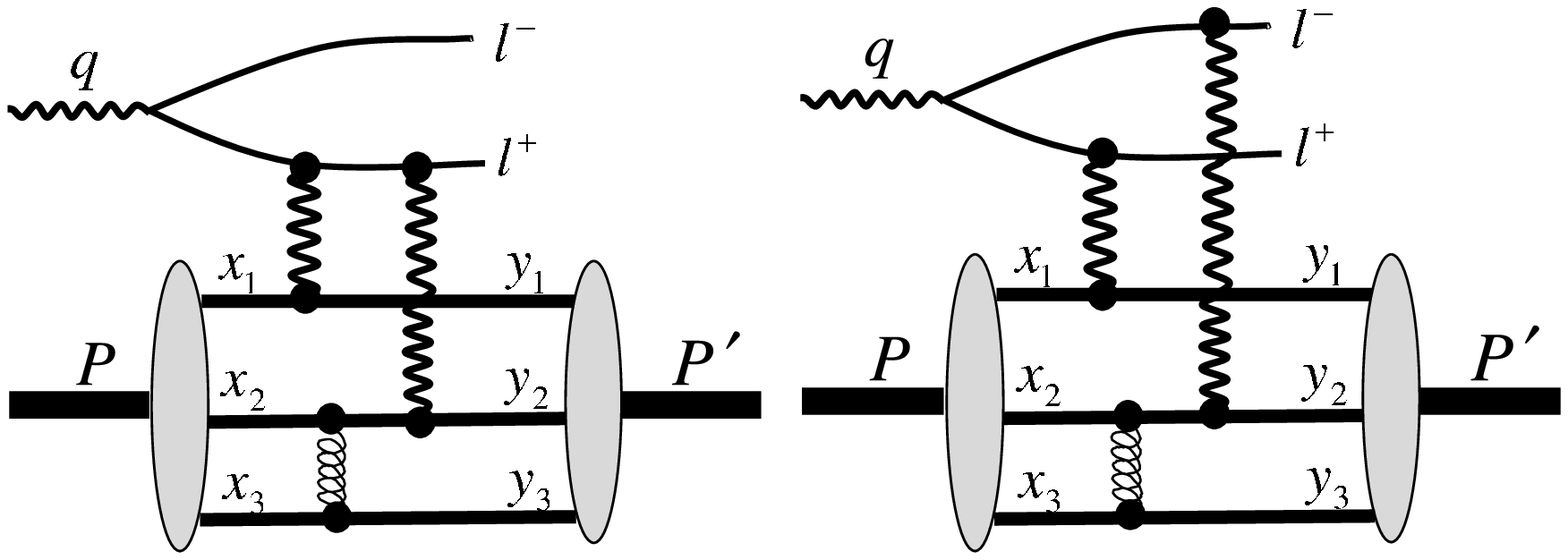}%
\hspace*{0.5cm}
\includegraphics[width=4.7cm,angle=90]{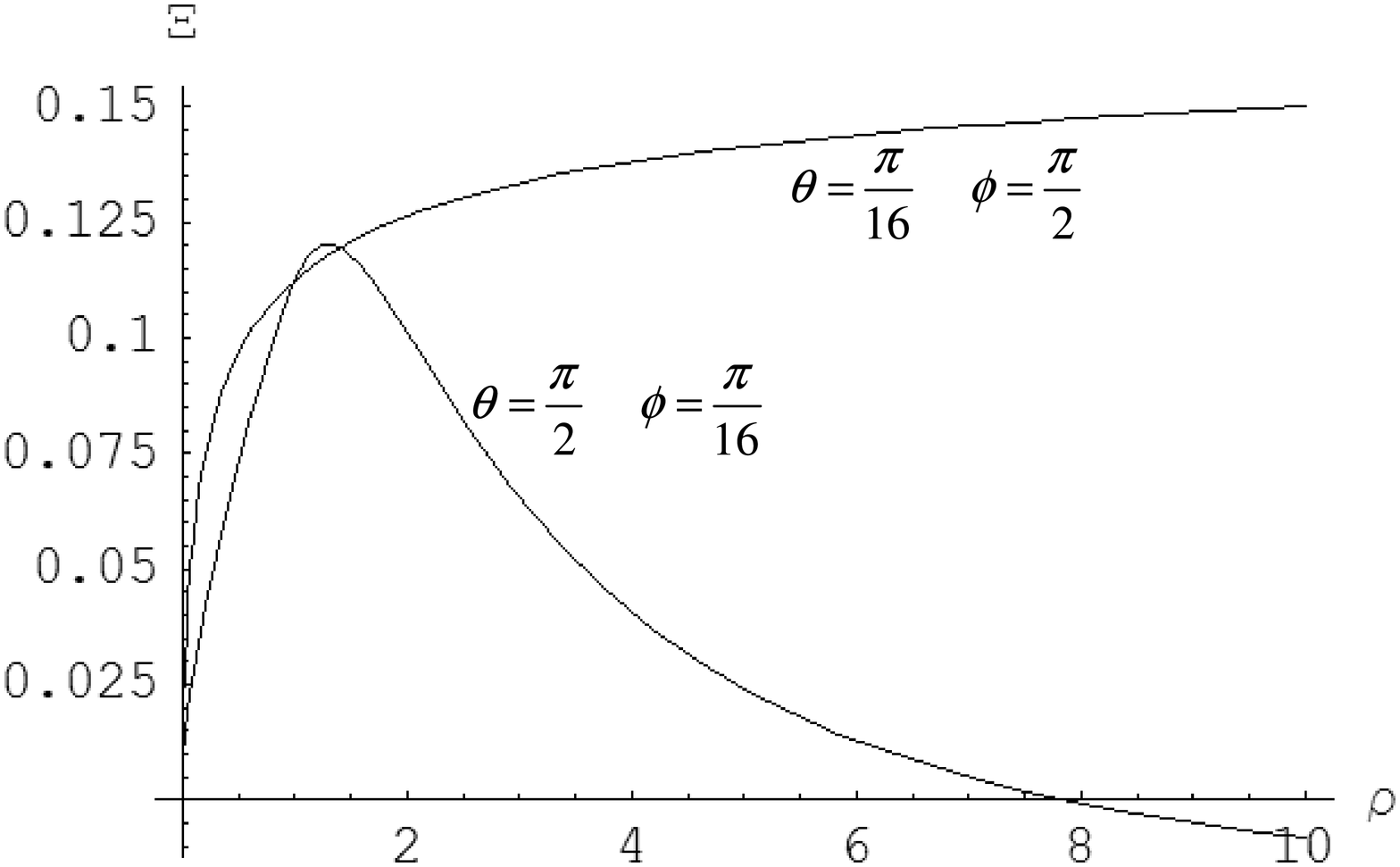}
\begin{minipage}{16.5cm}
\vspace*{0.5cm}
\caption{Typical TPE diagrams for lepton pair production
	involving two photons and one hard gluon {\bf (left)}.
	The graph shows the resulting lepton pair asymmetry
	at select angles {\bf (right)}.
	Figures taken from Ref.~\cite{Hoo06}.}
\label{fig:HB}
\end{minipage}
\end{center}
\end{figure}

Two-photon effects also manifest in the production of lepton-antilepton
pairs from a hadronic target by real photons.  The interference of one-
and two-photon exchange amplitudes leads to a charge asymmetry term that
can be calculated explicitly in the large-$t$ limit.  This was studied 
using pQCD by Hoodbhoy \cite{Hoo06} in the region of high center of mass
energy $s$ ($s \gg -t \gg M^2$).  Typical diagrams involving 2 photons
and 1 hard gluon exchange are shown in Fig.~\ref{fig:HB}, analogous
to the corresponding diagrams in elastic $ep$ scattering discussed
previously.  The lepton pair asymmetry for select angles in the lepton
pair center of mass is also shown as a function of $\rho\equiv -t/M^2$.
Note that Hoodbhoy's calculation and observations about the leading
contribution involving different quarks actually predates the work
of Refs.~\cite{Bor09, Kiv09}.

\newpage
\subsection{\it Dispersion relations}
\label{ssec:DRgg}

One of the sources of model dependence in the calculation of the
real parts of the TBE amplitudes arises from the intermediate
state in the box diagrams being off-shell.
For the nucleon elastic contribution, for example, this can be
parametrized through the off-shell dependence of the nucleon
electromagnetic form factors, where one of the nucleons is off its
mass shell, $W^2 = (p+q_1)^2 \not= M^2$, see Fig.~\ref{fig:TPE}.
In this case the vertex form factors are in general functions
of {\it two} variables, $q_1^2$ and $W^2$, which in the limit
$W^2 \to M^2$ reduce to the on-shell values.
Outside of a field-theoretical calculation of the amplitudes from
first principles, however, it is not possible to determine the
(unphysical) dependence on $W^2$ independently of a specific model.
This model dependence invariably introduces additional uncertainties
into the calculation.

One way to avoid the off-shell ambiguities is through the use of
dispersion relations \cite{Kro26}, which are based on general principles 
such as unitarity and analyticity of scattering amplitudes, and can in
principle relate real and imaginary parts of amplitudes for on-shell
processes.
The conservation of total probability in a scattering process
implies that the scattering matrix, ${\cal S}$, is unitary,
${\cal S}^\dagger {\cal S} = 1$.
In general the S-matrix element
${\cal S}_{fi} = \langle f| {\cal S} |i \rangle$
between initial state $i$ (with total momentum $p_i$) and final state
$f$ (total momentum $p_f$) can be written in terms of the invariant
amplitude as
\be
{\cal S}_{fi}\
=\ \delta_{fi} + i (2\pi)^4\, \delta^4(p_f-p_i)\, {\cal M}_{fi}.
\ee
Unitarity then requires that
\be
2\, \Im\, {\cal M}_{fi}\
=\ \sum_n\, \int d\rho\, {\cal M}_{nf}^*\, {\cal M}_{ni},
\label{eq:optical}
\ee
where $d\rho$ is the covariant phase-space factor for a complete
set of intermediate states $n$.  In the limit of forward elastic
scattering, $i=f$, the product of amplitudes in (\ref{eq:optical})
becomes diagonal, and can be expressed as a directly measurable
cross section.  This allows loop integrations to be evaluated in
terms of on-shell states, so that empirical {\it data} can be used
as input in the calculation of the imaginary part.

The Cauchy integral formula allows the amplitude ${\cal M}_{fi}(s)$
to be determined as a function of a real $s$-channel variable $s$
from the structure of its singularities in the complex plane.
The real and imaginary parts of ${\cal M}_{fi}(s)$ can be related by
\be
\Re\, {\cal M}_{fi}(s)\
=\ {1 \over \pi}\, {\cal P}\int_{-\infty}^\infty ds'\,
   { \Im\, {\cal M}_{fi}(s') \over s' - s },
\ee
where ${\cal P}$ denotes the principal value of the integral.

The dispersion approach has been applied in various studies in the
literature of TPE corrections to elastic $ep$ scattering, starting
with some early attempts in the late 1950s \cite{Dre57, Dre59, Wer61}
to compute ${\cal O}(e^4)$ corrections to elastic $ep$ cross sections.
More recently, Borisyuk and Kobushkin \cite{Bor08} used dispersion
relations to compute the contribution to TPE amplitudes arising from
nucleon elastic intermediate states, in terms of elastic on-shell
nucleon form factors parametrized as a sum of monopoles \cite{Blu03}.
Interestingly, the results of the dispersive calculations were found
to be very similar numerically to the direct loop computation
\cite{Blu05, Bor06}, especially at low $Q^2$.  This suggests that the
prescription of taking on-shell values for the half off-shell form
factors in the direct calculations provided a very good approximation.
At larger $Q^2$ ($Q^2 \gtorder 6$~GeV$^2$) the agreement between
the two approaches deteriorates, possibly hinting at the need to
account for the off-shell dependence of the form factors; at such
$Q^2$ values the reliability of the hadronic approach itself may
be more questionable.  In Ref.~\cite{Bor11} the authors applied the
same method to also compute the TPE corrections to the pion form
factor (see Sec.~\ref{ssec:pionFF} below), again finding very
similar results to the direct loop calculations \cite{Blu10}.

Elastic scattering from hadrons at forward angles was analyzed
some time ago within a dispersive approach by Bernab\'{e}u and
collaborators \cite{Ber80, Pen81, Bor86, Bor87}, and more recently
by Gorchtein \cite{Gor07}.  We sketch here the relevant details
(in the notation of this article), noting the overlap with the
discussion of forward-angle parity-violating $\gamma Z$ interference
in Sec.~\ref{ssec:DRgZ}.
In the forward angle limit ($q_1 \approx -q_2$ in Fig.~\ref{fig:TPE})
the imaginary part of the TPE amplitude, by virtue of the intermediate
state being on-shell, is related via the optical theorem
(\ref{eq:optical}) to the inclusive electromagnetic structure
functions of the nucleon,
\be
2\, \Im\, {\cal M}_{\gamma\gamma}\
=\ 4\pi M\,e^4\, \int{d^3 l \over (2\pi)^3 2 E_l}
   { 1 \over Q_1^4 }\, 
   L_{\mu\nu}\, W^{\mu\nu}\, ,
\label{eq:Imgg}
\ee
where $l = k-q_1$ is the momentum of intermediate state electron,
$E_l = \sqrt{(\bm{k}-\bm{q}_1)^2 + m_e^2}$ its energy, and
$Q_1^2 = -q_1^2$ is the virtuality of exchanged photon.  The factor
$4\pi M$ arises from the definition of the hadronic tensor.
The leptonic tensor in (\ref{eq:Imgg}) is given by
$L_{\mu\nu}
 = \ubar_e(k')\, \gamma_\mu(\vslash{l}+m_e)\gamma_\nu\, u_e(k)$,
where the scattered lepton momentum has been set to $k'=k-q$,
with $q$ kept small but finite.
%
%
The nucleon hadronic tensor can be written in terms of the
inclusive $F_1$ and $F_2$ structure functions as
\be
M W^{\mu\nu}\
=\ - g^{\mu\nu} F_1(W^2,Q_1^2)\
+\ {p^\mu p^\nu \over p \cdot q_1} F_2(W^2,Q_1^2),
\label{eq:Wmunu_def}
\ee
dropping terms proportional to $q_{\mu,\nu}$ which vanish when
contracted with $L_{\mu\nu}$.
Using Eq.~(\ref{eq:delta_gg}) the relative correction from TPE to
the elastic cross section can then be written as
\be
\Im\, \delta_{\gamma\gamma}(E)\
= -{\alpha\, t \over (2ME)^2}
   \int_{M^2}^s dW^2
   \int_0^{Q^2_{1,\rm max}} {dQ_1^2 \over Q_1^2}
   \left[ F_1
        + { s \left( Q^2_{1,\rm max}-Q_1^2 \right) \over
                     Q_1^2 \left( W^2 - M^2 + Q_1^2 \right) } F_2
   \right],
\label{eq:Imdelgg}
\ee
where the upper limit on the $Q_1^2$ integration is
$Q_{1,\rm max}^2 = 2ME(1-W^2/s)$, with $s = M^2+2 M E$ the total
center of mass energy squared, and $t=q^2$.  Note that the factor
$t$ in Eq.~(\ref{eq:Imdelgg}) implies that formally this contribution
vanishes at $t=0$.

The real part of $\delta_{\gamma\gamma}(E)$ at a given electron energy
$E$ is then evaluated from the imaginary part (\ref{eq:Imdelgg}) using
the fixed-$t$ dispersion relation
\be
\Re\, \delta_{\gamma\gamma}(E)\
=\ {1 \over \pi}\, {\cal P}\int_{-\infty}^\infty dE'\,
   { \Im\, \delta_{\gamma\gamma}(E') \over E' - E }.
\label{eq:DR1_gg}
\ee
Note that the contribution to the integral from the region $E' < 0$ 
corresponds to the crossed-box diagram in Fig.~\ref{fig:TPE}.
Invoking the crossing symmetry property discussed earlier in 
Eq.~(\ref{eq:cross}), one can write
\be
\Re\, \delta_{\gamma\gamma}(E)\
=\ {2 E \over \pi}
   {\cal P} \int_0^\infty dE' {1 \over E'^2-E^2}\,
   \Im\, \delta_{\gamma\gamma}(E').
\label{eq:DR2_gg}
\ee

Bernab\'{e}u {\it et al.} \cite{Ber80, Pen81, Bor86, Bor87} applied
this formalism to the case of elastic electron and muon scattering
from the spin-0 $^4$He and $^{12}$C nuclei at low energy, focusing in
particular on the continuum contributions.  The electron--positron
charge asymmetry for $^4$He, which depends on the interference of
the one- and two-photon exchange amplitudes, was found to be of the
order of 1\%, and about 4--5 smaller for $\mu^\pm$ \cite{Bor86}.
The effects of the finite size of the nucleus were found to be small
for kinetic energies up to $\approx 50$~MeV, but increased to
$\approx 30\%$ of the total forward amplitude at energies
$\approx 100$~MeV \cite{Bor87}, and somewhat more significant for
muon scattering.  The results of these studies were subsequently used
to constrain the subtraction constant in the dispersion relation for
the two-photon exchange amplitude at nonzero values of the momentum
transfer squared $t$.

More recently, Gorchtein \cite{Gor07} applied the forward dispersion
formalism to study the effect of TPE on the $e^+p$ to $e^-p$ cross
section ratio at higher energies.
Using phenomenological structure functions in the deep-inelastic
scattering region \cite{Cve01}, the ratio of $e^+p$ to $e^-p$
cross sections is found to be larger than unity, with effects
ranging from $\ltorder 1.5\%$ for incident energy $E=3$~GeV
to $\ltorder 2.5\%$ for $E \approx 10$~GeV.

Overall, the dispersion approach provides a unique method to
compute amplitudes from empirical inputs, in principle free of
model dependent ambiguities associated with off-shell effects
of intermediate states, particular at high mass $W^2$.
On the other hand, this virtue is at times somewhat negated by
the need to know the inputs at all energies, which are often
not well determined at very high energy.
Uncertainties in the high energy behavior of the input cross
sections can in some cases be as large as the off-shell
uncertainties in the direct, non-dispersive approaches.
Furthermore, at scattering angles away from the forward limit
the imaginary parts of TPE amplitudes are no longer related
to inclusive structure functions, but are given by nonforward
virtual Compton scattering amplitudes, for which there are
considerably fewer phenomenological constraints.
The dispersive framework therefore provides a useful
{\it complement} to the other methods discussed in this section.

%% file: section5.tex
\section{Impact of two-photon exchange on observables}
\label{sec:impact}

Having outlined the theoretical developments in the computation of
TPE corrections to elastic scattering, in this section we review the
implications of the corrections for observables.  Starting with the
most topical case of TPE effects on the proton electric to magnetic
form factor ratio (Sec.~\ref{ssec:impact_LT}), we also consider the
impact that TPE corrections have on electron--proton to positron--proton
elastic cross sections (Sec.~\ref{sec:impact_positron}), on the global
analysis of form factor data (Sec.~\ref{ssec:global}), as well as on
polarization observables (Sec.~\ref{ssec:impact_PT} and \ref{ssec:normal}).
To simplify the discussion, unless otherwise indicated, we restrict the
hadronic calculations to the nucleon elastic intermediate states only
(which provides a reasonable approximation to the total hadronic
contribution, see Sec.~\ref{sssec:resonance}).

\begin{figure}[htb]
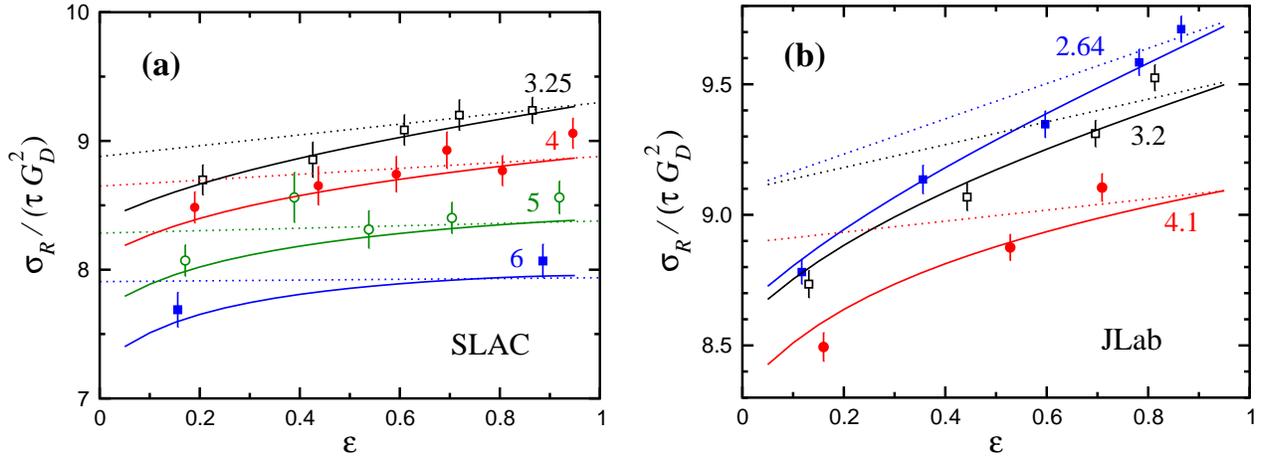

\begin{center}
\includegraphics[height=6cm]{Figs/fig21a.eps}%
\hspace*{0.5cm}\includegraphics[height=6cm]{Figs/fig21b.eps}
\begin{minipage}{16.5cm}
\caption{Comparison of the Born (dotted) and TPE (solid) calculations
	to measurements of $\sigma_R/(\tau G_D^2)$ for several values
	of $Q^2$:
{\bf (a)} SLAC data~\cite{And94} at $Q^2=3.25-6$~GeV$^2$;
{\bf (b)} Jefferson Lab data~\cite{Qat05} at $Q^2=2.64-4.1$~GeV$^2$.
        The Born cross section is evaluated using the form factors
	from Ref.~\cite{Bra02} which use PT extractions of $\ge$.
	The curves in (b) have been shifted by
	(+1.0\%, +2.1\%, +3.0\%) for $Q^2=(2.64, 3.2, 4.1)$~GeV$^2$.   
	Figure adapted from Ref.~\cite{Blu05}.}
\label{fig:sigR}
\end{minipage}
\end{center}
\end{figure}

\subsection{\it Cross sections and Rosenbluth separations}
\label{ssec:impact_LT}

The effect of the TPE corrections on the Rosenbluth separations,
calculated within the hadronic formalism of Sec.~\ref{ssec:nucleon},
is illustrated in Fig.~\ref{fig:sigR}.  The data points are the
reduced cross sections $\sigma_R$, scaled by the square of the dipole
form factor (\ref{eq:GD}) multiplied by $\tau$, from SLAC \cite{And94}
and the Jefferson Lab Super-Rosenbluth experiment \cite{Qat05}.
Here the Born-level prediction (dotted line) is obtained using the
form factor parametrization of Ref.~\cite{Bra02} in which $\ge$ is
fitted to the polarization transfer data, while the full result
(solid) includes the TPE corrections.  The TPE contributions yield a
significant increase of the slope, with some nonlinearity evident at
small $\eps$.  The corrected results are clearly in better agreement
with the data, although some residual difference between the Rosenbluth
and polarization measurements remain at the higher $Q^2$ values.

To obtain a rough estimate of the influence of these corrections on
the electric to magnetic proton form factor ratio $R = \mugegm$,
Blunden {\it et al.} \cite{Blu03, Blu05} considered a simplified
analysis in which the effective $\eps$ slope was approximated by a
linear fit to the full TPE corrections.  Such an analysis may not
provide an accurate estimate at very low $\eps$ or high $Q^2$ where
strong deviations from linearity arise; however, it is still
instructive to obtain an estimate of the effect on $R$ by taking
the slope over several ranges of $\eps$.
Blunden~\etal~\cite{Blu05} fit the relative correction to the
Born cross section $(1+\overline\delta)$ to a linear function
of $\eps$, of the form $a + b\, \eps$, for each value of $Q^2$
at which the ratio $R$ is measured.
The corrected reduced cross section in Eq.~(\ref{eq:sigmaR})
then becomes \cite{Blu05}
\be
\sigma_R\
\approx\ a\, \tau G_M^2(Q^2)
	\left[ 1 + {\eps \over \mu^2 \tau}
           \left( R^2 \left[ 1 + \eps b/a \right]
                + \mu^2 \tau b/a
           \right)
	\right],
\label{eq:sigRcor}
\ee
where
\be
R^2 = { \widetilde{R}^2 - \mu^2 \tau b/a
        \over 1 + \bar\eps b/a }
\label{eq:Rtrue}
\ee
is the ``true'' form factor ratio, corrected for TPE effects, and
$\widetilde{R}$ is the ``effective'' ratio, contaminated by TPE.
Note that in Eqs.~(\ref{eq:sigRcor}) and (\ref{eq:Rtrue}) the term
quadratic in $\eps$ has effectively been linearized by averaging
$\eps$ over the range fitted, $\eps \to \bar\eps$.
The approximation in Eq.~(\ref{eq:sigRcor}) is reasonable provided
$b/a \ll 1$, which is more valid at high $Q^2$ values.

\begin{figure}[t]
\begin{center}
\begin{minipage}{8cm}
\hspace*{-0.5cm}
\includegraphics[height=7cm]{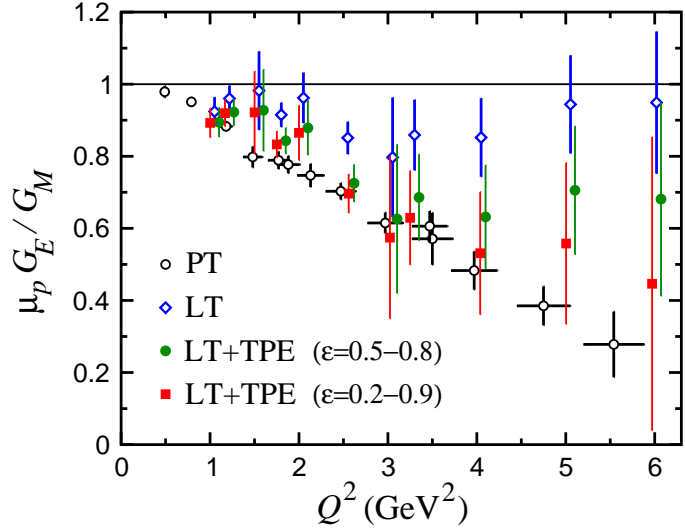}
\end{minipage}
\begin{minipage}{16.5cm}
\caption{The ratio of proton electric and magnetic form factors
	$\mu_p G_E/G_M$ measured using LT separation (open diamonds)
	\cite{Arr03} and polarization transfer (PT) (open circles)
	\cite{Jon00}.
        The LT points corrected for TPE are shown assuming a linear
	slope for $\eps=0.2-0.9$ (filled squares) and $\eps=0.5-0.8$
	(filled circles) (offset for clarity).
	Figure adapted from Ref.~\cite{Blu05}.}
\label{fig:GEMp}
\end{minipage}
\end{center}
\end{figure}

Considering two ranges for $\eps$, namely a large range $\eps=0.2-0.9$
and a more restricted range $\eps=0.5-0.8$, the resulting shift in $R$
is shown in Fig.~\ref{fig:GEMp}, together with the polarization transfer
data.  The linear $\eps$ approximation to the TPE correction should be
better for the latter, even though in practice experiments typically
sample values of $\eps$ near its extrema.

The effect of TPE on $R$ is clearly significant.  In particular,
the corrections have the proper sign and magnitude to resolve a
large part of the discrepancy between the LT and PT techniques.
While the early analysis of Ref.~\cite{Blu03} using simple monopole
form factors found a shift similar to that in for the $\eps=0.5-0.8$
range in Fig.~\ref{fig:GEMp}, which resolves around 1/2 of the
discrepancy, the nonlinearity at small $\eps$ makes the effective
slope somewhat larger if the $\eps$ range is taken between 0.2 and 0.9.
The magnitude of the effect in this case is sufficient to bring the
LT and PT points almost to agreement.  A more detailed analysis,
where the TPE correction is applied directly to the experimentally 
measured cross sections, is discussed in Sec.~\ref{sec:amt}.

Note that for the high $Q^2$ points, the value of $\mugegm$ decreases but
the uncertainty on the ratio increases when TPE corrections are applied.
This is because the quantity directly constrained by the experiment is 
the slope of the reduced cross section, which is related to $(\mugegm)^2$, 
rather than $\mugegm$.  So while the absolute uncertainty on the slope, 
and thus $(\mugegm)^2$, is unchanged, the uncertainty that this 
translates to for $\mugegm$ depends on the value, and increases when 
$\mugegm$ decreases.

The results in Fig.~\ref{fig:GEMp} are representative of most
calculations of the TPE corrections within the conventional hadronic
approach.  Qualitatively similar results are also obtained in the
GPD-based approach \cite{Che04, Afa05}, whose applicability is limited,
however, at low $Q^2$ and $\eps$ values ({\it cf.} Fig.~\ref{fig:BKpQCD}).
Note that since the hard electron--quark scattering amplitude gives a
{\it negative} slope in $\eps$, Fig.~\ref{fig:emu}, agreement with data
suggests an important role played by soft physics.

\subsection{\it $e^+p$/$e^-p$ ratios}
\label{sec:impact_positron}

As previewed in Sec.~\ref{sec:positron}, direct experimental evidence
for TPE effects can be obtained by comparing elastic $e^+p$ and $e^-p$
cross sections.  The ratio of these is defined as
\be
R^{e^+e^-}\
= { \sigma^{e^+p} \over \sigma^{e^-p} }\
\approx\
{ | {\cal M}_\gamma^{e^+} |^2
          + 2 \Re \left( {\cal M}_\gamma^{e^+ *}
                          {\cal M}_{\gamma\gamma}^{e^+}
                  \right)
\over
  | {\cal M}_\gamma^{e^-} |^2
          + 2 \Re \left( {\cal M}_\gamma^{e^- *}
                          {\cal M}_{\gamma\gamma}^{e^-}
                  \right) },
\ee
where $\sigma^{e^\pm p} \equiv d\sigma^{e^\pm p}/d\Omega$.
Whereas the electron Born amplitude ${\cal M}_\gamma^{e^-}$ changes
sign under the interchange $e^- \to e^+$, the TPE amplitude
${\cal M}_{\gamma\gamma}^{e^-}$ does not.  The interference of the
Born and TPE amplitudes therefore has the opposite sign for electron
and positron scattering, so that the cross section ratio can be
written
\be
R^{e^+e^-}\
\approx\ 1 - 2\, \overline\delta\, ,
\ee
where $\overline\delta$ is defined in Eq.~(\ref{eq:Delta_dif}).
The hadronic TPE calculation of $\overline\delta$ is illustrated
for a range of $Q^2$ values in Fig.~\ref{fig:del_gg}.
Since the finite part of the TPE contribution is negative over
most of the range of $\eps$, one would expect to see an enhancement
of the ratio of $e^+$ to $e^-$ cross sections.

Existing data on elastic $e^- p$ and $e^+ p$ scattering are sparse,
although some constraints exist from early measurements at SLAC
\cite{Bro65, Mar68}, Cornell \cite{And66}, DESY \cite{Bar67} and
Orsay \cite{Bou68}, as discussed in Sec.~\ref{sec:positron}.
The data are predominantly at low $Q^2$ and at forward scattering
angles, corresponding to large $\eps$ ($\eps \gtorder 0.7$), where
the TPE contributions to the cross section are small ($\ltorder 1\%$).
Nevertheless, the overall trend in the data reveals a small
enhancement in $R^{e^+e^-}$ at the lower $\eps$ values, as shown
in Fig.~\ref{fig:epluseminus}.

\begin{figure}[htb]
\begin{center}
\includegraphics[width=8.1cm]{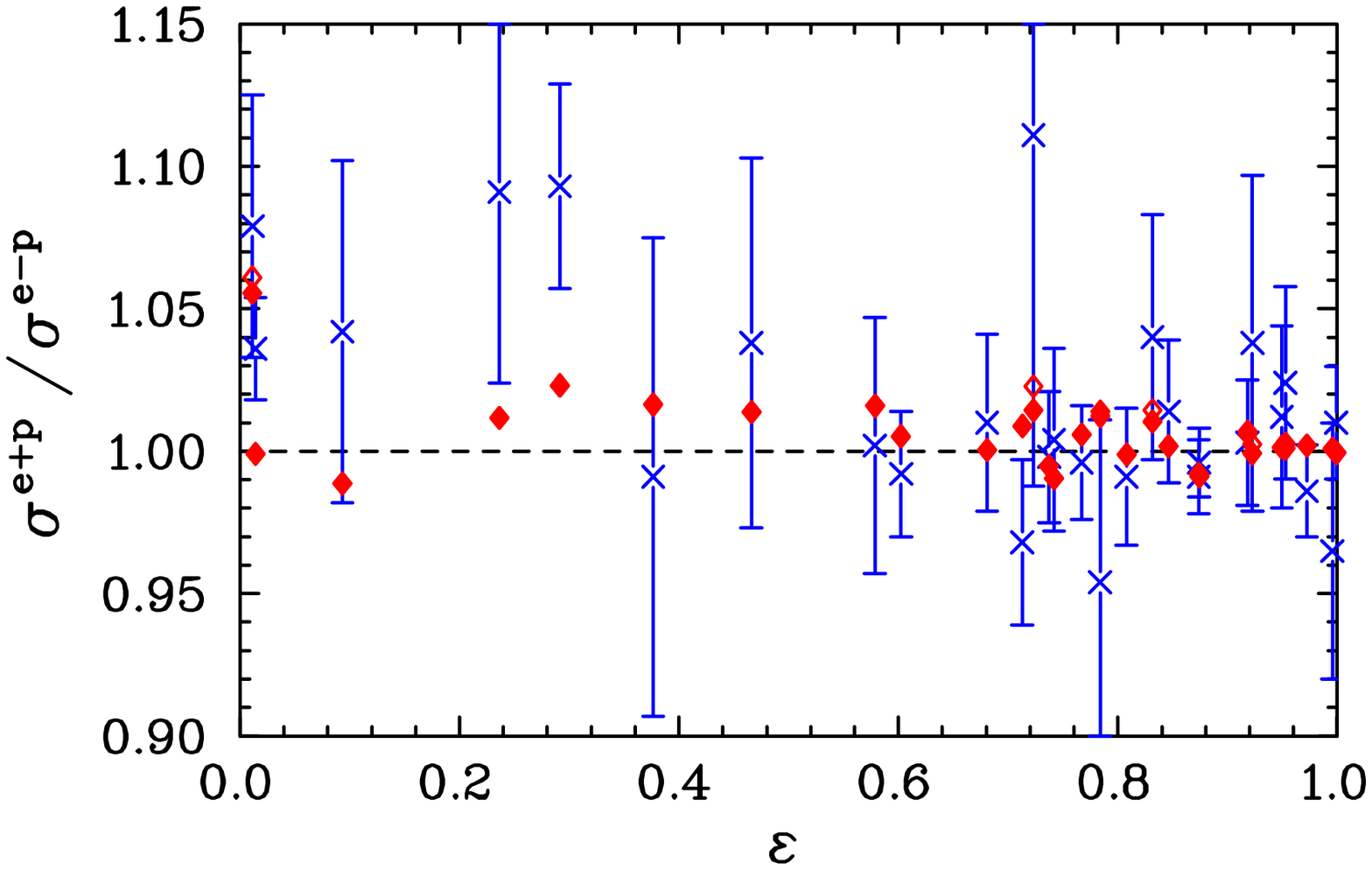}
\hspace*{0.2cm}
\includegraphics[width=7.8cm]{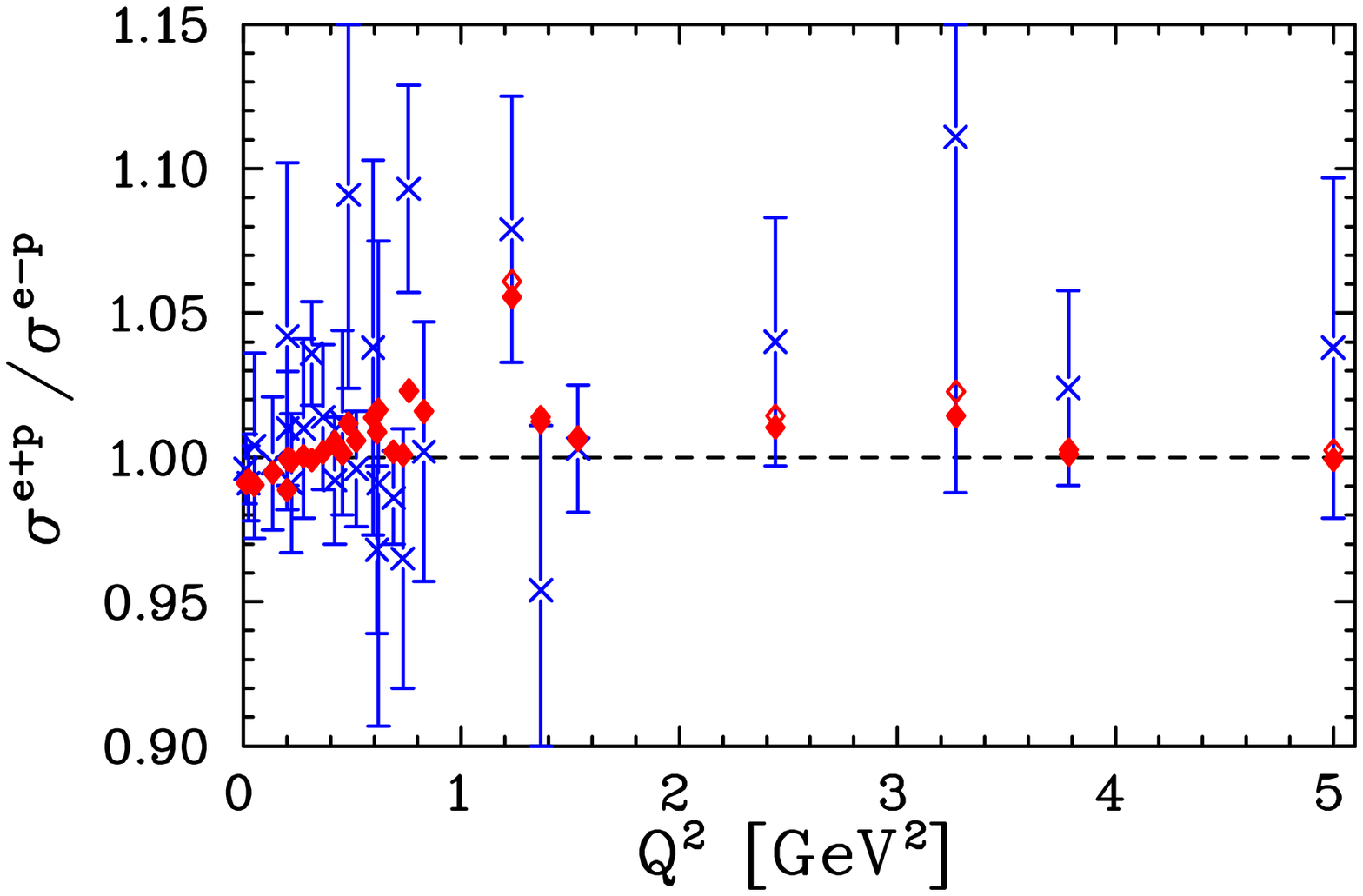}
\begin{minipage}{16.5cm}
\caption{Measured ratio of elastic $e^+ p$ to $e^-p$ cross sections
	(crosses), compared to the hadronic TPE calculations
	\cite{Blu05} (diamonds), presented as a function of $\eps$
	{\bf (left)} and $Q^2$ {\bf (right)}.}
\label{fig:Ree}
\end{minipage}
\end{center}
\end{figure}

The measurements of $R^{e^+e^-}$ are compared to the hadronic TPE
calculations \cite{Blu05} in Fig.~\ref{fig:Ree}, which generally
predict a slight enhancement in the ratio at the experimental
kinematics.  Overall, the TPE corrections agree reasonably well with
the data, although the experimental uncertainties are quite large,
especially where there are indications of a nonzero TPE effect.
Interestingly, the GPD-based calculations \cite{Che04, Afa05}
predict a suppression of $R^{e^+e^-}$ at large $\eps$ in the
$Q^2 \sim$~few GeV$^2$ range.  Better quality data, particularly
at backward angles, where an enhancement of up to $\sim 10\%$ is
predicted, are needed for a more definitive test of the TPE mechanism.

In the last few years, three new experiments have been initiated to
compare $e^+p$ and $e^-p$ scattering.  The first took data in 2009
at the VEPP-3 ring in Novosibirsk \cite{VEPP}, which uses an internal
target in a positron/electron storage ring to extract the ratio at
$Q^2=1.6$~GeV$^2$ and $\eps \approx 0.4$ \cite{Nik10}.
The analysis yields a raw $e^+p/e^-p$ ratio of $1.056 \pm 0.011$,
but corrections for charge-dependent bremsstrahlung, estimated to be
$\sim 3\%$, have to be applied \cite{Nik11}.
The completed experiment is statistics limited, with estimated
systematic uncertainties of 0.3\%~\cite{Nik11}, and plans for a
follow-up measurement with higher statistics at similar or somewhat
lower $Q^2$ are being evaluated.

Another experiment~\cite{e07005}, using a mixed beam of $e^+$ and $e^-$
generated via pair production from a secondary photon beam, recently
completed data taking at Jefferson Lab.  In this experiment, $e^- p$
and $e^+ p$ elastic scattering are measured simultaneously, using
detection of both the scattered lepton and struck proton to reconstruct
the initial lepton energy.  Cross sections can be measured for
$0.5 < Q^2 < 2.0$~GeV$^2$ and $\eps \gtorder 0.2$, and the analysis
is currently underway.  The use of a lepton beam with a large range of
energies allows for a mapping out of the $\eps$ dependence at fixed
$Q^2$ values over the kinematic range of the experiment.

Finally, the OLYMPUS experiment~\cite{olympus} will use the DORIS lepton
storage ring at DESY to make measurements with a fixed lepton energy,   
yielding $e^+p/e^-p$ ratios at several points from $Q^2=2.2$~GeV$^2$,     
$\eps=0.35$, to $Q^2=0.6$~GeV$^2$, $\eps=0.9$.  The experiment is
currently being installed and is scheduled to run in 2012.

\subsection{\it Polarization measurements}
\label{ssec:impact_PT}

While the results of the modern TPE calculations give a clear
indication of a sizable correction to the LT-separated data,
the obvious question which arises is to what extent does TPE
affect the polarization transfer results themselves.
The expectation is that since the PT measurements involve ratios
of cross sections, most of the radiative effects, including TPE,
should cancel.

The polarization transfer experiment involves the scattering of
longitudinally polarized electrons from an unpolarized proton
target, with the detection of the polarization of the recoil proton,
$\vec{e} + p \to e + \vec{p}$.  (The analogous process whereby a
polarized electron scatters elastically from a polarized proton
leaving an unpolarized final state gives rise to essentially the
same information.)  The calculation of the TPE corrections to the
cross sections with longitudinally or transversely polarized recoil
protons follows that of Sec.~\ref{ssec:nucleon}, with the spin
traces evaluated using the explicit expression for the spin-vectors
of the incident electron and recoil proton in Eqs.~(\ref{eq:s}) and
(\ref{eq:s_e}).

In the standard radiative corrections using the Mo-Tsai prescription
\cite{Tsa61, Mo69}, the corrections to the transverse and longitudinal
polarization are identical, so that no additional corrections beyond
hard bremsstrahlung are necessary \cite{Max00P}.
For the TPE corrections, in analogy with the unpolarized case,
Eq.~(\ref{eq:delta_gg}), the spin-dependent TPE corrections to the
longitudinal ($\overline\delta_L$) and transverse ($\overline\delta_T$)
cross sections are defined as the finite parts of the TPE contributions
relative to the IR expression from Mo and Tsai \cite{Tsa61, Mo69} in
Eq.~(\ref{eq:deltaIRMoT}), which are independent of polarization,
\be
\overline\delta_{L,T}\
=\ \delta_{L,T}\ -\ \delta_{\rm IR}{\rm (MoT)}\, .
\ee
In the polarization transfer measurements of the form factor ratio
$\mugegm$, one usually measures the ratio of the longitudinal or
transverse cross section to the unpolarized cross section, $P_L$ or   
$P_T$, respectively.  The TPE correction to the polarization transfer
ratio can therefore be incorporated as \cite{Blu05}
\be
{ P_{L,T} \over P_{L,T}^{\, 0} }\
=\ { 1 + \overline\delta_{L,T} \over 1 + \overline\delta },
\label{eq:polcorr}
\ee
where $P_{L,T}^{\, 0}$ are the polarized Born cross sections, and the
correction to the unpolarized cross section $\overline\delta$ is given
in Eq.~(\ref{eq:Delta_dif}).  Polarized target measurements make
similar comparisons of beam-target cross section asymmetries for
different target spin orientations, yielding two different combinations
of longitudinal and transverse spin asymmetries.

\begin{figure}[ht]
\vspace*{1cm}
\begin{center}
\begin{minipage}{8cm}
\includegraphics[height=6.3cm]{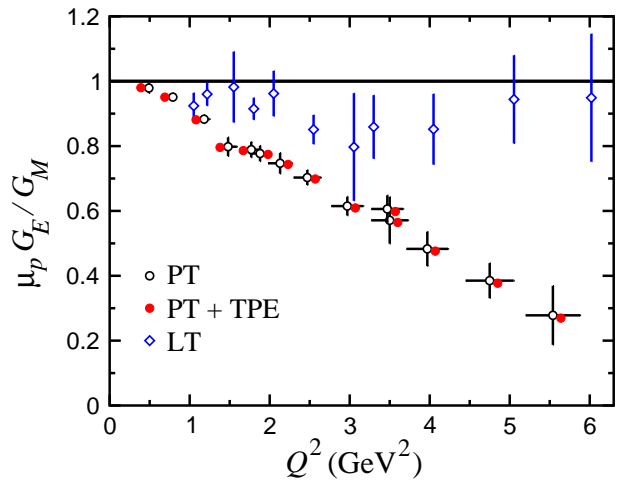}%
\end{minipage}
\begin{minipage}{16.5cm}
\caption{Proton electric to magnetic form factor ratio obtained from the
	PT measurements \cite{Jon00}, with (solid circles) and without
	(open circles) TPE corrections.  The corrected values have been
	offset for clarity.  The LT-separated ratio (open diamonds) from
        Fig.~\ref{fig:GEMp} is shown for comparison.
        Figure adapted from Ref.~\cite{Blu05}.}
\label{fig:GEMpt}
\end{minipage}
\end{center}  
\end{figure}

Taking $R = -\mu_p\sqrt{\tau(1+\eps)/2\eps}\, (P_T/P_L)$ to be the
corrected (``true'') electric to magnetic form factor ratio
(see Eq.~(\ref{eq:Rtrue})), the experimentally measured PT ratio is
\be
\widetilde R\ =\ R
  \left( {1 + \overline\delta_T \over 1 + \overline\delta_L}
  \right).
\label{eq:shiftedR}
\ee
Inverting Eq.~(\ref{eq:shiftedR}), the shift in $R$ is illustrated in
Fig.~\ref{fig:GEMpt}, relative to the uncorrected results and the LT
separated data.  Clearly the effect of TPE on the form factor ratio is
a very small, $\ltorder 3\%$ suppression at the larger $Q^2$ values,
which is well within the experimental uncertainties.
Note that the shift in $R$ in Eq.~(\ref{eq:shiftedR}) does not include
corrections due to hard photon bremsstrahlung, which are part of the
standard radiative corrections.  Since these would increase both the
numerator and denominator in Eq.~(\ref{eq:shiftedR}), the correction
in Fig.~\ref{fig:GEMpt} represents an upper limit on the shift in $R$.

%

\begin{figure}[ht]
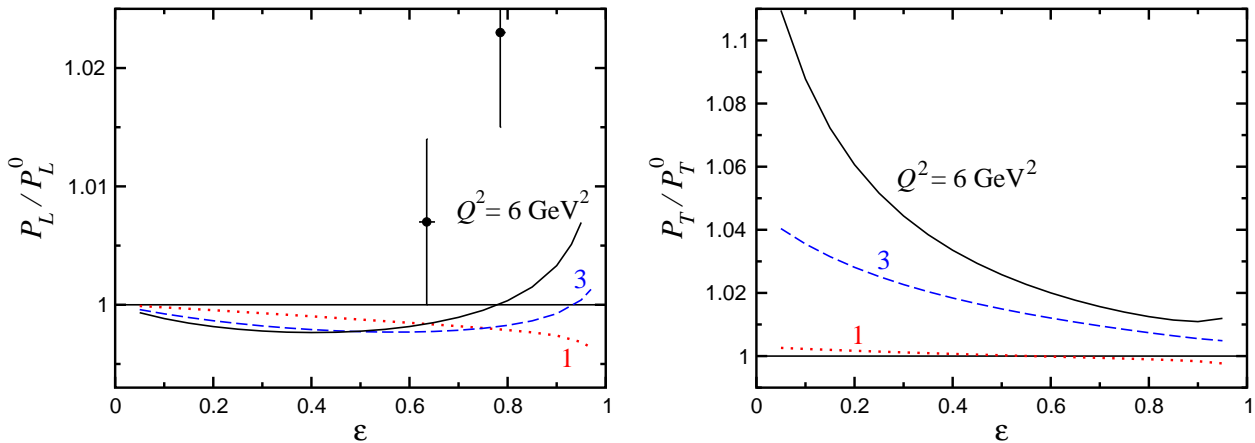

\begin{center}
\begin{minipage}{8cm}
\vspace*{1cm}
\hspace*{-4.5cm}\includegraphics[height=5.8cm]{Figs/fig25a.eps}%
\hspace*{0.5cm}\includegraphics[height=5.8cm]{Figs/fig25b.eps}%
\end{minipage}
\begin{minipage}{16.5cm}
\caption{Ratio of the finite part (with respect to the Mo-Tsai IR
	contribution (\ref{eq:deltaIRMoT})) of the Born+TPE correction
        relative to the Born term, for
        {\bf (left)} longitudinal and   
	{\bf (right)} transverse
        recoil proton polarization, at
        $Q^2 = 1$ (dotted), 3 (dashed) and 6~GeV$^2$ (solid).
	The longitudinal polarization data are from Ref.~\cite{Mez11},
	and have an overall normalization uncertainty of 0.01.
	Figure adapted from Ref.~\cite{Blu05}.}
\label{fig:delLT}
\end{minipage} 
\end{center}
\end{figure}

Although the TPE effects on the PT ratio $R$ are very small, this is
mostly because those measurements were typically performed at large
$\eps$ ($\eps \approx 0.7-0.8$) \cite{Jon00}, where the TPE corrections
are minimal.  At lower $\eps$, or for $\eps \sim 1$, the effects on
the polarizations can be significant.  This can be seen in
Fig.~\ref{fig:delLT}, which shows the longitudinal and transverse
polarizations from nucleon elastic intermediate states relative to the
Born terms.  The correction $\overline\delta_L$ to the longitudinal
cross section is approximately the same as the correction
$\overline\delta$ to the unpolarized cross section, so that TPE
correction to the longitudinal polarization $P_L$ is extremely small.
In fact, at extreme backward angles ($\eps=0$) the unpolarized and
longitudinal corrections $\overline\delta$ and $\overline\delta_L$ must
be identical \cite{Blu05}, and there is no TPE contribution to $P_L$.
In contrast, the combined effects of an increasing $\overline\delta_T$
at large $Q^2$ and $\overline\delta$ becoming more negative mean that
the correction to the transverse polarization $P_T$ is enhanced at 
backward angles, and grows with $Q^2$, as Fig.~\ref{fig:delLT}
illustrates.

In a recent Jefferson Lab experiment the ratios $R$ and $P_L/P_L^{\, 0}$
were measured at three $\eps$ values for a fixed $Q^2=2.5$~GeV$^2$ 
\cite{Mez11}.  In the Born approximation both of these should be
independent of $\eps$, so that any observed angular dependence would be
an indication of TPE contributions.  The results for the longitudinal
polarization ratio in Fig.~\ref{fig:delLT}~(left) are consistent with
no TPE effect at $\eps=0.635$, but indicate an $\approx 2\%$ enhancement
at the larger $\eps=0.785$ value (note that an overall normalization
uncertainty of 0.01 applies to the data).  While a rise as $\eps \to 1$
is predicted at larger $Q^2$, the magnitude of the effect is difficult
to reconcile with the calculation \cite{Blu05} and possibly suggests the
need to include higher-mass intermediate states in the TPE amplitude.
The results in the GPD-based approach \cite{Che04, Afa05} give
a similarly small effect, but with opposite sign compared with
the hadronic calculation \cite{Blu05}.

Data taken on the PT ratio $R$ at $Q^2=2.5$~GeV$^2$ are also
consistent with no significant $\eps$ dependence over the
range $0.15 \ltorder \eps \ltorder 0.8$, and hence in the
transverse ratio $P_T/P_T^{\, 0}$.  The hadronic calculation with
elastic nucleon intermediate states predicts an $\approx 2-3\%$
enhancement, which once again may indicate a role played by the
higher-mass resonances; the $\Delta(1232)$ contribution, for example,
may cancel some of the rise at $\eps \to 0$ (Sec.~\ref{sssec:Delta}).
It is interesting to observe that the GPD-based model predicts a
large {\it decrease} in the ratio $R$ (or transverse polarization)
at low $\eps$, opposite to the hadronic model.  Further measurements
of the $\eps$ dependence of the $P_T/P_T^{\, 0}$ ratio over a range
of $Q^2$ values would be very helpful in constraining the magnitude
and sign of the TPE effects.

Finally, two recent analyses \cite{Bor11a, Gut11} have attempted to
extract from the new data individual contributions to TPE amplitudes.
Because data exist for a limited set of $Q^2$ values, and because the
effect on the cross section is only inferred from the discrepancy
between LT and PT measurements, such analyses at present require
assumptions about the $\eps$ dependence of the amplitudes.
Although different decompositions of the amplitudes are used in
Refs.~\cite{Bor11a, Gut11}, the extracted amplitudes are found to
be similar when the different definitions are accounted for.

\subsection{\it Global form factor analysis}
\label{ssec:global}

With the realization that the Rosenbluth separations of elastic cross
section data were significantly affected by TPE, came the need to ensure
that for reliable extractions of the nucleon form factors the TPE
corrections must be incorporated in the analysis.  Initial efforts
focused on minimizing the impact of TPE in the extractions, based on
the assumption that TPE was the likely explanation for the LT/PT form
factor discrepancy, and that all observables (cross sections and
polarization components) were affected at a similar level.
Given that there are now calculations of TPE corrections that can explain
the discrepancy and are consistent with all experimental constraints on
TPE contributions, more recent efforts have focused on correcting the data
using calculated TPE contributions in the combined analysis of cross section
and polarization measurements.

\subsubsection{Phenomenological analyses of Rosenbluth and polarization data}

Because the Rosenbluth extractions at high $Q^2$ are very sensitive
to any small, angle dependent systematic effects, it was generally
believed that the polarization measurements provided a more reliable
way to extract $\ge$ at large $Q^2$, where it has minimal contribution
to the cross section.  Section~\ref{sec:estimates} discussed various
attempts to use the discrepancy to extract the TPE contributions to
the data.  However, when combining these measurements to extract the
form factors, rather than the TPE, the exact result is sensitive to
the way these inconsistent measurements are combined.

Many early attempts to extract form factors using both unpolarized
cross section and polarization transfer measurements simply assumed
that $\gm$ could be reliably extracted from the Rosenbluth results,
and $\ge$ could best be extracted using polarization measurements
of $\gegm$, combined with the Rosenbluth extractions of $\gm$.
This corresponds to the assumption, illustrated in
Fig.~\ref{fig:delta_sigma}, that whatever was deficient in the
cross section measurements led to a linear $\eps$ dependence which
had no impact at $\eps=0$, and thus the value of $\gm$ extracted
from Rosenbluth separations was unmodified.
However, if the discrepancy is caused by TPE contributions, then it is
$\eps=1$ where these contributions must be negligible, and thus there
will be both a large change in $\ge$ and a smaller change in $\gm$.
While the size of the discrepancy suggests that the impact on $\gm$
would be on the order of $2-3\%$, this is large compared to the
precision with which $\gm$ extractions were quoted and thus needs
to be taken into account.

In practice, different combined analyses made assumptions that led to
somewhat different results.  Because the polarization measurements were
only available with precision at high $Q^2$, some analyses~\cite{Kel02}
used $\gm$ from LT separations, while taking $\ge$ from LT separations
at low $Q^2$ and polarization measurements at high $Q^2$.  A slightly
different approach was taken in Ref.~\cite{Bra02}.  Rather than fitting
the extracted values of $\ge$ and $\gm$ from a mix of cross section
and polarization measurements, the slope of the reduced cross section
was fixed from polarization measurements, and $\gm$ varied to best
match the reduced cross section over the entire $\eps$ range.  This
corresponds to taking the slope predicted by polarization transfer in 
Fig.~\ref{fig:delta_sigma} and varying the normalization so that it
gives the best reproduction, on average, of the cross section measurement.
In this case, the slope will match the data at a weighted average value
of $\eps$, typically reproducing the cross section at a point in
$0.5 < \eps < 0.8$.  This is closer to what one expects from TPE,
where the agreement would occur at $\eps=1$, but still yields a
value of $\gm$ that is too small.

One issue of concern is the fact that the TPE-corrected form factors
will no longer provide a good parametrization of the form factors 
unless TPE contributions are explicitly included.  
This scenario can be seen from Fig.~\ref{fig:delta_sigma}, where $\gm$
is determined from the extrapolation of the cross section measurements
to $\eps=0$, and then the polarization measurements used to give the
$\eps$ dependence of the reduced cross section.
The form factors thus obtained, if treated in the Born approximation,
will yield the correct cross section only for scattering at 180$\deg$,
and produce the maximum deviation from the measured cross section at
small angles, where a large fraction of the high-precision measurements
is taken.  If these form factors are used as input in the analysis of
other measurements, such as when calculating the $ep$ elastic cross
section in the analysis of quasi-elastic proton knockout from a nucleus
\cite{Arr04a, Arr04c, Dut03} and the deuteron elastic form factors
\cite{Ada07}, then this error in the cross section can yield incorrect
results.

Note that this is not related to assumptions or errors in the
extraction.  The fact that TPE contributions are larger than
previously thought means that it is not sufficient to take the
Born form factors when one needs precise knowledge of the elastic
cross section; the additional TPE contributions must be included
along with the traditional radiative corrections to obtain the
observed cross section.  Because different analyses use different
approaches to correct for the TPE contributions, care must be taken
to apply \textit{consistent} TPE corrections when extracting the
form factors and obtaining the cross sections from these.
One way to avoid these issues is to provide both an extraction of
the TPE-corrected form factors and a fit to the total (Born+TPE)
elastic cross section~\cite{Arr04a, AMT07, Arr07b}.  This provides
a parametrization of the cross section that includes the observed
TPE contributions.

\subsubsection{Extraction of form factors including TPE}
\label{sec:amt}

Given that modern TPE calculations appear able to resolve most or all of
the form factor discrepancy, the next step is to apply these corrections
to the experimental observables and see if the results are consistent.
If so, a combined extraction of the form factors from all available data
can be performed.

\begin{figure}[thb]
\begin{center}
\includegraphics[width=6.25cm]{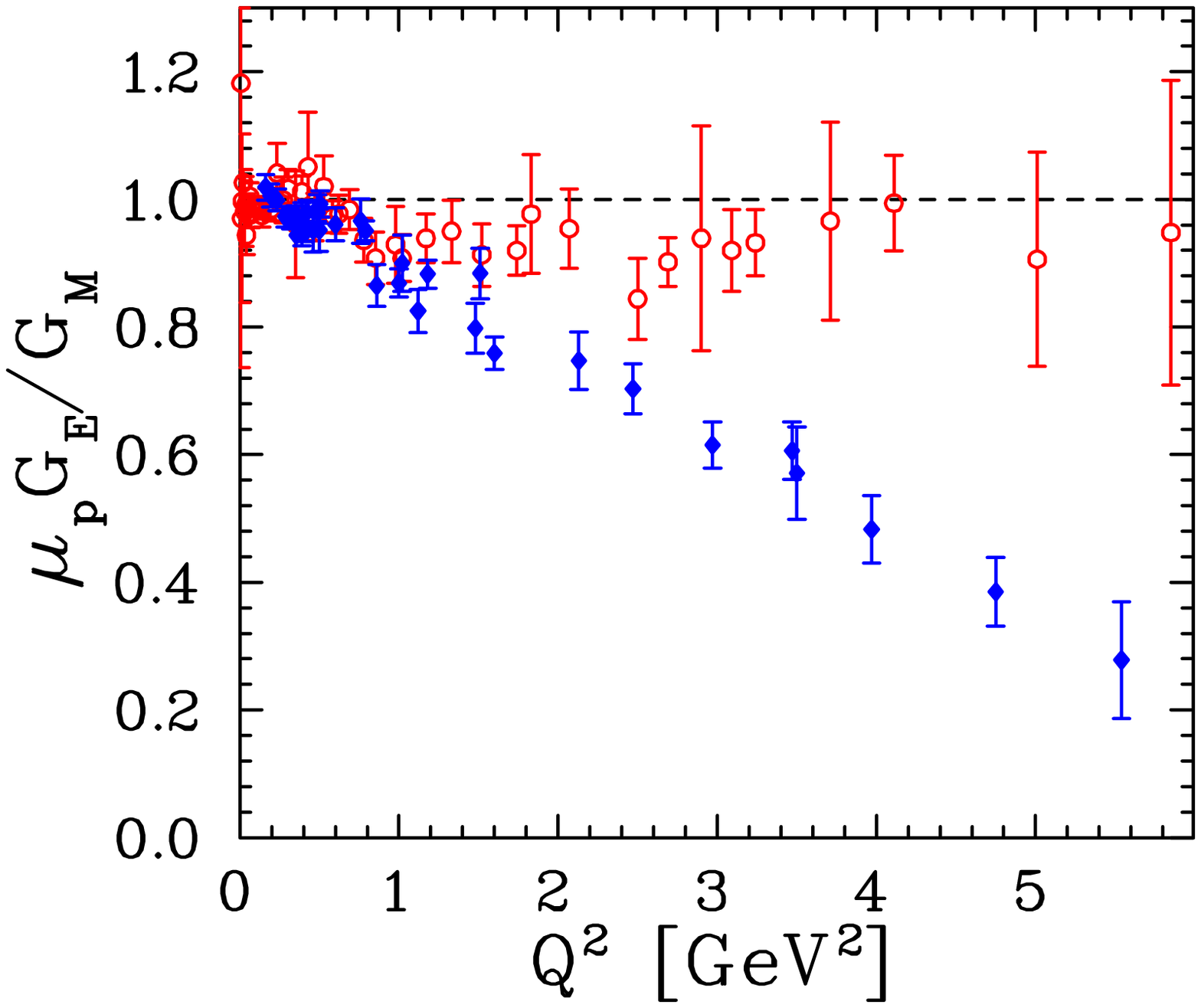}\hspace*{-0.2cm}
\includegraphics[width=5.2cm]{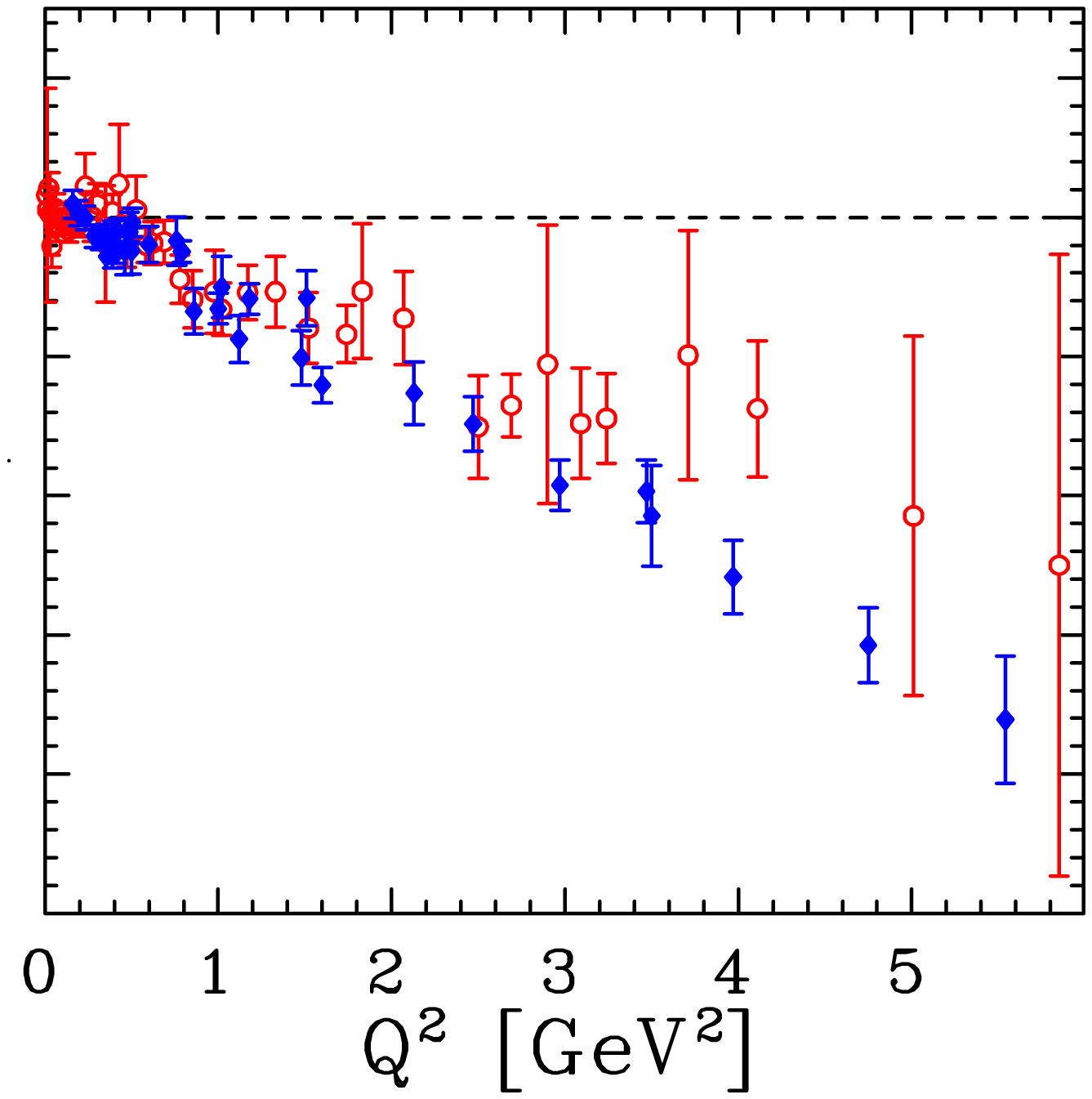}\hspace*{-0.2cm}
\includegraphics[width=5.2cm]{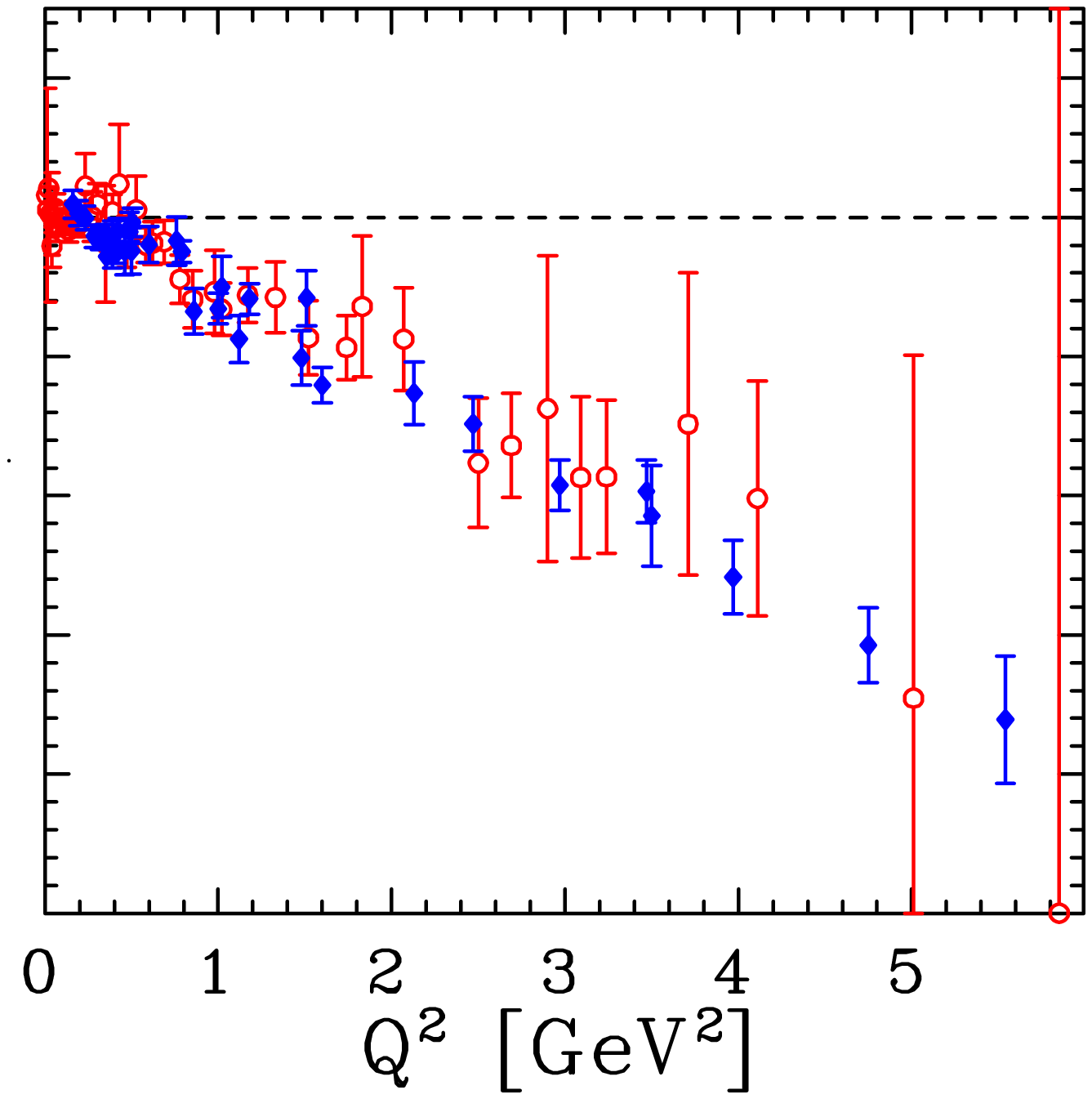}
\begin{minipage}{16.5cm}
\caption{Comparison of polarization measurements (filled diamonds)
	and LT separations (open circles) with no TPE corrections
	{\bf (left)}, TPE corrections from Ref.~\cite{Blu05}
	{\bf (center)}, and with the additional high-$Q^2$ correction
	applied in Ref.~\cite{AMT07} {\bf (right)}.}
\label{fig:AMT_compare}
\end{minipage}
\end{center}
\end{figure}

Such an analysis was performed in Ref.~\cite{AMT07} using the hadronic
TPE calculations with nucleon elastic intermediate states~\cite{Blu05},
as outlined in Sec.~\ref{ssec:nucleon}.  The hadronic TPE corrections
have the advantage that they are expected to be reliable for all
$\eps$ values at low $Q^2$, which is the kinematic region where
many of the extremely high precision measurements have been taken.
At higher $Q^2$ values, the calculations is expected to be less
complete, and it is important to test the corrections against the
observed discrepancy.
For $Q^2 \ltorder 2-3$~GeV$^2$, the TPE corrections to the cross
section bring the form factor ratio extracted from Rosenbluth separations
into excellent agreement with the polarization transfer measurements
(middle panel of Fig.~\ref{fig:AMT_compare}).  At higher $Q^2$ there
is a small residual, systematic disagreement.  Here a small additional
phenomenological correction was applied, linear in $\eps$ and with
magnitude comparable with estimates \cite{Kon05, Che04, Kon07} of
higher-mass intermediate state contributions (right panel of
Fig.~\ref{fig:AMT_compare}).

For the combined analysis of cross section and polarization measurements,
TPE corrections were applied to the extracted cross sections.  Most
experiments assumed a 1--1.5\% uncertainty due to radiative corrections,
with the dominant contribution coming from TPE.  Clearly, this was an
underestimate of the uncertainty when no TPE corrections were applied,
and was taken to be an appropriate uncertainty after applying the
hadronic correction of the TPE effects.  For the additional TPE
contribution associated with higher-mass intermediate states, 100\%
of the correction was applied as an additional uncertainty to the cross
section to reflect the impact of the poorly constrained TPE corrections at
high $Q^2$.  While the TPE calculation~\cite{Blu05} provides predictions
for the impact on the polarization transfer measurement, most of the
data are at large $\eps$, where the impact is extremely small compared
to the statistical uncertainties of the measurements.  In addition,
although the hadronic and partonic calculations yield similar results for
the correction to the unpolarized cross sections, they yield corrections
to the polarization measurements with opposite sign at large $Q^2$.
The analysis \cite{AMT07} therefore did not to include any TPE
corrections to the polarization measurement, as either calculation
would have had an extremely small impact on the final result.

\begin{figure}[thb]
\begin{center}
\includegraphics[width=8cm,angle=0]{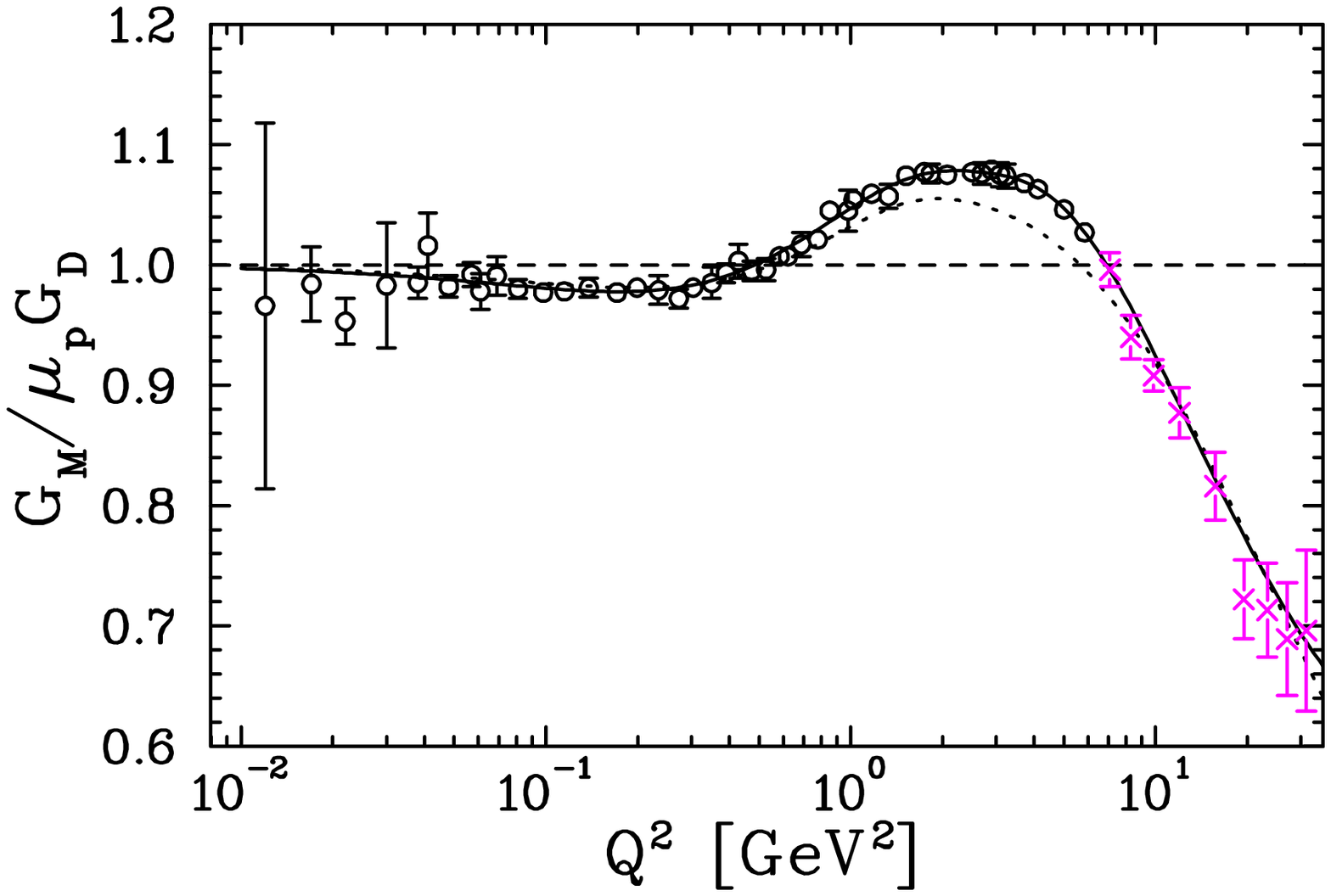}\hspace*{0.3cm}
\includegraphics[width=8cm,angle=0]{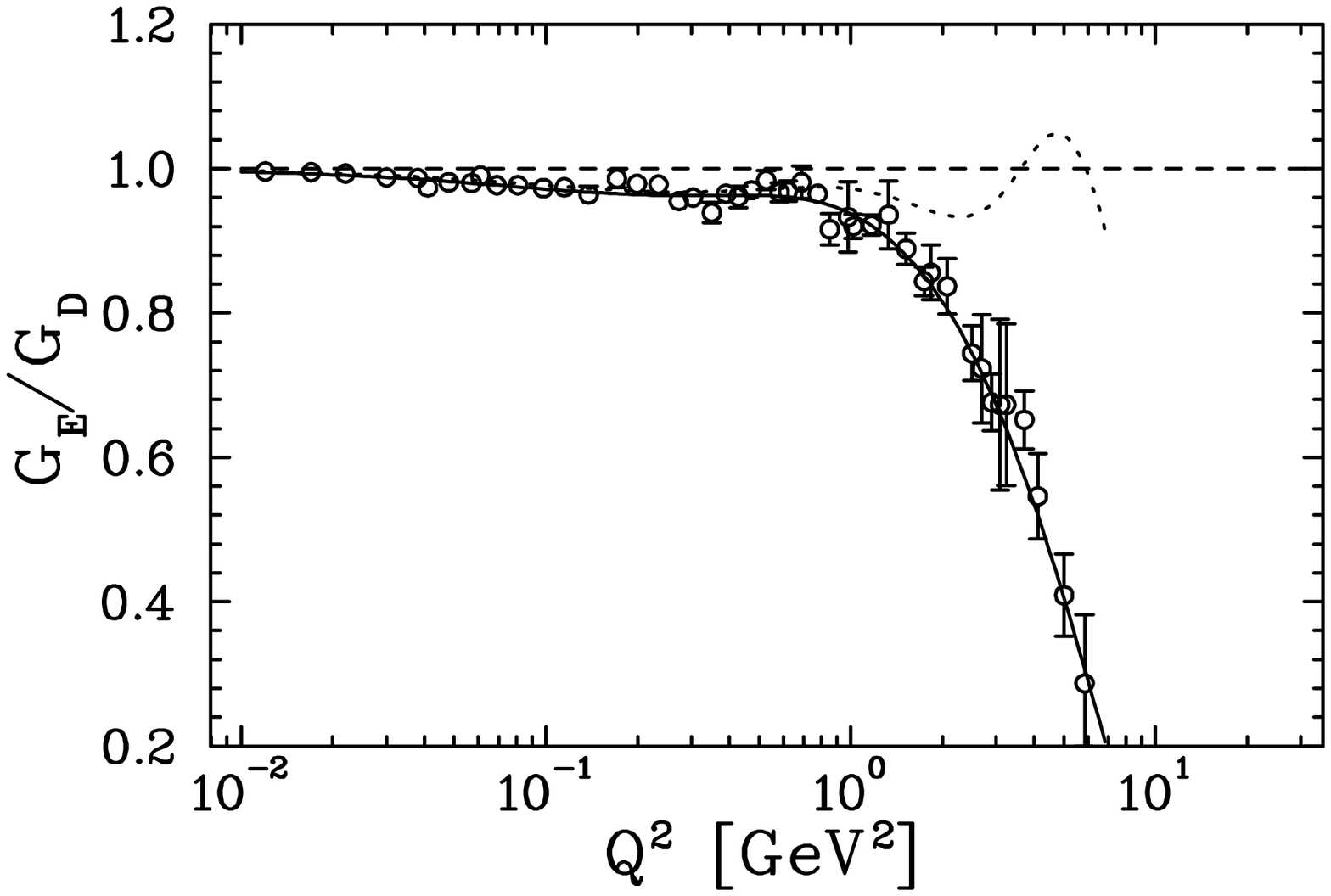}
\begin{minipage}{16.5cm}
\caption{Proton $\gm$ {\bf (left)} and $\ge$ {\bf (right)} form factors
	from the global fit in Ref.~\cite{AMT07}, scaled by the dipole
	form factor (\ref{eq:GD}).  Note that above $Q^2=6$~GeV$^2$
	there were no direct measurements of $\ge$ and $\gm$ was
	extracted from the cross section data with an additional
	uncertainty related to the uncertainty in $\ge$.  The solid
	line is the final fit, while the points are from direct
	extractions from data in small $Q^2$ bins.  The short-dashed
	line shows the fit from the cross sections with no TPE
	corrections applied.}
\label{fig:AMT_final}
\end{minipage}
\end{center}
\end{figure}

The extracted form factors from the combined analysis of polarization
measurements and TPE-corrected cross sections is shown in 
Fig.~\ref{fig:AMT_final}.  Since this publication, there have been
updated polarization results at high $Q^2$~\cite{Puc10, Puc11} and very
low $Q^2$~\cite{Ron07, Zha11, Ron11, Ven10}, as well as an extensive set 
of cross section measurements at low $Q^2$~\cite{Ber10}.  The global fit
of Ref.~\cite{AMT07} has been updated in Refs.~\cite{Zha11, Ven10} to
include the new polarization measurements, and inclusion of the new
cross section measurements~\cite{Ber10} and a detailed evaluation of
the uncertainties is in progress~\cite{Zha11b}.

\subsubsection{Impact on the extracted charge and magnetization radii
	of the proton}

The proton charge radius is related to the low $Q^2$ behavior of the
charge form factor.  In both electron scattering and atomic physics
extractions, the root-mean-square radius $R_E$ is defined in terms
of the slope of the form factor at $Q^2=0$,
\be
\ge(Q^2)\ =\ 1\ -\ {Q^2\, R_E^2 \over 6}\ +\ \cdots;\ \ \ \ \ \ \
R_E^2\ =\ -6 \left. {dG_E \over dQ^2} \right|_{Q^2=0} .
\ee
Because the TPE corrections are finite and have a significant $Q^2$
dependence in the limit $Q^2 \to 0$, as seen in Fig.~\ref{fig:del_gg},
they can impact the electron scattering extractions of the charge
radius.  These corrections were first included in the second Born
approximation by Rosenfelder~\cite{Ros00}, who found that they
increase the charge radius by about 0.01~fm.  More recently, the full
hadronic calculations were evaluated~\cite{Blu05S} and compared to the
Coulomb distortion correction~\cite{Ros00}.  The additional impact of
the TPE was very small, yielding an additional increase of 0.002~fm.
Since $\ge$ dominates the cross section at low $Q^2$, it can be extracted
reliably without requiring measurements at extremely large scattering
angles.  Therefore, it is not entirely surprising that the charge radius
is relatively insensitive to the difference between the second Born
approximation and the full hadronic TPE calculation which is most
important at larger scattering angles, as shown in Fig.~\ref{fig:2born}.

In contrast, the magnetic form factor becomes increasingly difficult
to extract from cross section measurements at very low $Q^2$.
In a Rosenbluth separation, one must extrapolate to $\eps=0$, where
the reduced cross section is proportional to $Q^2$ for $Q^2 \to 0$.
This makes the extraction sensitive to small $\eps$-dependent
corrections, enhancing the sensitivity to the TPE corrections, as well
as the difference between the second Born approximation and the full
calculation.  There is a significant difference between two recently
extracted values of the proton magnetization radius.  The first obtains
an RMS magnetic radius of 0.777$\pm$0.017~fm~\cite{Ber10} from a large
body of new, high-precision cross section measurements.  The second
finds 0.867$\pm$0.020~fm~\cite{Zha11} from a fit combining previous
cross section measurements with new low-$Q^2$ polarization
measurements~\cite{Zha11, Ron11}.  The latter analysis includes
hadronic TPE corrections for all cross section measurements,
while the former includes the Coulomb distortion correction of
Ref.~\cite{Mck48}, corresponding to the $Q^2=0$ limit of the
soft-photon approximation.  This is an overestimate of the correction
at all measured $Q^2$ values, and entirely neglects the $Q^2$
dependence of the correction, which is important in the extraction
of the magnetic radius. An estimate of the impact of a more complete
TPE correction~\cite{Arr11b} suggests that this could have a large
impact on the form factor and magnetic radius extracted in
Ref.~\cite{Ber10}.

\subsection{\it Normal asymmetries}
\label{ssec:normal}

The TPE exchange process gives rise to a nonzero contribution to
the elastic cross section for a recoil proton polarized normal
to the scattering plane. 
By time reversal invariance this is also equivalent to scattering
(unpolarized) electrons from a target polarized normal to the
scattering plane \cite{DeR71}.
Normal polarization observables vanishes in the Born approximation,
and their measurement directly accesses the imaginary part of the
TPE amplitude.
A detailed discussion of the formalism and experimental status of
normal asymmetry measurements can be found in Ref.~\cite{Car07}.
%

The target normal asymmetry $A_N$ is defined as \cite{Car07, DeR71,
Afa05C}
\be
A_N\ =\ { \sigma^\uparrow - \sigma^\downarrow \over
	  \sigma^\uparrow - \sigma^\downarrow },
\label{eq:Anormal_def}
\ee
where $\sigma^{\uparrow(\downarrow)}$ is the cross section for
unpolarized electrons scattering from a proton target with spin
parallel (antiparallel) to the direction normal to the scattering
plane, defined by the spin vector
$\zeta_N = \bm{k} \times \bm{k}'/|\bm{k} \times \bm{k}'|$
(see Eq.~(\ref{eq:s})).%
\footnote{Equivalently, by time reversal invariance, $A_N$ would be the
asymmetry for scattering unpolarized electrons from an unpolarized
proton target, with the recoil proton polarized normal to the
scattering plane.}
At order $\alpha$ in the electromagnetic coupling, the target normal
asymmetry is given by \cite{DeR71}
\be
A_N\ =\ { 2\, \Im\left( {\cal M}_\gamma^*\, {\cal M}_{\gamma\gamma} \right)
	\over |{\cal M}_\gamma|^2 },
\label{eq:Anormal}
\ee
providing a direct measure of the imaginary part of
${\cal M}_{\gamma\gamma}$.
%

\begin{figure}[ht]
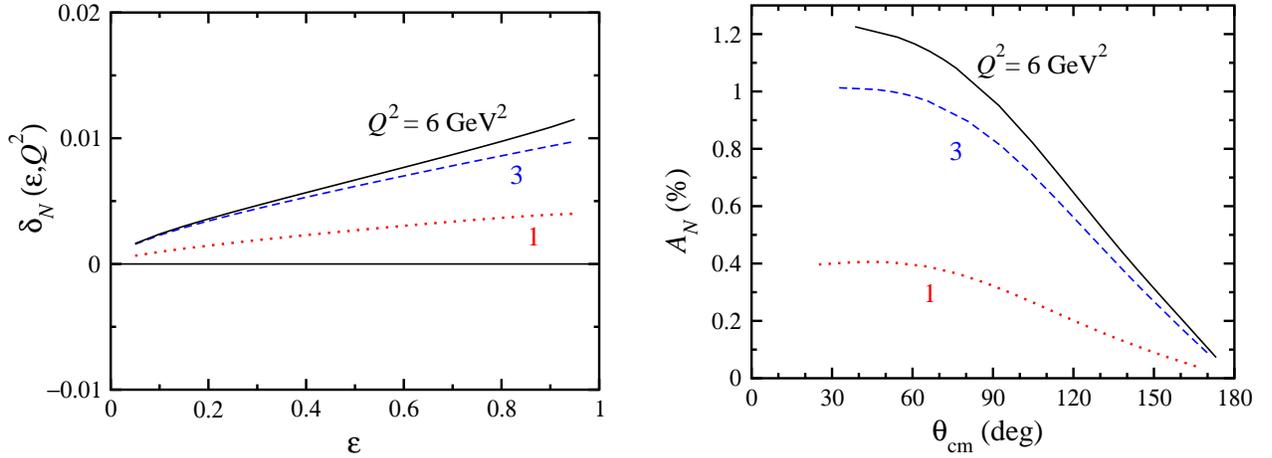

\vspace*{1cm}
\begin{center}
\includegraphics[height=6cm]{Figs/fig28a.eps}
\hspace*{0.5cm}
\includegraphics[height=6cm]{Figs/fig28b.eps}
\begin{minipage}{16.5cm}
\caption{
{\bf (Left)}
	Ratio of the TPE normal polarization correction to the
	unpolarized Born contribution as a function of $\eps$,
	for $Q^2 = 1$ (dotted), 3 (dashed) and 6~GeV$^2$ (solid).
{\bf (Right)}
	Target normal polarization asymmetry, as a percentage,
        as a function of the center of mass scattering angle,
        $\theta_{\rm cm}$, for $Q^2=1$ (dotted), 3 (dashed) and
        6~GeV$^2$ (solid).
	Figures adapted from Ref.~\cite{Blu05}.}
\label{fig:Anormal}
\end{minipage}  
\end{center}
\end{figure}

In Fig.~\ref{fig:Anormal}~(left) the ratio $\overline\delta_N$ of
the TPE normal contribution from nucleon elastic intermediate states
relative to the unpolarized Born cross section is shown as a
function of $\eps$ for several values of $Q^2$.
Note that there is no IR contribution to the normal polarization.
The normal polarization contribution is very small numerically,
$\delta_N \ltorder 1\%$, and has a very weak $\eps$ dependence.
In contrast to the TPE longitudinal and transverse polarization
corrections $\delta_L$ and $\delta_T$ in Sec.~\ref{ssec:impact_PT},
the normal polarization ratio is smallest at low $\eps$, becoming
larger with increasing $\eps$.

The target normal asymmetry $A_N$ in Eq.~(\ref{eq:Anormal}) is shown
in Fig.~\ref{fig:Anormal}~(right) as a function of the center of mass
scattering angle, $\theta_{\rm cm}$, for several values of $Q^2$.
The asymmetry is relatively small, of the order of 1\% at low
$\theta_{\rm cm}$ for $Q^2 \sim 3$~GeV$^2$, but grows with $Q^2$
\cite{Blu05}.
The magnitude of the elastic contribution to $A_N$ is relatively model
independent (see also Refs.~\cite{Afa05, DeR71}), as it is determined
mostly by the (on-shell) proton electromagnetic form factors, which
are reasonably well determined.
Estimates of contributions from higher-mass intermediate states
have been made within the GPD approach \cite{Afa05}, and suggest
that these may be as significant as the elastic at forward angles
$\theta_{\rm cm}$, but very small at backward angles.

Unfortunately no data exist on target normal asymmetries for
elastic scattering on a normal-polarized target.
Experiments have been performed at Jefferson Lab to measure the
target normal asymmetry of the neutron on a $^3$He target for
deep-inelastic scattering~\cite{e07013} and quasi-elastic neutron
knockout~\cite{e08005}, and the data are currently under analysis.
Note that the observation of nonzero effects here would not necessarily
signal TPE effects, as there are likely to be nuclear final state
interaction effects which can also produce a nonzero asymmetry.
A recent search for target normal asymmetries in deep-inelastic
scattering at HERMES~\cite{Air10} did not find a TPE signal within
the $\sim$10$^{-3}$ uncertainties of the measurement.

Although not directly relevant to the elastic form factor extraction,
the observation of nonzero target normal asymmetries would provide
direct evidence of TPE in elastic scattering.
Knowledge of the imaginary part of the TPE amplitude could be used
to constrain models of Compton scattering, or as input into dispersion
relations to obtain the real part of the TPE amplitude from the
imaginary part (see Sec.~\ref{ssec:DRgg}).

The imaginary part of the TPE amplitude can also be accessed by
measuring the electron beam asymmetry for electrons polarized normal
to the scattering plane scattering from unpolarized targets.  
The corresponding beam normal asymmetry $B_N$ is then defined
analogously to Eq.~(\ref{eq:Anormal_def}), with the electron spin
parallel or antiparallel to the normal polarization vector $\zeta_N$.
Since the beam normal asymmetry involves flipping the helicity of
the electron, is it zero in the limit $m_e \to 0$.  Several such
measurements have been made in elastic $ep$ scattering~\cite{Wel01,
Maa05, Arm07} in connection with measurements of parity-violating
elastic scattering.  The most recent result~\cite{And11} includes
both $ep$ and $ed$ scattering at backward angles, and the results
indicate the importance of including inelastic contributions in the
intermediate state.
%

%% file: section6.tex
\section{Two-photon exchange in other reactions}
\label{sec:other}

While the effect of TPE has been most dramatically illustrated for 
extraction of proton electric and magnetic form factors using the LT 
separation method, TPE has also been studied for its effect on the 
extraction of other observables, ranging from elastic neutron and
transition form factors, to the pion and nuclear form factors.

\subsection{\it Neutron form factors}

\begin{figure}[ht]
\vspace*{0.5cm}
\begin{center}
\includegraphics[height=6cm]{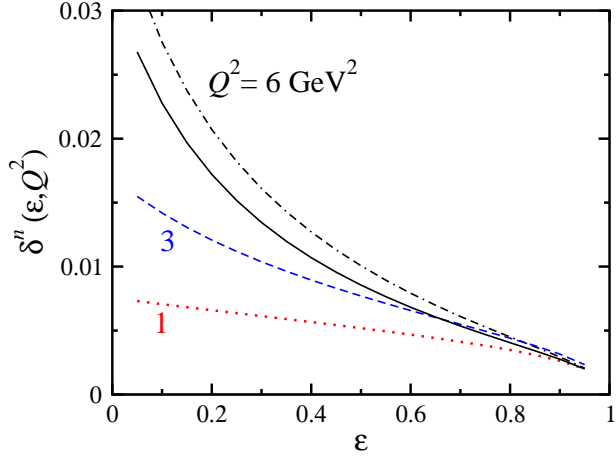}
\begin{minipage}{16.5cm}
\caption{TPE correction $\delta^n$ to the unpolarized electron--neutron
        elastic scattering cross section at $Q^2=1$ (dotted),
        3 (dashed) and 6~GeV$^2$ (solid and dot-dashed).
        The dot-dashed curve corresponds to the form factor
        parametrization of Ref.~\cite{Bos95}, while the others
	are from Ref.~\cite{Mer96}.
	Figure adapted from Ref.~\cite{Blu05}.}
\label{fig:del_n}
\end{minipage}
\end{center}
\vspace*{-.5cm}
\end{figure}

Because the magnitude of the electric form factor of the neutron is
relatively small compared to that of the proton, especially at low
$Q^2$, the effects of TPE may be even more pronounced for $\gen$ than
for $\gep$.  Within the hadronic framework of Sec.~\ref{sec:TPE}, the
relative TPE correction to the neutron elastic cross section is shown
in Fig.~\ref{fig:del_n} as a function of $\eps$ for $Q^2 = 1$, 3 and
6~GeV$^2$ \cite{Blu05}, using input neutron form factors from
Ref.~\cite{Mer96}.  For comparison the correction at $Q^2=6$~GeV$^2$
is also computed with the parametrization from Ref.~\cite{Bos95},
with the difference between these indicative of the model dependence
of the calculation.

Since there is no IR-divergent contribution to the TPE correction
for the neutron, the {\it total} correction $\delta^n$ is displayed
in Fig.~\ref{fig:del_n}.  As discussed in Sec.~\ref{ssec:emu},
for $Q^2 \to 0$ the TPE correction depends only on $F_1(0)$, and is
independent of the anomalous magnetic form factor $F_2$.  In this
limit the TPE correction for the neutron therefore vanishes.

\begin{figure}[bth]
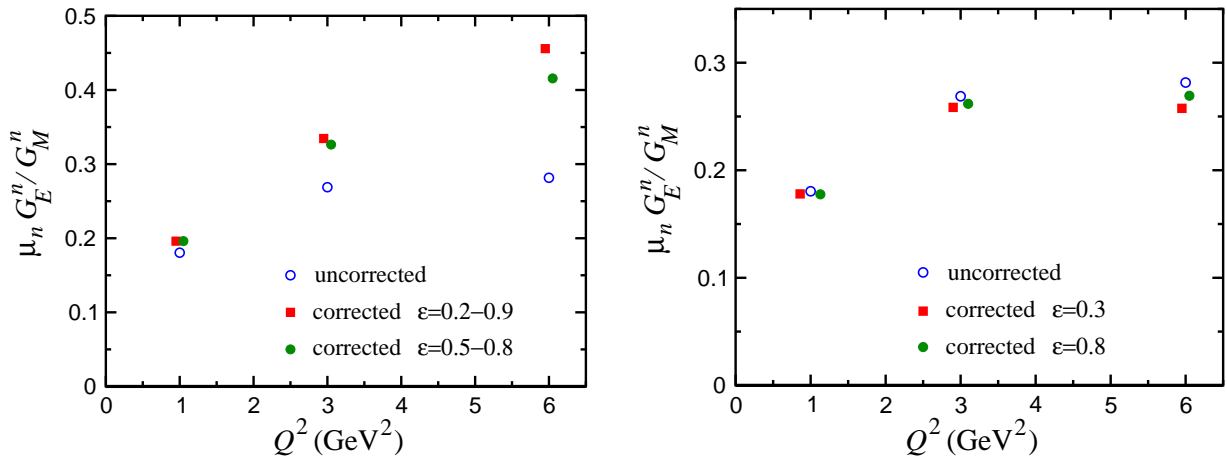

\vspace*{1cm}
\begin{center}
\hspace*{-0.5cm}\includegraphics[height=6cm]{Figs/fig30a.eps}
\hspace*{0.5cm}\includegraphics[height=6cm]{Figs/fig30b.eps}
\begin{minipage}{16.5cm}
\caption{{\bf (Left)}
	Effect of TPE on the ratio of neutron form factors
	$\mugegmn$ using LT separation.
        The uncorrected points (open circles) are shifted by taking
	linear fits to the TPE correction in Fig.~\ref{fig:del_n}
	over the quoted $\eps$ ranges.
	{\bf (Right)}
	TPE effect on polarization transfer measurements.
	The corrected points correspond to $\eps=0.3$
        (filled squares) and $\eps=0.8$
        (filled circles) (offset for clarity).
	Figure taken from Ref.~\cite{Blu05}.}
\label{fig:GEMn}
\end{minipage}
\end{center}
\end{figure}

The most notable difference with respect to the proton
(Fig.~\ref{fig:del_gg}) is the sign and slope of the TPE correction.
In particular, the magnitude of the neutron correction is around 3~times 
smaller than for the proton, and the negative slope arises from the 
negative anomalous magnetic moment of the neutron $\mu_n$ \cite{Blu05}.  
While the correction for the neutron is smaller than that for the proton,
the value of $\gen$ is small at all $Q^2$ values, so the effects of TPE
on an LT separation will be magnified.
Figure~\ref{fig:GEMn}~(left panel) shows the impact that a TPE
correction would have on a Rosenbluth extraction of $\mugegmn$.
The ``uncorrected'' values are taken from a global fit~\cite{Mer96},
and the filled circles show the modification after applying the TPE 
corrections from Fig.~\ref{fig:del_n}, assuming that the initial value 
came from an LT separation covering two different $\eps$ ranges.  
Inclusion of the TPE effect would improve the significance of the
upper limits on $\gen$ from the SLAC Rosenbluth extraction~\cite{Lun93}
up to $Q^2=4$~GeV$^2$, as well as yielding a small shift in $\gmn$.

While the TPE corrections to the form factor ratio from LT separations
are significant at large $Q^2$, the neutron $\gen$ form factor is 
typically extracted from polarization transfer experiments.
To compare the TPE effects on the ratio $\mugegmn$ extracted using the
polarization transfer method, we take the same ``uncorrected'' starting
values for $\mugegmn$, and show the impact of the TPE corrections at
two values of $\eps$, namely $\eps=0.3$ or 0.8 (right panel of
Fig.~\ref{fig:GEMn}).
%
%
The shift in the form factor ratio is considerably smaller than that
from the LT method, and below the present experimental uncertainties,
but nevertheless represents a $3-4\%$ suppression at $Q^2=3$~GeV$^2$
and $5-10\%$ at $Q^2=6$~GeV$^2$.

For kinematics typical of neutron form factor experiments at
Jefferson Lab, the ratio $\gegmn$ was recently measured
in experiment E93-038 \cite{Mad03} at $Q^2=1.45$~GeV$^2$ for
$\eps \approx 0.9$, at which the TPE correction was $\approx 2.5\%$.
In the subsequent extension E04-110 \cite{Mad04} at
$Q^2 \approx 4.3$~GeV$^2$, the TPE correction for
$\eps \approx 0.82$ is expected to be around 3\%. 
While small, these corrections will be important to take into
account in order to achieve precision at the several percent level.
Two-photon exchange effects will also need to be taken into account
when extracting the neutron magnetic form factor $\gmn$ from cross
section data, if precision at the $\sim 1-2\%$ level is sought.
In particular, in measurements of the ratio of neutron to proton
cross sections, the neutron cross section is obtained by multiplying
the ratio by the total measured proton cross section which depends on 
both the Born form factors and TPE corrections.  The extracted neutron
cross section must then be corrected for TPE effects in the neutron to
obtain the Born form factors.

\subsection{\it Electroproduction of resonances}

Beyond elastic final states, inelastic electroproduction channels
provide additional information on hadron structure, from mapping out
the spectrum of states in QCD, to probing the spatial distributions
of hadrons.  Of particular interest is the reaction whereby the
produced final state is in the $\Delta(1232)$ resonance region, which
has been recently studied both experimentally \cite{Tva06, Lia04}
and theoretically \cite{Pas06}.

At the Born level, $\Delta$ electroproduction is parametrized by the
same electromagnetic $N \to \Delta$ transition amplitudes that enter
the calculation of the $\Delta$ intermediate state contribution to the
TPE corrections in elastic $ep$ scattering in Sec.~\ref{sssec:Delta}.
It therefore provides an important consistency check on the role of
the $\Delta$ resonance in electron scattering. Kondratyuk and Blunden
\cite{Kon06} computed the TPE corrections to the unpolarized cross
section for $\Delta$ production in electron--proton collisions,
including both nucleon and $\Delta$ contributions in the intermediate
states in the one-loop diagrams.  As in the case of elastic scattering,
the TPE effects from the  intermediate nucleon and $\Delta$ have
opposite signs in most  kinematical regimes, and are pronounced
even at low energies.

It is straightforward to show that the sum of the box and crossed-box
amplitudes with an intermediate $\Delta$ is gauge invariant by itself.
For the nucleon intermediate state, an additional $\gamma \gamma N \Delta$
contact term has to be added to the box and crossed-box diagrams.
Such a term is required for gauge invariance, and can be constructed
by the standard procedure of minimal substitution~\cite{Kon00}.

\begin{figure}[htb]
\vspace*{-.5cm}
\begin{center}
\begin{minipage}{12cm}
\hspace*{-2.1cm}\includegraphics[scale=0.75]{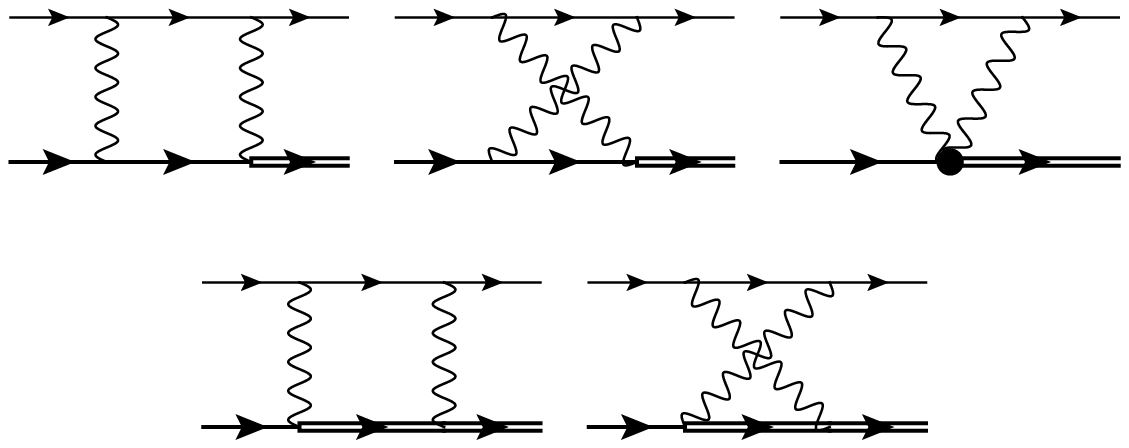}
\hspace*{0.5cm}\includegraphics[scale=0.38]{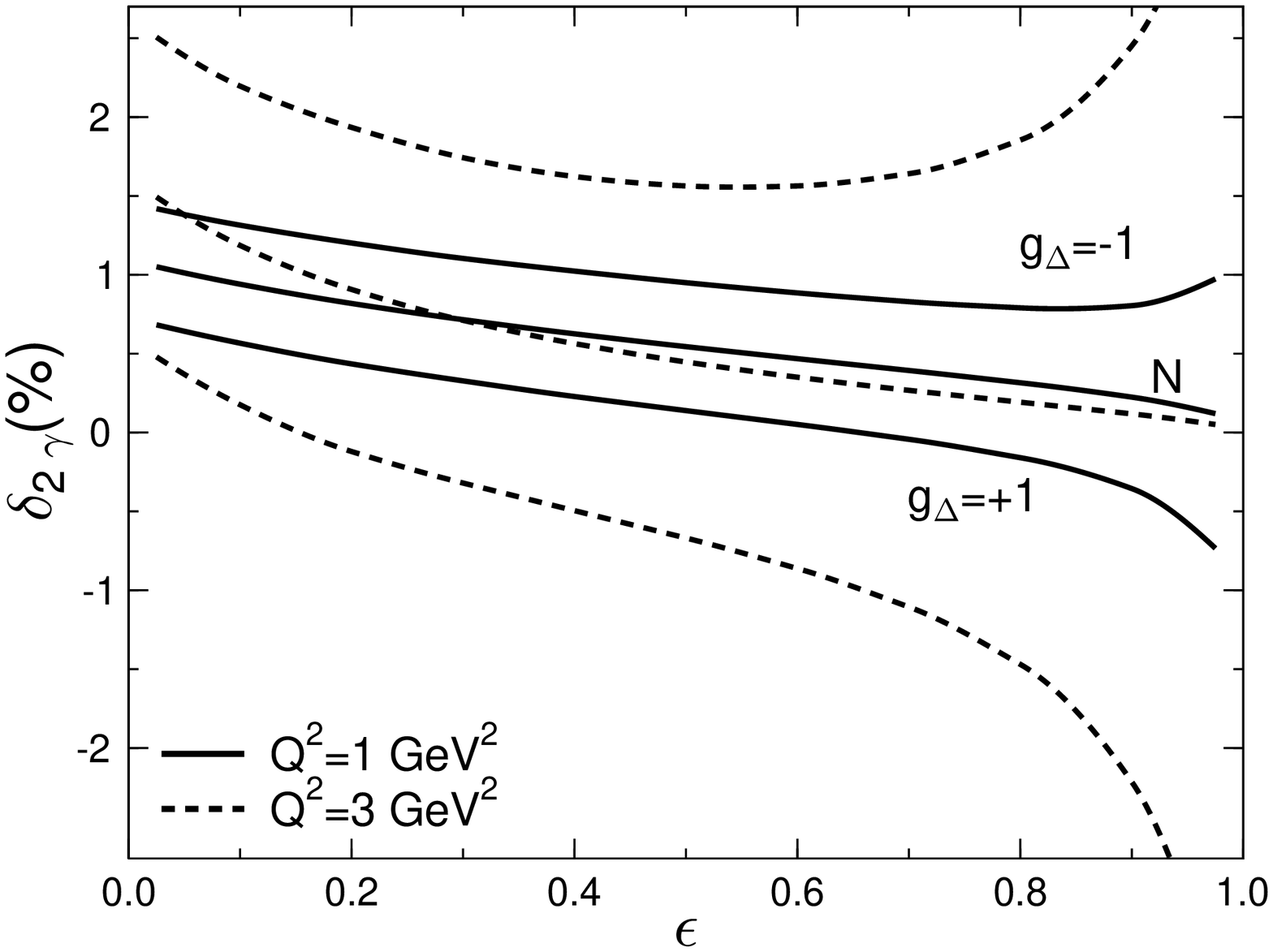}
\end{minipage}
\begin{minipage}{16.5cm}
\caption{
{\bf (Left)}
	Two-photon exchange graphs for the $e p \to e \Delta$ reaction.
	The loop diagram with the $\gamma \gamma N \Delta$ contact term
	(denoted by the black circle) ensures gauge invariance of the
	calculation.
{\bf (Right)}
	Two-photon exchange correction to the unpolarized cross section
	for $\Delta$ production in electron-proton collisions, calculated
	for $Q^2=1$~GeV$^2$ (solid) and $Q^2=3$~GeV$^2$ (dashed). 
	The upper, middle and lower pairs of lines are labeled by the
	values of the $\gamma \Delta \Delta$ coupling constant $g_{\Delta}$ 
	(``N" corresponding to $g_\Delta=0$, \ie~to the absence of an
	intermediate $\Delta$ state).
	Figures adapted from Ref.~\cite{Kon06}.}
\label{fig:dpol}
\end{minipage}
\end{center}
\end{figure}

An important theoretical ingredient of the calculation is the
$\gamma \Delta \Delta$ vertex form factor.  Various forms of this vertex
feature prominently in the studies of electromagnetic interactions of the
deuteron~\cite{Are78} and three-nucleon bound states~\cite{Del04}, as
well as in the recent extraction~\cite{Pas05} of the $\Delta$ magnetic
dipole moment.

The TPE corrections in $\Delta$ electroproduction are, as expected, 
generally much smaller than the Born contribution \cite{Kon06},
but can be important for a precise analysis of $ep$ scattering in the 
resonance region.  Results for $Q^2=1$~GeV$^2$ and $Q^2=3$~GeV$^2$ are 
shown in Fig.~\ref{fig:dpol}.  Kondratyuk and Blunden~\cite{Kon06} find
a pronounced $\eps$ dependence  of the TPE correction to the value and 
sign of the (dominant) $\gamma \Delta \Delta$ magnetic coupling constant. 
A current  analysis~\cite{Tva06} of experimental Rosenbluth separations 
in $ep$ scattering provides strong constraints on the nonlinearity in 
$\eps$, although additional high-precision data would allow more definite 
conclusions to be reached.

For further progress in the evaluation of higher-order effects in
electron-nucleon collisions, a more detailed knowledge of the
$\gamma \Delta \Delta$ vertex is needed.  Being formulated in terms of 
hadronic degrees of freedom, this model is somewhat complementary to the 
approach of Ref.~\cite{Pas06}, where TPE effects were calculated using 
the formalism of generalized parton distributions.

\subsection{\it Timelike form factors}
\label{ssec:timelike}

Electromagnetic form factors in the space-like ($Q^2>0$) and time-like 
($Q^2<0$) regions both yield information on the structure of hadrons. 
The form factors are real in the space-like region, while the time-like 
form factors have a phase structure reflecting the final-state 
interactions of the outgoing hadrons, and are therefore complex.

\begin{figure}[ht]
\begin{center}
\includegraphics[width=8cm]{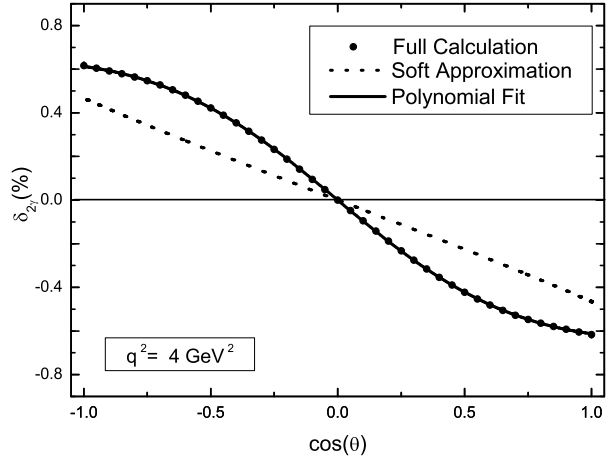}
\begin{minipage}{16.5cm}
\caption{Angular dependence of the TPE contribution to the differential
	cross section for $e^+e^- \to p\bar{p}$.
	Figure taken from Ref.~\cite{Che08}.}
\label{fig:Chen}
\end{minipage}
\end{center}
\end{figure}

The interference of one- and two-photon exchange can be studied in the 
process $e^+e^- \to p\bar{p}$, and is related by crossing symmetry to 
the elastic $ep$ interaction.  The interference manifests as an angular 
asymmetry in the differential cross section.  This was studied recently
by Chen~\etal~\cite{Che08} in a hadronic model with essentially the same 
ingredients as the $ep$ calculations.  The corrections are again at the 
few percent level, and are shown in Fig.~\ref{fig:Chen} for
$Q^2=4$~GeV$^2$.
A search for TPE effects in BABAR $e^+ e^- \to p \bar{p}$ data
\cite{Aub06} showed no indication of the expected forward-backward
asymmetry within the precision of the data~\cite{Tom08}.

In addition to the unpolarized differential cross section, 
Chen~\etal~\cite{Che08} considered the single-spin polarization 
observable $P_y$, and the double-spin polarization observables $P_x$ 
and $P_z$, which require the incoming electron to be polarized.
They suggested that $P_z$ in  particular should be considered in 
future experiments looking for TPE effects.

\subsection{\it Pion form factor}
\label{ssec:pionFF}

As the lightest bound state of quarks and antiquarks, the pion plays
a unique role in QCD, and determining its electromagnetic structure
is of great importance for understanding the realization of chiral
symmetry in nature.
Extractions of the pion form factor $F_\pi(Q^2)$ in the space-like
region from measurements of the pion electroproduction reaction
$e p \to e \pi^+ n$ have recently provided data on the $Q^2$
dependence of $F_\pi$ up to values of $Q^2 \sim 2.5$~GeV$^2$ 
\cite{Hor06, Tad07, Hub08}, and higher $Q^2$ measurements are planned
to $Q^2 \sim 6$~GeV$^2$.  The role of TPE in electromagnetic scattering 
from the pion was recently investigated by several authors \cite{Blu10, 
Bor11, Don10W, Kai11}.  The analysis of TPE from a spin-0 target is,
in fact, considerably simpler than that for spin-1/2 targets.

The form factor of the pion is defined through the matrix element
of the pion current,
\be
\langle \pi(p') | J^\mu(0) | \pi(p) \rangle\
=\ (p + p')^\mu\, F_\pi(Q^2).
\ee
In the Born approximation the amplitude for an electron scattering
from a $\pi^+$ is given by
\be
{\cal M}_\gamma^\pi\
=\ -{e^2 \over q^2}\ \ubar_e(k')\, \gamma_\mu\, u_e(k)\,
   (p+p')^\mu\, F_\pi(Q^2),
\ee
and the Born cross section in the target rest frame is
\be
{ d\sigma^{e\pi} \over d\Omega }\
=\ \sigma_{\rm Mott}\,  F_\pi^2(Q^2) \, ,
\ee
where $\sigma_{\rm Mott}$ is given in Eq.~(\ref{eq:Mott}).

The TPE amplitude for the box diagram has the form \cite{Blu10, Bor11}
\be
{\cal M}_{\gamma\gamma}^\pi\
=\ -ie^4 \int {d^4 q_1 \over (2\pi)^4}\
   L_{\mu\nu} H^{\mu\nu}_\pi\,
   \Delta_F(q_1,\lambda)\, \Delta_F(q_2,\lambda),
\label{eq:Mpigg}  
\ee
where the leptonic tensor $L_{\mu\nu}$ is given in 
Eq.~(\ref{eq:Lmunu_bc}), and the crossed-box is obtained using
crossing symmetry.
The pion hadronic tensor $H^{\mu\nu}_\pi$ in principle contains
contributions from all hadronic excitations in the intermediate state.
For the dominant pion elastic contribution, one has
\be
H^{\mu\nu}_\pi\
=\ (2p+q_1+q)^\mu\, F_\pi(Q_1^2)\,
   \Delta_F(p+q_1,m_\pi)\,
   (2p+q_1)^\nu\, F_\pi(Q_2^2).
\ee
The pion form factor is then modified according to
$F^2_\pi(Q^2)\ \to\ F^2_\pi(Q^2)\, (1 + \delta^\pi)$,
where, as for the nucleon TPE correction in Eq.~(\ref{eq:delta_gg}),
the relative correction $\delta^\pi$ is given by
\be
\delta^\pi\
=\ { 2 \Re \left( {\cal M}_\gamma^{\pi *}\, {\cal M}_{\gamma\gamma}^\pi 
	   \right)
     \over |{\cal M}_\gamma^\pi|^2}\, .
\ee

\begin{figure}[tb]
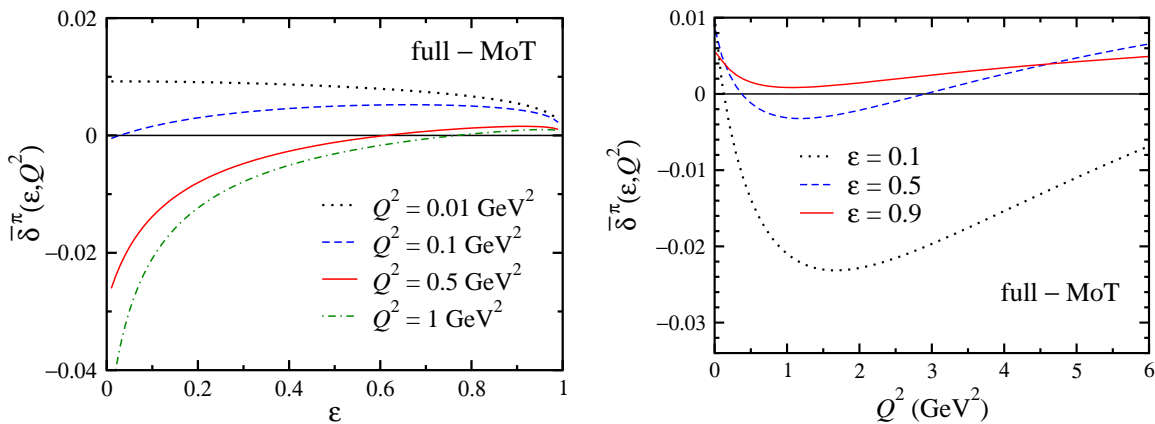

\begin{center}
\includegraphics[height=5.5cm]{Figs/fig33a.eps}%
\hspace*{0.5cm}\includegraphics[height=5.5cm]{Figs/fig33b.eps}
\begin{minipage}{16.5cm}
\caption{Two-photon exchange correction $\overline\delta^\pi$ to
	the pion form factor squared, relative to the Mo-Tsai (MoT)
	contribution \cite{Tsa61, Mo69}, as a function of $\eps$ for
	various $Q^2$ {\bf (left)}, and of $Q^2$ for various $\eps$
	{\bf (right)}.
	Figures adapted from Ref.~\cite{Blu10}.}
\label{fig:pi_epsMoT}
\end{minipage}
\end{center}
\end{figure}

As in the case of the nucleon electromagnetic form factors,
experimental analyses of pion form factor data typically use
radiative corrections computed by Mo and Tsai in the soft-photon
approximation \cite{Tsa61, Mo69} (see Sec.~\ref{ssec:general}).
The effect of the IR-finite, structure-dependent contribution
is illustrated in Fig.~\ref{fig:pi_epsMoT}, where the difference
between the full TPE correction and the Mo-Tsai prescription
\cite{Tsa61, Mo69},
$\overline\delta^\pi = \delta^\pi - \delta_{\rm IR}(\rm MoT)$,
is shown as a function of $\eps$ and $Q^2$.
Here a monopole parametrization for the ``bare'' pion form
factor in Eq.~(\ref{eq:Mpigg}) is used,
\be
F_\pi(Q^2)\ =\ \left( 1 + {Q^2 \over \Lambda_\pi^2} \right)^{-1},
\ee
with the cut-off parameter $\Lambda_\pi = 770$~MeV corresponding
to the $\rho$-meson mass.
The sensitivity of the results to the input pion form factor was
studied in Ref.~\cite{Blu10}, using an alternative parametrization
\cite{Mel03} which gives a better description of the available data
and builds in gauge invariance constraints for the $Q^2 \to 0$ limit
and perturbative QCD expectations for the $Q^2 \to \infty$ behavior.
The differences at low $Q^2$ are negligible, but become noticeable at
high $Q^2$, although do not affect the results qualitatively.

At low $Q^2$ ($Q^2 \sim 0.01$~GeV$^2$) the TPE correction is positive
and of the order of 1\% at backward angles (small $\eps$), decreasing
to zero as $\eps \to 1$.
For larger $Q^2$ the correction becomes more negative up to
$Q^2 \sim 1-2$~GeV$^2$, especially at backward angles,
but changes sign at intermediate $\eps$.
Unlike for $ep$ scattering, the $e\pi$ cross section vanishes
in the extreme backward limit ($\eps \to 0$).
Above $Q^2 \sim 2$~GeV$^2$ the correction grows once again,
reaching $\sim 1\%$ at $Q^2 = 10$~GeV$^2$ \cite{Blu10}.
Interestingly, the TPE correction is most positive at very small
$Q^2$ ($Q^2 \ll 1$~GeV$^2$) and at large $Q^2$ ($Q^2 \gg 1$~GeV$^2$),
reaching its minimum values at $Q^2 \sim 1-2$~GeV$^2$.

\begin{figure}[t]
\begin{center}
\hspace*{4cm}
\includegraphics[height=20cm,angle=270]{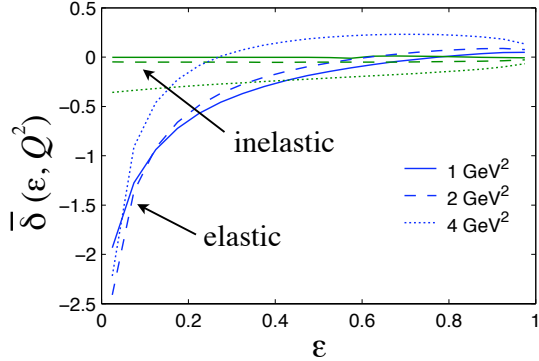}
\begin{minipage}{16.5cm}
\vspace*{-17cm}
\caption{Elastic and inelastic ($\rho$ and $b_1(1235)$) contributions
	to the TPE correction $\overline\delta^\pi$ (in percent)
	relative to the Maximon-Tjon result \cite{Max00T}, as a
	function of $\eps$ for $Q^2 = 1$, 2 and 4~GeV$^2$.
	Figure adapted from Ref.~\cite{Bor11}.}
\label{fig:piBK}
\end{minipage}
\vspace*{-8cm}
\end{center}
\end{figure}

The contributions from higher-mass intermediate states to the pion
hadronic tensor were considered in Ref.~\cite{Bor11} within a dispersion
relations approach.
Because the mass difference between the pion and the next excited
resonant state, the $\rho$ meson, is $\sim 5$ times as large as the
pion mass, one would not expect large contributions from excited
hadronic intermediate states.  
This was indeed confirmed in explicit calculations of the $\rho$ and
$b_1(1235)$ meson contributions, which were found to be negligible
for $Q^2 \ltorder 2$~GeV$^2$, as Fig.~\ref{fig:piBK} illustrates.
At larger $Q^2$ ($\gtorder 4$~GeV$^2$) the inelastic contributions
become comparable to the elastic for $\eps \gtorder 0.2$; however,
in the region at low $\eps$ where the TPE effect is greatest, these
are still significantly smaller than the elastic components.

In contrast to the proton form factor case, where the TPE effects
give large corrections to the elastic form factors extracted from
LT separated cross sections at large $Q^2$ \cite{AMT07}, the TPE
corrections to the pion form factor are relatively small.
This stems from the fact that electron scattering from a scalar target
is described by a single form factor, with no LT separation necessary.
On the other hand, since $F_\pi$ is extracted via LT separation
of the pion electroproduction cross section, TPE with one photon
attached to the pion and the other to the initial proton or final
neutron could modify the longitudinal cross section, and may need
to be considered.
Furthermore, the effects on the pion form factor from nonresonant
contributions in the intermediate state have not yet been evaluated,
and may also need to be considered in future analyses.

\subsection{\it Electron--nucleus scattering}

\subsubsection{TPE in deuteron form factors}
\label{sssec:D}

The TPE corrections to the (spin-1) deuteron ($D$) elastic
electromagnetic form factors and to the $e^+e^- \to D\bar{D}$ process
have been discussed in Refs.~\cite{Don06, Don09, Don09C, Kob10, Don10}.
The reaction amplitude contains six generalized form factors, but only
three linearly independent combinations of them (generalized charge
$G_C^D$, quadrupole $G_Q^D$, and magnetic $G_M^D$ form factors) 
contribute to the cross section~\cite{Don06}.

Dong and collaborators~\cite{Don09, Don09C, Don10} estimated all
the TPE corrections to the conventional form factors of the deuteron
using an effective Lagrangian approach.  In this model, the two photons 
couple to one of the two nucleons in the deuteron.  Dong found the TPE
corrections to be small (less than 1\% for $Q^2<2$~GeV$^2$), and most
significant for $G_M^D$~\cite{Don09}.  He also suggested that two of
the additional form factors could be tested in measurements of the
double and single polarization observables \cite{Don09}.  The effect
makes a sizable contribution at backward angles to the polarization
observable $P_y$ of $eD$ scattering, which vanishes in the Born
approximation \cite{Don10}.

Kobushkin \etal~\cite{Kob10} also discuss elastic electron-deuteron
scattering beyond the Born approximation, including contributions
where the two photons can also interact with different nucleons.
They conclude that TPE {\em may} give a large contribution to elastic
$eD$ scattering, but point out that their estimates have large
uncertainties.  The most important source of uncertainty comes from
the short-range part of the deuteron wave function.  It is suggested
that experimental study of TPE in elastic $eD$ scattering for $Q^2$
in the few GeV$^2$ range can give important information about the
deuteron structure at short distances.

\subsubsection{$^3$He elastic form factors}
\label{sssec:3He}

The hadronic formalism of Sec.~\ref{sec:TPE} has also been applied
to the case of elastic scattering from the spin-1/2 $^3$He nucleus
\cite{Blu05}, where contributions of unexcited $^3$He intermediate 
states were computed.  Of course these are likely to constitute only
a part of the entire TPE effect, as contributions from break-up channels 
may also be important.  Nevertheless, a comparison of the size and 
magnitude of the TPE corrections from elastic intermediate states in 
$^3$He relative to those in the proton illustrates a number of features 
of TPE.

The expressions used to evaluate the TPE contributions to the $^3$He
correction $\delta^{^3{\rm He}}$ are similar to those for the nucleon
in Sec.~\ref{ssec:nucleon}, but with some important differences.
The charge of the nucleus is $Z e$, with $Z=2$ for $^3$He, and the
mass $M_{^3{\rm He}}$ is approximately 3 times larger than the
nucleon mass, while the anomalous magnetic moment is
$\kappa_{^3{\rm He}} = -4.185$.
The internal photon--${}^3$He form factors are also softer than the 
corresponding proton form factor (the charge radius of the $^3$He 
nucleus is $\approx 1.88$~fm), and have zeros at $Q^2 \approx 0.45$
and 0.7~GeV$^2$ for the charge and magnetic form factors, respectively 
\cite{Amr94}.

\begin{figure}[ht]
\vspace*{0.5cm}
\begin{center}
\includegraphics[height=6cm]{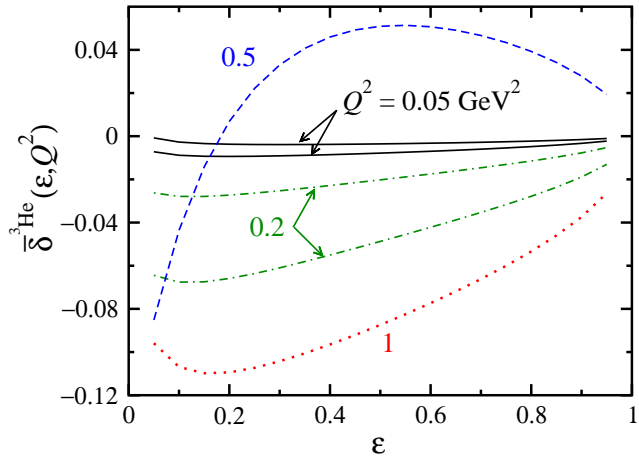}
\begin{minipage}{16.5cm}
\caption{TPE contribution to the unpolarized electron--$^3$He
	cross section, with the $^3$He elastic intermediate state,
	as a function of $\eps$, for $Q^2 = 0.05$ (solid), 0.2
	(dot-dashed), 0.5 (dashed) and 1~GeV$^2$ (dotted).
        A parametrization of the form factor from Ref.~\cite{Amr94}
        is used in all cases, except for the upper solid and
	dot-dashed curves, which uses a dipole with mass
	$\Lambda_{^3{\rm He}} = 0.37$~GeV.
	Figure adapted from Ref.~\cite{Blu05}.}
\label{fig:del_He3}
\end{minipage}
\end{center}
\end{figure}

The TPE correction $\overline\delta^{^3{\rm He}}$, relative to the
Mo-Tsai IR-divergent result, is shown in Fig.~\ref{fig:del_He3} as
a function of $\eps$ for several values of $Q^2$, using the $^3$He
form factors from Ref.~\cite{Amr94}.
As for the proton, the correction is negative at low $Q^2$, and
generally increases in magnitude with increasing $Q^2$.
%
%
The dramatic change of sign in $\overline\delta^{^3{\rm He}}$
at $Q^2 = 0.5$~GeV$^2$ stems from the presence of the zeros in   
the form factors at $Q^2 \sim 0.5$~GeV$^2$.

To estimate the model dependence of the results, Blunden {\it et al.}
also calculated $\overline\delta^{^3{\rm He}}$ assuming a dipole shape
with a cut-off mass $\Lambda_{^3{\rm He}} = 0.37$~GeV, fitted to the
$^3$He radius, which gives a reasonable approximation at low $Q^2$
($\ltorder 0.05$~GeV$^2$).
The results with this form factor are about a factor 2 smaller in
magnitude than for the form factor from Ref.~\cite{Amr94}.
At larger $Q^2$ the dipole shape is a less reliable representation
of the $^3$He form factor, making it difficult to estimate the model
dependence of $\delta^{^3{\rm He}}$.
The results in Fig.~\ref{fig:del_He3} illustrate the potential
relevance of TPE effects for future $^3$He form factor measurements,
which may reveal interesting TPE effects at larger $Q^2$ \cite{e04018}.

%% file: section7.tex
\section{Parity-violating electron scattering}
\label{sec:z}

In addition to the exchange of one or more virtual photons between
the electron and nucleon, the Standard Model allows the scattering
to take place via the exchange of a neutral $Z$ boson.
Because the $Z$ boson mass is some two orders of magnitude larger
than the proton mass, the weak exchange process is strongly
suppressed relative to the electromagnetic reaction.
Nevertheless, asymmetries sensitive to the $\gamma Z$ interference
amplitude, which are of the order of several parts per million (ppm),
have been measured in modern accelerator facilities.
These have been used to probe the strangeness content of the proton,
through measurements of the strange electric and magnetic form factors
\cite{HAP04,HAP06,HAP06a,HAP07,G005,G010,SAM97,MAM04,MAM05,MAM09},
as well as to measure the weak charge of the proton \cite{Qwk05}.

The $\gamma Z$ interference term is isolated by polarizing the
incident electron and measuring the difference between right-
and left-handed electrons scattering from unpolarized protons.
A parity-violating (PV) asymmetry can then be defined in terms
of the differential cross sections as
\be
A_{\rm PV}\ =\ {\sigma_+ - \sigma_- \over \sigma_+ + \sigma_-} ,
\label{eq:APVdef}
\ee
where $\sigma_h$ is the cross section for a right-hand
(helicity $h=+1$) or left-hand (helicity $h=-1$) polarized electron.
The numerator in the asymmetry is sensitive to the interference of
the vector and axial-vector currents, and hence violates parity.
The natural size of the PV asymmetry is $\sim Q^2/M_Z^2$, which
for $Q^2 \sim 0.1$~GeV$^2$ is of the order $10^{-5} - 10^{-6}$
--- values that can be routinely measured at current facilities
such as Jefferson Lab or MAMI at Mainz \cite{Bei05}.

In view of the large TPE effects on electromagnetic form factors
discussed in the proceeding sections, the question naturally arises
of what effect the exchange of two bosons ($\gamma$ or $Z$) may have
on PV asymmetries.  Because both the strange form factors and the
proton weak charge are numerically small quantities, these contributions
could affect their extraction significantly.

One should note that the ``standard electroweak radiative corrections''
already include a contribution from two-boson exchange (TBE) effects.  
These are usually taken from the classic calculation of Marciano and 
Sirlin \cite{Mar83, Mar84}, and include $\gamma Z$, $ZZ$, and $WW$
exchange box and crossed-box diagrams.  Because of the large masses
of the $Z$ and $W$ bosons, the box diagrams contain contributions for
virtual four-momenta of all mass scales.  For the $ZZ$ and $WW$ boxes,
the high mass scales dominate, and a calculation in terms of the quark
structure of the nucleon is reliable.  However, for the $\gamma Z$
diagram both high and low mass scales contribute.  The high-mass scale,
calculated in terms of quarks, accounts for most of the effect, but
the low momentum contribution in terms of a hadronic picture is not 
negligible.  The quark-based calculation is discussed in more detail
in Sec.~\ref{ssec:DRgZ}.

In this section we review the TBE corrections to PV asymmetries arising
from the interference between single $Z$ boson and $\gamma\gamma$
exchange amplitudes (which we denote by ``$Z(\gamma\gamma)$''),
and between the one-photon exchange and $\gamma Z$ interference
amplitudes (denoted by ``$\gamma(Z\gamma)$'').
Because PV electron scattering experiments are typically performed at
$Q^2$ values $\ltorder 1$~GeV$^2$, the hadronic formalism described
in Sec.~\ref{sec:TPE} is the more natural implementation of hadronic
structure effects.  For a detailed examination of the GPD-based
approach see Refs.~\cite{Car07,Afa05C}.

\subsection{\it Parity-violating asymmetries}
\label{ssec:APV}

In the Born approximation, the weak neutral current amplitude
is given by
\be
{\cal M}_Z\
=\ -{2 g^2 \over (4 \cos\theta_W)^2}\
   {1 \over M_Z^2-q^2}\
   j_{Z \mu}\ J_Z^\mu\
\approx\ -\left({G_F \over \sqrt{2}}\right)\ j_{Z \mu}\ J_Z^\mu\, ,
\label{eq:MZ}
\ee
where $g = e/\sin\theta_W$ is the weak coupling constant,
$M_Z$ is the $Z$ boson mass, and the strength of the effective
four-fermion interaction is given by the Fermi constant
$G_F = \pi\alpha/(\sqrt{2} M_Z^2 \sw\cw)$.
At tree level the weak mixing angle $\theta_W$ is related to the weak
boson masses by $\sw = 1 - M_W^2/M_Z^2$, where $M_W$ is the $W$ boson
mass.
The weak leptonic current is given by a sum of vector and axial-vector
terms \cite{PDG10},
\be
j_{Z \mu}\
=\ \ubar_e(k',h)\, \gamma_\mu (g^e_V - g^e_A \gamma_5)\, u_e(k,h)\, ,\ee
where the vector ($g^e_V$) and axial-vector ($g^e_A$) couplings of the
electron to the weak current are defined as
\be
g^e_V\ =\ -{1\over 2}\left(1 - 4 \sw\right)\, ,\ \ \ \
g^e_A\ =\ -{1\over 2}\, ,
\label{eq:geVA_def}
\ee
respectively.
Note that some of the TPE literature
\cite{Tjo08,Tjo09,Zho10,Bei05,Mus94,Zho07,Nag09,Gor09,Gor10,Gor11}
uses definitions whereby the vector and axial charges are scaled
by a factor 2 relative to those in Eq.~(\ref{eq:geVA_def}), so
that care should be taken when comparing formulas for amplitudes.

The matrix elements of the weak hadronic neutral current is given by
\be
J_Z^\mu\
=\ \ubar_N(p')\, \Gamma_Z^\mu(q)\, u_N(p)\, ,
\ee
where the current operator is parametrized by three weak nucleon
form factors,
\be
\Gamma_Z^\mu(q)\
=\ \gamma^\mu\, F_1^{ZN}(Q^2)\
+\ \frac{i \sigma^{\mu\nu} q_\nu}{2 M}\, F_2^{ZN}(Q^2)\
+\ \gamma^\mu \gamma_5\, G_A^{ZN}(Q^2)\ .
\label{eq:JZ}
\ee
Alternatively, the current operator can be expressed in terms of the
Sachs weak form factors $G_{E,M}^{ZN}$, in analogy with the
electromagnetic nucleon form factors in Eq.~(\ref{eq:GEMdef}),
\be
G_E^{ZN}(Q^2)\ =\ F_1^{ZN}(Q^2)\ -\ \tau F_2^{ZN}(Q^2)\, ,\ \ \ \ \ \
G_M^{ZN}(Q^2)\ =\ F_1^{ZN}(Q^2)\ +\ F_2^{ZN}(Q^2)\, .
\label{eq:GZN}
\ee

The differential cross section including $\gamma$ and $Z$ exchange
is given by Eq.~(\ref{eq:sigma0}), where the amplitude is now a sum
of the $\gamma$ and $Z$ Born contributions,
\be
\left| {\cal M} \right|^2\
=\ \left| {\cal M}_\gamma + {\cal M}_Z \right|^2\
=\ \left| {\cal M}_\gamma \right|^2
+ 2 \Re \left( {\cal M}^*_\gamma\, {\cal M}_Z \right)
+ \left| {\cal M}_Z \right|^2\ .
\label{eq:Mamp}
\ee
At the kinematics of interest in this review, the purely weak 
contribution $|{\cal M}_Z|^2$ is small compared with the other terms
and can be neglected.  The purely electromagnetic contribution cancels
in the numerator of the asymmetry in Eq.~(\ref{eq:APVdef}), so that
$A_{\rm PV}$ is sensitive to the parity-violating part of
$2\Re \left( {\cal M}^*_\gamma {\cal M}_Z \right)$, involving the
interference of ${\cal M}_\gamma$ with the product of vector and
axial-vector currents in ${\cal M}_Z$.
The vector-vector and axial-axial parts of ${\cal M}_Z$ also cancel
in the asymmetry.  The denominator is dominated by the electromagnetic
term, $|{\cal M}_\gamma|^2$.
Evaluating the asymmetry explicitly, one has for PV asymmetry of the
nucleon
\be
A^N_{\rm PV}\
=\ -\left( {G_F Q^2 \over 4\pi\alpha\sqrt{2}} \right)
   { 1 \over \sigma_R^{\gamma N} }
   \left\{ -2 g_A^e
	   \left( \eps\, G_E^{\gamma N} G_E^{ZN}
		+ \tau\, G_M^{\gamma N} G_M^{ZN}
	   \right)\
       +\ 2 g_V^e\, \eps'\, G_M^{\gamma N} G_A^{ZN}
   \right\},
\label{eq:APV}
\ee
where $\eps' = \sqrt{\tau (1+\tau)(1-\eps^2)}$, and $\sigma_R^{\gamma N}$
is the reduced $\gamma N$ cross section defined in Eq.~(\ref{eq:sigmaR}).

For a proton target, assuming isospin symmetry (see, however,
Refs.~\cite{Lew07,Xia08}), the weak vector form factors $G_{E,M}^{Zp}$
can be related to the electromagnetic form factors of the proton (neutron)
$G_{E,M}^{\gamma p(n)}$ by
\be
G_{E,M}^{Zp}\
=\ (1-4 \sw)\, G_{E,M}^{\gamma p} - G_{E,M}^{\gamma n} - G_{E,M}^s,
\label{eq:FZ}   
\ee
where $G_{E,M}^s$ are the contributions from strange quarks, and
\be
\hspace*{2cm}
G_E^{Zp}(0)\ \equiv\ Q_W^p\ =\ 1-4 \sw
\hspace*{2cm} {\rm [Born\ approximation]}
\ee
is the weak charge of the proton.
Because $Q_W^p$ is numerically small, the overall contribution to
$G_{E,M}^{Zp}$ from the proton electromagnetic form factors is
suppressed.
The weak axial-vector form factor of the proton is given by
$G_A^{Zp} = -G_A^p + G_A^s$, where 
$G_A^p(0) \equiv g_A = 1.267$ is the axial-vector charge,
and $G_A^s$ is the axial-vector strange quark contribution,
related to the spin of the proton carried by strange quarks
as measured in deep-inelastic scattering \cite{Kap88, Bas06}.
In terms of the proton vector ($A_V$), axial-vector ($A_A$) and
strange ($A_s$) contributions, the asymmetry can then be written as
\be
A^p_{\rm PV}
= - \left( {G_F Q^2 \over 4\pi\alpha\sqrt{2}} \right)
    \left( A_V + A_A + A_s \right),
\label{eq:APVp}
\ee
where
\bea
A_V &=& -2 g^e_A\,
\left[ (1 - 4\sw)
     - {1 \over \sigma_R^{\gamma p}}
       \left( \eps\, G_E^{\gamma p}\, G_E^{\gamma n}
	    + \tau\, G_M^{\gamma p}\, G_M^{\gamma n}
       \right)
\right],
\label{eq:A_V}						\\
A_A &=& 2 g^e_V\,
{\eps' \over \sigma_R^{\gamma p}}\, G_A^{Zp}\, G_M^{\gamma p}, \\
\label{eq:A_A}
A_s &=& 2 g^e_A\,
{1 \over \sigma_R^{\gamma p}}
\left( \eps G_E^{\gamma p}\, G_E^s
     + \tau G_M^{\gamma p}\, G_M^s
\right),
\label{eq:A_s}
\eea
where $\sigma_R^{\gamma p}$ is the reduced cross section for
a proton target.

\subsection{\it Two-boson exchange corrections}

Beyond the Born approximation, the PV asymmetry, Eq.~(\ref{eq:APV}),
receives corrections from higher-order radiative effects, including
both the electromagnetic two-photon exchange contributions discussed in
Sec.~\ref{sec:TPE} and corrections involving $\gamma$--$Z$ boson loops.
There are several ways in which the PV asymmetry can be represented
in the presence of higher-order radiative corrections.
The approach pioneered by Marciano and Sirlin \cite{Mar83, Mar84}
parametrizes the electroweak radiative effects in terms of parameters 
$\rho$ and $\kappa$, such that the weak charge of the proton in the 
presence of higher-order corrections becomes
\be
Q_W^p\ \to\ \rho\, (1 - 4\kappa\sw).
\ee
Including the standard radiative corrections, such as vacuum polarization,
vertex corrections, and hadron structure-independent two-boson exchange
contributions, one has $\rho = 0.9877$ and $\kappa = 1.0026$ \cite{PDG10}.
The TBE amplitudes discussed here give additional corrections to $\rho$
and $\kappa$ which are denoted in Refs.~\cite{Tjo08, Zho10, Zho07, Nag09}
by $\Delta\rho$ and $\Delta\kappa$, respectively.

An alternative parametrization is in terms of isoscalar and isovector
weak radiative corrections for the vector form factors, and a similar
set of corrections for the axial-vector form factors.
In this case the vector part of the PV asymmetry is written
\be
A_V\ =\ -2 g^e_A
\left[ (1-4\sw) (1+R_V^p)
     - {1 \over \sigma_R^{\gamma p}}
       \left( \eps\, G_E^{\gamma p} G_E^{\gamma n}
	    + \tau\, G_M^{\gamma p} G_M^{\gamma n}
       \right) (1+R_V^n)
\right],
\ee
where the proton and neutron radiative corrections are given,
to first order in $\rho-1$ and $\kappa-1$, by
\be
R_V^p = \rho - 1 - (\kappa-1)\ {4\sw \over 1-4\sw}\, ,\ \ \ \
R_V^n = \rho - 1\, .
\ee
The radiative corrections to the strange part of the asymmetry
(\ref{eq:A_s}) enter via a multiplicative factor
\be
A_s\ \to\ A_s (1+R_V^{(0)})\ ,
\ee
where the isoscalar radiative correction $R_V^{(0)} = \rho - 1$.
For the axial asymmetry $A_A$, the generalization of the tree level
axial form factor, $G_A^{Z p} \to \widetilde{G}_A^{Z p}$, implicitly
contains higher-order radiative corrections for the proton axial
current, as well as the hadronic anapole contributions
\cite{Mus94, You06}.

\begin{figure}[t]
\begin{center}
\begin{minipage}{8cm}
\includegraphics[angle=270,scale=0.4,clip=true,bb=42 0 320 600]{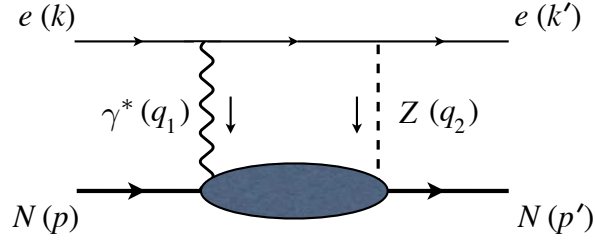}
\end{minipage}
\begin{minipage}{16.5cm}
\caption{An example of a two-boson exchange contribution to
	parity-violating elastic scattering, due to $\gamma Z$
	exchange (given by the amplitude ${\cal M}_{\gamma Z}$).
	Other contributions include the amplitude with the $\gamma$
	and $Z$ interchanged (denoted by ${\cal M}_{Z \gamma}$),
	and the crossed-box diagrams.}
\label{fig:TBE}
\end{minipage}
\end{center}
\end{figure}

More explicitly, the corrections to the cross section arising from
the $\gamma\gamma$ and $\gamma Z$ TBE contributions can be obtained
from Eq.~(\ref{eq:Mamp}) by the replacements
\be
{\cal M}_\gamma\
\to\ {\cal M}_\gamma + {\cal M}_{\gamma\gamma}\, ,\ \ \ \
{\cal M}_Z
\to {\cal M}_Z + {\cal M}_{\gamma Z} + {\cal M}_{Z \gamma}\, ,
\ee
where ${\cal M}_{\gamma Z}$ and ${\cal M}_{Z \gamma}$ correspond
to the two different orderings of the $\gamma$ and $Z$ exchanges
in the box and crossed-box diagrams, Fig.~\ref{fig:TBE}.

Because the coupling of the $Z$ boson to the electron and to the proton
contains both vector and axial-vector components, the interference
amplitude ${\cal M}_{\gamma Z}$ (and similarly ${\cal M}_{Z \gamma}$)
contains
(parity-conserving) vector electron--vector hadron and
  axial electron--axial hadron terms from the $Z$ exchange, along with
(parity-violating) vector electron--axial hadron and
  axial electron-vector hadron terms
(the $\gamma$ exchange contributions are pure of course pure
vector-vector).
The parity-violating asymmetry $A_{\rm PV}$ selects only the
vector-axial vector part of the amplitude,
${\cal M}_{\gamma Z}^{\rm (PV)}$, which can be decomposed into the
vector hadron contribution
${\cal M}_{\gamma Z}^{V \rm (PV)}$ and the axial-vector contribution
${\cal M}_{\gamma Z}^{A \rm (PV)}$.
These have the crossing symmetry properties under the interchange
$s \leftrightarrow u$ (or $E \leftrightarrow -E'$),
\bea
{\cal M}_{\gamma Z}^{V \rm (PV) xbox}(u,t)
&=& - \left. {\cal M}_{\gamma Z}^{V \rm (PV) box}(s,t)
      \right|_{s\to u},
\label{eq:PVcrossV}				\\
{\cal M}_{\gamma Z}^{A \rm (PV) xbox}(u,t)
&=& + \left. {\cal M}_{\gamma Z}^{A \rm (PV) box}(s,t)
      \right|_{s\to u},
\label{eq:PVcrossA}
\eea
and similarly for ${\cal M}_{Z \gamma}^{\rm (PV)}$.

The relative corrections from the $Z(\gamma\gamma)$ and
$\gamma(\gamma Z)$ interference terms to the PV asymmetry
can then be identified as
\bea
\delta_{Z (\gamma\gamma)}
&=& { 2 \Re \left( {\cal M}_Z^{\rm (PV)*} {\cal M}_{\gamma\gamma} \right)
      \over
      2 \Re \left( {\cal M}_Z^{\rm (PV)*} {\cal M}_\gamma \right)}\, , \\
\label{eq:dZ(gg)}
\delta_{\gamma (Z\gamma)}
&=& { 2 \Re \left( {\cal M}_\gamma^* {\cal M}_{\gamma Z}^{\rm (PV)}
                 + {\cal M}_\gamma^* {\cal M}_{Z \gamma}^{\rm (PV)} \right)
      \over
      2 \Re \left( {\cal M}_\gamma^* {\cal M}_Z^{\rm (PV)} \right)}\, ,
\label{eq:dg(gZ)}
\eea
with the full PV asymmetry, including TBE corrections, given by
\be
A_{\rm PV}\ =\ (1+\delta) A_{\rm PV}^0\
\equiv\ \left({1 + \delta_{Z (\gamma\gamma)} + \delta_{\gamma(Z\gamma)}
              \over 1 + \delta_{\gamma (\gamma\gamma)} }
        \right) A_{\rm PV}^0\, ,
\label{eq:delta_APV}
\ee
where $A_{\rm PV}^0$ is the Born level asymmetry in
Eq.~(\ref{eq:APV}).
The electromagnetic TPE correction $\delta_{\gamma(\gamma\gamma)}$
from Eq.~(\ref{eq:delta_gg}) is typically only a few percent 
\cite{Blu03, Blu05, Kon05}, so that the full correction $\delta$ can
be written approximately as
\be
\delta\ \approx\ \delta_{Z(\gamma\gamma)}
	       + \delta_{\gamma(Z\gamma)}
	       - \delta_{\gamma(\gamma\gamma)}\ .
\label{eq:delta_approx}
\ee

The amplitudes ${\cal M}_{\gamma\gamma}$, ${\cal M}_{\gamma Z}$ and
${\cal M}_{Z \gamma}$ contain contributions from both nucleon elastic
and inelastic intermediate states.
At forward scattering angles the contributions from the high-mass
inelastic continuum can be described with the help of dispersion
relations, discussed in Sec.~\ref{ssec:DRgZ} below.
For non-forward angles, the elastic and inelastic terms must be
computed directly, along the lines of the $\gamma\gamma$ corrections
in Sec.~\ref{sec:TPE}.

\subsubsection{Nucleon elastic contributions}
\label{sssec:gZN}

The calculation of the $\gamma Z$ correction with a nucleon intermediate
state to the $\vec{e} p \to e p$ cross section, where the incident
electron is polarized with helicity $h$, mirrors that of the
$\gamma\gamma$ amplitude ${\cal M}_{\gamma\gamma}$ in Eq.~(\ref{eq:Mgg}),
with the replacement of the $\gamma N N$ vertex function
$\Gamma_\gamma^\mu$ by $\Gamma_Z^\mu$ in Eq.~(\ref{eq:JZ})
\cite{Tjo09, Zho10, Zho07},
\be
{\cal M}_{\gamma Z}^{N}\
= -i\, e^2\, \left({G_F\over \sqrt{2}}\right)\, M_Z^2\,
\int {d^4 q_1\over (2\pi)^4}\
     L_{\mu\nu}^{\gamma Z}\, H^{\mu\nu}_{(\gamma Z) N}\,
     \Delta_F(q_1,h)\, \Delta_F(q_2,M_Z).
\label{eq:MgZN}
\ee
Here the leptonic tensor $L_{\mu\nu}^{\gamma Z}$ for the box diagram
is given by
\be
L_{\mu\nu}^{\gamma Z}\
=\ \ubar_e(k',h)\, 
   (g_V^e \gamma_\mu - g_A^e \gamma_\mu \gamma_5)\,
	  S_F(k-q_1,m_e)\, \gamma_\nu\ u_e(k,h).
\label{eq:Lmunu_gZ}
\ee
The hadronic $\gamma Z$ interference tensor $H^{\mu\nu}_{(\gamma Z) N}$
for the nucleon elastic intermediate state is given by
\be
H^{\mu\nu}_{(\gamma Z) N}\
=\ \ubar_N(p')\,
   \Gamma_Z^\mu(q_2)\, S_F(p+q_1,M)\, \Gamma_\gamma^\nu(q_1)\,
   u_N(p).
\label{eq:HgZN}
\ee
Similar expressions hold for the conjugate amplitude
${\cal M}_{Z \gamma}^{N}$.
Only the PV parts of the amplitudes ${\cal M}_{Z \gamma}^{N}$ and
${\cal M}_{\gamma Z}^{N}$ contribute to the PV asymmetry
$A_{\rm PV}$.

While the electromagnetic nucleon form factors used in the $\gamma NN$
vertex functions are relatively well constrained by data, for the
weak $ZNN$ vertex the form factors are generally less well determined.  
For the weak vector current, the form factors can be directly related
to the $\gamma NN$ form factors through the conservation of the vector
current (CVC).  
The axial-vector current, on the other hand, is not conserved, however,
some constraints on the axial form factor have been obtained from
elastic neutrino scattering data.
Using a dipole fit,
	$G_A^{Zp}(Q^2) = -g_A / ( 1 + Q^2/\Lambda_{N (A)}^2 )^2$,
one finds that the mass parameter $\Lambda_{N (A)}^2 \approx 1$~GeV.
Since one of the main purposes of the PV experiments is to extract
strange contributions to form factors by comparing the measured
asymmetry with the predicted zero-strangeness asymmetry, in the
computation of the TBE corrections one can set the strange form
factors to zero, $F_{1,2}^s = 0 = G_A^s$.

\begin{figure}[t]
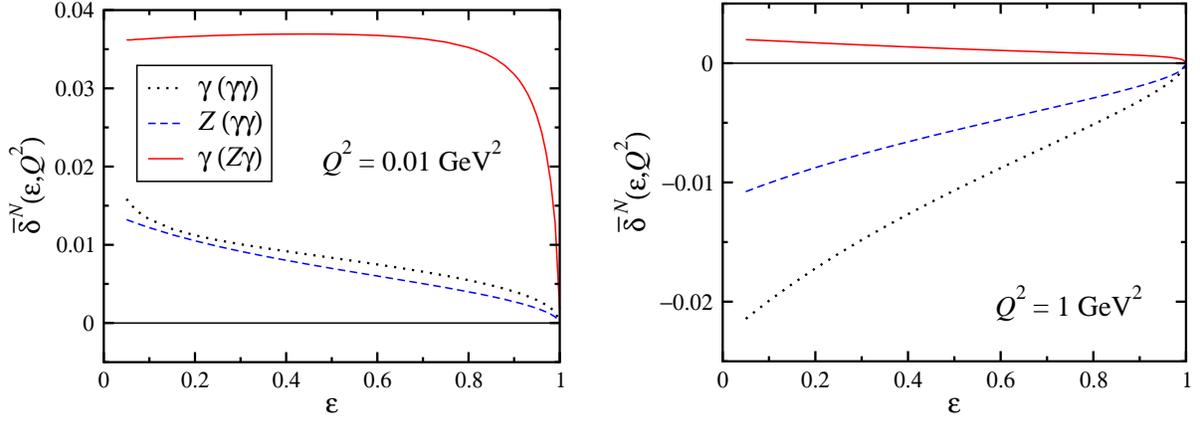

\begin{center}
\includegraphics[height=5.5cm]{Figs/fig37a.eps}
\hspace*{0.5cm}\includegraphics[height=5.5cm]{Figs/fig37b.eps}
\begin{minipage}{16.5cm}
\caption{TBE corrections with nucleon intermediate states, for the
       $\gamma(\gamma\gamma)$ (dotted),
       $Z(\gamma\gamma)$ (dashed) and
       $\gamma(Z\gamma)$ (solid) contributions
       at $Q^2=0.01$ {\bf (left)} and 1~GeV$^2$ {\bf (right)}.
       The finite part of the correction is defined with
       respect to the Mo-Tsai IR contribution,
       $\overline\delta^N = \delta^N - \delta_{\rm IR}(\rm MoT)$
       \cite{Tsa61, Mo69}.
       The $\gamma(\gamma\gamma)$ correction enters with the opposite
       sign in the asymmetry, Eq.~(\ref{eq:delta_approx}).
       Figure adapted from Ref.~\cite{Tjo09}.}
\label{fig:delN}
\end{minipage}
\end{center}
\end{figure}

The contributions from the nucleon intermediate states to the TBE
correction $\delta^N$ are illustrated in Fig.~\ref{fig:delN} as a
function of $\eps$ for $Q^2 = 0.01$ and 1~GeV$^2$.
At low $Q^2$ the $\gamma(\gamma\gamma)$ and $Z(\gamma\gamma)$
contributions are very similar, and therefore partially cancel
in the asymmetry $A_{\rm PV}$.  Consequently the asymmetry is
determined mostly by the $\gamma(Z\gamma)$ component.
At larger $Q^2$ ($\gtorder 1$~GeV$^2$) the $\gamma(Z\gamma)$
component decreases in magnitude, while the $\gamma(\gamma\gamma)$
and $Z(\gamma\gamma)$ pieces become large and more negative 
\cite{Blu03, Blu05, Tjo09}.
The dependence of the total correction $\delta^N$ on the input form
factors was found \cite{Blu05, Tjo09} to be very weak for all $\eps$,
and only becomes appreciable at large $Q^2$ ($Q^2 \gtorder 1$~GeV$^2$).
The correction at $Q^2 = 0.01$~GeV$^2$ also appears relatively flat
over the range $0.1 \ltorder \eps \ltorder 0.8$, before dropping
rapidly as $\eps \to 1$.
At large $Q^2$ the total TBE correction becomes more strongly $\eps$
dependent, decreasing in magnitude at forward scattering angles but
increasing at backward angles ($\eps \to 0$).

\subsubsection{$\Delta$ intermediate states}

Contributions to TBE amplitudes from the excitation of the $\Delta(1232)$
resonance were recently considered by Tjon {\it et al.} \cite{Tjo09} 
and Nagata {\it et al.} \cite{Nag09}.  Evaluation of the relevant TBE 
amplitudes follows that of the electromagnetic contribution in 
Sec.~\ref{sssec:Delta}, extending the formalism of Kondratyuk {\it et al.}
\cite{Kon05} to the weak sector.

For the $ZN\Delta$ vertex both vector and axial-vector contributions enter.
The vector transitions are required by CVC to have the same form as the 
$\gamma N\Delta$ in Eq.~(\ref{eq:gDN}),
\bea
\Gamma_{ZN \to \Delta}^{\alpha\mu(V)}(p_\Delta,q)
&=& {1 \over 2 M_\Delta^2}\sqrt{2\over 3}
\Big\{
  g_1^V(Q^2)
  \left[ g^{\alpha \mu} \fslash{q}\,\fslash{p}_\Delta
    - \fslash{q}\, \gamma^\alpha p_\Delta^\mu
    - \gamma^\alpha \gamma^\mu q\cdot p_\Delta
    + \fslash{p}_\Delta\,\gamma^\mu q^\alpha
  \right]                               \nn\\
& & \hspace*{-2.5cm}
+\ g_2^V(Q^2)
  \left[ q^\alpha p_\Delta^\mu - g^{\alpha\mu} q\cdot p_\Delta
  \right]\
+\ {g_3^V(Q^2) \over M_\Delta}
  \left[ q^2 \left( \gamma^\alpha p_\Delta^\mu
                  - g^{\alpha\mu} \fslash{p}_\Delta
             \right)
       + q^\mu \left( q^\alpha \fslash{p}_\Delta
                    - \gamma^\alpha q\cdot p_\Delta
               \right)
  \right]
\Big\} \gamma_5
\label{eq:ZDN_V}
\eea
where again the factor $\sqrt{2/3}$ is associated with the
$N \to \Delta$ weak isospin transition.
Using CVC and isospin symmetry, the vector $ZN\Delta$ form factors
can be related to the $\gamma N \Delta$ form factors by \cite{Tjo09}
\be
g_i^V(Q^2) = 2 (1 - 2\sw)\, g_i(Q^2)\, ,\ \ \ i=1,2,3
\ee
where the $Q^2$ dependence of the electromagnetic $\gamma N \Delta$
form factor is parametrized as in Sec.~\ref{sssec:Delta}.

For the axial-vector vertex, nonconservation of the axial-vector
current implies the existence of an addition form factor.
One can use the partially conserved axial current (PCAC) hypothesis
to relate two of the form factors, leaving a similar expression to
that in Eq.~(\ref{eq:ZDN_V}),
\bea
\Gamma_{ZN \to \Delta}^{\alpha\mu(A)}(p_\Delta,q)
&=& {1 \over 2 M_\Delta^2}\sqrt{2\over 3}
\Big\{
  g_1^A(Q^2)
  \left[ g^{\alpha \mu} \fslash{q}\,\fslash{p}_\Delta
    - \fslash{q}\, \gamma^\alpha p_\Delta^\mu
    - \gamma^\alpha \gamma^\mu q\cdot p_\Delta
    + \fslash{p}_\Delta\,\gamma^\mu q^\alpha
  \right]                               \nn\\
& & \hspace*{-2.5cm}
+\ g_2^A(Q^2)
  \left[ q^\alpha p_\Delta^\mu - g^{\alpha\mu} q\cdot p_\Delta
  \right]\
+\ {g_3^A(Q^2) \over M_\Delta}
  \left[ q^2 \left( \gamma^\alpha p_\Delta^\mu
                  - g^{\alpha\mu} \fslash{p}_\Delta
             \right)
       + q^\mu \left( q^\alpha \fslash{p}_\Delta
                    - \gamma^\alpha q\cdot p_\Delta
               \right)
  \right]
\Big\}.
\label{eq:ZDN_A}
\eea
Note that here the weak isospin transition factor has been absorbed
into the definition of the couplings \cite{Lal05, Lal06}.

The axial form factors are less well determined, but some constraints
have been extracted from analysis of $\nu$ scattering data.
In a recent analysis, Lalakulich {\it et al.} \cite{Lal05, Lal06}
parametrized the $\nu N \to \mu \Delta$ cross sections from bubble
chamber experiments at low $Q^2$ in terms of phenomenological form
factors.  The available data can be described by the form factors
\cite{Tjo09, Lal06, Lal07}
$g_1^A(Q^2) = 0$,
$g_2^A(Q^2) = (Q^2/4 M^2)\, g_3^A(Q^2)$,
%
with the couplings at $Q^2=0$ determined in Ref.~\cite{Lal06},
and the $Q^2$ dependence given by a dipole form having a cut-off
mass of $\Lambda_{\Delta (A)} = 1.0$~GeV.
Note that Ref.~\cite{Nag09} uses a different basis of form factors
than in Eqs.~(\ref{eq:ZDN_V}) and (\ref{eq:ZDN_A}); for the relation
between these see \cite{Tjo09}.  As for the electromagnetic case,
the vertex with an incoming $\Delta$ can be obtained from
Eq.~(\ref{eq:Gconj}).

The $\gamma Z$ interference amplitude for the box
diagram with a $\Delta$ intermediate state can then be written 
\cite{Tjo09, Nag09}
\be
{\cal M}_{\gamma Z}^{\Delta}\
= -i\, e^2\, \left({G_F\over \sqrt{2}}\right)\, M_Z^2\,
   \int {d^4 q_1\over (2\pi)^4}\
   L_{\mu\nu}^{\gamma Z}\, H^{\mu\nu}_{(\gamma Z) \Delta}\,
   \Delta_F(q_1,0)\, \Delta_F(q_2,M_Z),
\label{eq:MgZD}
\ee
where the leptonic tensor $L_{\mu\nu}^{\gamma Z}$ is the same
as in Eq.~(\ref{eq:Lmunu_gZ}).
The hadronic interference tensor $H^{\mu\nu}_{(\gamma Z) \Delta}$
for the $\Delta$ intermediate state is now
\be
H^{\mu\nu}_{(\gamma Z) \Delta}\
=\ \ubar_N(p')\,
   \Gamma_{\Delta\to ZN}^{\mu\alpha}(p+q_1,-q_2)\,
   S_{\alpha\beta}(p+q_1,M_\Delta)\,
   \Gamma_{\gamma N\to\Delta}^{\beta\nu}(p+q_1,q_1)\,
   u_N(p),
\label{eq:HgZD}
\ee
where $\Gamma_{\Delta\to Z N}^{\mu\alpha}$ is the sum of the vector
and axial-vector vertices in Eqs.~(\ref{eq:ZDN_V}) and (\ref{eq:ZDN_A}).
The corresponding amplitude ${\cal M}_{\gamma Z}^{\Delta}$ can be
derived in a similar manner.

\begin{figure}
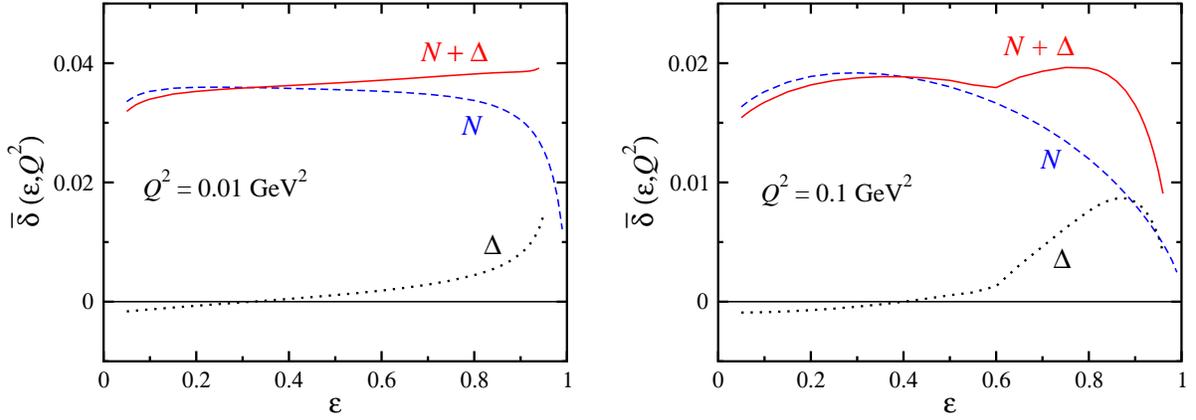

\begin{center}
\includegraphics[height=5.5cm]{Figs/fig38a.eps}
\hspace*{0.5cm}\includegraphics[height=5.5cm]{Figs/fig38b.eps}
\begin{minipage}{16.5cm}
\caption{Total finite parts of the TBE corrections $\overline\delta$,
       relative to the Mo-Tsai contribution \cite{Tsa61, Mo69}, with
       nucleon (dashed) and $\Delta$ (dotted) intermediate states,
       as well as the sum (solid), at
       $Q^2=0.01$~GeV$^2$ {\bf (left)} and 0.1~GeV$^2$ {\bf (right)}.
       Figure adapted from Ref.~\cite{Tjo09}.}
\label{fig:delND}
\end{minipage}
\end{center}
\end{figure}

The total TBE contribution from $\Delta$ intermediate states,
relative to the Mo-Tsai IR result \cite{Tsa61, Mo69}, is shown
in Fig.~\ref{fig:delND} as a function of $\eps$ for $Q^2 = 0.01$
and 0.1~GeV$^2$.
At low $Q^2$ values the $\Delta$ contribution is dominated by the
$\gamma(Z\gamma)$ term.  The negligible $\gamma\gamma$ interference
with the Born $\gamma$ or $Z$ exchange reflects the vanishing of the
two-photon exchange correction in the $Q^2 \to 0$ limit.
For comparison we also show in Fig.~\ref{fig:delND} the nucleon
elastic contribution, and the sum of the nucleon and $\Delta$ terms.
The $\Delta$ correction is strongly suppressed at low $\eps$, but
grows with increasing $\eps$, becoming as important as the nucleon
elastic part near the forward limit, $\eps \to 1$.
It is interesting to contrast this with the role of the $\Delta$
in TPE corrections to electromagnetic scattering discussed in
Sec.~\ref{sssec:Delta}, where the $\Delta$ contribution was negligible
at forward angles but becomes important at backward angles, where
magnetic scattering is dominant.

The increased magnitude of the $\Delta$ contribution to PVES at forward
angles reflects the growth of the invariant center of mass energy for
fixed $Q^2$ as $\eps \to 1$ \cite{Tjo09, Nag09} --- since the $\Delta$
intermediate state amplitudes ${\cal M}_{\gamma\gamma}^{\Delta}$ and
${\cal M}_{\gamma Z}^{\Delta}$ have numerators with higher powers of
loop momenta than the corresponding nucleon amplitudes, $\delta^\Delta$
will grow faster with invariant energy than $\delta^N$.
One method to ensure consistency with the physical bounds at
asymptotically high energy requires is to use so-called ``sideways''
form factors, which depend on the center of mass energy as well as
on the mass of the virtual boson.
Alternatively, a dispersive approach, where the intermediate states
are on-shell, may be adopted, such as that at forward angles
discussed in Sec.~\ref{ssec:DRgZ}.
In the absence of a detailed analysis of the high-energy behavior
of the higher-mass contributions, the predictions of the $\Delta$
calculation should not be taken too literally at very large
$\eps \sim 1$.

\subsubsection{Effects on the strange form factors}

While the TBE contributions to the parity-violating asymmetry in elastic
$ep$ scattering are relatively small, there are two ways in which the
impact on the measurements can be enhanced.  Although the asymmetry does
not have large corrections, these measurements are typically looking
for contributions from strange quarks which are seen through very small
deviations from the asymmetry expected in the absence of strange quarks.
A few-percent correction to $A_{\rm PV}$ will therefore be significant
if the contribution from strange quarks is at the level of $5-10\%$.
In addition, the corrections discussed in the previous section show the
difference between the Born prediction for $A_{\rm PV}$ and the value
including TBE contributions, assuming that one is starting from the
correct electromagnetic form factors.  There is an additional effect
for measurements which used proton form factors extracted without
accounting for TPE corrections in determining the value of $A_{\rm PV}$
for the case with no strange quark contributions.

In Ref.~\cite{Arr07b}, the impact of TPE on the electromagnetic form
factors and the TPE corrections to the parity-violating asymmetry
were examined, but $\gamma Z$ box contributions were not included.
The direct TPE contributions to the asymmetry were found to be
$\ltorder 1\%$ for $Q^2 \le 3$~GeV$^2$, and smallest at large $\eps$,
where the highest precision measurements were planned.  However, the
impact of using proton form factors that did not have TPE corrections
applied was found to be much larger, well over 5\% at 3~GeV$^2$, with
the largest corrections in high-$\eps$ region where high precision
measurements were planned.
A prescription was given that allowed for approximate corrections
to be applied, without explicit inclusion of the TPE corrections.
But because this estimate did not include the $\gamma Z$ contributions
and the TPE-corrected form factors can now be reliably extracted,
this procedure is now most relevant if looking at previous extractions
of the strangeness contribution based on TPE-uncorrected form factors.
However, the use of TPE-uncorrected form factors yields the largest
TPE-related error in most of the earlier extractions, as the absolute 
corrections for the two-photon and $\gamma Z$ diagrams is relatively small.
However, because these corrections modify all of the extracted strange
form factor contributions in a highly-correlated fashion, even these
small corrections may become important in global analyses of the strange
quark contributions.

The full effects of the TBE corrections (two-photon and $\gamma Z$
exchange) on the $A_{\rm PV}$ asymmetry at kinematics corresponding
to experiments 
\cite{HAP04,HAP06,HAP06a,HAP07,G005,G010,SAM97,MAM04,MAM05,MAM09}
designed to measure the strange quark form factors of the nucleon
were considered in Refs.~\cite{Tjo09, Zho10, Zho07, Nag09}.
For the forward angle HAPPEX \cite{HAP04} and G0 \cite{G005}
measurements, the nucleon correction $\delta^N$ was found to be in
the vicinity of $\sim 0.1-0.2\%$, increasing to $\sim 1.0-1.5\%$
for the backward angle G0 \cite{G010} and the earlier SAMPLE
\cite{SAM97} measurements.  In contrast, the $\Delta$ contribution
$\delta^\Delta$ is almost negligible at backward angles, but becomes
more important at forward angles, where as discussed above its
evaluation is more questionable.

The impact of these corrections on the strange form factors is
difficult to gauge without performing a full reanalysis of the data,
since in general different electroweak parameters and form factors
are used in the various experiments.
Nevertheless, following Zhou {\em et al.} \cite{Zho07} attempts have
been made \cite{Tjo08, Zho10, Nag09} to quantify the effect from the
TBE corrections on the combination $G_E^s + \beta G_M^s$ measured in
the PV experiments, where $\beta$ depends on the kinematics of a
particular experiment.
Zhou {\em et al.} assume the experimental asymmetry
$A_{\rm PV}^{\rm exp}$ can be written as \cite{Zho07}
\be
A_{\rm PV}^{\rm exp}(\rho,\kappa)
= A_{\rm PV}^{0}(\rho',\kappa') (1 + \delta^N + \delta^\Delta),
\ee
where $A_{\rm PV}^0$ is the Born asymmetry (\ref{eq:APV}),
and $\rho' = \rho - \Delta\rho$ and $\kappa' = \kappa - \Delta\kappa$
remove the existing (hadron structure-independent) two-boson exchange
corrections contained in $\rho$ and $\kappa$ \cite{Mar83, Mar84},
From the calculated TBE corrections one can then determine
$A_{\rm PV}^{0}$ and extract the strange asymmetry $A_s$ (\ref{eq:A_s}).
Defining the relative correction $\delta_G$ to the strange form factors
by   $\overline{G}_E^s + \beta \overline{G}_M^s
      = (G_E^s + \beta G_M^s)(1+\delta_G)$,
where $\overline{G}_{E,M}^s$ is extracted from the experimental
asymmetry and $G_{E,M}^s$ is obtained from the Born asymmetry,
one has \cite{Zho07}
\be
\delta_G
= {1 \over A_{\rm PV}^{\rm exp} - A_V - A_A}
  \left\{ A_{\rm PV}^{\rm exp}
	  \left( {\Delta\rho\over\rho} - \delta \right)
	+ {G_F Q^2\over 2\sqrt{2}\pi\alpha} \rho\sw \Delta\kappa
	- A_A {\Delta\rho\over\rho}
  \right\}.
\label{eq:delG}
\ee
Because the magnitude of the strange form factors is generally rather
small \cite{You06, Liu07}, dividing by the strange asymmetry in
Eq.~(\ref{eq:delG}) can give a quite large relative correction
$\delta_G$.  The size of the correction was found to be typically
within 10--20\% for most PV experiments \cite{Zho10}.

The model dependence of these corrections was explored in
Ref.~\cite{Tjo08}, where an estimate was made of the induced difference
in the strange asymmetry extracted using various input electroweak
form factors.  Comparing results with empirical \cite{AMT07} and
monopole form factors (as used in Ref.~\cite{Zho07}), the effects
ranged from around 15\% for the HAPPEX kinematics \cite{HAP04, HAP07}
to over 30\% for the PVA4 kinematics \cite{MAM04}.
One should caution, however, that these values are indicative only,
and a more detailed reanalysis of the strange form factor data
including TBE effects {\it a priori} would be needed to reach more
robust conclusions.

In addition, the fractional contributions may not reflect the
importance of the corrections because the strange quark contributions
are related to the difference between the measured asymmetry and the
expected asymmetry in the absence of strange quark contributions.
Thus, a large fractional correction could simply indicate that the
strangeness contribution is extremely small.  It may therefore be
more useful to compare the size of the TBE corrections to the
uncertainty in the $A_{\rm PV}$ measurements.

A preliminary attempt at incorporating TBE corrections in a global
strange form factors analysis was made recently by Young \cite{You09}.
Using PV scattering data below $Q^2 = 0.3$~GeV$^2$ the TBE corrections
in the proton were found to modify the strange electric and magnetic
form factors by
\be
G_E^s = 0.0025(182)\ \longrightarrow\ 0.0023(182)\, ,\ \ \ \
G_M^s = -0.011(254)\ \longrightarrow\ -0.020(254)\, ,
\ee
at a scale $Q^2=0.1$~GeV$^2$.
The effect on the strange magnetic form factor is an almost factor
two increase in the magnitude; while significant, this is still well
within the current experimental uncertainty.
The shift in the strange electric form factor is somewhat smaller.
This is mostly because the $G_E^s$ form factor is determined largely by
the $^4$He data, and the TBE effects here have not yet been computed.
Overall, though, the conclusion appears to be that TBE effects provide
relatively mild corrections to strange quark form factors.
On the other hand, a significantly larger effect from TBE has been
found in nearly forward PV electron scattering at very low $Q^2$,
which we discuss next.

\subsection{\it $\gamma Z$ corrections to the proton weak charge}
\label{ssec:DRgZ}

In parallel with the discussion of Sec.~\ref{ssec:DRgg}, we consider
the forward scattering limit, where $k'=k-q$, with $t=q^2$ kept small
but finite.
The parity-violating proton asymmetry in Eq.~(\ref{eq:APVp}) is related
to the weak charge of the proton $Q_W^p$ \cite{Mus94},
\be
A_{\rm PV}\ \to\ \frac{G_F}{4 \pi \alpha \sqrt{2}}\, t\,Q_W^p\, .
\label{eq:Apvt0}
\ee
Including electroweak radiative corrections, the proton weak charge
is defined at zero electron energy $E$ and zero momentum transfer as
\cite{Erl03}   
\be
Q_W^p\ =\ (1 + \Delta\rho + \Delta_e) (1 - 4 \sw(0) + \Delta_e')\
       +\ \Box_{WW}\ +\ \Box_{ZZ}\ +\ \Box_{\gamma Z}(0),
\label{eq:Qwp}
\ee
where $\sw(0) = 0.23867(16)$ is the weak mixing angle at zero momentum,
and the corrections $\Delta\rho$, $\Delta_e$ and $\Delta_e'$ are given
in Ref.~\cite{Erl03}.
The contributions $\Box_{WW}$ and $\Box_{ZZ}$ from the $WW$ and $ZZ$
box and crossed-box diagrams can be computed perturbatively,
while the $\gamma Z$ interference correction $\Box_{\gamma Z}(E)$
in addition depends on physics at long distance scales
\cite{Mus94, Mar83, Mar84, Erl03, Erl05}.  The current best
theoretical estimate from Ref.~\cite{Erl05} is $Q_W^p=0.0713(8)$.
An explicit energy-dependence is shown for $\Box_{\gamma Z}(E)$ in
Eq.~(\ref{eq:Qwp}), in anticipation of a significant variation of this
quantity for electron scattering in the GeV range. The other radiative
corrections in Eq.~(\ref{eq:Qwp}) are not expected to show such a
dependence.

Note that the absolute correction $\Box_{\gamma Z}$ to the proton
weak charge is related to the relative correction $\delta_{\gamma Z}$
in Eq.~(\ref{eq:dg(gZ)}) by
\be
\Box_{\gamma Z}(0)\
\equiv\ Q_W^p\, \delta_{\gamma Z}\
=\ Q_W^p\,
 { \Re \left( {\cal M}_\gamma^*
	      \left[ {\cal M}_{\gamma Z}^{\rm (PV)}
	           + {\cal M}_{Z \gamma}^{\rm (PV)}
	      \right]
       \right)
   \over
   \Re \left( {\cal M}_\gamma^* {\cal M}_Z^{\rm (PV)} \right) }.
\label{eq:gZbox}
\ee
Corrections from the interference of ${\cal M}_Z^{\rm (PV)}$
with the TPE amplitude under the replacement
${\cal M}_\gamma \to {\cal M}_\gamma + {\cal M}_{\gamma\gamma}$
(namely, $\delta_{Z(\gamma\gamma)}$ in Eq.~(\ref{eq:dZ(gg)})) vanish
in the forward limit, and therefore do not affect the asymmetry.

A precise determination of the proton weak charge $Q_W^p$ can provide
an important test of the Standard Model through verification of the
predicted running of the weak mixing angle from the $Z$ boson pole
to low energies \cite{Erl05}.
The ${\rm Q_{weak}}$ experiment at Jefferson Lab \cite{Qwk05},
which will complete data taking in 2012, was designed to measure
$Q_W^p$ to a higher level of precision than previously possible.
In combination with constraints from atomic parity violation
\cite{Por09}, the ${\rm Q_{weak}}$ measurement aims to either
discover evidence for new physics beyond the Standard Model that
leads to parity violation in electron scattering, or raise the
limit on its mass scale to above 2~TeV, complementing direct
searches at the LHC \cite{Erl05, You07}.

As discussed above, the $\gamma Z$ correction in general has
contributions from the
vector electron--axial vector hadron coupling of the $Z$ boson
  ($\Box_{\gamma Z}^{A}$) and from the
axial electron--vector hadron coupling of the $Z$
  ($\Box_{\gamma Z}^{V}$),
$\Box_{\gamma Z} = \Box_{\gamma Z}^{A} + \Box_{\gamma Z}^{V}$.
The vector hadron contribution $\Box_{\gamma Z}^{V}$ vanishes in
the limit of zero energy, but is finite at $E > 0$.
The axial hadron correction $\Box_{\gamma Z}^{A}$, which is dominant
at very low electron energies relevant to atomic parity-violation
experiments, was estimated some time ago by Marciano and Sirlin
\cite{Mar83, Mar84} in terms of a free quark model-inspired loop
calculation.

At forward angles, the correction $\Box_{\gamma Z}$ can be computed
from its imaginary part using standard forward dispersion relations
analogous to Eq.~(\ref{eq:DR1_gg})~\cite{Gor09},
\be
\Re\, \Box_{\gamma Z}(E)
= {1 \over \pi}
    {\cal P} \int_{-\infty}^\infty dE'\,
    {\Im\, \Box_{\gamma Z}(E') \over E'-E}.
\label{eq:DR}
\ee
The imaginary part of $\Box_{\gamma Z}$ depends on the PV $ep \to eX$
cross section, which can be expressed in terms of interference
electroweak structure functions.
The integration over negative energies in Eq.~(\ref{eq:DR}) corresponds
to the crossed $\gamma Z$ box diagram, and can be related to the box
diagram using the crossing symmetry properties of the $\gamma Z$
amplitudes in Eq.~(\ref{eq:PVcrossV}) and (\ref{eq:PVcrossA}).
For the vector hadron and axial hadron loop corrections one has,
in analogy with Eq.~(\ref{eq:DR2_gg}),
\bea
\Re\, \Box_{\gamma Z}^{V}(E)
&=& {2E \over \pi}
    {\cal P} \int_0^\infty dE' {1 \over E'^2-E^2}\,
    \Im\, \Box_{\gamma Z}^{V}(E'),
\label{eq:DRV}					\\
\Re\, \Box_{\gamma Z}^{A}(E)
&=& {2 \over \pi}
    {\cal P}  \int_0^\infty dE' {E' \over E'^2-E^2}\,
    \Im\, \Box_{\gamma Z}^{A}(E').
\label{eq:DRA}
\eea
It is evident from Eq.~(\ref{eq:DRV}) that at zero energy the vector
hadron correction vanishes, $\Re\, \Box_{\gamma Z}^{V}(0)=0$.

From the optical theorem, the imaginary part of PV $\gamma Z$ exchange
amplitude can be written in terms of the cross section for all possible 
final hadronic states,
\be
2\,\Im\, \left( {\cal M}_{\gamma Z}^{\rm (PV)}
	   + {\cal M}_{Z \gamma}^{\rm (PV)}
      \right)
=\ 4 \pi M\, e^2 \left({-2 G_F \over \sqrt{2}}\right)\
   \int{d^3l \over (2\pi)^3 2 E_l}
   \left( {1\over Q_1^2} \right)
   {1 \over 1+Q_1^2/M_Z^2}\,
   L_{\mu\nu}^{\gamma Z}\, W^{\mu\nu}_{\gamma Z}\, ,
\label{eq:ImBoxdef}
\ee
where $l = k-q_1$, $E_l = \sqrt{(\bm{k}-\bm{q}_1)^2 + m_e^2}$,
and $Q_1^2 = -q_1^2$.
The $\gamma Z$ interference hadronic tensor can be parametrized in
terms of three interference electroweak structure functions,
\be
M W^{\mu\nu}_{\gamma Z}\
=\ - g^{\mu\nu} F_1^{\gamma Z}
   + {p^\mu p^\nu \over p\cdot q_1} F_2^{\gamma Z}
   - i \eps^{\mu\nu\lambda\rho}
     {p_\lambda q_{1\rho} \over 2 p \cdot q_1} F_3^{\gamma Z}.
\label{eq:Wmunu_gZ}
\ee
The structure functions $F_{1,2}^{\gamma Z}$ are analogous to the
electromagnetic structure functions in Eq.~(\ref{eq:Wmunu_def})
and contribute to the vector hadron correction $\Box_{\gamma Z}^{V}$,
while the $F_3^{\gamma Z}$ structure function contributes to the
axial hadron correction $\Box_{\gamma Z}^{A}$.
They are in general functions of the exchanged boson virtuality $Q_1^2$
and of the invariant mass $W$ of the exchanged boson and proton, or
alternatively of the Bjorken variable $x = Q_1^2/(W^2-M^2+Q_1^2)$.

In analogy with Eq.~(\ref{eq:Imdelgg}), the imaginary parts of the
vector hadron and axial hadron contributions to the $\gamma Z$ box
diagrams can be written as \cite{Gor09, Gor11, Sib10, Ris10, Blu11}
\bea
\Im\, \Box_{\gamma Z}^{V}(E)
&=& {\alpha \over (2ME)^2}
    \int_{M^2}^s dW^2
    \int_0^{Q^2_{1,\rm max}} {dQ_1^2 \over 1+Q_1^2/M_Z^2}			
    \left[ F_1^{\gamma Z}
         + { s \left( Q^2_{1, \rm max}-Q_1^2 \right) \over
		      Q_1^2 \left( W^2 - M^2 + Q_1^2 \right) }
	   F_2^{\gamma Z}
    \right],
\label{eq:ImBoxV}					\\
\Im\, \Box_{\gamma Z}^{A}(E)
&=& \hat{v}_e {\alpha \over (2ME)^2}
    \int_{M^2}^s dW^2
    \int_0^{Q^2_{1, \rm max}} {dQ_1^2 \over 1+Q_1^2/M_Z^2}
    \left[ {2ME \over W^2-M^2+Q_1^2} - {1 \over 2} \right]
    F_3^{\gamma Z},
\label{eq:ImBoxA}
\eea
where
$\hat{v}_e \equiv (1-4\hat{s}^2)$, with
$\hat{s}^2 \equiv \sin^2{\theta_W(M_Z^2)} = 0.23116(13)$ 
in the $\overline{\rm MS}$ scheme \cite{PDG10}.
%
%
Note that Eqs.~(\ref{eq:ImBoxV}) and (\ref{eq:ImBoxA}) assume that
the scale dependence of $\alpha$ and $\sin^2\theta_W$ is negligible
(see, however, Ref.~\cite{Blu11}).

In evaluating the imaginary parts of $\Box_{\gamma Z}^{V,A}$
it is convenient to split the integration into three regions:
(i) elastic with $W^2=M^2$;
(ii) resonance with $(M+m_\pi)^2 \le W^2 \ltorder 4$~GeV$^2$; and
(iii) deep inelastic scattering (DIS) region, with $W^2>4$~GeV$^2$.
The elastic contributions, including both the $\gamma Z$ and $Z \gamma$
orderings, can be written in terms of the elastic electroweak form
factors defined in Sec.~\ref{ssec:APV},
\bea
F_1^{\gamma Z (\rm el)}(W^2,Q_1^2)
&=& 2M^2 \tau\, G_M^{\gamma p}\, G_A^Z\, \delta(W^2-M^2),
\label{eq:F1el}				\\
F_2^{\gamma Z (\rm el)}(W^2,Q_1^2)
&=& {4M^2 \tau \over 1+\tau}\,
    \left( G_E^{\gamma p}\, G_E^{Zp}\,
	   +\, \tau\, G_M^{\gamma p}\, G_M^{Zp}
    \right)\, \delta(W^2-M^2),
\label{eq:F2el}				\\
F_3^{\gamma Z (\rm el)}(W^2,Q_1^2)
&=& -4M^2 \tau\, G_M^{\gamma p}\, G_A^Z\, \delta(W^2-M^2),
\label{eq:F3el}
\eea
where here $\tau = Q_1^2/4M^2$.
Note that with dipole parametrizations of form factors the integrals
(\ref{eq:DRV}), (\ref{eq:DRA}), (\ref{eq:ImBoxV}) and (\ref{eq:ImBoxA})
can be performed analytically, providing a useful cross-check of the
numerical calculations.
One can verify that the results in fact coincide exactly with the
direct loop calculations \cite{Tjo08, Zho07} of the $\gamma Z$
corrections in Sec.~\ref{sssec:gZN}, which do not use dispersion
relations, and in which the intermediate nucleon is off-shell.
The result for $\Re\, \Box_{\gamma Z}^{A}(E)$ also agrees at $E=0$
with the classic calculation of Marciano and Sirlin \cite{Mar83}, if the
parameters are adjusted to correspond to those of Ref.~\cite{Mar83}.
To simplify notation, in the following we denote
$\Re\, \Box_{\gamma Z}^{V,A}$ by $\Box_{\gamma Z}^{V,A}$,
since these are the quantities of interest for the observables.
Numerical results for the elastic contribution,
$\Box_{\gamma Z}^{A \rm (el)}(E)$, are given in Table~\ref{tab:boxa}
at several energies $E$.

\begin{table}[bt]
\begin{center}
\begin{minipage}{16.5cm}
\caption{\label{tab:boxa}
   Contributions to the real part of $\Box_{\gamma Z}^{A}(E)$
   from different kinematical regions~\cite{Blu11}. The quoted
   uncertainty on the total accounts for model-dependence as
   well as uncertainties in parameters. The energy of the
   ${\rm Q_{weak}}$ experiment is 1.165~GeV.}
\end{minipage}
\begin{tabular}{llll}
\\ \hline
& \multicolumn{3}{c}{$E$ (GeV)}\\
\cline{2-4}
& 0.0 & 1.165 & 3.0\\
\hline
elastic 				 & 0.00064 & 0.00005 & 0.00001 \\
resonance 			 & 0.00023 & 0.00012 & 0.00002 \\
DIS ($Q^2>1$~GeV$^2$) & 0.00327 & 0.00328 & 0.00330 \\
DIS ($Q^2<1$~GeV$^2$) & 0.00024 & 0.00025 & 0.00027 \\
\cline{1-4}
Total 			  	 & 0.0044(4) & 0.0037(4) & 0.0036(4)\\
\hline
\end{tabular}
\end{center}
\end{table}

In the nucleon resonance region, while there is an abundance
of electroproduction data, there are no direct measurements of
the interference structure functions $F_{1,2,3}^{\gamma Z}$.
For vector transitions to isospin $I=3/2$ states, such as the
$\Delta(1232)$ resonance, conservation of the vector current and
isospin symmetry require the weak isovector transition form factors
to be equal to the electromagnetic ones multiplied by $(1+Q_W^p)$.
For isospin $I=1/2$ resonances, which contain contributions from
isovector and isoscalar currents, using SU(6) quark model wave
functions one can verify that for the most prominent $I=1/2$
states the magnitudes of the $Z$-boson transition couplings are
equal to the respective photon couplings to within a few percent
\cite{Gor11, Sib10}.
Estimates of the resonance region contribution to the box diagram,
$\Box_{\gamma Z}^{V \rm (res)}$, have been made recently
\cite{Gor09, Gor11, Sib10, Ris10} by parametrizing resonance region
structure function data from SLAC, Jefferson Lab and elsewhere
in terms of resonant and nonresonant background components
\cite{Sib10, Chr10}.
The model dependence of relating the vector $\gamma Z$ resonance
structure functions to the electromagnetic structure functions
has been studied in Ref.~\cite{Gor11}.

For the axial-vector resonance contributions,
$\Box_{\gamma Z}^{A \rm (res)}$, one can use parametrizations
of the transition form factors in neutrino scattering from
Lalakulich {\it et al.} \cite{Lal05, Lal06}, with modified
isospin factors appropriate to $\gamma Z$.
These form factors have been fit to the Jefferson Lab pion
electroproduction data (vector part) and pion production data
in $\nu$ and $\bar\nu$ scattering at ANL, BNL and Serpukhov
(axial-vector part) up to $Q_1^2 = 3.5$~GeV$^2$.
At larger $Q_1^2$ the resonance contributions are suppressed by the
$Q_1^2$ dependence of the transition form factors, which is stronger 
for the dominant $\Delta(1232)$ resonance than for the higher-mass
resonances \cite{Lal06}.
The resulting resonance contribution,
$\Box_{\gamma Z}^{A \rm (res)}(0)$, is smaller than the elastic term at 
$E=0$, but decreases less rapidly with increasing energy, as shown in 
Table~\ref{tab:boxa}.

The contributions from high $W$ can be computed by analyzing the
high-$Q_1^2$ ($Q_1^2 > Q_0^2$) and low-$Q_1^2$ ($Q_1^2 < Q_0^2$)
regions separately, with $Q_0^2$ taken to be $\approx 1$~GeV$^2$.
Structure functions in the high-$W$, high-$Q_1^2$ region of
deep-inelastic scattering can be related to leading twist parton
distribution functions (PDFs), and at leading order in $\alpha_s$
are given by
\bea
F_2^{\gamma Z (\rm DIS)}
&=& x \sum_q 2\, e_q\, g_V^q\, ( q + \bar q )\
 =\ 2x F_1^{\gamma Z (\rm DIS)},
\label{eq:F2q}				\\
x F_3^{\gamma Z (\rm DIS)}
&=& x \sum_q 2\, e_q\, g_A^q\, ( q - \bar q ),
\label{eq:F3q}
\eea
where $q$ and $\bar q$ are the quark and antiquark PDFs,
and $g_V^u = 1/2 - (4/3) \sw$ and $g_V^d = -1/2 + (2/3) \sw$
are the weak vector charges for $u$ and $d$ quarks, respectively.
Note that in the limit $2 g_V^q \to e_q$ the interference structure
functions $F_2^{\gamma Z (\rm DIS)} \to F_2^{\gamma (\rm DIS)}$,
where
\be
F_2^{\gamma (\rm DIS)}\ =\ x \sum_q e_q^2\, ( q + \bar q ).
\ee
In the case of a free quark target one has
$F_3^{\gamma Z} = (5/3)\,x\,\delta(1-x)$, which gives a
contribution to the axial hadron $\gamma Z$ correction
\be
\Box_{\gamma Z}^{A\, \rm (free\ quark)}\
=\ \hat{v}_e\,{5\alpha\over 2\pi} \,
   \left( \ln{M_Z^2 \over M^2} + \frac{3}{2} \right).
\label{eq:BoxgZms}
\ee
This reproduces the perturbative result in the free quark model
of Ref.~\cite{Mar83} at $E=0$, where $M$ is interpreted as
a hadronic mass scale.

As observed in Ref.~\cite{Gor09}, for instance, for three quark
flavors the sum over electroweak charges
$\sum_q 2\, e_q\, g_V^q = {2 \over 3} (1+Q_W^p) \approx \sum_q e_q^2$,
so that in the high-$W$ region dominated by sea quarks, the quark
distributions are approximately flavor independent, in which case
$F_{1,2}^{\gamma Z} \approx F_{1,2}^\gamma$.
In addition, while the Callan-Gross relation between the $F_1$ and
$F_2$ structure functions is assumed in Eq.~(\ref{eq:F2q}) at high
$Q_1^2$, at finite $Q_1^2$ the contribution from $F_1$ is obtained
from $F_2$ and the ratio
$R_{LT} \equiv \sigma_L/\sigma_T
	     = (1+4M^2x^2/Q_1^2) F_2/(2 x F_1) - 1$
of the longitudinal and transverse cross sections \cite{Whi90}.

One can simplify the expressions for $\Box_{\gamma Z}^{V,A}$ in
the DIS region by firstly interchanging the order of the integrations
in Eqs.~(\ref{eq:DRV}), (\ref{eq:DRA}) and (\ref{eq:ImBoxV}),
(\ref{eq:ImBoxA}) to perform the integrals over the energy analytically
\cite{Ris10, Blu11}.  The resulting integrands can be expanded at high
$Q_1^2$ and low $E$ in powers of $x^2/Q_1^2$, yielding a series whose
coefficients are moments of structure functions \cite{Blu11},
\bea
\Box_{\gamma Z}^{V\, \rm (DIS)}(E)
&=& {\alpha \over \pi}\, 2 M E
    \int_{Q_0^2}^\infty {dQ_1^2 \over Q_1^4 (1+Q_1^2/M_Z^2)}
    \biggl[ M_2^{(2)}
          + \frac{2}{3} M_1^{(2)}
          + {2M^2 \over 3Q_1^4} (E^2-Q_1^2) M_2^{(4)}	\nn\\
& & \hspace*{5.5cm}
          +\ {2M^2 \over 5Q_1^4} (4E^2 - 5Q_1^2) M_1^{(4)} + \ldots
    \biggr],
\label{eq:momV}						\\
\Box_{\gamma Z}^{A\, \rm (DIS)}(E)
&=& \hat{v}_e {3\alpha \over 2\pi}
    \int_{Q_0^2}^\infty {dQ_1^2 \over Q_1^2 (1+Q_1^2/M_Z^2)}
    \biggl[ M_3^{(1)}
          + {2M^2 \over 9Q_1^4} (5E^2 - 3Q_1^2) M_3^{(3)} + \ldots
    \biggr].
\label{eq:momA}
\eea
Here the moments of the structure functions are defined as
\be
M_i^{(n)}(Q_1^2)\
\equiv\ \int_0^1 dx\,x^{n-2}\, {\cal F}_i^{\gamma Z}(x,Q_1^2),\quad
i=1,2,3,
\label{eq:momdef}
\ee
where
${\cal F}_i^{\gamma Z}
= \left\{ xF_1^{\gamma Z}, F_2^{\gamma Z}, xF_3^{\gamma Z} \right\}$.
Note that the upper limit $x_{\rm max}$ on the $x$-integrals in 
Eqs.~(\ref{eq:momV}) and (\ref{eq:momA}) has been approximated by 1;
the resulting error was found to be less than $10^{-4}$ for
$Q_1^2 > 1$~GeV$^2$ \cite{Blu11}.
The large-$x$ contributions to $M_i^{(n)}(Q_1^2)$ become more
important for large $n$; however, the higher moments are
suppressed by increasing powers of $1/Q_1^2$.
In practice, the integrals in Eqs.~(\ref{eq:momV}) and (\ref{eq:momA})
are dominated by the lowest moments, with the $1/Q_1^2$ corrections
being relatively small in DIS kinematics.
For the axial-vector hadron part, the lowest moment,
$M_3^{(1)}(Q_1^2)$, is the $\gamma Z$ analog of the Gross--Llewellyn
Smith (GLS) sum rule \cite{Gro69} for $\nu N$ DIS, which at leading
order counts the number of valence quarks in the nucleon.
The corresponding quantity for $\gamma Z$ is
$\sum_q 2 e_q\, g_A^q = 5/3$, so that at next-to-leading order   
in the $\overline{\rm MS}$ scheme
\be
M_3^{(1)}(Q_1^2)\
=\ \frac{5}{3} \left( 1 - {\alpha_s(Q_1^2) \over \pi} \right).
\ee
The lowest moment contribution to Eq.~(\ref{eq:momA}) is therefore
\be
\Box_{\gamma Z}^{A\, \rm (MS)}(0)\
\approx\ \hat{v}_e {3\alpha \over 2\pi}
	 \int_{Q_0^2}^\infty dQ_1^2\,
	 {M_3^{(1)}(Q_1^2) \over Q_1^2 (1+Q_1^2/M_Z^2)},
\label{eq:MS}
\ee
which is identical to the Marciano-Sirlin result \cite{Mar83} for
the high energy contribution to the box diagram.

For $Q_1^2 < Q_0^2$ a partonic description of the structure functions
is not valid.  In particular, since the integrals over $Q_1^2$ in 
Eqs.~(\ref{eq:ImBoxV}) and (\ref{eq:ImBoxA}) extend down to $Q_1^2=0$,
and the upper limit on the $x$-integral, $x_{\rm max}$, is also limited 
by $Q_1^2$, one requires the behavior of the structure functions in
the limit of both low $x$ and low $Q_1^2$.
In the case of the vector $F_2^{\gamma Z}$ structure function,
conservation of the two vector currents requires
$F_2^{\gamma Z} \sim Q_1^2$ as $Q_1^2 \to 0$.
In computing the high-$W$ contributions to $\Box_{\gamma Z}^{V}$
several authors have used Regge inspired parametrizations
\cite{Cve01, Cap94, Kai99} of the electromagnetic structure functions,
and approximating $F_2^{\gamma Z} \approx F_2^\gamma$ at small $x$.

By contrast, $F_3^{\gamma Z}$ depends on both vector and axial-vector
currents, so that no analogous current conservation constraint exists.
In the absence of data on $F_3^{\gamma Z}$ in the low-$x$, low-$Q_1^2$
region, Blunden {\it et al.} considered models for its possible $x$ and
$Q_1^2$ dependence such that $F_3^{\gamma Z}$ at $x_{\rm max}$ should
not diverge as $Q_1^2 \to 0$, and should match the partonic structure
function at $Q_1^2=Q_0^2$.  This model-dependence is reflected in the
quoted uncertainty in Table~\ref{tab:boxa}.

\begin{figure}[htb]
\begin{center}
\vspace*{0.5cm}
\includegraphics[height=7cm]{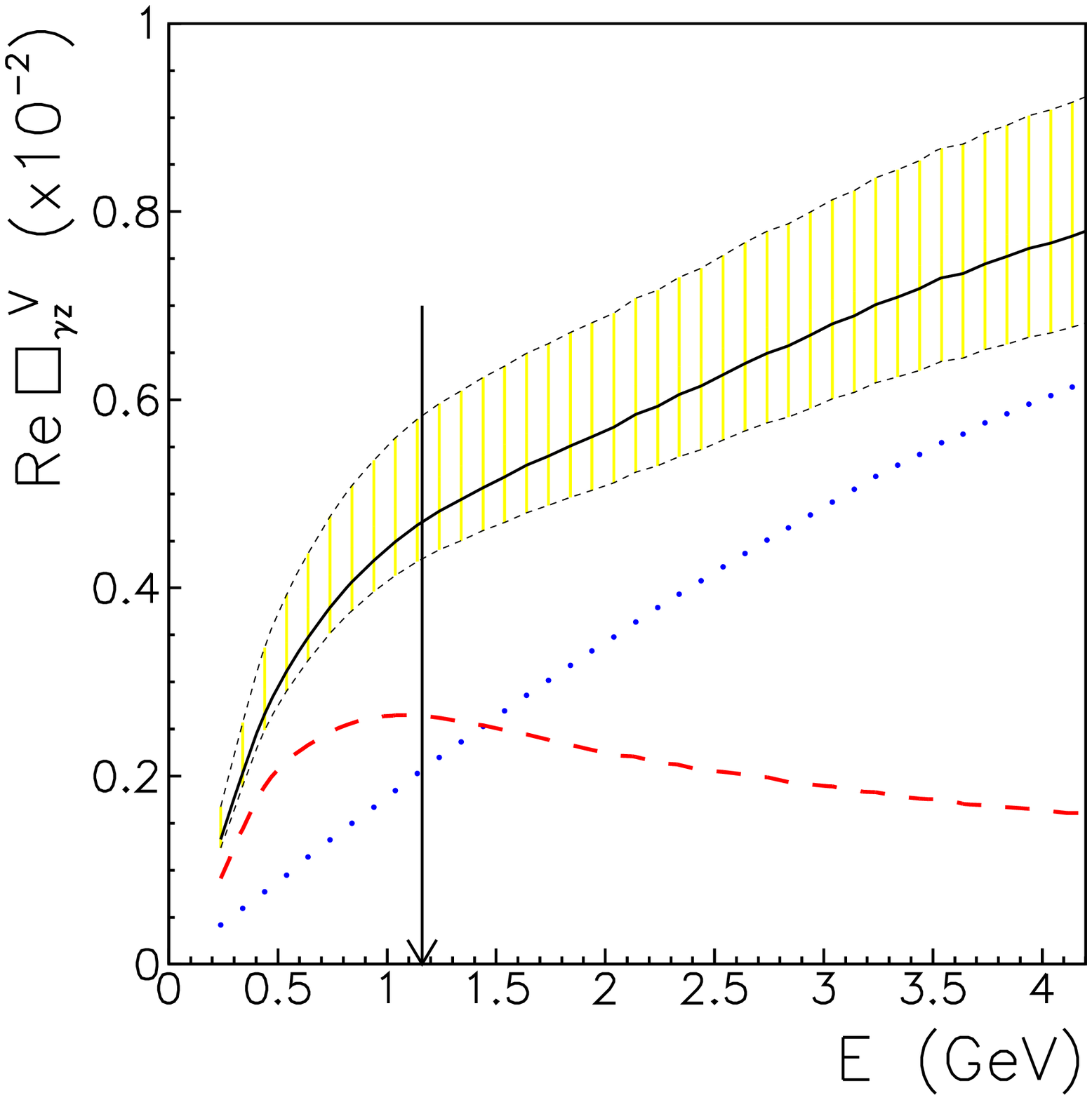}%
\hspace*{0.5cm}\includegraphics[height=6.5cm]{Figs/fig39b.eps}
\begin{minipage}{16.5cm}
\caption{$\gamma Z$ box corrections to $Q_W^p$ for the
{\bf (left)}
	vector hadron $\Box_{\gamma Z}^{V}$ contribution showing the
	resonant (dashed) and nonresonant (dotted) components, and the
        sum (solid, and shaded);
{\bf (right)}
        axial hadron $\Box_{\gamma Z}^{A}$ contribution (labeled ``A'')
	together with the sum of axial and vector hadron corrections
	(``V+A''), and the $E=0$ result from Refs.~\cite{Mar83,Erl03}
	(``MS'', extended to finite $E$ for comparison).
        The vertical lines at $E=1.165$~GeV indicate the energy
	of the ${\rm Q_{weak}}$ experiment.
        Figures taken from Refs.~\cite{Sib10, Blu11}.}
\label{fig:boxtotal}
\end{minipage}
\end{center}
\end{figure}

The total vector and axial hadron corrections
$\Box_{\gamma Z}^{V, A}(E)$ are shown in Fig.~\ref{fig:boxtotal}
as a function of the incident electron energy $E$.
For the vector hadron correction $\Box_{\gamma Z}^{V}$,
Fig.~\ref{fig:boxtotal} (left), most of the strength ($\sim 80\%$) 
comes from relatively low energies, below 4~GeV, where the $Q_1^2$
range extends to $\sim 6$~GeV$^2$, and $W$ to $\sim 3$~GeV.
This coincides precisely with the region that a wealth of very accurate
electroproduction data exists from Jefferson Lab \cite{Lia04, Osi03,
Nic00, Mal09}.
The nonresonant contribution to $\Box_{\gamma Z}^{V}$ is small
at low energies, rising linearly with $E$ in this region.
The resonant part increases steeply to a maximum at $E \sim 1$~GeV,
before falling off like $1/E$ \cite{Gor09, Sib10}.
Sibirtsev {\it et al.} find the resonant and nonresonant contributions
to $\Box_{\gamma Z}^{V}$ to be 0.0026 and 0.0021, respectively, at
the energy relevant for the ${\rm Q_{weak}}$ experiment, $E=1.165$~GeV 
(although one should note that this separation is somewhat arbitrary,
as only the total cross section is physically meaningful).  The error
band, which grows with energy, is obtained from the uncertainty in the
fit parameters using a variational method \cite{Sib10}.

Using different structure function inputs, Rislow and Carlson
\cite{Ris10} find
$\Box_{\gamma Z}^{V}(1.165~\rm GeV) = 0.0057 \pm 0.0009$,
which is slightly higher than but consistent with the value from 
Sibirtsev {\it et al.} \cite{Sib10}.
Gorchtein {\it et al.} \cite{Gor11}, on the other hand, find
$\Box_{\gamma Z}^{V}(1.165~\rm GeV) = 0.0054 \pm 0.0020$,
which is again consistent with the other estimates but has a larger
uncertainty due to the larger range of input structure functions
considered there.

The axial hadron correction $\Box_{\gamma Z}^{A}$ in
Fig.~\ref{fig:boxtotal} (right) is dominated by the DIS
contribution, which has negligible $E$ dependence.
On the other hand, the resonance and low-$Q^2$ DIS
contributions dominate the uncertainties.
The total axial hadron correction $\Box_{\gamma Z}^{A}(E)$ is
$0.0044(4)$ at $E=0$, and $0.0037(4)$ at $E=1.165$~GeV.
This should be compared to the value $0.0052(5)$ used in
Ref.~\cite{Erl03}, which is assumed to be energy independent.
Combined with the correction to $\Box_{\gamma Z}^{V}$,
this shifts the theoretical estimate for $Q_W^p$ from $0.0713(8)$
to $0.0705(8)$, with a total additional energy dependent correction
of $0.0040^{+0.0011}_{-0.0004}$ at $E=1.165$~GeV.

\begin{figure}[htb]
\begin{center}
\hspace*{4.5cm}
\includegraphics[height=9cm,angle=90]{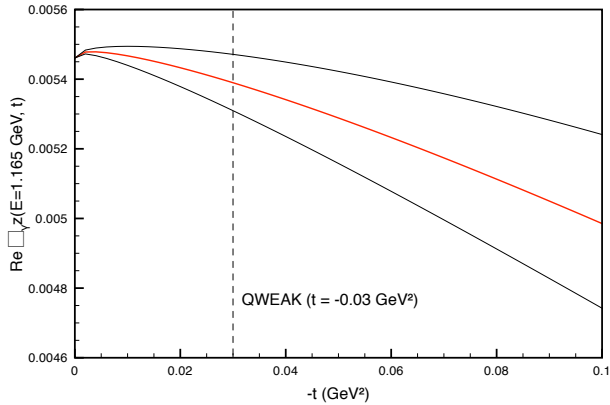}
\hspace*{-13cm}
\begin{minipage}{16.5cm}
\caption{Dependence of the vector hadron correction $\Box_{\gamma Z}^{V}$
	on the momentum transfer squared $t$ at the ${\rm Q_{weak}}$
	energy $E=1.165$~GeV.  The outer curves indicate the uncertainty.
	Figure taken from Ref.~\cite{Gor11}.}
\label{fig:box_tdep}
\end{minipage}
\end{center}
\end{figure}

The $t$ dependence of the $\gamma Z$ corrections, from the forward limit
$-t \equiv Q^2 = 0$ to the ${\rm Q_{weak}}$ value of $-t = 0.03$~GeV$^2$,
was considered by Gorchtein {\it et al.} \cite{Gor09, Gor11} using the
phenomenological {\it ansatz}
\be
\Box_{\gamma Z}(E,t)
=\ \Box_{\gamma Z}(E,0)\, {\exp(-B|t|/2) \over F_1^{\gamma p}(t)},
\ee
where $F_1^{\gamma p}(t)$ is the Dirac proton form factor, and
$B = (7 \pm 1)$~GeV$^{-2}$ is extracted from analysis of data forward
Compton scattering data on $^4$He nuclei \cite{Ale87}.
Although the estimate is considered by the authors to be exploratory,
it suggests that the dispersion correction decreases by only about 2\%
from its $t=0$ value \cite{Gor11}.

The corrections $\Box_{\gamma Z}^{V, A}$ are important for the
interpretation of the ${\rm Q_{weak}}$ experiment, given its projected
uncertainty of $\pm 0.003$ \cite{Qwk05}.
It is also critical to the physical interpretation of the experiment
which is expected to constrain possible sources of parity violation
from beyond the Standard Model at a mass scale of $\gtorder 2$~TeV
\cite{You07}.
The uncertainties in the corrections can be reduced with future
parity-violating structure function measurements at low $Q^2$,
such as those planned at Jefferson Lab \cite{PVDIS, PVDIS12}.
The high precision determination of $Q_W^p$ would then allow more
robust extraction of signals for new physics beyond the Standard Model.

%% file: section8.tex
\section{Conclusions and outlook}
\label{sec:conc}

The renewed interest in the role of two-photon exchange in
electron--hadron scattering sprouted from the challenge to resolve
a major discrepancy between new polarization transfer measurements
of the proton's electric to magnetic form factor ratio and earlier
results using Rosenbluth separations of elastic cross sections,
which has been confirmed now by several independent experiments.
In the intervening decade since the first polarization transfer
experiments were performed at Jefferson Lab, the study of TPE has
flourished into an area rich in phenomenology, with many developments
in experiment and theory that have significantly altered how we think
about the basic process of electromagnetic scattering from hadrons.
The most likely candidate that has emerged to explain the form factor
discrepancy is two-photon exchange, in particular the hadron
structure dependent parts of the $\gamma\gamma$ box and crossed-box
diagrams neglected in the standard treatments of radiative corrections.

In this article we have reviewed the developments in the study of TPE
over the past decade, focusing on the hadronic framework for radiative
corrections applicable for most measurements extending into the
$Q^2 \sim$~few~GeV$^2$ range.  This period has witnessed major advances
in theoretical calculations of TPE amplitudes, which we have outlined
in this review, together with their implications for observables.
We have surveyed the effects of TPE corrections on various elastic
$ep$ scattering observables, including Rosenbluth separations of
cross sections, polarization measurements, and $e^+p/e^-p$ ratios,
as well as the Born-forbidden normal beam and target asymmetries.
Two-photon effects also find their way into other observables, such
as neutron form factors, resonance electroproduction, the pion form
factor, and elastic electron--nucleus (deuteron and $^3$He) cross
sections, which we have summarized here.  The theoretical advances
in TPE computations have furthermore allowed for the first time the
{\it a priori} inclusion of TPE corrections in a global fit of form
factor data, which has recently been performed.

Extending the discussion into the weak sector, we reviewed the
calculation of corrections to parity-violating elastic scattering
cross sections and asymmetries from $\gamma Z$ interference diagrams.
These turn out to have relatively minor impact on the extraction
of the strange nucleon electromagnetic form factors, but play an
important role in measurements of the weak charge of the proton at
forward angles, which gives direct access to the weak mixing angle,
$\sin^2\theta_W$.

At present, the strongest experimental constraints on the size and
kinematic dependence of TPE corrections come from the discrepancy
between form factors extracted from elastic cross section and
polarization measurements, assuming one attributes the discrepancy
to TPE.  Measurements of beam normal single spin asymmetries also
provide important tests of TPE calculations, but are sensitive to the
imaginary parts of TPE amplitudes which are not directly probed in
form factor measurements.  The most direct experimental information
comes from comparisons of positron and electron scattering from the
proton at large $\eps$, and the limit on nonlinear contributions in
$\eps$ to polarization observables.  Unfortunately, at present these
provide only limited evidence for TPE contributions, and serve mainly
to constrain the magnitude of the effects.

As we have shown in this review, modern calculations of TPE corrections
can resolve most of the observed form factor discrepancy and are
consistent with all other experimental constraints; however, it will be
important to test their validity over as large a range of observables
and reactions as possible.  Including the TPE corrections and applying
an estimate of their uncertainties at high $Q^2$ yields form factors in
which TPE effects are not the dominant source of uncertainty \cite{AMT07}.

In future an important goal will be to obtain {\it direct} experimental 
verification that the form factor discrepancy is indeed explained by 
TPE effects.  This will involve an experimental program consisting
of several efforts, including measurement of the $e^+ p/e^- p$ cross 
section ratio and detailed examination of the $\eps$ dependence of 
polarization observables, particularly since in some cases theoretical 
predictions for these vary considerably.  An important opportunity may
be presented by the development of {\it positron} beams at a future
12~GeV upgraded Jefferson Lab \cite{JPOS09}.
In addition, measurement of various beam and target normal asymmetries
will provide direct evidence for Born-forbidden effects and constrain
input for dispersion relation analyses.

Further theoretical work will seek better control of higher-mass
intermediate state contributions to the box diagrams, especially at
higher $Q^2$ values.  Along this direction, it will be important to
go beyond the resonance approximation to include the nonresonant
background in the intermediate state spectrum.  This naturally
becomes difficult to do within any specific model, and a more
effective approach may be to utilize dispersion relation to obtain
TPE amplitudes from the imaginary parts of the Compton scattering
{\it data}.  This is particularly appealing at forward angles, where
the amplitudes can be related to inclusive structure functions.

This approach is proving to be especially relevant for estimating
$\gamma Z$ interference corrections to the proton weak charge at
low $Q^2$, currently being measured in the Q$_{\rm weak}$ experiment
at Jefferson Lab.  Extending this program into the non-forward
region will be an important challenge, but one which will allow
for a reliable global analysis of the parity-violating elastic
scattering data to extract strange nucleon form factors correcting
for the effects of two-boson exchange.

Finally, we should note that TPE corrections are indeed small
in general, and for a large range of experiments are at the
${\cal O}(\alpha)$ (or $\sim 1\%$) level that has historically been
assumed.  It is only certain cases, typically at backward angles for
two-photon exchange or forward angles for $\gamma Z$ corrections,
where they tend to be larger, at the $\sim$~few~\% level; $\gep$ is
a very special case where even a few percent effect in the cross
section has a large impact on the Rosenbluth $\gegm$ extraction.
Two-photon exchange corrections therefore do not necessarily require
revising methodologies for electron scattering in general, so much
as reminding us of the need to take greater care in estimating their
possible impact on observables that may be particularly susceptible
to their effects.

\section*{Acknowledgments}

We are grateful to our collaborator, the late John~A.~Tjon, for his
many fundamental contributions to the study of radiative corrections
--- his footprint runs throughout this review.
We also thank A.~Afanasev, S.~Kondratyuk, R.~E.~Segel, A.~Sibirtsev,
I.~Sick, E.~Tomasi-Gustafsson, A.~W.~Thomas and S.~N.~Yang for helpful
discussions and collaborations that contributed to this article.
This work was supported by NSERC (Canada) and the U.S.\ Department of
Energy, Office of Nuclear Physics under contract DE-AC02-06CH11357, and
contract DE-AC05-06OR23177, under which Jefferson Science Associates,
LLC operates Jefferson Lab.

%% file: references.tex

\end{document}

%% file: TBE.bbl
\begin{thebibliography}{99}
\itemsep -2pt


\bibitem{Per07}
C.~E.~Perdrisat, V.~Punjabi and M.~Vanderhaeghen,
        \Journal{\PPNP}{59}{694.}{2007}

\bibitem{Arr07a}
J.~Arrington, C.~D.~Roberts and J.~M.~Zaontti,
        \Journal{\JPG}{34}{S23.}{2007}

\bibitem{Arr11}
J.~Arrington, C.~W.~de~Jager and C.~E.~Perdrisat,
	{\it J. Phys.: Conf. Ser.} 299 (2011) 011001.


\bibitem{Pes95}
M.~E.~Peskin and D.~V.~Schroeder,
	{\it An Introduction to Quantum Field Theory},
	Addison-Wesley (1995).

\bibitem{Blu03}
P.~G.~Blunden, W.~Melnitchouk and J.~A.~Tjon,
        \Journal{\PRL}{91}{142304.}{2003}
        
\bibitem{Blu05}
P.~G.~Blunden, W.~Melnitchouk and J.~A.~Tjon,
        \Journal{\PRC}{72}{034612.}{2005}

\bibitem{Kon05} 
S.~Kondratyuk, P.~G.~Blunden, W.~Melnitchouk and J.~A.~Tjon,
        \Journal{\PRL}{95}{172503.}{2005}

\bibitem{Tjo08}
J. A. Tjon and W. Melnitchouk,
        \Journal{\PRL}{100}{082003.}{2008}

\bibitem{Blu10}
J. A. Tjon, P. G. Blunden and W. Melnitchouk,
	\Journal{\PRC}{81}{018202.}{2010}

\bibitem{Ros50}
M.~N.~Rosenbluth,
        \Journal{\PREV}{79}{615.}{1950}

\bibitem{Wal94}
R.~C.~Walker {\em et al.},
        \Journal{\PRD}{49}{5671.}{1994}

\bibitem{Arr03}
J.~Arrington,
        \Journal{\PRC}{68}{034325.}{2003}

\bibitem{Chr04}
M.~E.~Christy {\em et al.},
        \Journal{\PRC}{70}{015206.}{2004}

\bibitem{Qat05}
I.~A.~Qattan {\em et al.},
        \Journal{\PRL}{94}{142301.}{2005}

\bibitem{Max00P}
L.~C.~Maximom and W.~C.~Parke,
        \Journal{\PRC}{61}{045502.}{2000}

\bibitem{Jon00}
M.~K.~Jones {\it et al.},
        \Journal{\PRL}{84}{1398.}{2000}

\bibitem{Gay01}
O.~Gayou {\it et al.},
        \Journal{\PRC}{64}{038202.}{2001}

\bibitem{Gay02}
O.~Gayou {\it et al.},
        \Journal{\PRL}{88}{092301.}{2002}

\bibitem{Pun05}
V.~Punjabi {\it et al.},
        \Journal{\PRC}{71}{055202.}{2005}

\bibitem{Mac06}
G.~MacLachlan {\it et al.},
	\Journal{\NPA}{764}{261.}{2006}

\bibitem{Ron07}
G.~Ron {\it et al.},
        \Journal{\PRL}{99}{202002.}{2007}

\bibitem{Puc10}
A.~J.~R.~Puckett {\it et al.},
        \Journal{\PRL}{104}{242301.}{2010}
        
\bibitem{Zha11}
X.~Zhan {\it et al.},
        arXiv:1102.0318 [nucl-ex].

\bibitem{Puc11}
A.~J.~R.~Puckett {\it et al.},
        arXiv:1102.5737 [nucl-ex].
        
\bibitem{Ron11}
G.~Ron {\it et al.},
	arXiv:1103.5784 [nucl-ex].

\bibitem{Cra06}		
C.~B.~Crawford {\it et al.},
	\Journal{\PRL}{98}{052301.}{2007}

\bibitem{Jon06}		
M.~K.~Jones {\it et al.},
	\Journal{\PRC}{74}{035201.}{2006}

\bibitem{Jai03}
P.~Jain and J.~P.~Ralston,
        \Journal{\it Pramana}{61}{987.}{2003}
        
\bibitem{Bel03}
A.~V.~Belitsky, X.~Ji and F.~Yuan,
        \Journal{\PRL}{91}{092003.}{2003}


\bibitem{Bos95}
P.~E.~Bosted,
        \Journal{\PRC}{51}{409.}{1995}
 
\bibitem{Lit70}
J.~Litt \etal,
        \Journal{\PLB}{31}{40.}{1970}

\bibitem{Pri71}
L.~E.~Price {\it et al.},
        \Journal{\PRD}{4}{45.}{1971}

\bibitem{Bar73}
W.~Bartel {\it et al.},
        \Journal{\NPB}{58}{429.}{1973}

\bibitem{And94}
L.~Andivahis \etal,
        \Journal{\PRD}{50}{5491.}{1994}

\bibitem{Qat06}
I.~A.~Qattan,
        Ph.D. Thesis, Northwestern University, nucl-ex/0610006.

\bibitem{Arr04a}
J.~Arrington,
        \Journal{\PRC}{69}{022201(R).}{2004}

\bibitem{Tsa61}
Y.~S.~Tsai,
        \Journal{\PREV}{122}{1898.}{1961}
        
\bibitem{Mo69}
L.~W.~Mo and Y.~S.~Tsai,
        \Journal{\RMP}{41}{205.}{1969}
        
\bibitem{Mck48}
W.~A.~McKinley and H.~Feshbach,
        \Journal{\PREV}{74}{1759.}{1948}

\bibitem{Lew56}
R.~R.~Lewis \etal,
        \Journal{\PREV}{102}{537.}{1956}

\bibitem{Dre57}
S.~D.~Drell and M.~Ruderman,
        \Journal{\PREV}{106}{561.}{1957}

\bibitem{Dre59}
S.~D.~Drell and S.~Fubini,
        \Journal{\PREV}{113}{741.}{1959}

\bibitem{Wer61}
N.~R.~Werthammer and M.~A.~Ruderman,
        \Journal{\PREV}{123}{1005.}{1961}

\bibitem{Cam67}
J.~A.~Campbell,
        \Journal{\NPB}{1}{283.}{1967}

\bibitem{Cam69a}
J.~A.~Campbell,
        \Journal{\PREV}{180}{1541.}{1969}

\bibitem{Gre69}
G.~K.~Greenhut,
        \Journal{\PREV}{184}{1860.}{1969}

\bibitem{Bro70}
R.~W.~Brown {\em et al.},
        \Journal{\PRD}{1}{1432.}{1970}

\bibitem{Ent01}
R.~Ent {\it et al.},
        \Journal{\PRC}{64}{054610.}{2001}

\bibitem{Max00T}
L.~C.~Maximon and J.~A.~Tjon,
        \Journal{\PRC}{62}{054320.}{2000}
        
\bibitem{Ros00}
R.~Rosenfelder,
        \Journal{\PLB}{479}{381.}{2000}

\bibitem{Arr04c}
J.~Arrington and I.~Sick,
        \Journal{\PRC}{70}{028203.}{2004}

\bibitem{Gui03}  
P.~A.~M.~Guichon and M.~Vanderhaeghen,
        \Journal{\PRL}{91}{142303.}{2003} 
        
\bibitem{Rek04}
M.~P.~Rekalo and E.~Tomasi-Gustafsson,
        \Journal{\NPA}{742}{322.}{2004}  

\bibitem{Afa03}
A.~Afanasev, S.~Brodsky and C.~Carlson,  
        presented at the Oct. 2003 meeting of the DNP.

\bibitem{Che04}
Y.~C.~Chen, A.~Afanasev, S.~J.~Brodsky, C.~E.~Carlson and M.~Vanderhaeghen,
        \Journal{\PRL}{93}{122301.}{2004}
        
\bibitem{Bel08}
M.~A.~Belushkin, H.-W.~Hammer and U.-G.~Miessner
        \Journal{\PLB}{658}{138.}{2008}

\bibitem{Alb09}
W.~M.~Alberico, S.~M.~Bilenky, C.~Giunti and K.~M.~Gracyzk,
	\Journal{\JPG}{36}{115009}{2009}

\bibitem{Tom05}
E.~Tomasi-Gustafsson and G.~I.~Gakh,
        \Journal{\PRC}{72}{015209.}{2005}

\bibitem{Che07}
Y.-C.~Chen, C.-W.~Kao and S.-N.~Yang,
        \Journal{\PLB}{652}{269.}{2007}

\bibitem{Bor07a}
D.~Borisyuk and A.~Kobushkin,
	\Journal{\PRC}{76}{022201.}{2007}

\bibitem{Qat11}
I.~A.~Qattan and A.~Alsaad
        \Journal{\PRC}{83}{054307}{2011}

\bibitem{Gra11}
K.~M.~Graczyk,
	arXiv:1106.1204 [hep-ph].

\bibitem{Arr05}
J.~Arrington,
        \Journal{\PRC}{71}{015202.}{2005}

\bibitem{Bor07}
D.~Borisyuk and A.~Kobushkin,
	\Journal{\PRC}{75}{038202.}{2007}

\bibitem{Bor11a}
D.~Borisyuk and A.~Kobushkin,
	\Journal{\PRD}{83}{057501.}{2011}

\bibitem{Gut11}
J.~Guttmann, N.~Kivel, M.~Meziane and M.~Vanderhaeghen,
        arXiv:1012.0564 [hep-ph].

\bibitem{Mez11}
M.~Meziane {\it et al.},
	\Journal{\PRL}{106}{132501.}{2011}

\bibitem{You62}
D.~Yount and J.~Pine,
        \Journal{\PREV}{128}{1842.}{1962}
        
\bibitem{Bro65}                 
A.~Browman, F.~Liu and C.~Schaerf,
	\Journal{\PREV}{139}{B1079.}{1965}

\bibitem{And66}                 
R.~L.~Anderson {\em et al.},
	\Journal{\PRL}{17}{407.}{1966}

\bibitem{Bar67}                 
W.~Bartel {\em et al.},
	\Journal{\PLB}{25}{242.}{1967}

\bibitem{And68}
R.~L.~Anderson {\it et al.}
        \Journal{\PREV}{166}{1336.}{1968}
        
\bibitem{Bou68}                 
B.~Bouquet {\em et al.},
	\Journal{\PLB}{26}{178.}{1968}

\bibitem{Mar68}
J.~Mar \etal,
        \Journal{\PRL}{21}{482.}{1968}

\bibitem{Cam69b}
L.~Camilleri \etal,
        \Journal{\PRL}{23}{149.}{1969}

\bibitem{Arr04b}
J.~Arrington,
        \Journal{\PRC}{69}{032201(R).}{2004}

\bibitem{VEPP}
J.~Arrington {\em et al.},
	{\em Two-photon exchange and elastic scattering of 
	electrons/positrons on the proton},
	proposal for an experiment at VEPP-3 (2004),
	nucl-ex/0408020.

\bibitem{e07005}
Jefferson Lab Experiment E07-005,
	{\it Beyond the Born approximation: A precise comparison of
	positron-proton and electron-proton elastic scattering in CLAS},
	W.~Brooks {\em et al.}, spokespersons.

\bibitem{olympus}
M.~Kohl,
	{\it AIP Conf. Proc.} 1160 (2009) 19.

\bibitem{e05017}
Jefferson Lab Experiment E05-017,
	{\it A measurement of two-photon effects in upolarized
	electron-proton scattering},
	J.~Arrington, spokesperson.

\bibitem{Ber10}
J.~Bernauer {\it et al.},
        \Journal{\PRL}{105}{242001.}{2010}

\bibitem{Pao10}
M.~Paolone {\it et al.},
	\Journal{\PRL}{105}{072001.}{2010}

\bibitem{Tva06}
V.~Tvaskis \etal,
        \Journal{\PRC}{73}{025206.}{2006}

\bibitem{Rek99}
M.~P.~Rekalo, E.~Tomasi-Gustafsson and D.~Prout,
        \Journal{\PRC}{60}{042202.}{1999}

\bibitem{Abi07}
Z.~Abidin and C.~E.~Carlson,
        \Journal{\PRD}{77}{037301.}{2007}

\bibitem{Afa05} 
A.~V.~Afanasev, S.~J.~Brodsky, C.~E.~Carlson, Y.~C.~Chen and M.~Vanderhaeghen,
        \Journal{\PRD}{72}{013008.}{2005}


\bibitem{Hoo79}
G.~'t~Hooft and M.~Veltman,
        \Journal{\NPB}{153}{365.}{1979}
                
\bibitem{Sch49}
J.~Schwinger,
	\Journal{\PREV}{76}{790.}{1949}

\bibitem{Pas79}
G.~Passarino and M.~J.~Veltman,
        \Journal{\NPB}{160}{151.}{1979}

\bibitem{Wei09}
F.~Weissbach, K.~Hancken, D.~Trautmann and I.~Sick
	\Journal{\PRC}{80}{064605.}{2009}

\bibitem{Bys07}
Y.~M.~Bystritskiy, E.~A.~Kuraev and E.~Tomasi-Gustafsson
	\Journal{\PRC}{75}{015207.}{2007}

\bibitem{Hah99}
T.~Hahn and M.~Perez-Victoria,
	\Journal{Comput.~Phys.~Commun.}{118}{153.}{1999}

\bibitem{Bee90}
W.~Beenakker and A.~Denner,
        \Journal{\NPB}{338}{349.}{1990}
                
\bibitem{Mei63}
N.~T.~Meister and D.~R.~Yennie,
        \Journal{\PREV}{130}{1210.}{1963}
        
\bibitem{Tjo09} 
J.~A.~Tjon, P.~G.~Blunden and W.~Melnitchouk,
        \Journal{\PRC}{79}{055201.}{2009}

\bibitem{Mer96}
P.~Mergell, U.-G.~Meissner and D.~Drechsel,
        \Journal{\NPA}{596}{367.}{1996}

\bibitem{Bra02}
E.~J.~Brash {\em et al.},
	\Journal{\PRC}{65}{051001(R).}{2002}

\bibitem{Jon73}
H.~F.~Jones and M.~D.~Scadron,
	\Journal{\ANNP}{81}{1.}{1973}

\bibitem{Kon01}
S.~Kondratyuk and O.~Scholten,
	\Journal{\PRC}{64}{024005.}{2001}

\bibitem{Zho10}
H.~Q.~Zhou, C.~W.~Kao, S.~N.~Yang and K.~Nagata,
	\Journal{\PRC}{81}{035208.}{2010}

\bibitem{Pas04}
V.~Pascalutsa and J.~A.~Tjon,
	\Journal{\PRC}{70}{035209.}{2004}

\bibitem{Kon07}
S.~Kondratyuk and P.~G.~Blunden,
        \Journal{\PRC}{75}{038201.}{2007}

\bibitem{Kor00}
A.~Yu.~Korchin and O.~Scholten,
        \Journal{\PRC}{60}{015205.}{2000}

\bibitem{Pen02}
G.~Penner and U.~Mosel,
        \Journal{\PRC}{66}{055212.}{2002}

\bibitem{Kur10}
E.~A.~Kuraev and E.~Tomasi-Gustafsson,
	{\it Phys. Part. Nucl. Lett.} 7 (2010) 67.

\bibitem{Dal51}
R.~H.~Dalitz,
	{\it Proc. Roy. Soc. (London)} 206 (1951) 509.

\bibitem{Blu05S}
P.~G.~Blunden and I.~Sick,
	\Journal{\PRC}{72}{057601.}{2005}
                
\bibitem{Sic98}
I.~Sick and D.~Trautmann,
         \Journal{\NPA}{637}{559.}{1998}

\bibitem{Isk93}
I.~F.~Iskhakov,
	\Journal{\RPJ}{36}{582.}{1993}

\bibitem{Car07}
C.~Carlson and M.~Vanderhaeghen,
	\Journal{\it Ann. Rev. Nucl. Part. Sci.}{57}{171.}{2007}

\bibitem{Bor09}		
D.~Borisyuk and A.~Kobushkin,
	\Journal{\PRD}{79}{034001.}{2009}

\bibitem{Kiv09}
N.~Kivel and M.~Vanderhaeghen,
        \Journal{\PRL}{103}{092004.}{2009}

\bibitem{Hoo06}
P.~Hoodbhoy,
        \Journal{\PRD}{73}{054027.}{2006}

\bibitem{Kro26}		
R.~de~L.~Kronig,
        {\it Opt. Soc. Am.} 12 (1926) 547;
H.~A.~Kramers,
        {\it Atti Cong. Intern. Fisica (Transactions of Volta Centenary
        Congress), Como} 2 (1927) 545.

\bibitem{Bor08}		
D.~Borisyuk and A.~Kobushkin,
	\Journal{\PRC}{78}{025208.}{2008}

\bibitem{Bor06}		
D.~Borisyuk and A.~Kobushkin,
	\Journal{\PRC}{74}{065203.}{2006}

\bibitem{Bor11}
D.~Borisyuk and A.~Kobushkin,
	\Journal{\PRC}{83}{025203.}{2011}

\bibitem{Ber80}
J.~Bernab\'{e}u and J.~A.~Pe\~{n}arrocha,
	\Journal{\PRD}{22}{1082.}{1980}

\bibitem{Pen81}
J.~A.~Pe\~{n}arrocha and J.~Bernab\'{e}u,
        \Journal{\ANNP}{135}{321.}{1981}

\bibitem{Bor86}
J.~Bordes, J.~A.~Pe\~{n}arrocha and J.~Bernab\'{e}u,
        \Journal{\PLB}{173}{86.}{1986}

\bibitem{Bor87}
J.~Bordes, J.~A.~Pe\~{n}arrocha and J.~Bernab\'{e}u,
	\Journal{\PRD}{35}{3310.}{1987}

\bibitem{Gor07}
M.~Gorchtein,
	\Journal{\PLB}{644}{322.}{2007}

\bibitem{Cve01}
G.~Cvetic, D.~Schildknecht, B.~Surrow and M.~Tentyukov,
	\Journal{\EPJC}{20}{77.}{2001}


\bibitem{Nik10}
D.~M.~Nikolenko {\it et al},
        \Journal{\em Phys. Atom. Nucl}{73}{1322.}{2010}

\bibitem{Nik11}
D.~M.~Nikolenko {\it et al.},
        \Journal{PoS}{ICHEP2010}{164.}{2010}

\bibitem{Kel02}
J.~Kelly,
        \Journal{\PRC}{70}{068202.}{2002}

\bibitem{Dut03}
D.~Dutta {\it et al.},
        \Journal{\PRC}{68}{064603.}{2003}

\bibitem{Ada07}
C.~Adamuscin, L.~Bimbot, S.~Dubnicka, A.~Z.~Dubnickova and
E.~Tomasi-Gustafsson,
	\Journal{\PRC}{78}{025202.}{2008}

\bibitem{AMT07}
J.~Arrington, W.~Melnitchouk and J.~A.~Tjon,
        \Journal{\PRC}{76}{035205.}{2007}
        
\bibitem{Arr07b}
J.~Arrington and I.~Sick,
        \Journal{\PRC}{76}{035201.}{2007}

\bibitem{Ven10}
S.~Venkat, J.~Arrington, G.~A.~Miller and X.~Zhan,
        \Journal{\PRC}{83}{015203.}{2010}
        
\bibitem{Zha11b}
X.~Zhan and J.~Arrington,
        to be submitted to \PRC.
      
\bibitem{DeR71}
A.~De~R\'{u}jula, J.~M.~Kaplan and E.~de~Rafael,
	\Journal{\NPB}{35}{365.}{1971}

\bibitem{Arr11b}
J.~Arrington,
	submitted to \PRL.

\bibitem{Afa05C}
A.~V.~Afanasev and C.~E.~Carlson,
	\Journal{\PRL}{94}{212301.}{2005}

\bibitem{e07013}
Jefferson Lab Experiment E07-013,
	{\it Target normal single-spin asymmetry in inclusive
	DIS $n(e,e')$ with a polarized $^3${\rm He} target},
	T.~Averett {\it et al.}, spokespersons.

\bibitem{e08005}
Jefferson Lab Experiment E08-005,
	{\it Measurement of the target single-spin asymmetry $A_y$ in 
	the quasi-elastic $^3\stackrel{\longrightarrow}{\rm He}(e,e'n)$ 
	reaction},
	V.~A.~Sulkosky contact person.

\bibitem{Air10}
A.~Airapetian, \etal,
        \Journal{\PLB}{682}{351.}{2010}

\bibitem{Wel01}
S.~P.~Wells {\it et al.},
	\Journal{\PRC}{63}{064001.}{2001}

\bibitem{Maa05}
F.~E.~Maas {\it et al.},
	\Journal{\PRL}{94}{082001.}{2005}

\bibitem{Arm07}
D.~S.~Armstrong {\it et al.},
	\Journal{\PRL}{99}{092301.}{2007}

\bibitem{And11}
D.~Androic \etal,
	arXiv:1103.3667 [nucl-ex].


\bibitem{Lun93}
A.~Lung {\it et al.},
        \Journal{\PRL}{70}{718.}{1993}

\bibitem{Mad03}
R.~Madey {\em et al.},
        \Journal{\PRL}{91}{122002.}{2003}

\bibitem{Mad04}
Jefferson Lab Experiment E04-110,
        {\em The neutron electric form factor at $Q^2$=4.3~(GeV/c)$^2$ 
	from the reaction $^2{\rm H}(\vec e,e'\vec n) ^1{\rm H}$ via
	recoil polarimetry},
	R.~Madey, spokesperson.

\bibitem{Lia04}
Y.~Liang {\it et al.},
	arXiv:nucl-ex/0410027.

\bibitem{Pas06}
V.~Pascalutsa, C.~E.~Carlson and M.~Vanderhaeghen,
	\Journal{\PRL}{96}{012301.}{2006}

\bibitem{Kon06}
S.~Kondratyuk and P.~G.~Blunden,
	\Journal{\NPA}{778}{44.}{2006}

\bibitem{Kon00}
S.~Kondratyuk and O.~Scholten,
	\Journal{\NPA}{677}{396.}{2000}

\bibitem{Are78}
H.~J.~Weber and H.~Arenh{\"o}vel,
	\Journal{\PREP}{36}{277.}{1978}.

\bibitem{Del04}
A.~Deltuva \etal,
	\Journal{\PRC}{69}{034004.}{2004}

\bibitem{Pas05}
V.~Pascalutsa and M.~Vanderhaeghen,
	\Journal{\PRL}{94}{102003.}{2005}

\bibitem{Che08}
D.~Y.~Chen, H.~Q.~Zhou and Y.~B.~Dong,
        \Journal{\PRC}{78}{045208.}{2008}

\bibitem{Aub06}
B.~Aubert {\it et al.},
	\Journal{\PRD}{73}{012005.}{2006}

\bibitem{Tom08}
E.~Tomasi-Gustafsson, E.~A.~Kuraev, S.~Bakmaev and S.~Pacetti,
	\Journal{\PLB}{659}{197.}{2008}

\bibitem{Hor06}
T.~Horn {\it et al.},
        \Journal{\PRL}{97}{192001.}{2006}

\bibitem{Tad07}
V.~Tadevosyan {\it et al.},
        \Journal{\PRC}{75}{055205.}{2007}

\bibitem{Hub08}
G.~M.~Huber {\it et al.},
        \Journal{\PRC}{78}{045203.}{2008}

\bibitem{Don10W}
Y.~-B.~Dong and S.~D.~Wang,
	\Journal{\PLB}{684}{123.}{2010}

\bibitem{Kai11}
N.~Kaiser,
	\Journal{\JPG}{38}{025003.}{2011}

\bibitem{Mel03}      
W.~Melnitchouk,
        \Journal{\EPJA}{17}{223.}{2003}                

\bibitem{Don06}
Y.~B.~Dong,~C.W.~Kao, S.~N.~Yang and Y.~C.~Chen,
        \Journal{\PRC}{74}{064006.}{2006}

\bibitem{Don09}
Y.~B.~Dong,
        \Journal{\PRC}{80}{025208.}{2009}

\bibitem{Don09C}
Y.~B.~Dong and D.~Y.~Chen,
	\Journal{\PLB}{675}{426.}{2009}

\bibitem{Kob10}
A.~P.~Kobushkin, Ya.~D.~Krivenko-Emetov and S.~Dubni\v{c}ka,
        \Journal{\PRC}{81}{054001.}{2010}

\bibitem{Don10}
Y.~B.~Dong,
        \Journal{\PRC}{82}{068202.}{2010}

\bibitem{Amr94} 
A.~Amroun {\it et al.},
        \Journal{\NPA}{579}{596.}{1994}                 

\bibitem{e04018}
Jefferson Lab Experiment E04-018,
        {\it Elastic electron scattering off $^3${\rm He}
	and $^4${\rm He} at large momentum transfers},
        J.~Gomez, A.~Katramatou and G.~Petratos, spokespersons.


\bibitem{HAP04}		
K.~A.~Aniol {\it et al.},
	\Journal{\PRC}{69}{065501.}{2004}

\bibitem{HAP06}		
K.~A.~Aniol {\it et al.},
        \Journal{\PLB}{635}{275.}{2006} 

\bibitem{HAP06a}	
K.~A.~Aniol {\it et al.},
	\Journal{\PRL}{96}{022003.}{2006}

\bibitem{HAP07}		
A.~Acha {\it et al.},
	\Journal{\PRL}{98}{032301.}{2007}

\bibitem{G005}
D.~S.~Armstrong {\it et al.},
	\Journal{\PRL}{95}{092001.}{2005}

\bibitem{G010}
D.~Androic {\it et al.},
	\Journal{\PRL}{104}{012001.}{2010}

\bibitem{SAM97}
B.~Mueller {\it et al.},
	\Journal{\PRL}{78}{3824.}{1997}

\bibitem{MAM04}
F.~E.~Maas {\it et al.},
	\Journal{\PRL}{93}{022002.}{2004}

\bibitem{MAM05}
F.~E.~Maas {\it et al.},
	\Journal{\PRL}{94}{152001.}{2005}

\bibitem{MAM09}
S.~Baunack {\em et al.},
	arXiv:0903.2733 [nucl-ex].

\bibitem{Qwk05}
Jefferson Lab Experiment E05-020,
	{\it Qweak: A search for new physics at the TeV scale
	via a measurement of the proton's weak charge},
	R.~D.~Carlini {\em et al.}, spokespersons.

\bibitem{Bei05}
E.~J.~Beise, M.~L.~Pitt and D.~T.~Spayde,
        \Journal{\PPNP}{54}{289.}{2005}

\bibitem{Mar83}
W.~J.~Marciano and A.~Sirlin,
	\Journal{\PRD}{27}{552.}{1983}

\bibitem{Mar84}
W.~J.~Marciano and A.~Sirlin,
	\Journal{\PRD}{29}{75.}{1984}

\bibitem{PDG10}
K.~Nakamura {\it et al.},
        \Journal{\JPG}{37}{075021.}{2010}

\bibitem{Mus94}
M.~J.~Musolf {\em et al.},
	\Journal{\PREP}{239}{1.}{1994}

\bibitem{Zho07}
H.~Q.~Zhou, C.~W.~Kao and S.~N.~Yang,
	\Journal{\PRL}{99}{262001.}{2007}

\bibitem{Nag09}
K.~Nagata, H.~Q.~Zhou, C.~W.~Kao and S.~N.~Yang,
	\Journal{\PRC}{79}{062501.}{2009}

\bibitem{Gor09}
M.~Gorchtein and C.~J.~Horowitz,
	\Journal{\PRL}{102}{091806.}{2009}

\bibitem{Gor10}
M.~Gorchtein, C.~J.~Horowitz and M.~J.~Ramsey-Musolf,
	{\it AIP Conf. Proc.} 1265 (2010) 328.

\bibitem{Gor11}
M.~Gorchtein, C.~J.~Horowitz and M.~J.~Ramsey-Musolf,
	arXiv:1102.3910 [hep-ph].

\bibitem{Lew07}
R.~Lewis, 
        \Journal{\EPJA}{32}{409.}{2007}

\bibitem{Xia08}
Z.~-T.~Xia, W.~Zuo,
        \Journal{\PRC}{78}{015209.}{2008}

\bibitem{Kap88}
D.~B.~Kaplan and A.~Manohar,
        \Journal{\NPB}{310}{527.}{1988}

\bibitem{Bas06}
S.~D.~Bass, R.~J.~Crewther, F.~M.~Steffens and A.~W.~Thomas,
        \Journal{\PLB}{634}{249.}{2006}

\bibitem{You06}
R.~D.~Young, J.~Roche, R.~D.~Carlini and A.~W.~Thomas,
	\Journal{\PRL}{97}{102002.}{2006}

\bibitem{Lal05}
O.~Lalakulich and E.~A.~Paschos,
	\Journal{\PRD}{71}{074003.}{2005}

\bibitem{Lal06}
O.~Lalakulich, E.~A.~Paschos and G.~Piranishvili,
	\Journal{\PRD}{74}{014009.}{2006}

\bibitem{Lal07}
O.~Lalakulich, W.~Melnitchouk and E.~A.~Paschos,
	\Journal{\PRC}{75}{015202.}{2007}

\bibitem{Liu07}
J.~L.~Liu, R.~D.~McKeown and M.~J.~Ramsey-Musolf,
	\Journal{\PRC}{76}{025202.}{2007}

\bibitem{You09}
R.~D.~Young,
	private communication.

\bibitem{Erl03}
J.~Erler, A.~Kurylov and M.~J.~Ramsey-Musolf,
	\Journal{\PRD}{68}{016006.}{2003}

\bibitem{Erl05}
J.~Erler and M.~J.~Ramsey-Musolf,
	\Journal{\PRD}{72}{073003.}{2005}

\bibitem{Por09}
S.~G.~Porsev, K.~Beloy and A.~Derevianko,
	\Journal{\PRL}{102}{181601.}{2009}

\bibitem{You07}
R.~D.~Young, R.~D.~Carlini, A.~W.~Thomas and J.~Roche,
	\Journal{\PRL}{99}{122003.}{2007}

\bibitem{Sib10}
A.~Sibirtsev, P.~G.~Blunden, W.~Melnitchouk and A.~W.~Thomas,
	\Journal{\PRD}{82}{013011.}{2010}

\bibitem{Ris10}
B.~Rislow and C.~E.~Carlson,
	arXiv:1011.2397 [hep-ph].

\bibitem{Blu11}
P.~G.~Blunden, W.~Melnitchouk and A.~W.~Thomas,
	arXiv:1102.5334 [hep-ph].

\bibitem{Chr10}
M.~E.~Christy and P.~E.~Bosted,
	\Journal{\PRC}{81}{055213.}{2010}

\bibitem{Whi90}
L.~W.~Whitlow {\it et al.},
	\Journal{\PLB}{250}{193.}{1990}

\bibitem{Gro69}
D.~J.~Gross and C.~H.~Llewellyn Smith,
	\Journal{\NPB}{14}{337.}{1969}

\bibitem{Cap94}
A.~Capella, A.~Kaidalov, C.~Merino and J.~Tran Thanh Van,
	\Journal{\PLB}{337}{358.}{1994}

\bibitem{Kai99}
A.~B.~Kaidalov and C.~Merino,
	\Journal{\EPJC}{10}{153.}{1999}

\bibitem{Osi03}
M.~Osipenko {\it et al.},
	\Journal{\PRD}{67}{092001.}{2003}

\bibitem{Nic00}
I.~Niculescu {\it et al.},
	\Journal{\PRL}{85}{1186.}{2000}

\bibitem{Mal09}
S.~P.~Malace {\it et al.},
	\Journal{\PRC}{80}{035207.}{2009}

\bibitem{Ale87}
A.~S.~Aleksanian {\it et al.},
	\Journal{\SJNP}{45}{628.}{1987}

\bibitem{PVDIS}
Jefferson Lab Experiment E05-007,
	{\it PVDIS: Parity violation in deep-inelastic scattering},
	R.~Michaels~\etal, spokespersons.

\bibitem{PVDIS12}
Jefferson Lab Experiment E12-07-102,
	{\it Precision measurement of the parity-violating asymmetry
	in deep-inelastic scattering off deuterium using baseline
	12~GeV equipment in Hall~C},
	K.~Paschke~\etal, spokespersons.

\bibitem{JPOS09}
J.~Dumas, J.~Grames and E.~Voutier,
	{\it AIP Conf. Proc.} 1160 (2009) 120.

\end{thebibliography}
